\RequirePackage[displaymath]{lineno} % Display line numbers
\documentclass[aps,twpcolumn,showpacs,byrevtex,prl,reprint]{revtex4-2}
\usepackage{epsfig}
\usepackage{graphicx}% Include figure files
\usepackage{dcolumn}% Align table columns on decimal point
\usepackage{bm}% bold math
\usepackage{overpic}
\usepackage{subfigure}
\usepackage{float}
\usepackage{color}
\usepackage{amsmath}
\usepackage{mathcomp}
\usepackage{mathrsfs}
\usepackage{multirow}
\usepackage{rotating}
\usepackage{hyperref}
\usepackage{threeparttable}
\newcommand{\BESIIIorcid}[1]{\href{https://orcid.org/#1}{\hspace*{0.1em}\raisebox{-0.45ex}{\includegraphics[width=1em]{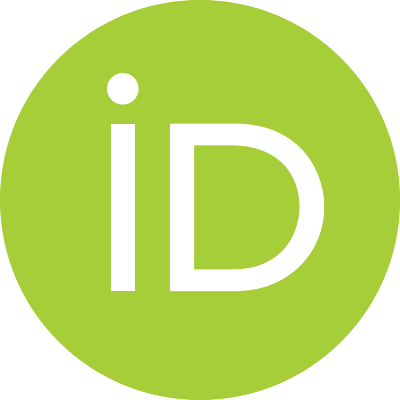}}}}

\begin{document}
\normalsize
\parskip=5pt plus 1pt minus 1pt

%\linenumbers
\title{\boldmath Observation of an Altered $a_{0}(980)$ Line shape in $D^{+} \rightarrow \pi^{+}\eta\eta$}
\vspace{-1cm}

\author{
\begin{small}
\begin{center}
  %% Saved at => 2025-01-01
M.~Ablikim$^{1}$\BESIIIorcid{0000-0002-3935-619X},
M.~N.~Achasov$^{4,b}$\BESIIIorcid{0000-0002-9400-8622},
P.~Adlarson$^{77}$\BESIIIorcid{0000-0001-6280-3851},
X.~C.~Ai$^{82}$\BESIIIorcid{0000-0003-3856-2415},
R.~Aliberti$^{36}$\BESIIIorcid{0000-0003-3500-4012},
A.~Amoroso$^{76A,76C}$\BESIIIorcid{0000-0002-3095-8610},
Q.~An$^{73,59,\dagger}$,
Y.~Bai$^{58}$\BESIIIorcid{0000-0001-6593-5665},
O.~Bakina$^{37}$\BESIIIorcid{0009-0005-0719-7461},
Y.~Ban$^{47,g}$\BESIIIorcid{0000-0002-1912-0374}
% ...
\\
\vspace{0.2cm}
(BESIII Collaboration)\\
\vspace{0.2cm}{\it
$^{1}$ Institute of High Energy Physics, Beijing 100049, People's Republic of China\\
% ...
$^{4}$ Budker Institute of Nuclear Physics SB RAS (BINP), Novosibirsk 630090, Russia\\
% ...
$^{36}$ Johannes Gutenberg University of Mainz, Johann-Joachim-Becher-Weg 45, D-55099 Mainz, Germany\\
$^{37}$ Joint Institute for Nuclear Research, 141980 Dubna, Moscow region, Russia\\
% ...
$^{47}$ Peking University, Beijing 100871, People's Republic of China\\
% ...
$^{58}$ Southeast University, Nanjing 211100, People's Republic of China\\
$^{59}$ State Key Laboratory of Particle Detection and Electronics, Beijing 100049, Hefei 230026, People's Republic of China\\
% ...
$^{73}$ University of Science and Technology of China, Hefei 230026, People's Republic of China\\
% ...
$^{76}$ University of Turin and INFN, (A)University of Turin, I-10125, Turin, Italy; (B)University of Eastern Piedmont, I-15121, Alessandria, Italy; (C)INFN, I-10125, Turin, Italy\\
$^{77}$ Uppsala University, Box 516, SE-75120 Uppsala, Sweden\\
% ...
$^{82}$ Zhengzhou University, Zhengzhou 450001, People's Republic of China\\
\vspace{0.2cm}
$^{\dagger}$ Deceased\\
% ...
$^{b}$ Also at the Novosibirsk State University, Novosibirsk, 630090, Russia\\
% ...
$^{g}$ Also at State Key Laboratory of Nuclear Physics and Technology, Peking University, Beijing 100871, People's Republic of China\\
% ...
} 
\end{center}
\vspace{0.4cm}
\end{small}
}
\affiliation{}
\vspace{-4cm}

\begin{abstract}
Using $20.3~{\rm fb}^{-1}$ of $e^+e^-$ collision data collected with the
BESIII detector at $\sqrt{s}=3.773~{\rm GeV}$, we perform the first amplitude
analysis of the decay $D^+\to\pi^+\eta\eta$. 
The intermediate process $D^+\to a_0(980)^+\eta$, $a_0(980)^+\to\pi^+\eta$, is the only component retained in the nominal amplitude model,
and its branching fraction is measured to be $(3.67\pm0.12_{\rm stat}\pm0.06_{\rm syst})\times10^{-3}$. 
The $\pi^+\eta$ mass spectrum associated with $a_0(980)^+\eta$ production exhibits
a line shape that differs substantially from those observed in
$D_{(s)}\to a_0(980)\pi$ and $D^0\to a_0(980)^-e^+\nu_e$ decays. We examine
several conventional descriptions of the $a_0(980)$ amplitude, including
Flatt\'e, dispersively modified Flatt\'e, $T$-matrix, and $K$-matrix
parameterizations. With reference $a_0(980)$ parameters, neither these models
nor their extensions including additional small resonant or non-resonant
amplitudes reproduce the observed line shape satisfactorily. When the
$a_0(980)$ parameters are allowed to float, satisfactory fits can be obtained,
but the pole mass is driven well above the $K\bar K$ threshold,
inconsistent with the near-threshold character of the $a_0(980)$. The results
reveal a tension between fit quality and the physical pole position in
conventional direct-production amplitude models. 
\end{abstract}
\clearpage
\newpage
\maketitle

\section{Introduction}
\label{sec:intro}

The interpretation of invariant-mass spectra is a central task in hadron spectroscopy. Resonance properties, such as the quantum numbers $I^GJ^{PC}$, pole positions, masses, and widths, provide essential information for identifying hadronic states. 
Among the light hadrons below $1~{\rm GeV}/c^2$, the pseudoscalar and vector mesons are well established, whereas the scalar mesons $f_0(500)$, $K_0^*(700)$, $f_0(980)$, and $a_0(980)$ remain of particular interest because their internal structures are still under discussion. 
Recent studies of charm-hadron decays involving the $a_0(980)$ have provided important experimental information on its production and line shape~\cite{BESIII:2019jjr, BESIII:2021aza, BESIII:2023htx, BESIII:2024tpv, BESIII:2024mbf}. In these channels, however, the $a_0(980)$ usually appears together with several other intermediate amplitudes, and the extracted $a_0(980)$ line shape can be affected by interference with them. 

The decay $D^+\to\pi^+\eta\eta$ provides a relatively clean environment for studying the $a_0(980)$ line shape. 
The upper kinematic boundary of the $\pi^+\eta$ invariant mass, $M(\pi^+\eta)$, is $1.322~{\rm GeV}/c^2$, allowing only a limited set of $\pi^+\eta$ resonant contributions. 
Besides the $a_0(980)$, the $a_2(1320)$ lies at the kinematic boundary and is strongly suppressed by phase space. 
Possible contributions in the $\eta\eta$ system are limited to isoscalar states such as $f_0(1370)$, $f_2(1270)$, and $f_0(1500)$~\cite{ParticleDataGroup:2024cfk}.
Considering the available phase space and the known decay branching fractions, these contributions are expected to be small in $D^+\to\pi^+\eta\eta$. 
Therefore, this decay is especially suitable for investigating the $D^+\to a_0(980)^+\eta$ intermediate process and the associated $a_0(980)$ line shape.

In this work, we perform the first amplitude analysis of $D^+\to\pi^+\eta\eta$ using $20.3~{\rm fb}^{-1}$ of $e^+e^-$ collision data collected with the BESIII detector at $\sqrt{s}=3.773~{\rm GeV}$~\cite{Ablikim:2013ntc, BESIII:2015equ, BESIII:2024lbn}. 
We examine a broad set of conventional amplitude models by combining several $a_0(980)$ line shape parameterizations---Flatt\'e~\cite{Flatte:1976xu}, dispersively modified Flatt\'e~\cite{Bugg:2008ig}, $T$-matrix~\cite{Ikeno:2021kzf}, and $K$-matrix~\cite{Anisovich:1997pe}---with possible additional cascade decay amplitudes.
This study provides a systematic test of how the choice of $a_0(980)$ parameterization affects both the description of the data and the extracted pole position.

% \section{The formulas for the lineshape of $a_0(980)$}
\section{$a_0(980)$ line shape parameterizations}
\label{sec:lineshape}

In this analysis, four descriptions of the $a_0(980)$ line shape are tested: the Flatt\'e parameterization, the dispersively modified Flatt\'e parameterization, the $T$-matrix formalism, and the $K$-matrix formalism.
These descriptions are introduced below and used in the amplitude fits.

\subsection{Flatt\'e parameterization}
Denoting $s\equiv M_{\pi\eta}^2$, the Flatt\'e formula is 
\begin{equation}
    P_{a_{0}(980)}(s) = \frac{1}{M_{0}^{2} - s - i\sum_{j}g_{j}^{2}\rho_{j}(s)},
\end{equation}
where $M_0$ is the bare mass parameter, $g_j$ is the coupling constant for channel $j$, and $\rho_j(s)$ is the corresponding phase space factor, defined as
\begin{align}
    \rho_{j}(s) &= 
    \begin{cases}
        \frac{1}{s}\sqrt{(s-s_{j}^+)(s-s_{j}^-)} & s \geq s_{j}^+ \\
        \frac{i}{s}\sqrt{(s_{j}^+ - s)(s - s_{j}^-)} & s_{j}^- \leq s < s_{j}^+ \\ 
        0 & s < s_{j}^-
    \end{cases},
    \label{eq:phasespacefactor}
\end{align}
where $s_j^\pm=(m_{j1}\pm m_{j2})^2$, with $m_{j1}$ and $m_{j2}$ being the masses of the two particles in the channel $j$. 
Here, for the $s < s_{j}^-$ region, $(s-s_{j}^+)(s-s_{j}^-)>0$, but the phase space actually is not allowed, so the $\rho_j(s)$ is nonphysical and set to be 0. 
The coupled channels considered are $\pi\eta$, $K\bar K$, and $\pi\eta'$. 
Unless otherwise stated, the Flatt\'e parameters are fixed to the values reported by CLEO~\cite{CLEO:2011upl}.

\subsection{Dispersively modified Flatt\'e parameterization}
%
% To solve the nonphysical result in the $s < s_{j}^-$ region, the Flatt\'e formalism is modified using the dispersion relationship. 
To avoid the spurious behavior associated with the naive continuation of the two-body phase-space factor below the $s_{j}^-$ region, and to incorporate the corresponding dispersive contribution to the real part of the self-energy, a dispersively modified Flatt\'e parameterization is employed.
The dispersively modified Flatt\'e parameterization is written as
\begin{equation}
    P_{a_{0}(980)}(s) = \frac{1}{M_0^2 - s - \sum_{j}g_{j}^2\Pi_{j}(s)},
\end{equation}
where $\Pi_j(s)$ is the channel-dependent dispersive function defined in Ref.~\cite{BESIII:2016tqo}; the same convention and subtraction scheme are used here. 
Its imaginary part is related to the corresponding two-body phase space, while its real part gives the dispersive threshold correction. 
Unless otherwise stated, the parameters in this description are fixed to the values reported by BESIII~\cite{BESIII:2016tqo}.

\subsection{$T$-matrix formalism}

In the $T$-matrix formalism, we follow the coupled-channel treatment of Ref.~\cite{Ikeno:2021kzf}. 
According to the quark-level diagrams considered in that work, the primary tree-level production of $D^+\to K^+\bar K^0\eta$ is absent. 
The source term is therefore assigned to the $\pi^+\eta$ channel, while the $K^+\bar K^0$ channel is generated through coupled-channel final-state interactions.
Since the two $\eta$ mesons in the final state are identical, the total amplitude is symmetrized under $\eta_a\leftrightarrow\eta_b$. 
Defining $s_a=m^2_{\pi^+\eta_a}$ and $s_b=m^2_{\pi^+\eta_b}$, the amplitude is written as~\cite{Ikeno:2021kzf}
\begin{eqnarray}
\begin{aligned}
\label{eq:tmatrix_amp}
    \mathcal{M}_{D^+ \to \pi^+ \eta \eta} &= g_1 [ 1 + G_{\pi\eta}(s_a) T_{\pi\eta, \pi\eta}(s_a) \\
    &+ G_{\pi\eta}(s_b) T_{\pi\eta, \pi\eta}(s_b)],
\end{aligned}
\end{eqnarray}
where $g_1$ is an overall normalization factor. 

The scattering matrix is obtained from the algebraic Bethe-Salpeter equation,
\begin{equation}
    T(s)=\left[1-V(s)G(s)\right]^{-1}V(s),
\end{equation}
where $V$ and $T$ are matrices in the coupled-channel space and $G(s)=\mathrm{diag}\left(G_{\pi\eta}(s),G_{K\bar K}(s)\right)$.
The coupled channels are chosen as (1) $\pi^+\eta$ and (2) $K^+\bar K^0$. Following Refs.~\cite{Ikeno:2021kzf,Song:2025ofe}, the potential matrix is
\begin{gather} 
\begin{aligned}
V_{11} &= -\frac{2m_{\pi}^2}{3f_{\pi\eta}^2}, \\
\quad 
V_{12} &= -\frac{3s - m_{K^+}^2 - m_{\bar{K}^0}^2 - m_{\eta}^2}{3\sqrt{3}f_{K\bar{K}}f_{\pi\eta}},\\
\quad 
V_{22} &= -\frac{s}{4f_{K\bar{K}}^2},
\end{aligned}
\end{gather}
where the effective decay constants are set to $f_{\pi\eta}=f_{K\bar K}=93~{\rm MeV}$~\cite{Song:2025ofe}.
For each channel $i$, the two-point loop function is defined as
\begin{equation}
    G_i(s) = i \int \frac{\mathrm{d}^4q}{(2\pi)^4} \frac{1}{(P-q)^2 - m_{i2}^2 + i\epsilon} \frac{1}{q^2 - m_{i1}^2 + i\epsilon}.
\end{equation}
where $P^2=s$, and $m_{i1}$ and $m_{i2}$ are the masses of the two particles in the channel $i$.
Following Refs.~\cite{Ikeno:2021kzf,Song:2025ofe}, we use a three-momentum cutoff regularization,
\begin{equation}
    G_i(s) = \frac{1}{4\pi^2}\int_0^{q_{\rm max}} \frac{q^2(\omega_{i1}+\omega_{i2})} {\omega_{i1}\omega_{i2} \left[s-(\omega_{i1}+\omega_{i2})^2+i\epsilon\right]}\,dq,
\end{equation}
where $\omega_{ij}=\sqrt{m_{ij}^2+q^2}$, $q$ is the three-momentum magnitude, and $q_{\rm max}=0.6~{\rm GeV}$ is the cutoff.

\subsection{$K$-matrix formalism}

In the $S$-wave $K$-matrix formalism, the production amplitude for the channel $a$ is written as
\begin{equation}
    F_a(s)=\sum_b\left[I-iK(s)\rho(s)\right]^{-1}_{ab}P_b(s),
    \label{eq:Kmatrix_F}
\end{equation}
where $I$ is the identity matrix, $\rho(s)$ is the diagonal matrix of phase space factors, and $P_b(s)$ is the production vector. 
Following Ref.~\cite{Anisovich:1997pe}, the coupled channels are chosen as (1) $\pi\eta$, (2) $K\bar K$, (3) $\pi\eta'$, and (4) multi-meson states. 
The corresponding $K$-matrix is parameterized as
\begin{equation}
    K_{ab}(s)=\sum_\alpha \frac{g_a^{(\alpha)}g_b^{(\alpha)}}{M_\alpha^2-s} + f_{ab}\frac{1.5~{\rm GeV}^2+s_1}{s+s_1}.
    \label{eq:Kmatrix}
\end{equation}

The phase space factors for the first three channels are defined using the same convention as in the Flatt\'e parameterization in Eq.~\ref{eq:phasespacefactor}. 
For the multi-meson channel, $i=4$, a smooth threshold form is adopted,
\begin{equation}
    \rho_{4}(s) =  
    \begin{cases}
        1 & s \geq s_{th}^b \\ 
        \left( \frac{1 - s_{th}^a/s}{1 - s_{th}^a/s_{th}^b} \right)^{\! 5/2} 
        & s_{th}^a \leq s < s_{th}^b \\
        0 & s < s_{th}^a
    \end{cases},
\end{equation}
where $s_{th}^a = (m_\eta+3m_\pi)^2$, $s_{th}^b = 2.25~\mathrm{GeV}^2$.

The production vector describes the coupling of the initial $D^+$ decay to the coupled channels. 
We consider the standard form
\begin{equation}
    P_a(s)=\sum_\alpha \frac{\beta_\alpha g_a^{(\alpha)}}{M_\alpha^2-s} + f_a^{\rm prod}\frac{1.5~{\rm GeV}^2+s_1^{\rm prod}} {s+s_1^{\rm prod}} .
    \label{eq:Pvector_smooth}
\end{equation}
The smooth production term is included only in the $\pi\eta$ channel; thus $f_a^{\rm prod}=0$ for $a\ne1$. 
As an alternative description of the smooth production term, we also consider a polynomial form for the $P$ vector ~\cite{CrystalBarrel:2019zqh},
\begin{equation}
    P_a(s)=\sum_\alpha \frac{\beta_\alpha g_a^{(\alpha)}}{M_\alpha^2-s} +\sum_{n=0}^{1}c_{a,n}s^n ,
    \label{eq:polyPvector}
\end{equation}
where $c_{a,n}=0$ for $a\ne1$.

Following Ref.~\cite{Anisovich:1997pe}, two bare poles are included in the $K$-matrix. 
For the first three channels, the couplings of each bare pole are constrained by
\begin{equation}
    g_1^{(\alpha)}=\frac{\cos\Theta}{\sqrt{2}}g^{(\alpha)},\quad
    g_2^{(\alpha)}=\frac{\sqrt{\lambda}}{2}g^{(\alpha)},\quad
    g_3^{(\alpha)}=\frac{\sin\Theta}{\sqrt{2}}g^{(\alpha)},
\end{equation}
where the subscript denotes the channel index, $\Theta\simeq41.4^\circ$ is the $\eta$-$\eta'$ mixing angle in the quark-flavor basis, and $\lambda=0.6$ accounts for the suppression of strange-quark-pair creation. 
The parameters of the $K$-matrix are listed in Table~\ref{tab:parsforKmatrix}.

Finally, for the decay $D^+\to(\pi^+\eta)_{S-\rm{wave}}\eta$ , the $K$-matrix amplitude is symmetrized over the two identical $\eta$ mesons as
\begin{equation}
    A_{D^+\to(\pi^+\eta)_{S-\rm{wave}}\eta} = F_1(s_a)+F_1(s_b).
    \label{eq:Kmatrix_Damp}
\end{equation}

\begin{table}[htbp]
\centering
\caption{$K$-matrix parameters quoted in Ref.\cite{Anisovich:1997pe}. Masses and couplings are given in $\mathrm{GeV}/c^{2}$, while units of $\mathrm{GeV}^2/c^{4}$ for s-related quantities are implied, and only $f_{11}$ is non-zero. }
\label{tab:parsforKmatrix}
\resizebox{\columnwidth}{!}{
    \renewcommand{\arraystretch}{1.5} 
    \setlength{\tabcolsep}{4pt} 
    \begin{tabular}{l c c@{\hspace{1.0cm}} c c}
    \hline\hline
     & \multicolumn{4}{c}{$a_0$-resonances without $K$-matrix background term} \\
    \hline
     & \multicolumn{2}{c}{Solution 1} & \multicolumn{2}{c}{Solution 2} \\
    \hline
     & $\alpha=1$ & $\alpha=2$ & $\alpha=1$ & $\alpha=2$ \\
    \hline
    $M_\alpha$ & $0.963$ & $1.630$ & $0.965$ & $1.654$ \\
    $g^{(\alpha)}$   & $0.879$ & $0.702$ & $0.901$ & $0.435$ \\
    $g_4^{(\alpha)}$ & $0.598$ & $0.511$ & $0.689$ & $0.687$ \\
    \hline
     & \multicolumn{4}{c}{$a_0$-resonances with $K$-matrix background term} \\
    \hline
     & \multicolumn{2}{c}{Solution 3} & \multicolumn{2}{c}{Solution 4} \\
    \hline
     & $\alpha=1$ & $\alpha=2$ & $\alpha=1$ & $\alpha=2$ \\
    \hline
    $M_\alpha$ & $0.944$ & $1.624$ & $0.939$ & $1.640$ \\
    $g^{(\alpha)}$   & $0.879$ & $0.702$ & $0.901$ & $0.435$ \\
    $g_4^{(\alpha)}$ & $0.651$ & $0.519$ & $0.653$ & $0.651$ \\
    & $f_{11}=0.529$ & $s_1=1.0$  & $f_{11}=0.731$ & $s_1=1.9$ \\    
    \hline\hline
    \end{tabular}
}
\end{table}

% The total real parameters to be fitted are:
% \begin{gather*}
%     \beta_1,\quad \beta_2, \quad f^{\mathrm{prod}}_{1}, \quad s_1^{\mathrm{prod}}
% \end{gather*}

%%=============================================================================================
\subsection{Discussion}

Using the Flatt\'e, dispersively modified Flatt\'e, and $T$-matrix descriptions introduced above, 
we compare the corresponding normalized $\pi^+\eta$ mass distributions for the 
$D^{+}\to a_{0}(980)^+\eta$, $a_{0}(980)^+\to\pi^+\eta$ decay chain. 

\begin{figure}[htbp]
\begin{center}
\begin{minipage}[b]{0.40\textwidth}
\epsfig{width=0.98\textwidth,clip=true,file=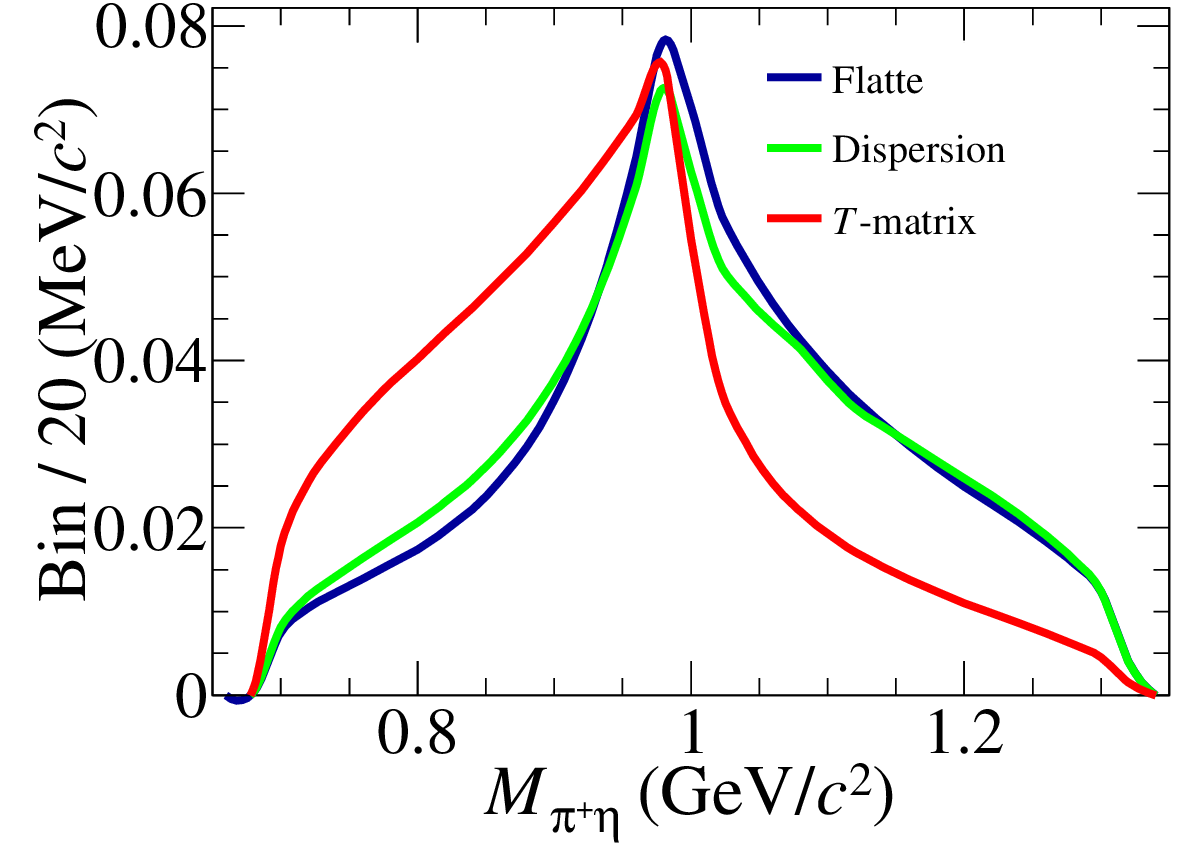}
\end{minipage}
\caption{Normalized $\pi^+\eta$ mass distributions obtained with the Flatt\'e (blue), dispersively modified Flatt\'e (green), and $T$-matrix (red) descriptions. Each distribution is normalized to unit area.}
\label{fig:a0lineshapegen}
\end{center}
\end{figure}

As shown in Fig.~\ref{fig:a0lineshapegen}, the Flatt\'e and dispersively modified Flatt\'e parameterizations give rather similar line shapes, while the $T$-matrix formalism leads to a visibly different distribution. 
This difference is expected because, in the implementation used here, the $T$-matrix amplitude describes the coupled-channel $\pi\eta$ final-state interaction, including both resonant and non-resonant scattering components generated within the model. 
In contrast, the Flatt\'e and dispersively modified Flatt\'e forms parameterize the $a_0(980)$ propagator with prescribed coupled-channel couplings and do not include an explicit non-resonant $\pi\eta$ scattering background.

The $K$-matrix result is not shown in this illustrative comparison because its production amplitude depends on the choice of the production vector. 
In particular, the coefficients $\beta_\alpha$ describe the relative production strengths of the $K$-matrix bare poles in the initial $D^+$ decay, while the smooth production term introduces an additional non-pole component. 
Therefore, unlike the Flatt\'e-type propagator forms, the $K$-matrix formalism does not define a unique line shape before the production parameters are specified in the amplitude fit.

\section{Event selection}

The BESIII detector and the upgraded multi-gap resistive plate chambers used in the time-of-flight systems are described in Refs.~\cite{detector,MRPC} in detail, respectively.
Since the $D^{+}$ and $D^{-}$ are produced together in pairs, the double tag (DT) method~\cite{MARK-III:1985hbd} is employed to suppress backgrounds. 
Six tag channels 
$D^{-} \to K^{+}\pi^{-}\pi^{-} (\pi^{0})$, $D^{-} \to K^{0}_{S}\pi^{-} (\pi^{0})$, $D^{-} \to K^{0}_{S}\pi^{-}\pi^{-}\pi^{+}$, and $D^{-} \to K^{+}K^{-}\pi^{-}$ are used.
Signal MC samples with $\psi(3770) \to D^{+}D^{-}$, $D^{-} \to \mathrm{tags}$ and $D^{+} \to \pi^{+}\eta\eta$ are generated, where the amplitude model from the fit to the data is used for the signal decay.
The tracking, particle identification (PID), $K_{S}^{0}$, $\pi^{0}$, and $\eta$ reconstruction are almost identical to those in Ref.~\cite{BESIII:2023exq}, except for the signal window of invariant mass for $\eta$ candidates, which is set to $0.45<M(\gamma\gamma)_{\eta}<0.65$~GeV$/c^{2}$ to improve the $\eta$ reconstruction efficiency. 
Two variables, $M_{\mathrm{BC}} = \sqrt{E_{\mathrm{beam}}^{2} - |\Vec{P}_{D^{\pm}}|^{2}}$ and 
$\Delta E = E_{D^{\pm}} - E_{\mathrm{beam}}$, are used to identify $D^{\pm}$ mesons,
where $(E_{D^{\pm}},\Vec{P}_{D^{\pm}})$ is the four-momentum of the $D^{\pm}$ meson and $E_{\mathrm{beam}}$ is the beam energy.
For both the tag and signal sides, any candidate with $M_{\mathrm{BC}}<1.83$~GeV$/c^{2}$ or $|\Delta E|>0.1$~GeV is first rejected. 
The candidate for each tag mode with $|\Delta E|$ closest to 0 is selected, while the signal candidate with 
$M_{\mathrm{BC}}$ closest to the $D$ nominal mass~\cite{ParticleDataGroup:2024cfk} is chosen. 
Backgrounds are studied with an inclusive Monte Carlo (MC) sample simulated with {\sc geant4}~\cite{sim},
which includes all known open-charm decays, charmless decays and initial-state radiative decays to the $J/\psi$ or $\psi(3686)$. All particle decays with known branching fractions are modeled with {\sc evtgen}~\cite{EvtGen}. 
The remaining unknown charmonium decays are generated with {\sc lundcharm}~\cite{Chen:2000tv}. 

For the tag channels, the $M_{\mathrm{BC}}$ signal windows are defined as $\pm 6$~MeV$/c^{2}$ around the known $D^{-}$ mass~\cite{ParticleDataGroup:2024cfk}; 
and the $\Delta E$ windows are set to 3.5 times the corresponding resolutions~\cite{BESIII:2023exq}. 
On the signal side, the $M_{\mathrm{BC}}$ signal window is required to be in the range $[1.860,~1.880]$~GeV$/c^{2}$.
Since the dominant background is from processes with no $\eta$ in the final state, for events used in the amplitude analysis a multi-variant analysis (MVA)~\cite{Hocker:2007ht} is employed in which a Gradient Boosted Decision Tree (BDTG) classifier is developed based on the inclusive MC sample. %
The BDTG takes five discriminating variables: the natural logarithm of the $\chi^{2}$ of constraining the two-photon pairs to the $\eta$ mass; 
the invariant masses and the cosine of the helicity angles of the photon decaying from $\eta_{1,2}$ with higher energy, 
where the subscripts $1$ and $2$ for $\eta$ represents a more or less energetic $\eta$.
Studies of the MC samples show that a requirement on the output of the BDTG retains 78.3\% signals and rejects 89.7\% backgrounds.
The possible distortion of the Dalitz-plot distribution due to this requirement is studied with MC samples and found to be negligible.
A further requirement of $|\Delta E|<0.040$~GeV is then imposed.
Furthermore, to ensure all candidate events fall into the physical region of phase space, a kinematic fit is performed, where aside from four-momentum conservation, 
the invariant masses of $M(\gamma\gamma)_{\eta}$ and $M(\pi^{+}\eta \eta)_{D}$ are constrained to individual values quoted from the PDG~\cite{ParticleDataGroup:2024cfk}.
Finally, a sample of 1624 candidates is retained with a purity of $(85.1 \pm 0.9)\%$.

\section{Amplitude analysis formalism}

The amplitude analysis is performed using accepted events in data based on an unbinned maximum likelihood fit. 
The logarithm of the likelihood is constructed as 
\begin{eqnarray}
\begin{aligned}
\ln\mathcal L= \sum_{i=1}^{N_{\rm evt}} \ln\!\left[ f_{\rm s}\widetilde S(p_i) +(1-f_{\rm s})\widetilde B(p_i) \right],
\end{aligned}
\end{eqnarray}
where $f_{\mathrm{s}}$ is the signal purity, $p$ is the four-momenta of the final particles, and $\tilde{S}(p)$ and $\tilde{B}(p)$ are the normalized probability density functions (PDFs) of the signal and background, respectively:
\begin{eqnarray}
\begin{aligned}
\tilde{S}(p)& = \frac{\epsilon(p)|\mathcal{M}(p)|^{2}R_{3}(p)}{\int{\epsilon(p)|\mathcal{M}(p)|^{2}R_{3}(p)\mathrm{d}^3p}},\\
\tilde{B}(p)& = \frac{\epsilon(p)B_{\epsilon}(p)R_{3}(p)}{\int{\epsilon(p)B_{\epsilon}(p)R_{3}(p)\mathrm{d}^3p}}.	
\end{aligned}
\end{eqnarray}
Here, $R_{3}(p)$ is the three-body phase space factor, and $\epsilon(p)$ is the efficiency function.
The $\mathcal{M}(p)$ is the total amplitude of the signal, and $B_{\epsilon}(p)$ is the efficiency corrected background PDF.
The $B_{\epsilon}(p)$ is calculated with 
\begin{gather}
    B_{\epsilon}(p) = B(p)/\epsilon(p),
\end{gather}
where the $B(p)$ is the function extracted from the inclusive MC sample with the signal removed and the $\epsilon(p)$ is the corresponding efficiency.

\subsection{Signal amplitude formalism}

The signal amplitude $\mathcal{M}(p)$ is modeled as the coherent sum of all intermediate-process and partial-wave contributions,
\begin{equation}
\mathcal{M}(p) = \sum_{\alpha}{c_{\alpha}e^{i\phi_{\alpha}}A_{\alpha}}, 
\end{equation}
where $\alpha$ labels a particular intermediate process or partial-wave contribution, and $c_{\alpha}$ and $\phi_{\alpha}$ are its magnitude and phase, respectively.

Because the final state contains two identical $\eta$ mesons, each contribution is coherently symmetrized under their exchange:
\begin{equation}
A_{\alpha}(p_{\pi},p_{\eta_a},p_{\eta_b}) =
A_{\alpha}^{(0)}(p_{\pi},p_{\eta_a},p_{\eta_b})+
A_{\alpha}^{(0)}(p_{\pi},p_{\eta_b},p_{\eta_a}),
\end{equation}
where $A_{\alpha}^{(0)}$ denotes the unsymmetrized contribution and is constructed as
\begin{equation}
A_{\alpha}^{(0)} = P_{\alpha}S_{\alpha}F^{R}_{\alpha}F^{D}_{\alpha},
\end{equation}
where $P_{\alpha}$ is the propagator for a resonant intermediate state. A relativistic Breit--Wigner function is used for all resonances except the $a_{0}(980)$. For the $\pi\eta$ $S$ wave dominated by the $a_{0}(980)$, the Flatt\'e, dispersively modified Flatt\'e, $T$-matrix, and $K$-matrix descriptions are considered and are all denoted by $P_{\alpha}$ in the amplitude notation. For the possible nonresonant amplitudes, no resonance propagator is included, and $P_{\alpha}$ is set to unity.

The spin factor $S_{\alpha}$ is constructed using the covariant tensor formalism~\cite{Zou:2002ar}. Consider the generic decay chain $D^{+}\to RP_{3}$, $R\to P_{1}P_{2}$, where $P_{1}$, $P_{2}$, and $P_{3}$ are spin-zero pseudoscalar particles. The first-order operators at the $D^{+}$ and $R$ vertices are
\begin{equation}
\begin{aligned}
\tilde{T}^{(1)\mu}_D   & = r^{\mu}_D - \frac{p_D \cdot r_D}{p^2_D}p^{\mu}_D,\\
\tilde{t}^{(1)}_{R\,\mu}& = r_{R\,\mu} - \frac{p_R \cdot r_R}{p^2_R}p^{\mu}_R,	
\end{aligned}
\end{equation}
where $p_D$ and $p_R$ are the four-momenta of the $D^{+}$ meson and the two-body subsystem $R$, respectively; $p_{1,2,3}$ are the four-momenta of the final-state particles; and $r_D=p_R-p_3$ and $r_R=p_1-p_2$. The corresponding second-order operators are
\begin{equation}
\begin{aligned}
\tilde{T}^{(2)\mu\nu}_D   & = \tilde{T}^{(1)\mu}_D\tilde{T}^{(1)\nu}_D 
                             - \frac{1}{3}\tilde{T}^{(1)}_D\cdot\tilde{T}^{(1)}_D(g^{\mu\nu}-\frac{p^{\mu}_D p^{\nu}_D}{p^{2}_D}),\\
\tilde{t}^{(2)}_{\mu\nu}(R)& = \tilde{t}^{(1)}_{R\,\mu}\tilde{t}^{(1)}_{R\,\nu} 
                             - \frac{1}{3}\tilde{t}^{(1)}_R\cdot\tilde{t}^{(1)}_R(g_{\mu\nu}-\frac{p_{R\,\mu} p_{R\,\nu}}{p^{2}_R}).	
\end{aligned}
\end{equation}
Here, $g_{\mu\nu}$ is the metric tensor. Owing to the limited phase space, only orbital angular momenta up to $L=2$ are considered. The spin factor for a contribution with orbital angular momentum $L$ is evaluated using the generic expression
\begin{equation}
S(L) = \begin{cases}
    1, &L=0\\
    \tilde{T}^{(1)\mu}_D\tilde{t}^{(1)}_{R\,\mu}, &L=1\\
    \tilde{T}^{(2)\mu\nu}_D\tilde{t}^{(2)}_{R\,\mu\nu}, &L=2.
\end{cases}
\end{equation}
For each contribution $\alpha$, $S_{\alpha}$ is obtained from this expression using its corresponding value of $L$.

The factors $F^{R}_{\alpha}$ and $F^{D}_{\alpha}$ are the Blatt--Weisskopf barrier factors for the decay of the two-body subsystem $R$ and for the $D^{+}$ decay vertex, respectively. For the relevant orbital angular momentum $L$, the generic barrier factor is
\begin{equation}
F(L,q,r_e) = \begin{cases}
    1, &L=0\\
    \sqrt{\frac{1+z_0}{1+z}}, &L=1\\
    \sqrt{\frac{9+3z_0+z_0^2}{9+3z+z^2}}, &L=2.
\end{cases}
\end{equation}
Here,
\begin{equation}
z=(qr_e)^2, \qquad z_0=(q_0r_e)^2,
\end{equation}
where $q_0$ is the value of $q$ when the invariant mass of $R$ is set to the reference mass $M_0$. For $F^{D}_{\alpha}$, $q$ is the common magnitude of the three-momenta of $R$ and $P_3$ in the $D^{+}$ rest frame, and the effective radius of the $D^{+}$ meson is set to $r_e=5.0~\mathrm{GeV}^{-1}$ in the nominal fit. For $F^{R}_{\alpha}$, $q$ is the common magnitude of the three-momenta of $P_1$ and $P_2$ in the $R$ rest frame, and the effective radius of $R$ is set to $r_e=3.0~\mathrm{GeV}^{-1}$.

For a resonant intermediate state, $M_0$ is its nominal mass, so that the corresponding barrier factor is normalized to unity at $q=q_0$. For the possible nonresonant amplitudes considered in this analysis, no natural resonance mass is available. The reference mass is therefore set to $M_0(f_2(1270))$ for $D^{+}\to(\eta\eta)_T\pi^{+}$ and to $M_0(a_2(1320))$ for $D^{+}\to(\pi^{+}\eta)_V\eta$ and $D^{+}\to(\pi^{+}\eta)_T\eta$. The subscripts $V$ and $T$ denote the vector and tensor nonresonant amplitudes, respectively. These reference values only fix the constant normalization of the barrier factors and do not imply the presence of the corresponding resonances. Because this constant normalization can be absorbed into the fitted complex coefficient, replacing the numerator of a nonresonant barrier factor by unity would give an equivalent parameterization.

\subsection{Fit fraction and branching fraction of intermediate process} 
The fit fraction (FF) of the $\alpha^{\rm th}$ amplitude is calculated with 
\begin{eqnarray}
\begin{aligned}
FF_{\alpha} = \frac{\int{|c_{\alpha}A_{\alpha}(p)|^{2}R_{3}(p)d^{3}p}}{\int{|M(p)|^{2}R_{3}(p)d^{3}p}}. 
\end{aligned}
\end{eqnarray}
Here, it is noted that there is no efficiency function being incorporated. 
The branching fraction of the $\alpha^{\rm th}$ intermediate process is 
\begin{eqnarray}
\begin{aligned}
\mathcal{B}_{\alpha} = \mathcal{B}(D^{+} \to \pi^{+}\eta\eta)FF_{\alpha},
\end{aligned}
\end{eqnarray}

\section{Signal Model Construction}
\label{sec:a0onlyconfirmation}

\begin{figure}[htbp]
\begin{center}
\begin{minipage}[b]{0.40\textwidth}
\epsfig{width=0.98\textwidth,clip=true,file=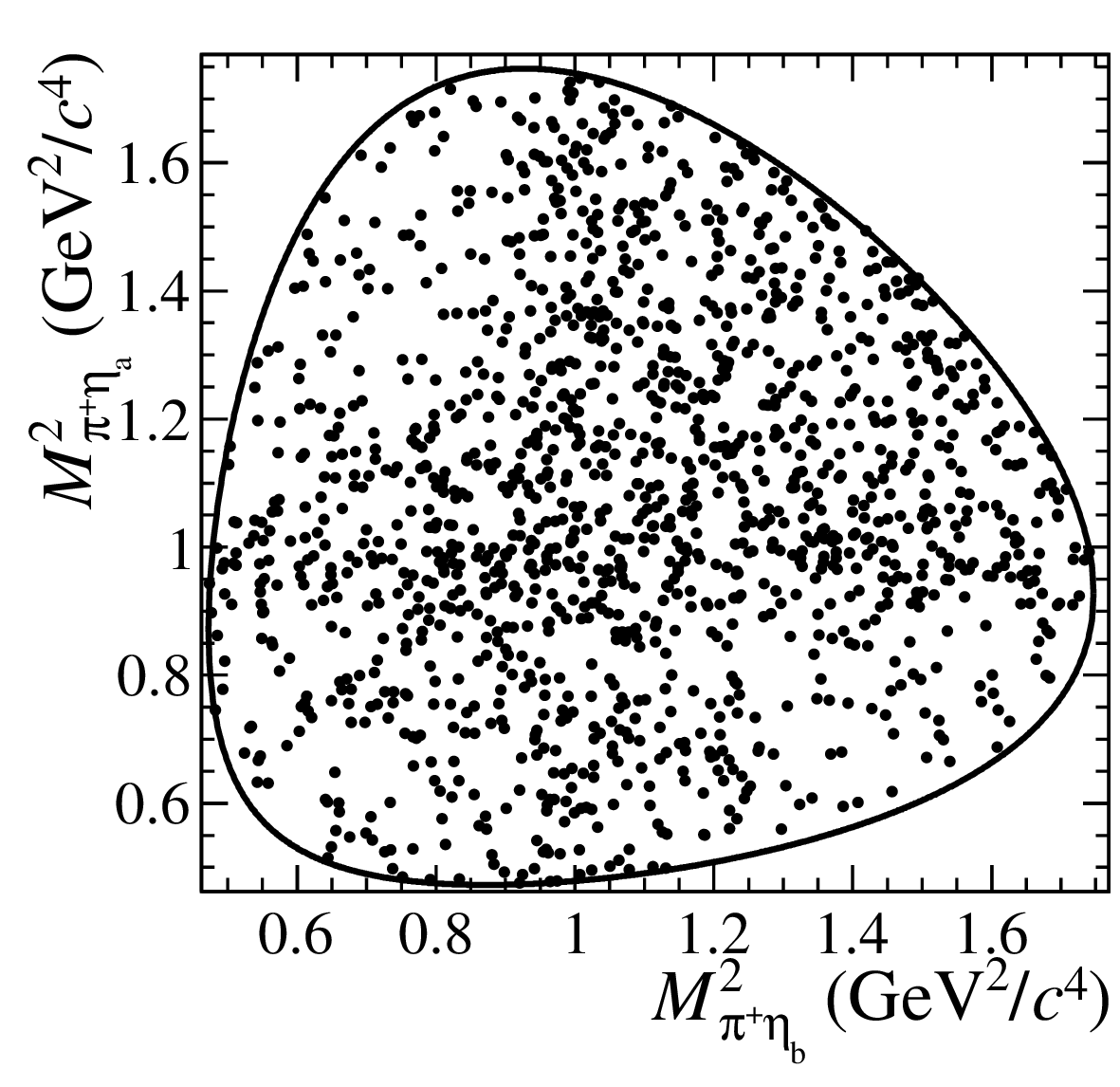}
\end{minipage}
\caption{Dalitz plot of $M^2(\pi^+\eta_a)$ versus $M^2(\pi^+\eta_b)$ for the selected $D^+\to\pi^+\eta\eta$ candidates. 
The subscripts $a$ and $b$ label the two identical $\eta$ candidates, whose assignment is randomized on an event-by-event basis.}
\label{fig:Dalitz}
\end{center}
\end{figure}

\begin{figure}[htbp]
\begin{center}
\begin{minipage}[b]{0.23\textwidth}
\epsfig{width=0.98\textwidth,clip=true,file=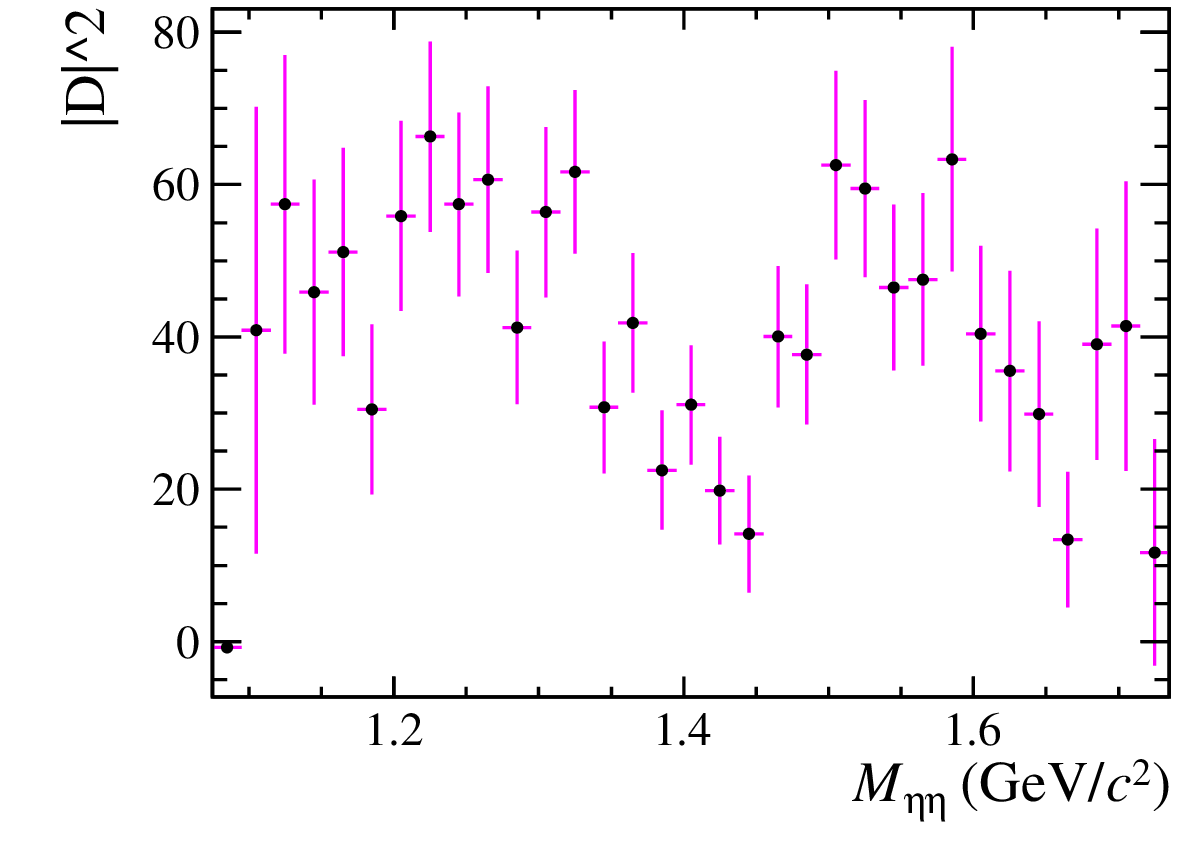}
\put(-85,65){$|S|^{2}$}
\end{minipage}
\begin{minipage}[b]{0.23\textwidth}
\epsfig{width=0.98\textwidth,clip=true,file=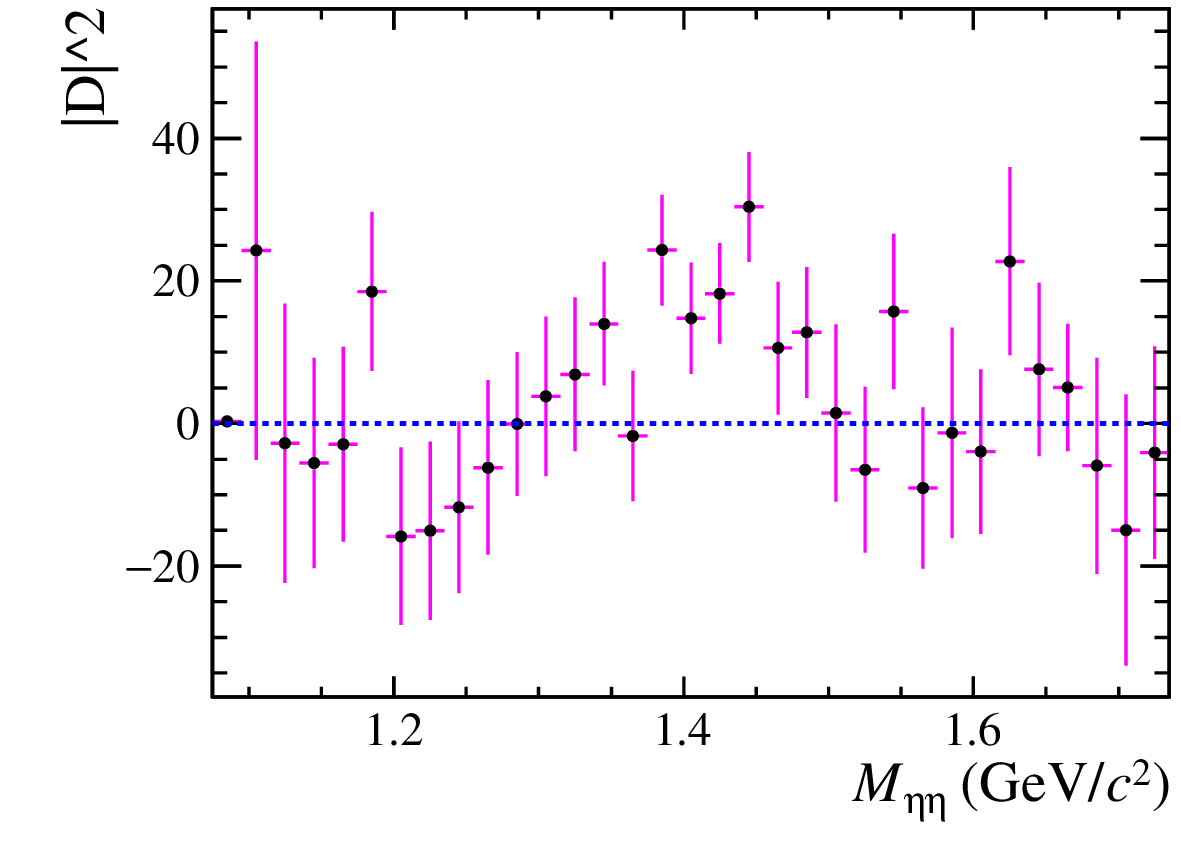}
\put(-85,65){$|D|^{2}$}
\end{minipage}
\caption{The Legendre-moment-weighted projections on $M(\eta\eta)$ for data are plotted for (left) $S-{\rm wave}$ and (right) $D-{\rm wave}$. }
\label{fig:metaetapw}
\end{center}
\end{figure}

In the Dalitz plot of $M^2(\pi^+\eta_a)$ versus $M^2(\pi^+\eta_b)$, shown in Fig.~\ref{fig:Dalitz}, a prominent band corresponding to the $a_0(980)^+$ resonance is observed in the $\pi^+\eta$ system, while no other clear structures are visible.  
Here, the subscripts $a$ and $b$ label the two identical $\eta$ candidates, whose assignment is randomized on an event-by-event basis.
Furthermore, Legendre-moment-weighted projections on $M(\eta\eta)$ are also performed to check whether there is any $f_{0}$ and $f_{2}$ states, 
which are shown in Fig.~\ref{fig:metaetapw}. The detail formalisms are given in App.~\ref{sec:Legendre}. 
Here, the background shapes obtained from the inclusive MC show good agreement with data, 
thus the background contribution in Legendre-moment-weighted projections are estimated from inclusive MC and subtracted from the data. 
Then efficiency and three-body phase space $M(\eta\eta)\rho_{D\to S\pi^{+}}\rho_{S\to\eta\eta}$ corrections are also considered, 
Where, the phase space factors can be calculated with Eq.~\ref{eq:phasespacefactor} and $M(\eta\eta)$ comes from the projections being 
shown for the invariant mass spectrum. 

\begin{figure}[htbp]
\begin{center}
\begin{minipage}[b]{0.23\textwidth}
\epsfig{width=0.98\textwidth,clip=true,file=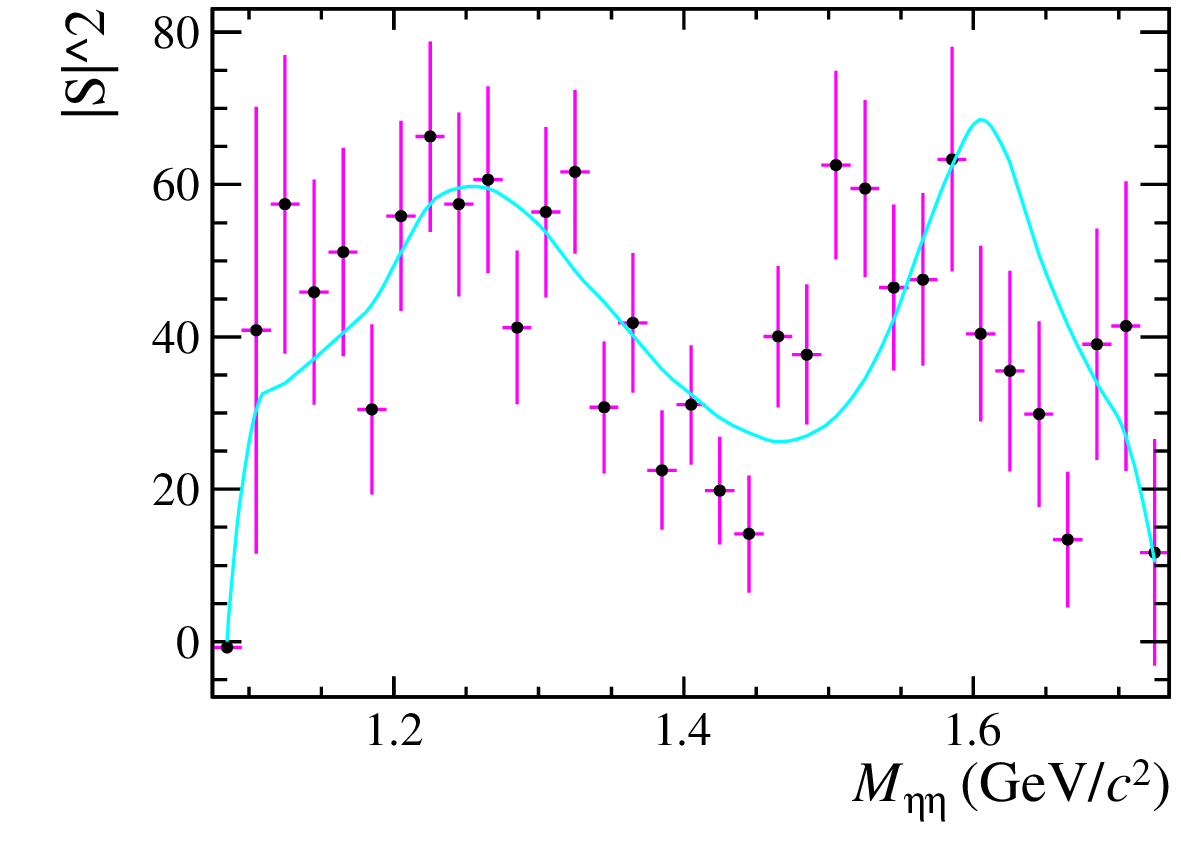}
%\put(-85,65){F}
\end{minipage}
\begin{minipage}[b]{0.23\textwidth}
\epsfig{width=0.98\textwidth,clip=true,file=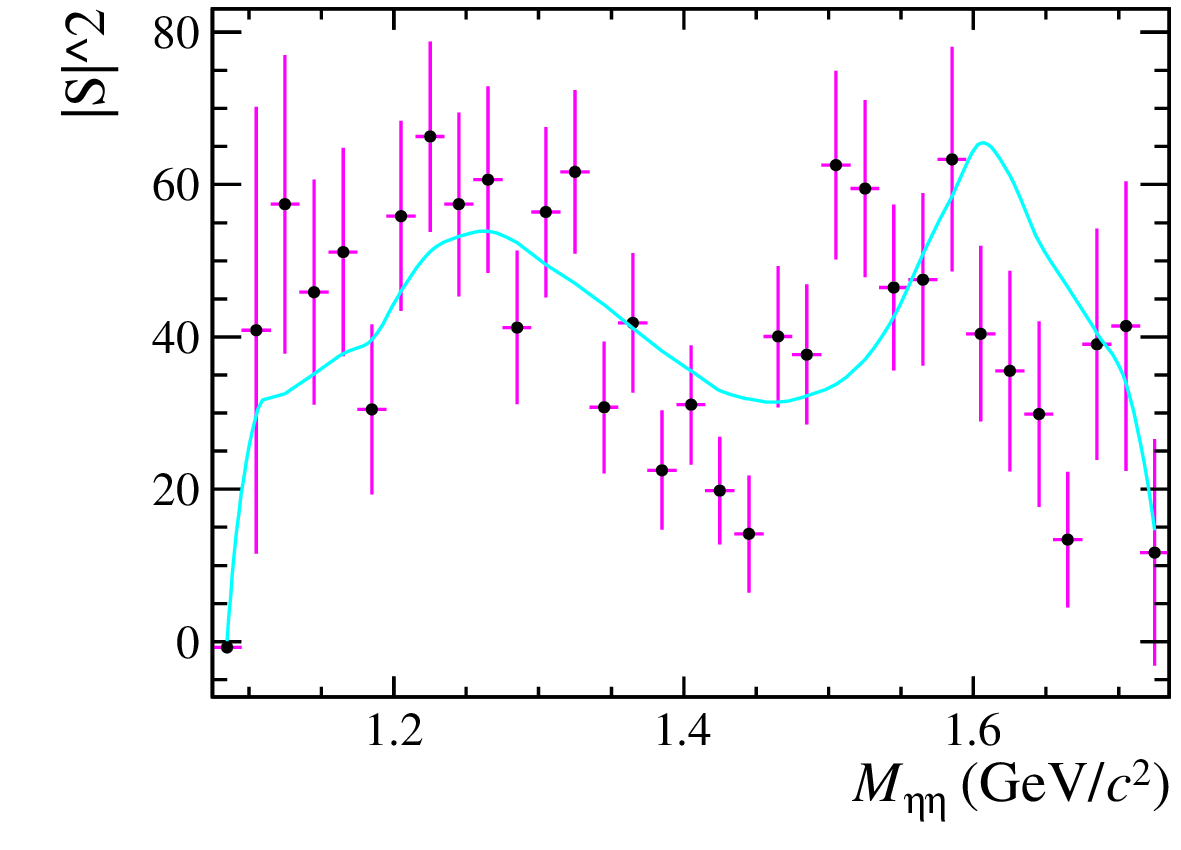}
%\put(-85,65){D}
\end{minipage}
\caption{The Legendre-moment-weighted projections on $M(\eta\eta)$ for (dots with errors) data and the expected line shape reflected from $a_{0}(980)$. }
\label{fig:meteetapwa0eta_noloop_nofpar}
\end{center}
\end{figure}

The $D-$wave is found to be around 0, while some possible enhancements are found for $S-$wave. 
To validate whether the possible enhancements are reflected by the $a_{0}(980)^{+}$, we also perform the same projections for MC samples generated with 
$D^{+} \to a_{0}(980)^{+}\eta$, $a_{0}(980)^{+} \to \pi^{+}\eta$, where, the $a_{0}(980)$ line-shape is described with (left) Flatt\'e and (right) dispersively modified Flatt\'e formalisms. 
The results are shown in Fig.~\ref{fig:meteetapwa0eta_noloop_nofpar}. 
From Fig.~\ref{fig:metaetapw} and Fig.~\ref{fig:meteetapwa0eta_noloop_nofpar}, the possible enhancement is from the $a_{0}(980)^{+}$ reflection;
there is no obvious structure for $M(\eta\eta)$. 
The $a_{0}(980)^{+}$ reflection almost describe the $(\eta\eta)_{S-{\rm wave}}$ in data. 
Therefore, the channel $D^+\to a_0(980)^+\eta$, $a_0(980)^+\to\pi^+\eta$ is expected to be the dominant channel. 

Motivated by this observation, the baseline signal model includes only the decay chain $D^+\to a_0(980)^+\eta$, $a_0(980)^+\to\pi^+\eta$. 
In each fit, this amplitude is chosen as the reference amplitude, with its magnitude and phase fixed to 1.0 and 0.0, respectively. 
The additional amplitudes listed in Table~\ref{tab:addamp} are tested individually. 
The statistical significance of each additional amplitude is evaluated from the change in the log-likelihood and the change in the number of free parameters between the fits with and without the amplitude. 

The tests are divided into four sets, according to the different $a_0(980)^+$ line shapes, described by:
\begin{enumerate}
\item[A.] the Flatt\'e parameterization;
\item[B.] the dispersively modified Flatt\'e parameterization;
\item[C.] the $T$-matrix formalism; and
\item[D.] the $K$-matrix formalism.
\end{enumerate}

\begin{table}[htbp]
\begin{center}
\caption{The additional amplitudes being considered, where the subscripts V and T represent the vector and tensor non-resonant states.}
\begin{tabular}{c|c} \hline 
 ~          & Amplitude \\ \hline
 I          & $D^{+} \to f_{0}(1370)\pi^{+}, f_{0}(1370) \to \eta\eta$ \\
 II         & $D^{+} \to f_{0}(1500)\pi^{+}, f_{0}(1500) \to \eta\eta$ \\
 III        & $D^{+} \to f_{0}(1710)\pi^{+}, f_{0}(1710) \to \eta\eta$ \\
 IV         & $D^{+} \to f_{2}(1270)\pi^{+}, f_{2}(1270) \to \eta\eta$ \\
 V          & $D^{+} \to f^{\prime}_{2}(1525)\pi^{+}, f^{\prime}_{2}(1525) \to \eta\eta$ \\ 
 VI         & $D^{+} \to f_{2}(1565)\pi^{+}, f_{2}(1565) \to \eta\eta$ \\
 VII        & $D^{+} \to f_{2}(1640)\pi^{+}, f_{2}(1640) \to \eta\eta$ \\
 VIII       & $D^{+} \to (\eta\eta)_{T} \pi^{+}$ \\
 IX         & Constant term \\
 X          & $D^{+} \to (\pi^{+}\eta)_{V} \eta$\\
 XI         & $D^{+} \to a_{2}(1320)^{+}\eta$ \\
 XII       & $D^{+} \to (\pi^{+}\eta)_{T}\eta$ \\
\hline
\end{tabular}
\label{tab:addamp}
\end{center}
\end{table}

In the test sets A and B, the $a_0(980)$ parameters are fixed to the values reported in Refs.~\cite{CLEO:2011upl} and~\cite{BESIII:2016tqo}, respectively. 
In the test set C, the $T$-matrix amplitude is used as described in Sec.~\ref{sec:lineshape}, without introducing additional free parameters.
In the test set D, the scattering part of the $K$-matrix formalism is fixed: the bare masses, channel couplings, and smooth-background parameters of the $K$-matrix are taken from Table~I of Ref.~\cite{Anisovich:1997pe}. 
The production-vector parameters, which describe the coupling of the initial $D^+$ decay to the $K$-matrix bare poles and smooth non-pole production terms, are treated as free fit parameters.
In the test sets A--D, the amplitude ``IX. Constant term'' represents a uniform nonresonant amplitude over phase space. 
When this term is included together with the dominant $a_0(980)^+\eta$ amplitude, the fitted solution exhibits a large constructive interference, exceeding 30\%. 
Such a large interference indicates a strong correlation between the constant term and the $a_0(980)^+\eta$ amplitude, which makes the separation of the two contributions unstable and model dependent. 
Therefore, the constant term is not included as an independent component in the nominal signal model.

In addition, we also employ a model-independent amplitude analysis for $(\eta\eta)_{S-{\rm wave}}$ instead of a simple constant for IX. 
The results are similar with the fits to the model with further considering the amplitude IX, large constructive interference is observed between 
$D^+ \to (\eta\eta)_{S-{\rm wave}}\pi^{+}$ and $D^{+} \to a_{0}(980)^{+}\eta$. 
The detail results can be found at Sec.~\ref{sec:MIPWAres}. 

\subsection{Test set A: Flatt\'e parameterization}
\label{sec:testsetA}

\begin{figure}[htbp]
\begin{center}
\begin{minipage}[b]{0.23\textwidth}
\epsfig{width=0.98\textwidth,clip=true,file=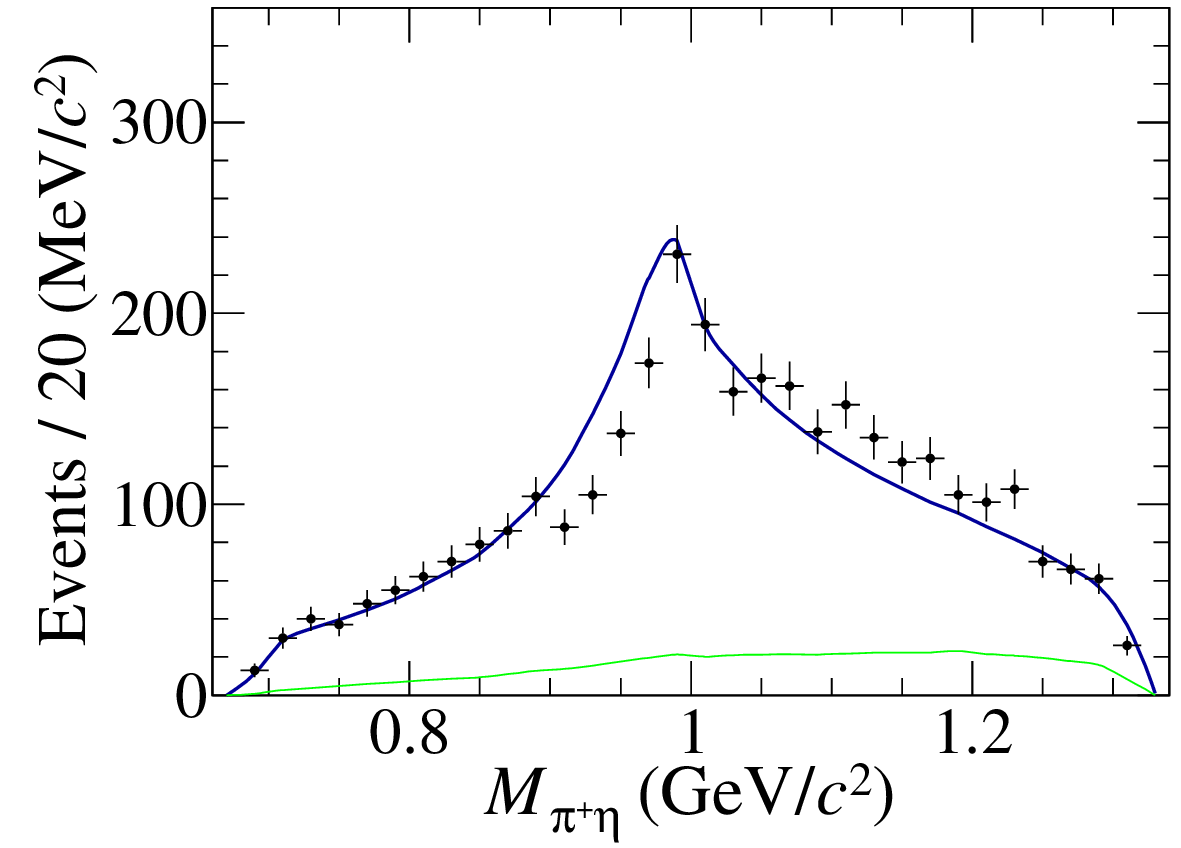}
%\put(-25,65){(a)}
\end{minipage}
\begin{minipage}[b]{0.23\textwidth}
\epsfig{width=0.98\textwidth,clip=true,file=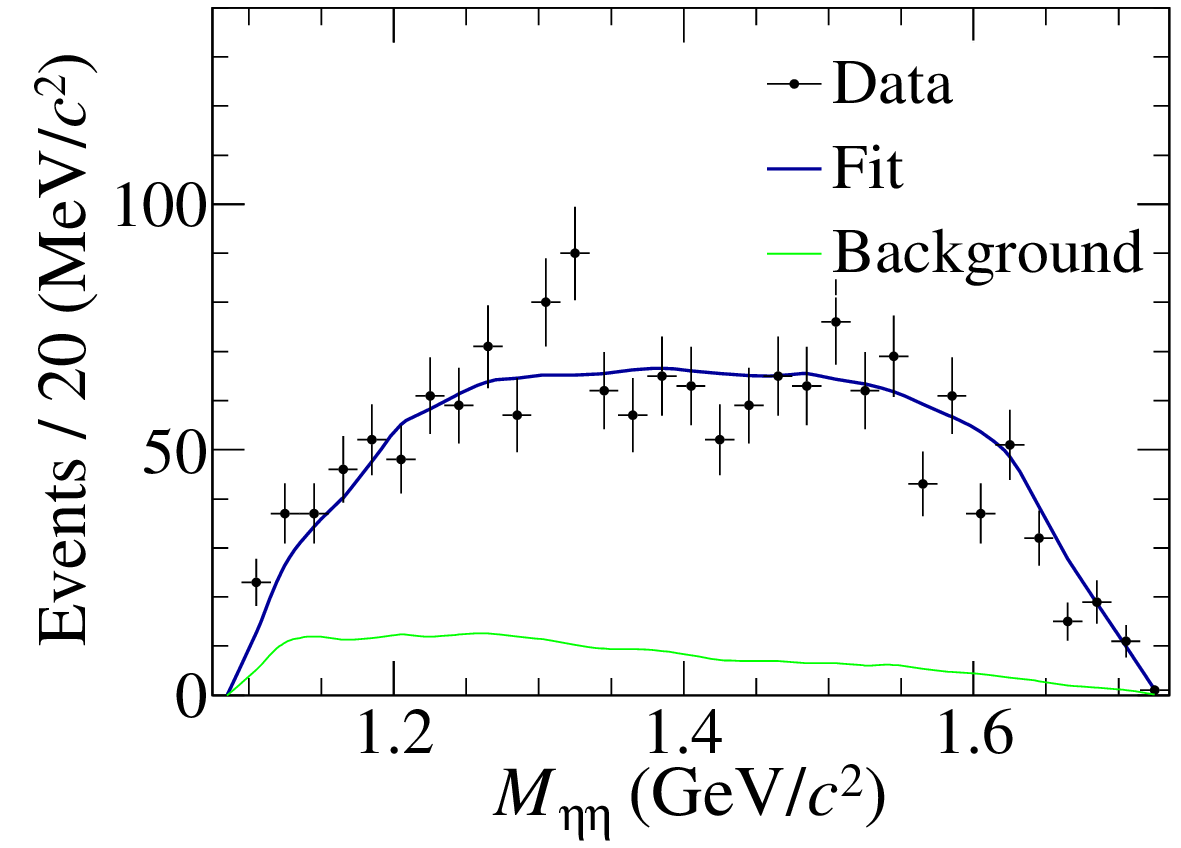}
%\put(-25,65){(b)}
\end{minipage}
\caption{The projections of (left) $M(\pi^{+}\eta)$ and (right) $M(\eta\eta)$ for the  baseline model with Flatt\'e parameterization for $P_{a_0(980)}$. 
The black dots with error bars are the data; the blue lines are the total fit; and the green lines are the background. }
\label{fig:Flattebaseline}
\end{center}
\end{figure}

The baseline model gives $\ln\mathcal{L}=105.3$ and $\chi^2/{\rm NDOF}=161.1/91$. 
Throughout this paper, the $M(\pi^+\eta)$ projection is obtained by filling both $M(\pi^+\eta_a)$ and $M(\pi^+\eta_b)$ combinations for each event, in order to reduce statistical fluctuations. 
The projections of $M(\pi^+\eta)$ and $M(\eta\eta)$ are shown in Fig.~\ref{fig:Flattebaseline}. 
The observed $a_0(980)$ line shape is not well reproduced by the baseline model.

The additional amplitudes listed in Table~\ref{tab:addamp} are then tested by adding them individually to the baseline model. 
The $\Delta\ln\mathcal{L}$ values, fit qualities, and significances relative to the baseline model are summarized in Table~\ref{tab:Flatteaddamp}.
% The values of $\ln\mathcal{L}$ increment, fit qualities, and significances relative to the baseline model are summarized in Table~\ref{tab:Flatteaddamp}.

\begin{table}[htbp]
\begin{center}
\caption{Fit results for models obtained by adding one additional amplitude to the Flatt\'e baseline model. The $\Delta\ln\mathcal{L}$ and significance are evaluated relative to the baseline model.}
% \caption{The fit qualities for the fits after adding one additional amplitude in the baseline model. The significance is calculated relative to the baseline model.}
\begin{tabular}{c|ccc} \hline 
Amplitude         & $\Delta\ln\mathcal{L}$ & $\chi^2/{\rm NDOF}$ & Significance ($\sigma$)\\ \hline
 I                & $+2.8$   & $146.1/89$  & 1.9                  \\
 II               & $+7.0$   & $157.3/89$  & 3.3                  \\
 III              & $+7.7$   & $141.8/89$  & 3.5                  \\
 IV               & $+7.7$   & $143.7/89$  & 3.5                  \\
 V                & $+14.6$  & $136.0/89$  & 5.0                  \\  %(inconsistent with the PDG result of D+ ->K-K+ pi+ and f2(1525)->K-K+ and f2(1525)->eta eta) && mass can not be restored
 VI               & $+17.7$   & $128.1/89$  & 5.6                  \\ % Significance with baseline + V + VI v.s. baseline + V. results 2.6sigma for 1525 and 3.6sigma for 1565
 VII              & $+18.1$   & $133.5/89$  & 5.7                  \\ %% the same with V. VI.
 VIII             & $+11.5$   & $134.0/89$  & 4.4                  \\
 IX               & $+28.0$   & $106.6/89$  & 7.2                  \\
 X                & $+2.1$    & $146.6/89$  & 1.5                  \\
 XI               & $+11.2$   & $152.1/89$  & 4.3                  \\
 XII              & $+4.8$    & $158.6/89$  & 2.7                  \\
\hline
\end{tabular}
\label{tab:Flatteaddamp}
\end{center}
\end{table}

\begin{figure}[htbp]
\begin{center}

%==================== f2'(1525) ====================
\begin{minipage}[b]{0.23\textwidth}
\centering
\epsfig{width=0.98\textwidth,clip=true,file=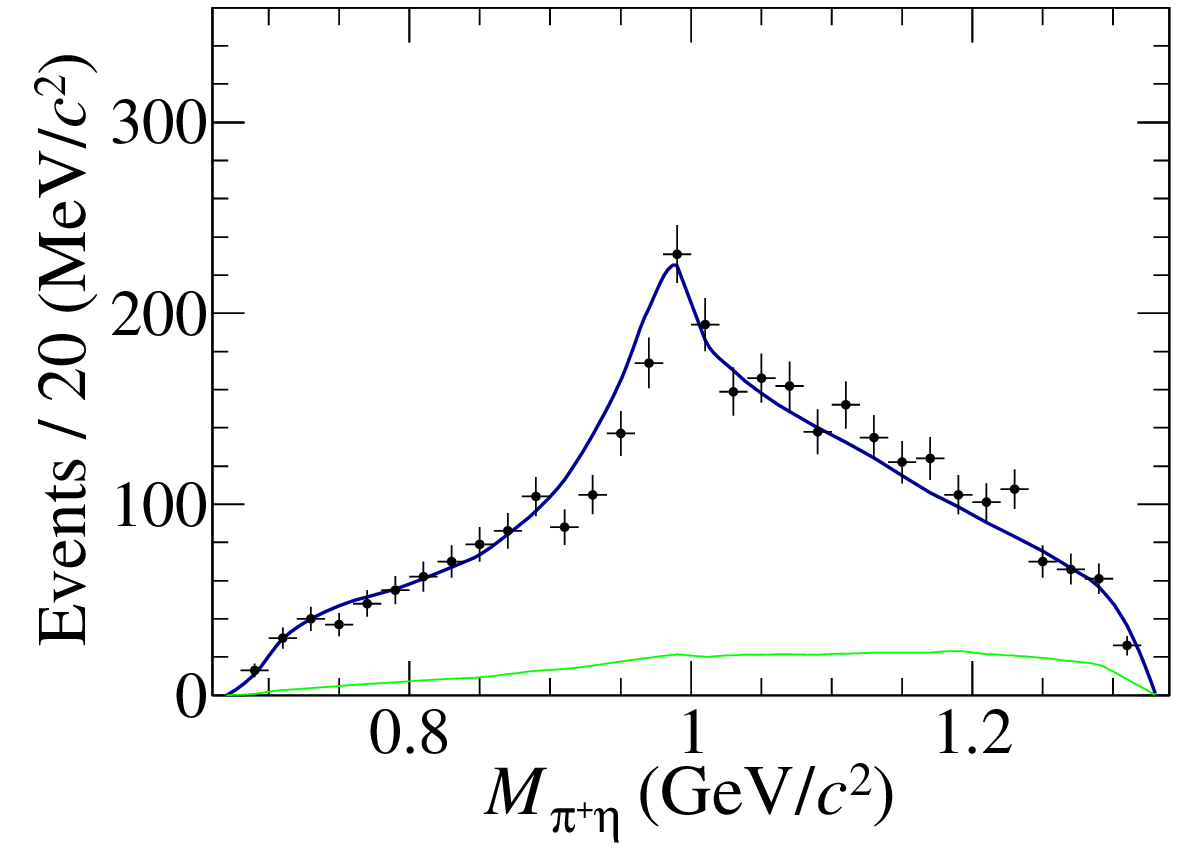}
\put(-85,65){V}
% \put(-80,65){VI.$f_{2}^{\prime}(1525)$}
\end{minipage}
\begin{minipage}[b]{0.23\textwidth}
\centering
\epsfig{width=0.98\textwidth,clip=true,file=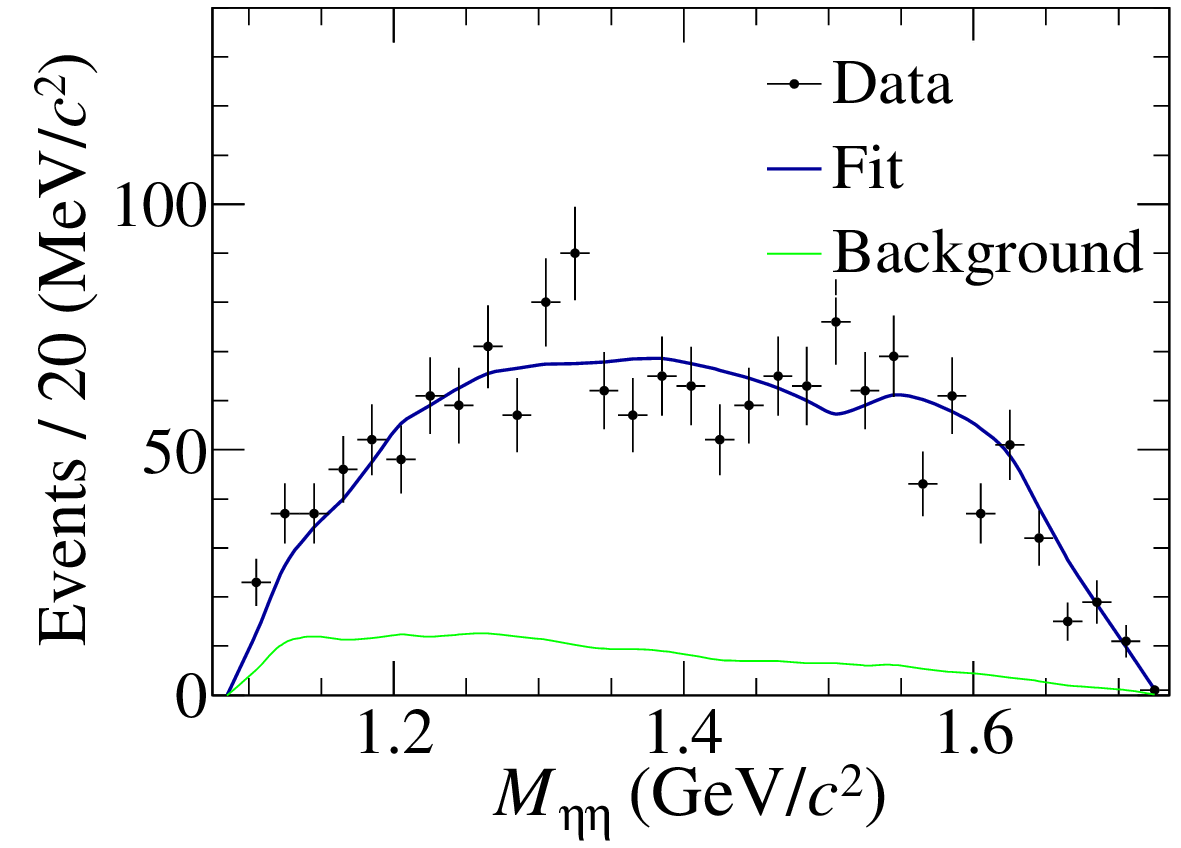}
%\put(-25,65){(b)}
\end{minipage}
%==================== f2(1565) ====================
\begin{minipage}[b]{0.23\textwidth}
\centering
\epsfig{width=0.98\textwidth,clip=true,file=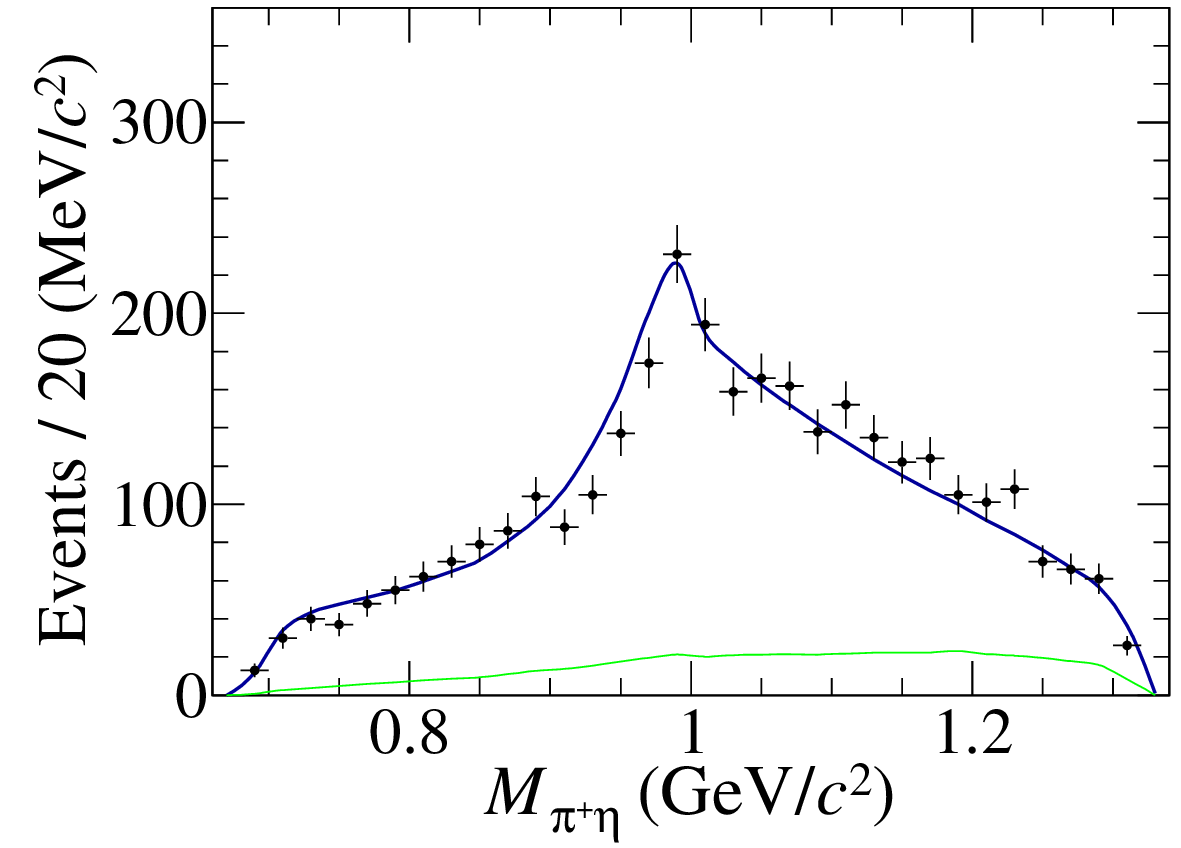}
\put(-85,65){VI}
% \put(-80,65){VII.$f_{2}(1565)$}
\end{minipage}
\begin{minipage}[b]{0.23\textwidth}
\centering
\epsfig{width=0.98\textwidth,clip=true,file=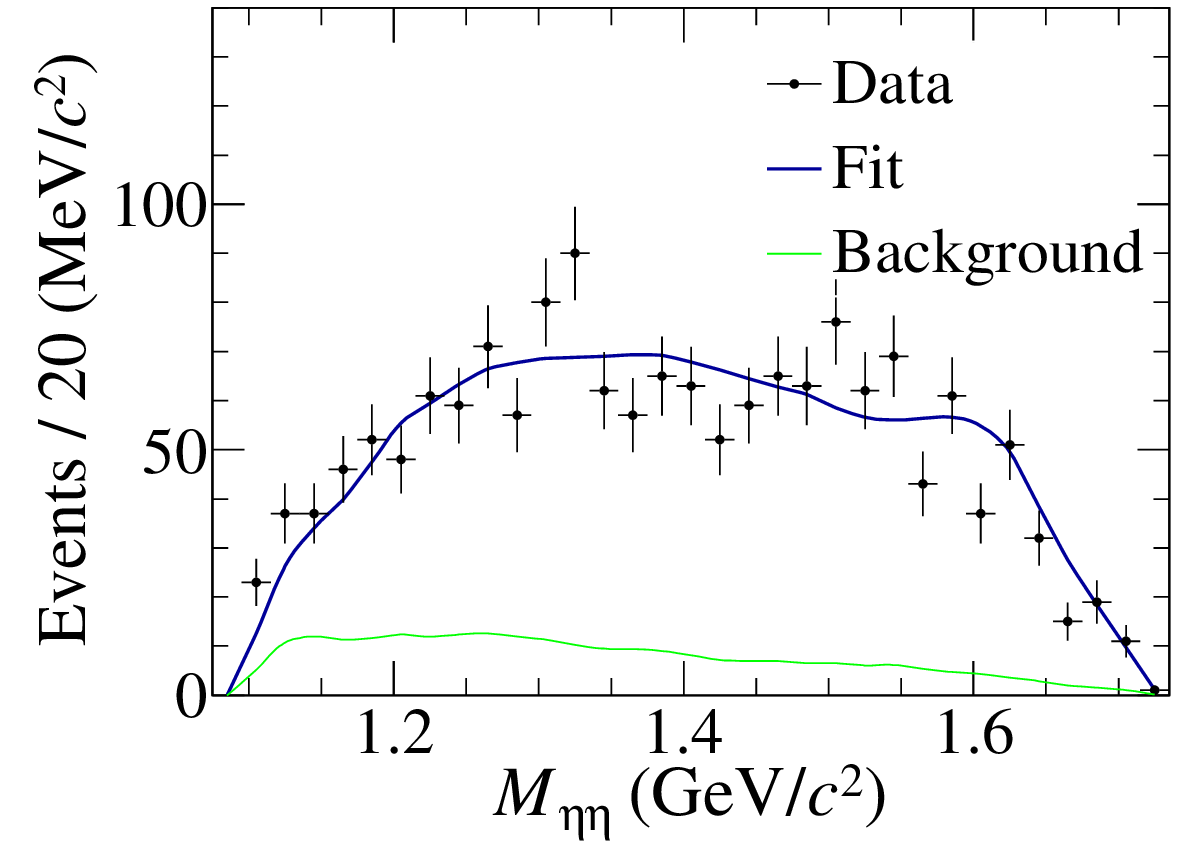}
%\put(-25,65){(d)}
\end{minipage}
%==================== f2(1640) ====================
\begin{minipage}[b]{0.23\textwidth}
\centering
\epsfig{width=0.98\textwidth,clip=true,file=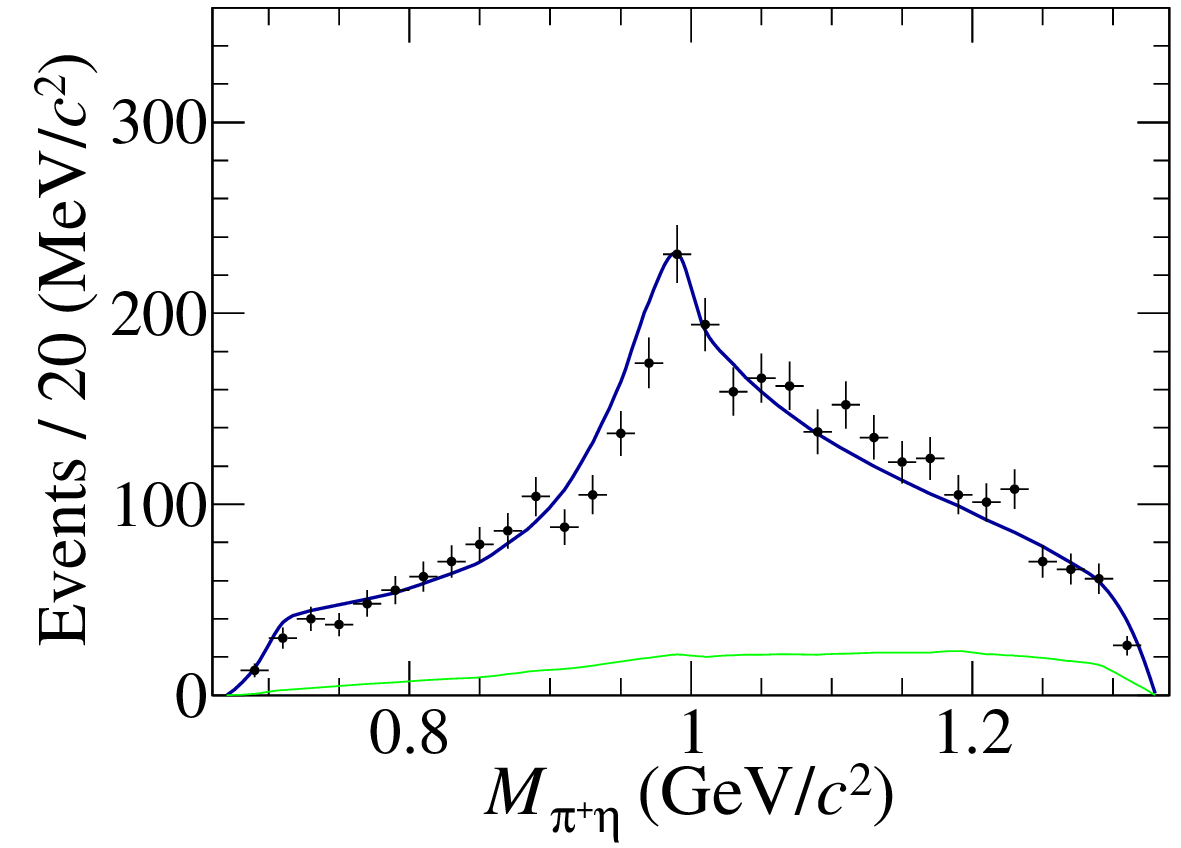}
\put(-85,65){VII}
% \put(-80,65){VIII.$f_{2}(1640)$}
\end{minipage}
\begin{minipage}[b]{0.23\textwidth}
\centering
\epsfig{width=0.98\textwidth,clip=true,file=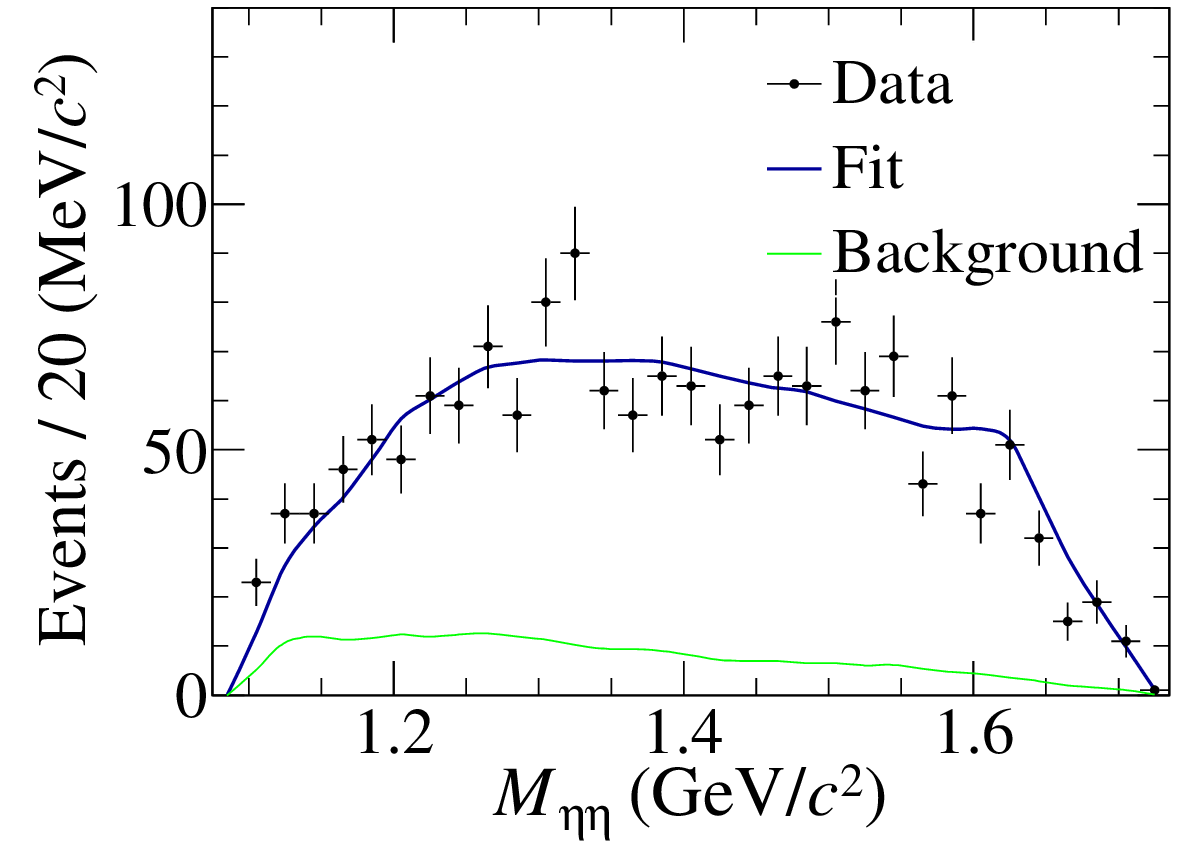}
%\put(-25,65){(f)}
\end{minipage}

\caption{The same as Fig.~\ref{fig:Flattebaseline}, but for the direct-production models including one additional
tensor amplitude V. $f_{2}(1270)$,  VI. $f_2^{\prime}(1525)$, and VII. $f_2(1640)$ with the Flatt\'e parameterization for $P_{a_0(980)}$. }
\label{fig:f2addamp_main}
\end{center}
\end{figure}

As given in Table~\ref{tab:Flatteaddamp}, adding a single additional amplitude improves the fit quality in some cases, but none of these models provides a satisfactory description of the data. 
The largest improvement is obtained for the constant term; however, as discussed above, this term is strongly correlated with the dominant $a_0(980)^+\eta$ amplitude and does not provide a stable independent contribution.
The single-amplitude tests give significances above $5\sigma$ for the tensor states $f_2'(1525)$, $f_2(1565)$, and $f_2(1640)$.
The projections for the three tensor amplitudes that give the largest improvements in the direct-production tests are shown in Fig.~\ref{fig:f2addamp_main}. 
The complete set of projections for all tested additional amplitudes in the Flatt\'e case is provided in Fig.~\ref{fig:Flatteaddamp} of Appendix~\ref{app:addamp_projections}.
However, no corresponding peak-like structures are observed in the $M(\eta\eta)$ projection. 
Since these tensor states populate the same high-$M(\eta\eta)$ region and have partially overlapping line shapes, their apparent significances in the single-amplitude tests can be enhanced by correlations with the dominant $a_0(980)^+\eta$ amplitude and with each other.

The interpretation of the $f_2'(1525)$ contribution is further constrained by existing information from $D^+\to K^-K^+\pi^+$. 
Using the branching fractions for $f_2'(1525)\to\eta\eta$ and $f_2'(1525)\to K\bar K$, together with the FFs obtained in the present fit, one would expect a corresponding contribution in $D^+\to K^-K^+\pi^+$ at the percent level. 
Such a contribution is not supported by previous Dalitz-plot analyses of $D^+\to K^-K^+\pi^+$~\cite{CLEO:2008msk,EvangelhoVieira:2015ujl}. 
This comparison suggests that the large significance obtained for $f_2'(1525)$ in the single-amplitude test is likely driven by correlations and should not be interpreted as evidence for an independent $f_2'(1525)\pi^+$ component.

A consistency check is also performed for the $f_2(1270)$ contribution. 
Using the PDG branching fractions $\mathcal{B}(D^+\to f_2(1270)\pi^+, f_2(1270)\to\pi^+\pi^-)$, $\mathcal{B}(f_2(1270)\to\pi\pi)$, $\mathcal{B}(f_2(1270)\to\eta\eta)$ and $\mathcal{B}(D^+\to\pi^+\eta\eta)$, the expected FF for $D^+\to f_2(1270)\pi^+$, $f_2(1270)\to\eta\eta$ is estimated to be $(1.1\pm0.2)\times10^{-3}$. 
When the FF is fixed to this value in the fit including amplitude IV, the significance of the $f_2(1270)$ contribution is reduced to below $3\sigma$.

\begin{figure}[htbp]
\begin{center}
\begin{minipage}[b]{0.23\textwidth}
\epsfig{width=0.98\textwidth,clip=true,file=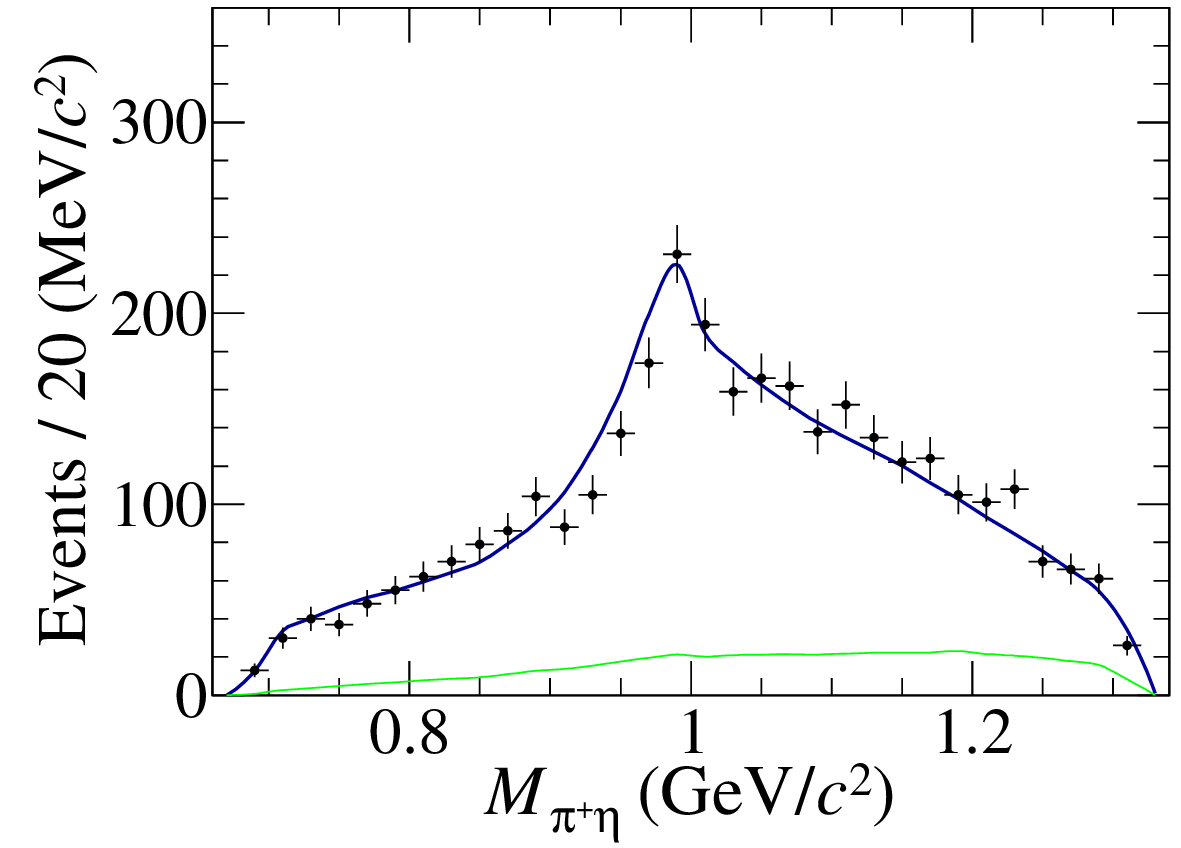}
\put(-85,65){V$+$VI}
\end{minipage}
\begin{minipage}[b]{0.23\textwidth}
\epsfig{width=0.98\textwidth,clip=true,file=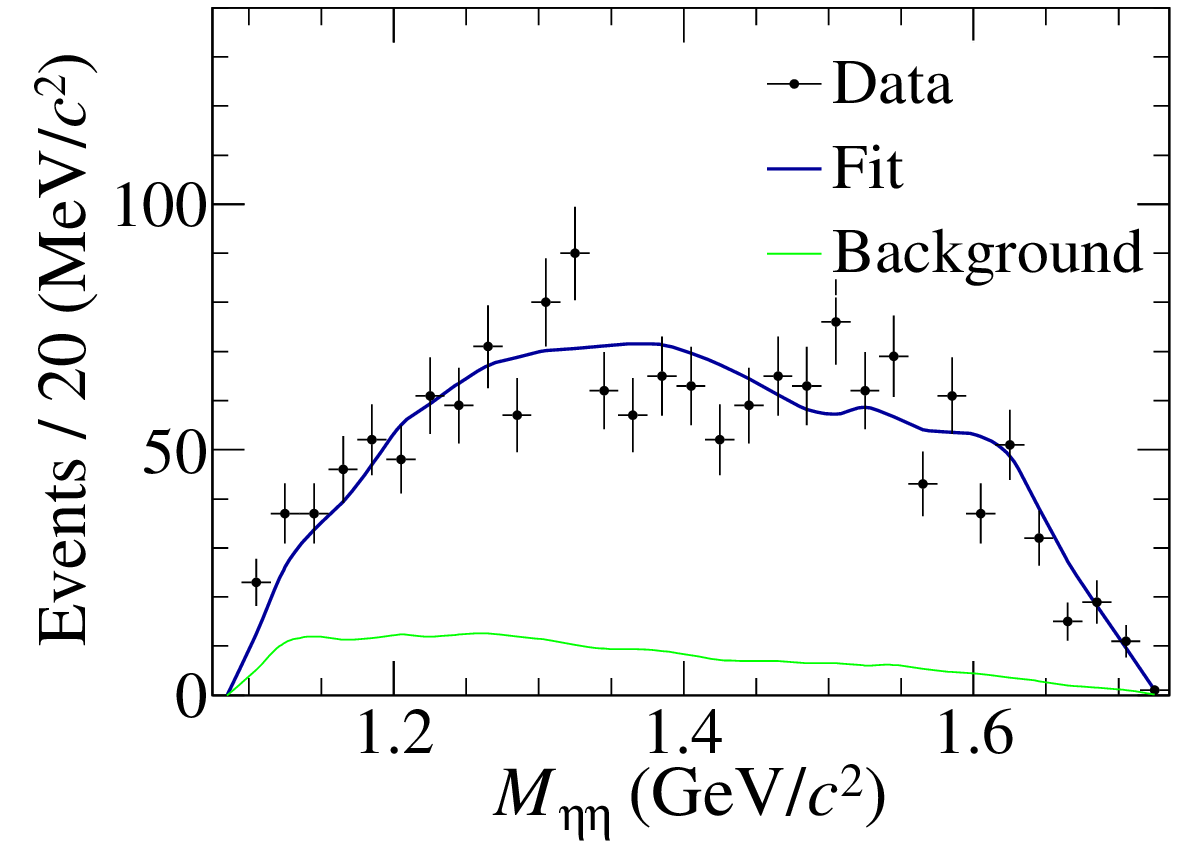}
%\put(-25,65){(b)}
\end{minipage}
\begin{minipage}[b]{0.23\textwidth}
\epsfig{width=0.98\textwidth,clip=true,file=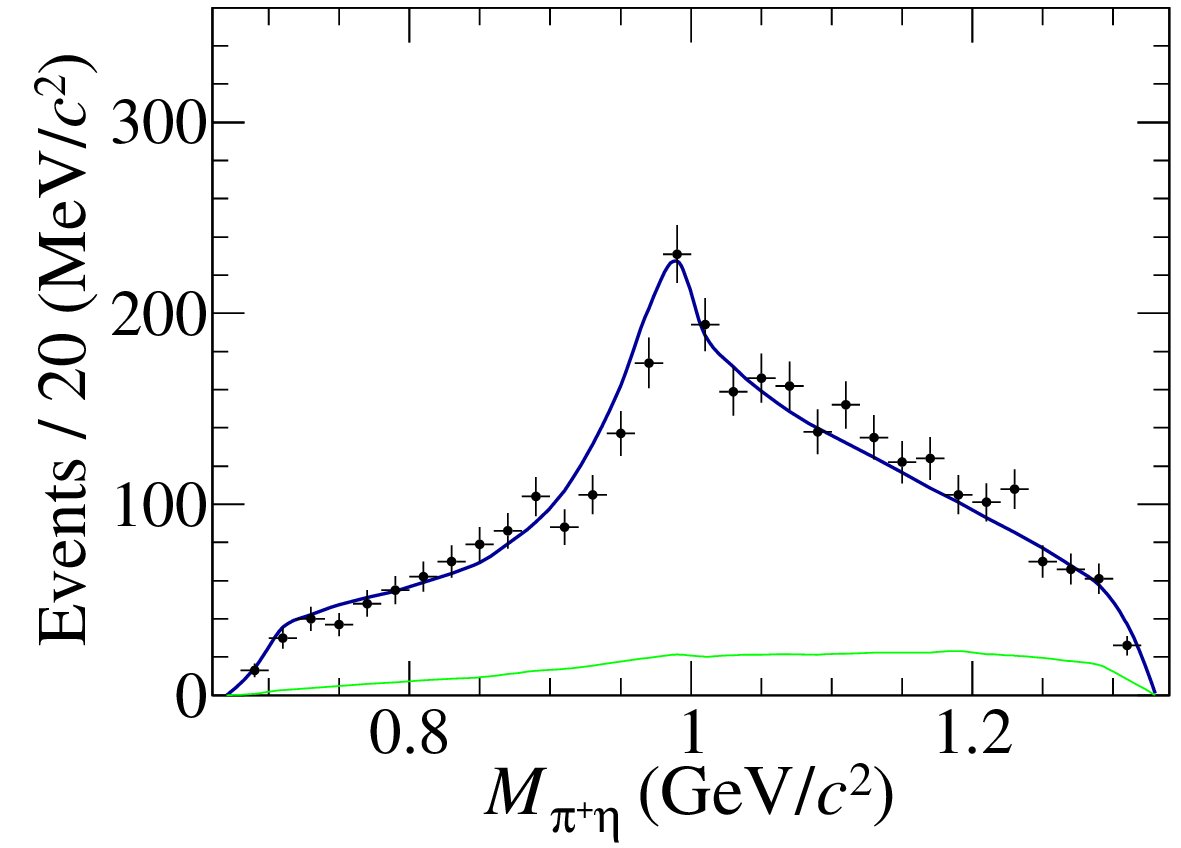}
\put(-85,65){V$+$VII}
\end{minipage}
\begin{minipage}[b]{0.23\textwidth}
\epsfig{width=0.98\textwidth,clip=true,file=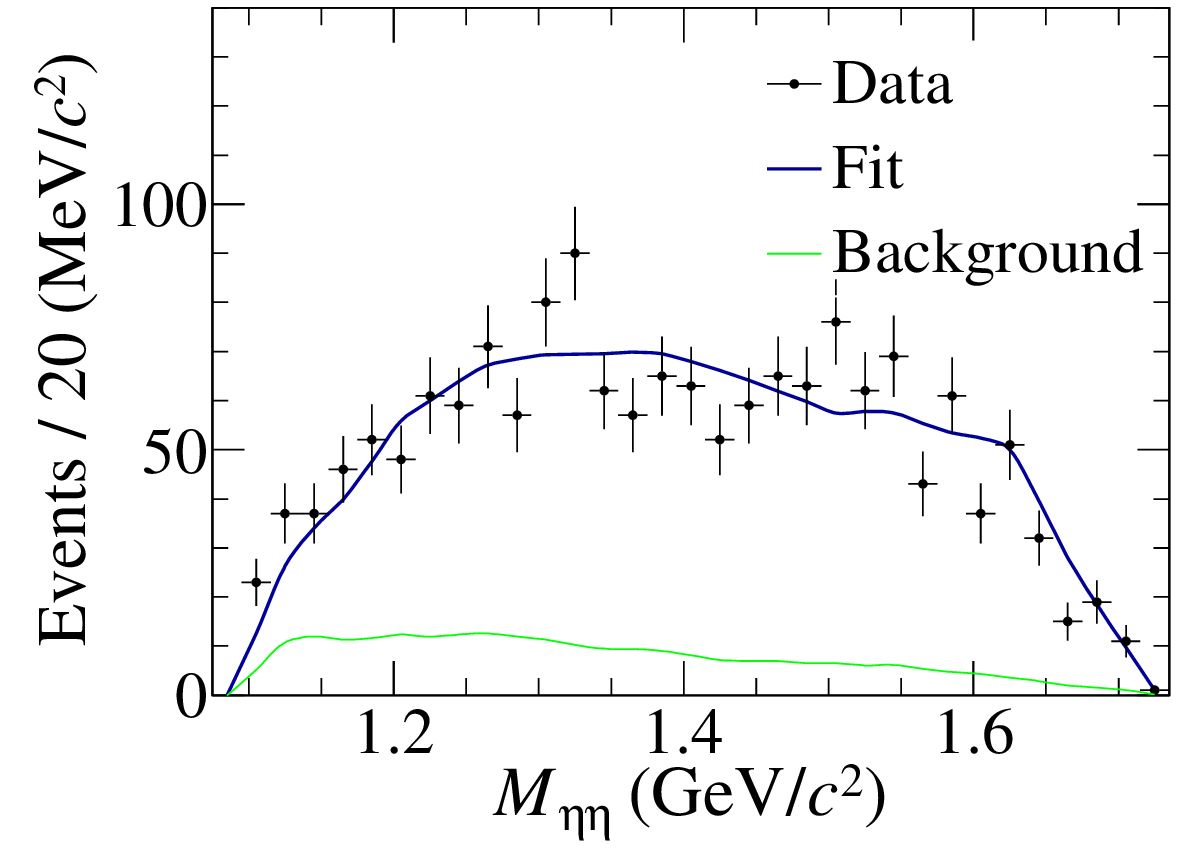}
%\put(-25,65){(d)}
\end{minipage}
\begin{minipage}[b]{0.23\textwidth}
\epsfig{width=0.98\textwidth,clip=true,file=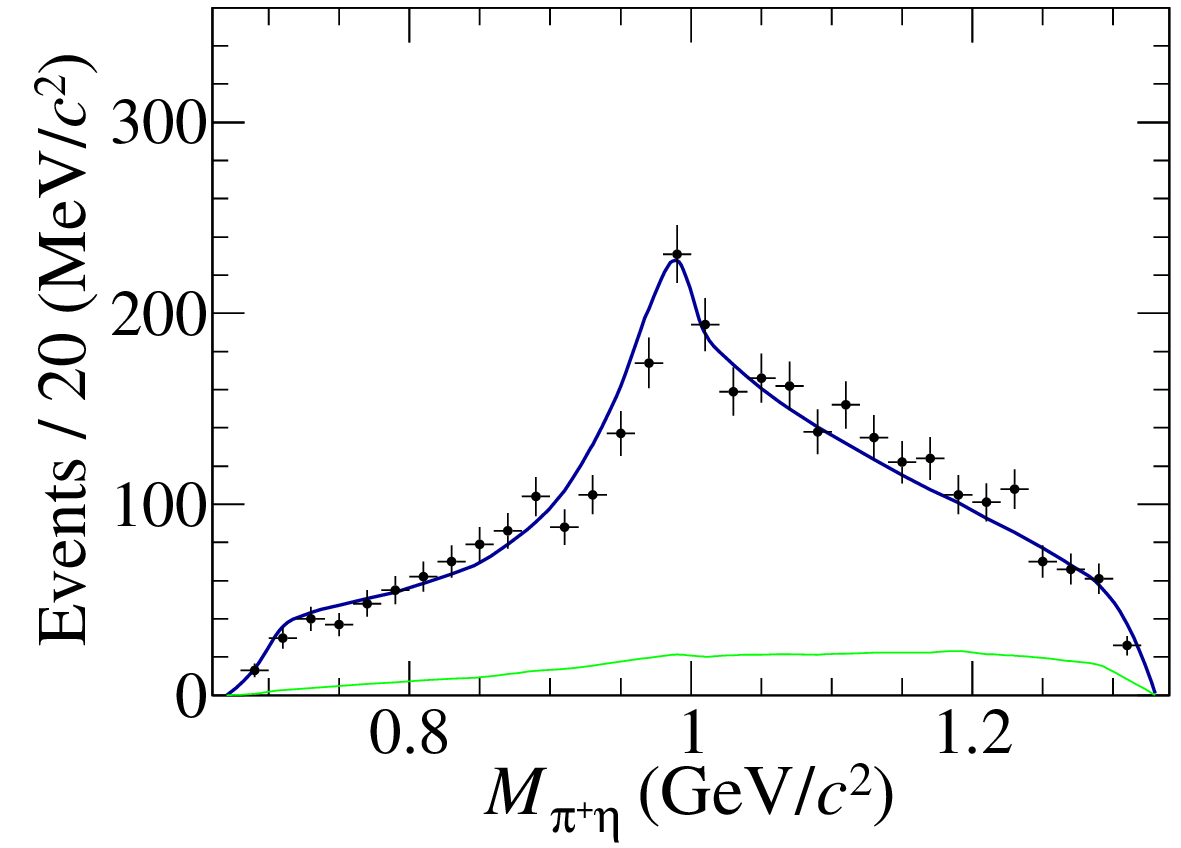}
\put(-85,65){VI$+$VII}
\end{minipage}
\begin{minipage}[b]{0.23\textwidth}
\epsfig{width=0.98\textwidth,clip=true,file=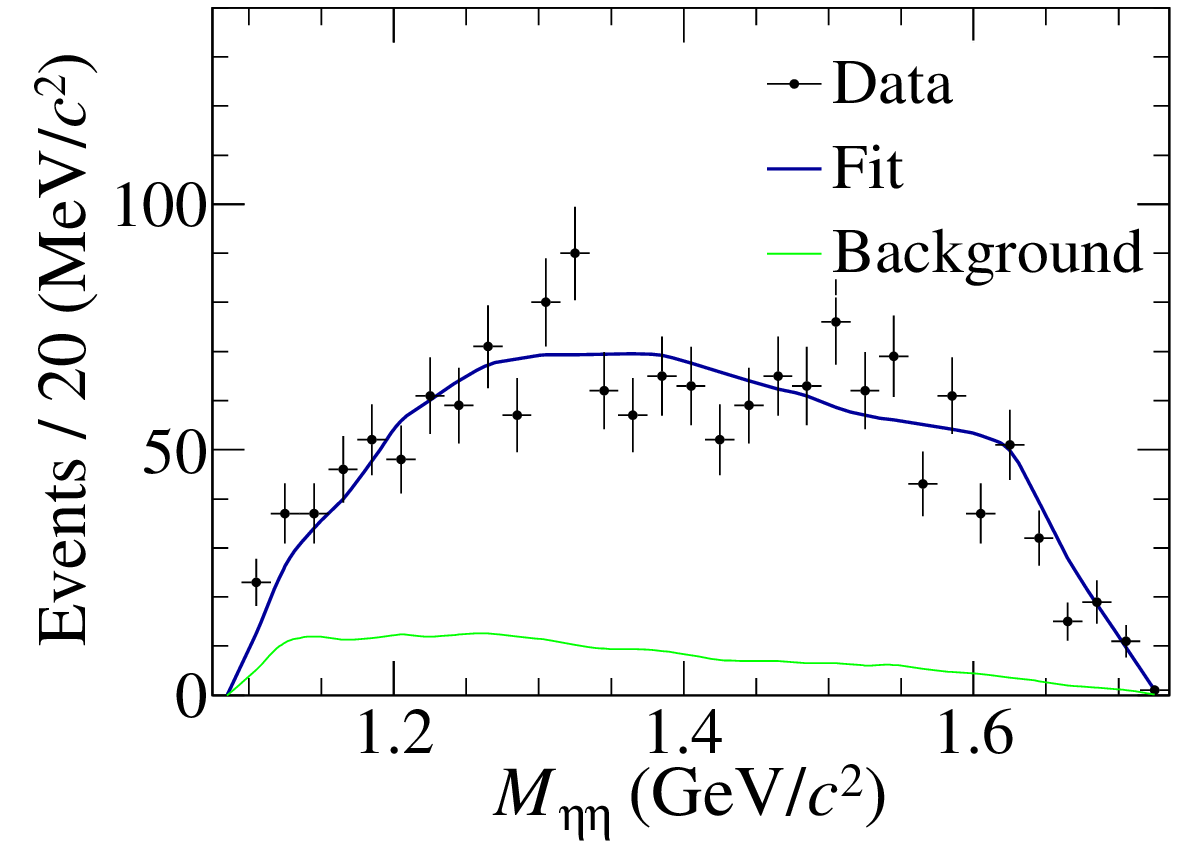}
%\put(-25,65){(f)}
\end{minipage}
\caption{The same as Fig.~\ref{fig:Flattebaseline}, but for the models after further considering (top) $f_{2}^{\prime}(1525)+f_{2}(1565)$, 
(middle) $f_{2}^{\prime}(1525) + f_{2}(1640)$, and (bottom) $f_{2}(1565)+f_{2}(1640)$.}
\label{fig:CLEOcpar2f2}
\end{center}
\end{figure}

To further examine correlations among the tensor components, three fits are performed by including two of the three tensor states $f_2'(1525)$, $f_2(1565)$, and $f_2(1640)$ simultaneously. 
The projections are shown in Fig.~\ref{fig:CLEOcpar2f2}, and the fit results are summarized in Table~\ref{tab:CLEOcpar2f2}. 
The significance of each tensor state is evaluated from the change in $\ln\mathcal{L}$ with respect to the corresponding fit containing only the other tensor state.

As shown in Table~\ref{tab:CLEOcpar2f2}, the incremental significances of the tensor states are substantially smaller than those obtained in the single-amplitude tests. 
When all three tensor states are included simultaneously, the fit gives $\chi^2/{\rm NDOF}=120.2/85$ and the significance of $f_{2}'(1525)$, $f_{2}(1565)$, 
and $f_{2}(1640)$ are only 2.2$\sigma$, 1.6$\sigma$ and 0.6$\sigma$, respectively, indicating only a marginal improvement. 
These results confirm that the apparent tensor contributions are strongly correlated and do not provide stable independent components.

\begin{table}[htbp]
\begin{center}
\caption{The fit qualities for the fits with models adding two additional $f_{2}$ states in the baseline model are listed. 
The significance is calculated with comparing with the baseline plus single $f_{2}$ states. }
\begin{tabular}{c|ccccc} \hline 
\multirow{2}{*}{Amplitude}   & \multirow{2}{*}{$\chi^2/{\rm NDOF}$} & \multicolumn{3}{c}{Significance ($\sigma$)} \\
            ~                &  ~                           &   V     &   VI   &  VII                     \\ \hline
 V + VI                      & $119.1/87$                   &  2.6    &  3.6   &  -                       \\
 V + VII                     & $126.6/87$                   &  1.9    &  -     & 3.1                      \\
 VI + VII                    & $127.7/87$                   &  -      &  1.1   & 1.3                      \\
\hline
\end{tabular}
\label{tab:CLEOcpar2f2}
\end{center}
\end{table}

Since the single-amplitude test with $f_2(1565)$ gives the best fit quality among the three tensor-state candidates, it is used as the reference model for further tests of additional components. 
Additional tests are performed by adding each of the remaining amplitudes, except for the constant term, to the model ``baseline+$f_2(1565)$''. 
The largest value of $\ln\mathcal{L}$ is 127.3, and the best fit quality is $\chi^2/{\rm NDOF}=113.4/87$, obtained with the model ``baseline+$f_2(1565)+f_0(1710)$''. 
Relative to the ``baseline+$f_2(1565)$'' model, the improvement corresponds to only $2.5\sigma$. 
Moreover, the improvement is mainly associated with a large interference between the $f_0(1710)$ and $a_0(980)^+\eta$ amplitudes. 
For the other tested models, $\ln\mathcal{L}$ remains below 127.0 and the fit quality is worse than $\chi^2/{\rm NDOF}=119.5/87$.

Based on these studies, we conclude that the discrepancy between the data and the baseline Flatt\'e model cannot be resolved by adding small conventional resonant or non-resonant amplitudes.

\subsection{Test set B: dispersively modified Flatt\'e parameterization}
\label{sec:testsetB}

\begin{figure}[htbp]
\begin{center}
\begin{minipage}[b]{0.23\textwidth}
\epsfig{width=0.98\textwidth,clip=true,file=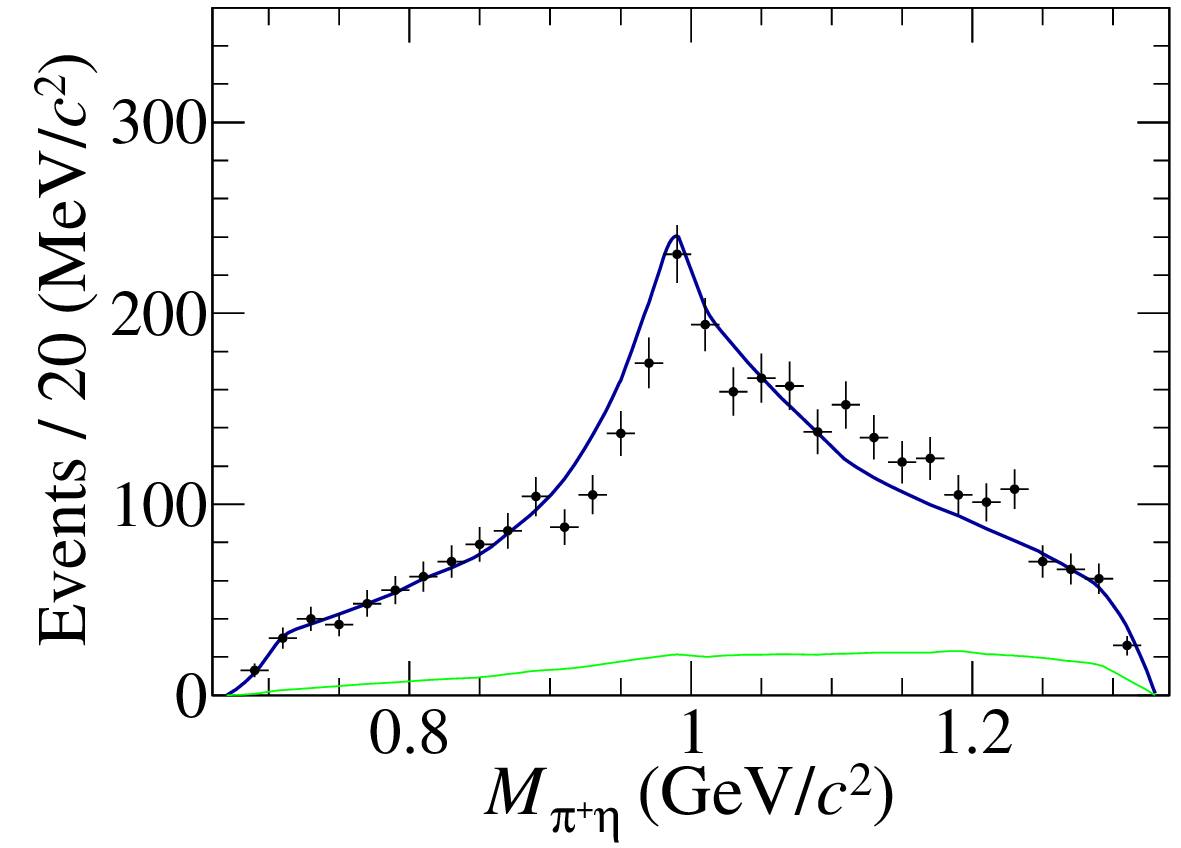}
%\put(-25,65){(a)}
\end{minipage}
\begin{minipage}[b]{0.23\textwidth}
\epsfig{width=0.98\textwidth,clip=true,file=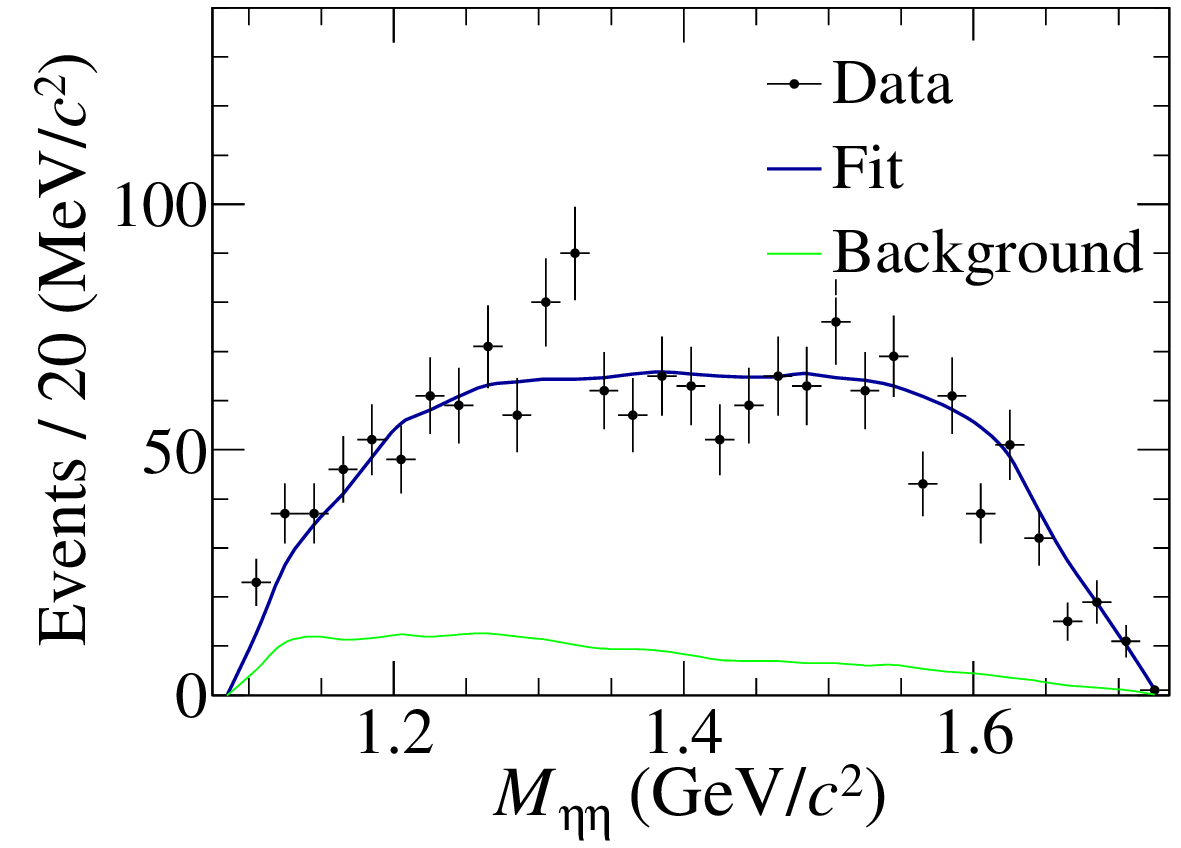}
%\put(-25,65){(b)}
\end{minipage}
\caption{The same as Fig.~\ref{fig:Flattebaseline}, but for the baseline model using the dispersively modified Flatt\'e parameterization.}
\label{fig:dispersivebase}
\end{center}
\end{figure}

As discussed in Sec.~\ref{sec:testsetA}, the discrepancy between the data and the Flatt\'e baseline model cannot be resolved by adding the conventional additional amplitudes listed in Table~\ref{tab:addamp}.  
The baseline fit gives $\ln\mathcal{L}=112.0$ and $\chi^2/{\rm NDOF}=162.7/91$. 
The corresponding projections are shown in Fig.~\ref{fig:dispersivebase}. 

The twelve additional amplitudes listed in Table~\ref{tab:addamp} are then tested by adding them individually to the dispersive baseline model. 
The corresponding projections are shown in Fig.~\ref{fig:dispersive} of Appendix~\ref{app:addamp_projections}. 
% The values of $\ln\mathcal{L}$, fit qualities, and significances relative to the baseline model are summarized in Table~\ref{tab:dispersive}. 
The $\Delta\ln\mathcal{L}$ values, fit qualities, and significances relative to the baseline model are summarized in Table~\ref{tab:dispersive}.
The largest improvement is obtained for model IX, the constant term. 
However, as in the test set A, this term produces a large interference with the dominant $a_0(980)^+\eta$ amplitude and is therefore not included as an independent component in the nominal signal model. 
For the remaining fits, the mismatch in the $M(\pi^+\eta)$ spectrum persists.

\begin{table}[htbp]
\begin{center}
\caption{Fit results for models obtained by adding one additional amplitude to the dispersive baseline model. The $\Delta\ln\mathcal{L}$ and significance are evaluated relative to the baseline model.}
\begin{tabular}{c|ccc} \hline 
Amplitude  & $\Delta\ln\mathcal{L}$      &   $\chi^{2}/\mathrm{NDOF}$ & Significance($\sigma$)     \\ \hline
I          & $+5.9$        &   $136.9/89$               & 2.9                     \\  
II         & $+2.6$        &   $156.1/89$               & 1.7                     \\
III        & $+13.1$       &   $125.9/89$               & 4.7                     \\
IV         & $+8.8$        &   $143.6/89$               & 3.7                     \\
V          & $+7.5$        &   $147.6/89$               & 3.4                     \\
VI         & $+11.8$       &   $138.6/89$               & 4.4                     \\
VII        & $+12.0$       &   $143.4/89$               & 4.5                     \\
VIII       & $+10.6$       &   $130.1/89$               & 4.2                     \\
IX         & $+21.1$       &   $105.4/89$               & 6.1                     \\
X          & $+4.7$        &   $142.5/89$               & 2.5                     \\
XI         & $+2.0$        &   $164.5/89$               & 1.4                     \\
XII        & $+0.1$        &   $162.6/89$               & 0.1                     \\
\hline
\end{tabular}
\label{tab:dispersive}
\end{center}
\end{table}

Compared with the Flatt\'e baseline model, the dispersive baseline model has a larger likelihood value. 
Consequently, the relative improvements obtained by adding the $f_0$ and $f_2$ amplitudes are reduced, and none of these resonant amplitudes reaches a significance above $5\sigma$. 
Among the resonant amplitudes, the largest improvement is obtained when the $f_0(1710)\pi^+$ amplitude is added. 
In this fit, the FF of the $f_0(1710)\pi^+$ contribution is $(1.3\pm0.6)\%$, while the associated interference fraction is $(-4.3\pm3.8)\%$. 
Since no clear corresponding structure is observed in the $M(\eta\eta)$ projection, this contribution is not considered to provide a stable independent component and is not adopted in the nominal signal model.

As in Sec.~\ref{sec:testsetA}, further tests are performed using the model ``baseline+$f_2(1565)$'' as the reference, because it gives the best fit quality among the single-amplitude resonant additions in this test set. 
Additional amplitudes, except for the constant term, are then added individually to this reference model. 
The largest likelihood value, $\ln\mathcal{L}=129.5$, is obtained for the model ``baseline+$f_2'(1525)+f_2(1565)$'', with $\chi^2/{\rm NDOF}=128.6/87$. 
Relative to the corresponding single-$f_2$ models, the incremental significances of the $f_2'(1525)$ and $f_2(1565)$ contributions are $2.9\sigma$ and $4.1\sigma$, respectively. 
This behavior is similar to that observed in the test set A and indicates that the apparent tensor contributions are strongly correlated rather than stable independent components.

Based on these studies, we conclude that the discrepancy between the data and the dispersive baseline model cannot be resolved by adding small conventional resonant or non-resonant amplitudes.

\subsection{Test set C: $T$-matrix formalism}
\label{sec:testsetC}

\begin{figure}[htbp]
\begin{center}
\begin{minipage}[b]{0.23\textwidth}
\epsfig{width=0.98\textwidth,clip=true,file=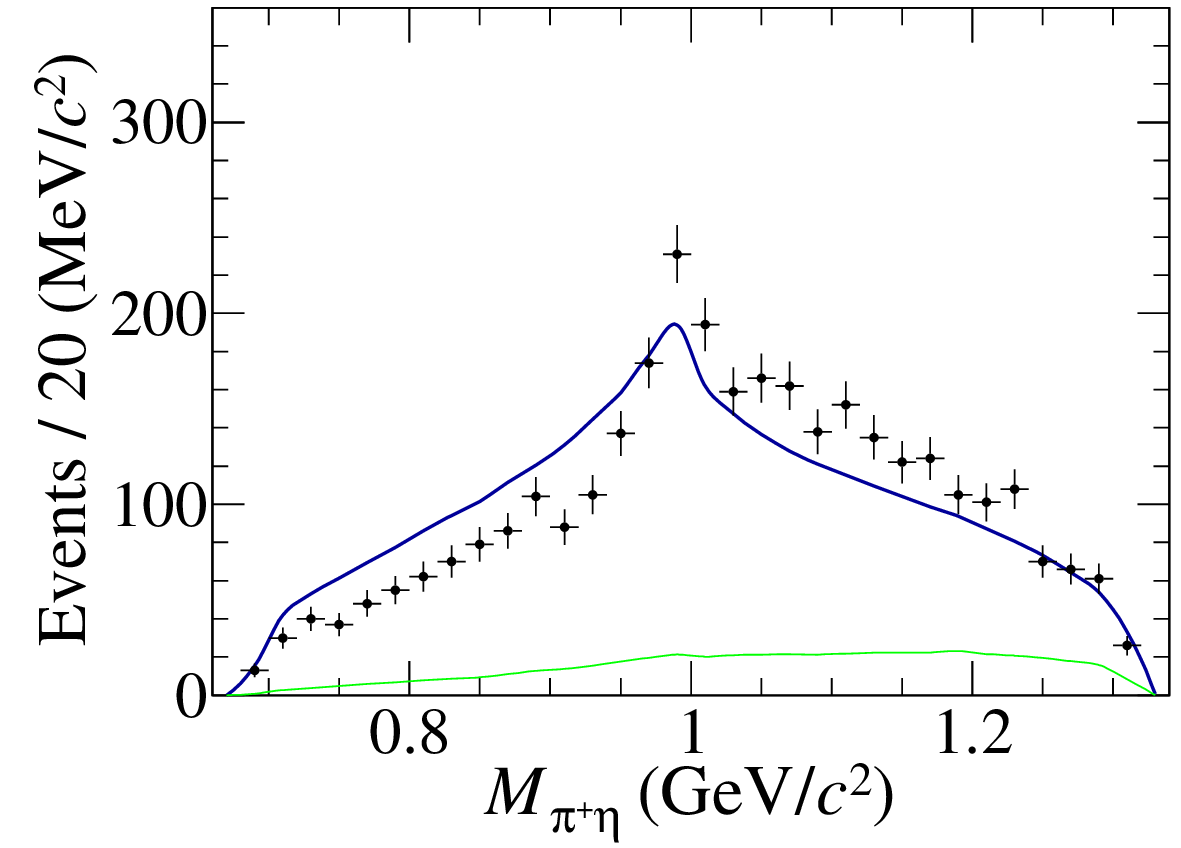}
%\put(-25,65){(a)}
\end{minipage}
\begin{minipage}[b]{0.23\textwidth}
\epsfig{width=0.98\textwidth,clip=true,file=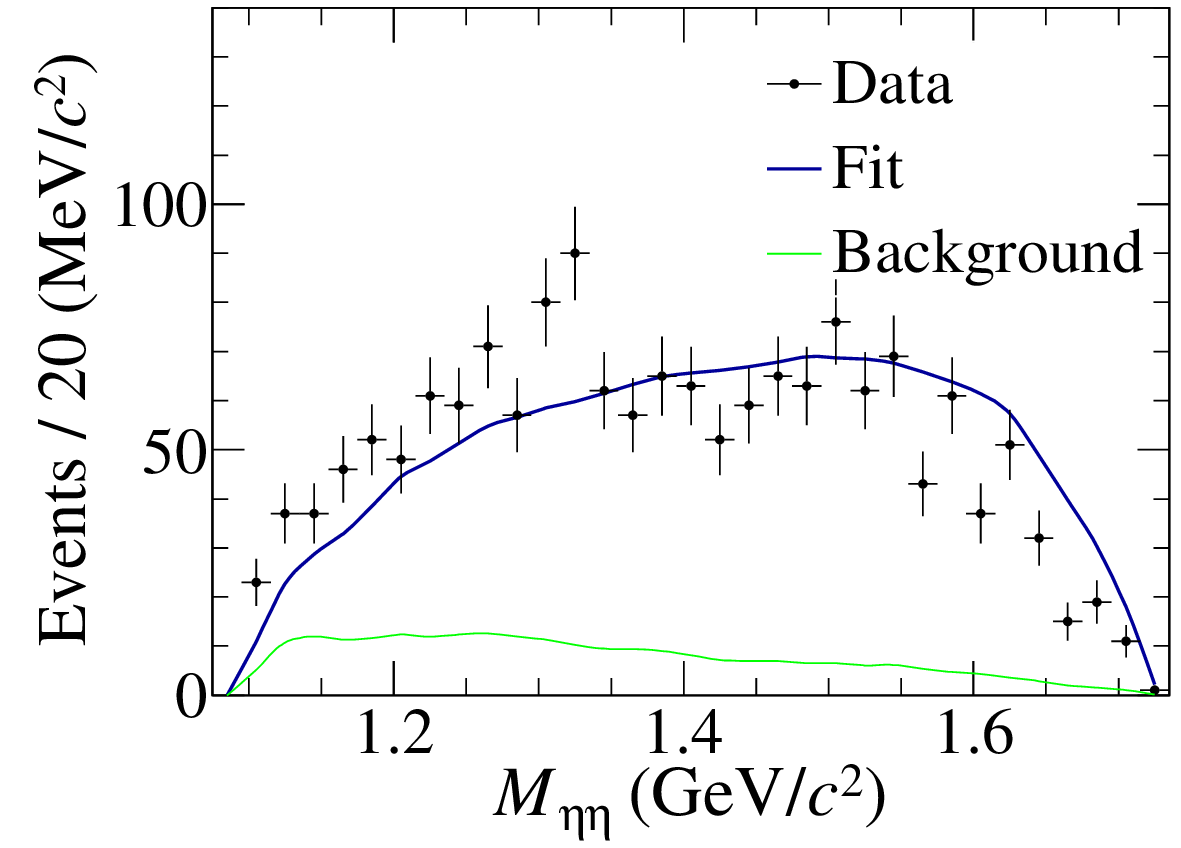}
%\put(-25,65){(b)}
\end{minipage}
\caption{The same as Fig.~\ref{fig:Flattebaseline}, but for the baseline model using the fixed $T$-matrix description of the $\pi\eta$ $S$-wave.}
\label{fig:TMatrix}
\end{center}
\end{figure}

In the test set C, the $\pi\eta$ $S$-wave dominated by the $a_0(980)$ contribution is described by the fixed $T$-matrix amplitude introduced in Sec.~\ref{sec:lineshape}. 
The baseline fit gives $\ln\mathcal{L}=67.9$ and $\chi^2/{\rm NDOF}=278.7/91$. 
The corresponding projections are shown in Fig.~\ref{fig:TMatrix}. 
Compared with the Flatt\'e and dispersively modified Flatt\'e baseline fits in Secs.~\ref{sec:testsetA} and~\ref{sec:testsetB}, this fit gives a substantially worse description of the data.

The additional amplitudes listed in Table~\ref{tab:addamp} are then tested by adding them individually to the $T$-matrix baseline model. 
The corresponding projections are shown in Fig.~\ref{fig:TMatrixadd} of Appendix~\ref{app:addamp_projections}, and the fit results are summarized in Table~\ref{tab:TMatrix}. 
Because the $T$-matrix baseline model gives a poor description of the data, several additional amplitudes lead to large apparent improvements in $\ln\mathcal{L}$. 
However, these improvements should not be interpreted as evidence for independent physical contributions 
since the fit quality of the corresponding baseline model are much worse than those using Flatt\'e and dispersively modified Flatt\'e parameterizations for $P_{a_0(980)}$. 
Furthermore, the large contributions given by these tests disagree with those presented in sections~\ref{sec:testsetA} and~\ref{sec:testsetB}, and the independent Dalitz plot analyses 
for the decay $D^+\to K^-K^+\pi^+$~\cite{CLEO:2008msk,EvangelhoVieira:2015ujl}.

\begin{table}[htbp]
\begin{center}
% \caption{Fit results for models obtained by adding one additional amplitude to the $T$-matrix baseline model.}
\caption{Fit results for models obtained by adding one additional amplitude to the $T$-matrix baseline model. The $\Delta\ln\mathcal{L}$ and significance are evaluated relative to the baseline model.}
\begin{tabular}{cccc} \hline 
Amplitude  & $\Delta\ln\mathcal{L}$  & $\chi^{2}/\mathrm{NDOF}$ & Significance\\ \hline
I          & $+38.9$ & $179.2/89$               & 8.5  \\  
II         & $+11.7$ & $251.5/89$               & 4.4  \\
III        & $+45.2$ & $172.0/89$               & 9.2  \\
IV         & $+25.8$ & $208.7/89$               & 6.8  \\
V          & $+13.7$ & $245.0/89$               & 4.8  \\
VI         & $+24.5$ & $223.9/89$               & 6.6  \\ 
VII        & $+26.6$ & $228.7/89$               & 7.0  \\
VIII       & $+26.1$ & $207.4/89$               & 6.9  \\
IX         & $+59.4$ & $130.4/89$               & 10.7 \\
X          & $+20.6$ & $223.3/89$               & 6.0  \\
XI         & $+9.3$  & $231.3/89$               & 3.9  \\
XII        & $+12.0$ & $234.5/89$               & 4.5  \\
\hline
\end{tabular}
\label{tab:TMatrix}
\end{center}
\end{table}

In particular, the models ``baseline+III'' and ``baseline+IX'' exhibit large interference with the dominant $\pi\eta$ $S$-wave amplitude. 
These solutions are therefore not considered stable independent descriptions of the data. 
For the remaining models, although the likelihood values are improved relative to the $T$-matrix baseline, the fit qualities remain poor and the mismatch in the $M(\pi^+\eta)$ spectrum persists. 
Since the baseline $T$-matrix description is already strongly disfavored, the large significances obtained for several additional amplitudes mainly reflect their ability to compensate for the poor $\pi\eta$ line shape description. 
They are therefore not interpreted as evidence for genuine additional intermediate states.

Additional tests including more than one extra amplitude show similar behavior: the improvements are mainly driven by strong correlations and interference effects rather than by well-resolved structures in the data. 
Therefore, the fixed $T$-matrix description, even when supplemented with conventional resonant or non-resonant amplitudes, does not provide a satisfactory description of the data.

%\clearpage
%\newpage

\subsection{Test set D: $K$-matrix formalism}
\label{sec:testsetD}

In this section, the $\pi\eta$ $S$-wave dominated by the $a_0(980)$ contribution is described with the $K$-matrix formalism. 
The scattering part of the $K$-matrix is fixed to the four parameter sets reported in Ref.~\cite{Anisovich:1997pe}, denoted as s1--s4 and listed in Table~\ref{tab:parsforKmatrix}. 
For each of the parameter set, the production-vector parameters are determined from the fit. 
In the baseline fits, the parameter $s_1^{\rm prod}$ is found to be consistent with zero within its uncertainty and becomes poorly constrained when allowed to float. 
It is therefore fixed to zero in all subsequent fits.

\begin{figure}[htbp]
\begin{center}
\begin{minipage}[b]{0.23\textwidth}
\epsfig{width=0.98\textwidth,clip=true,file=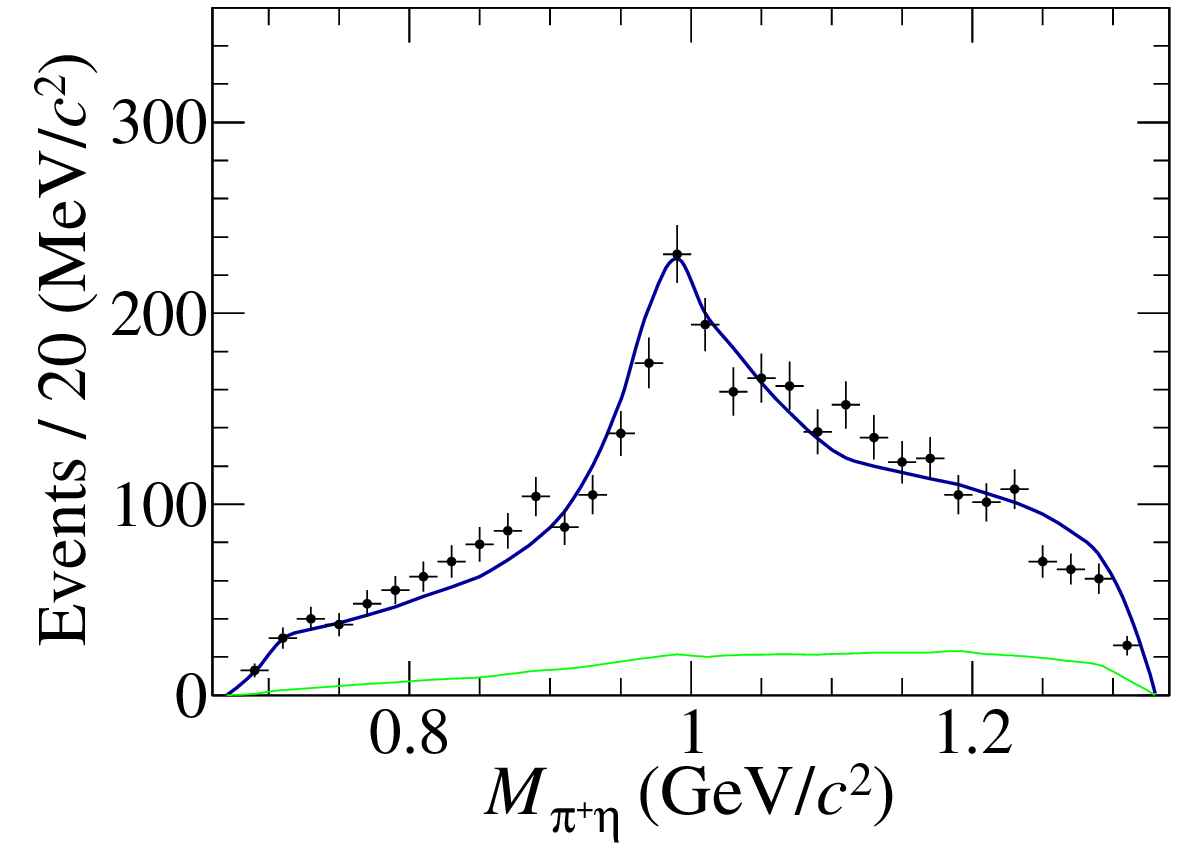}
\put(-75,65){s1}
%\put(-25,65){(a)}
\end{minipage}
\begin{minipage}[b]{0.23\textwidth}
\epsfig{width=0.98\textwidth,clip=true,file=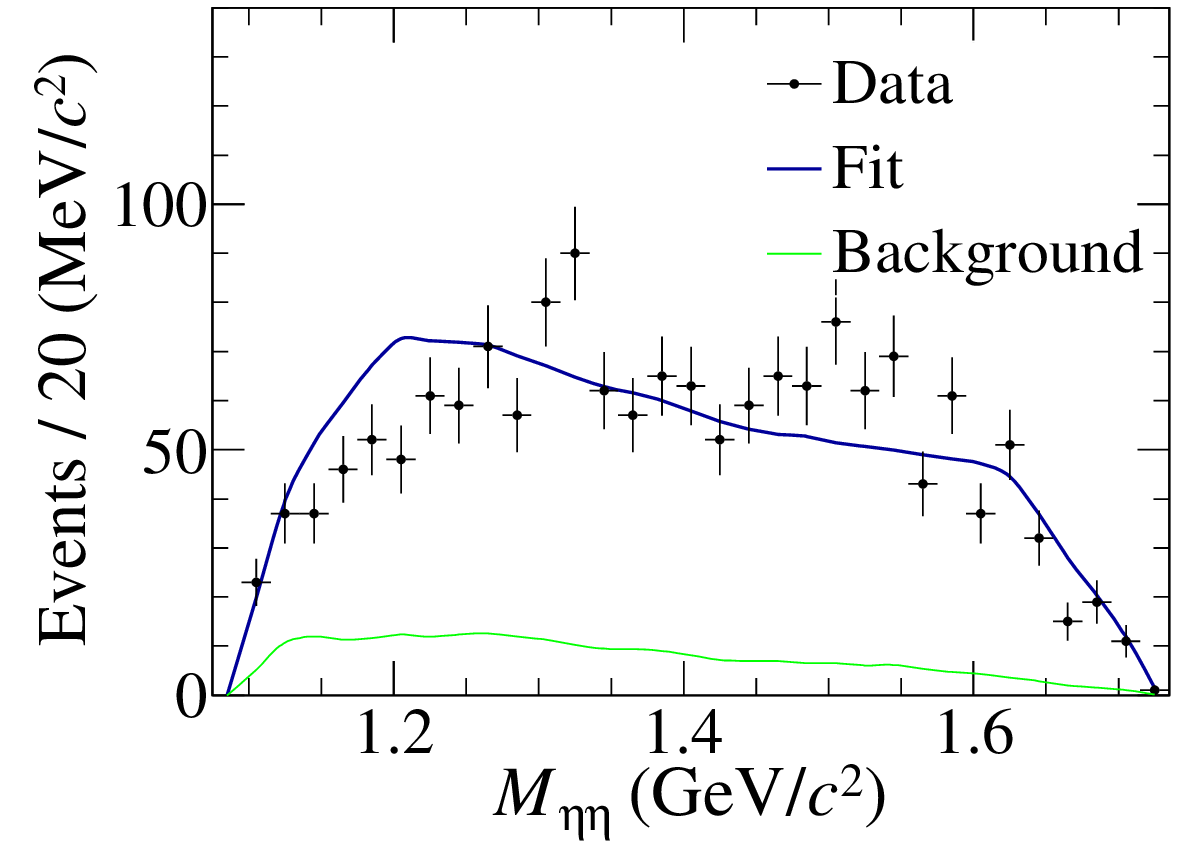}
%\put(-25,65){(b)}
\end{minipage}
\begin{minipage}[b]{0.23\textwidth}
\epsfig{width=0.98\textwidth,clip=true,file=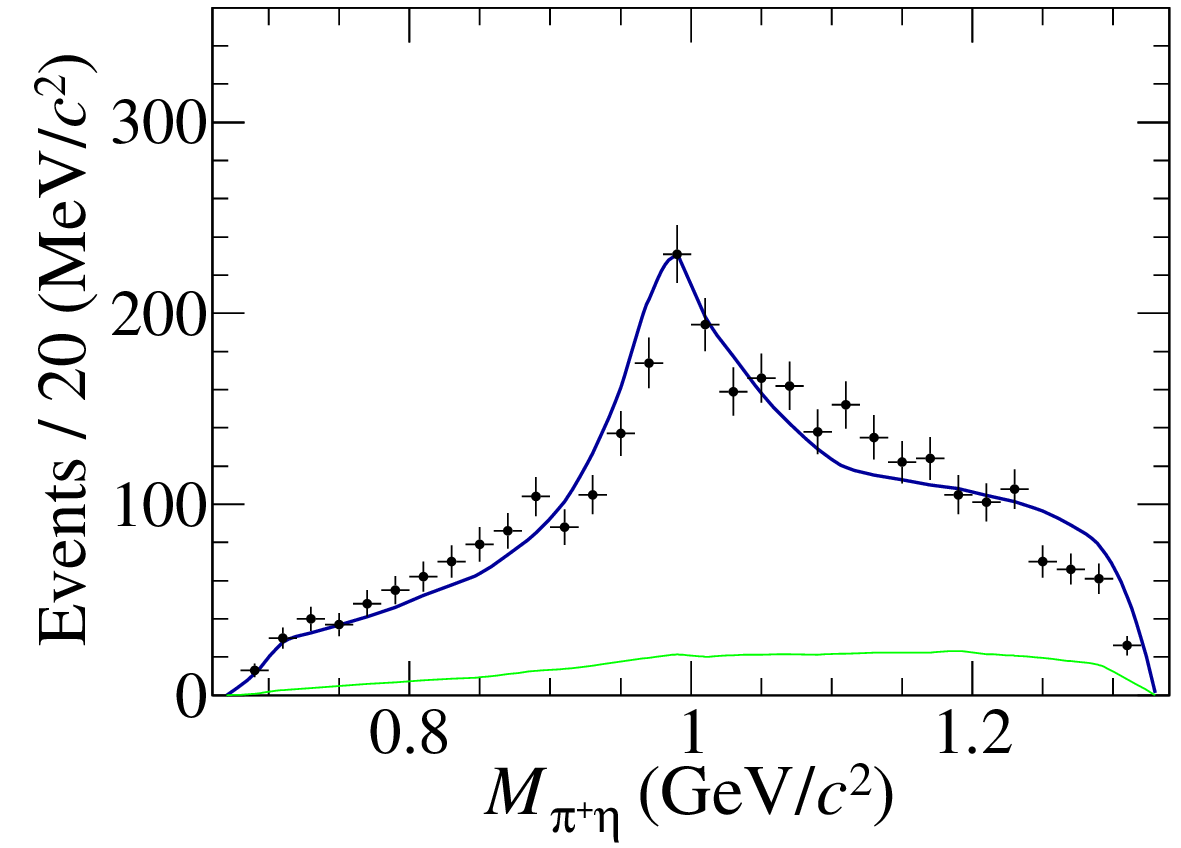}
\put(-75,65){s2}
%\put(-25,65){(c)}
\end{minipage}
\begin{minipage}[b]{0.23\textwidth}
\epsfig{width=0.98\textwidth,clip=true,file=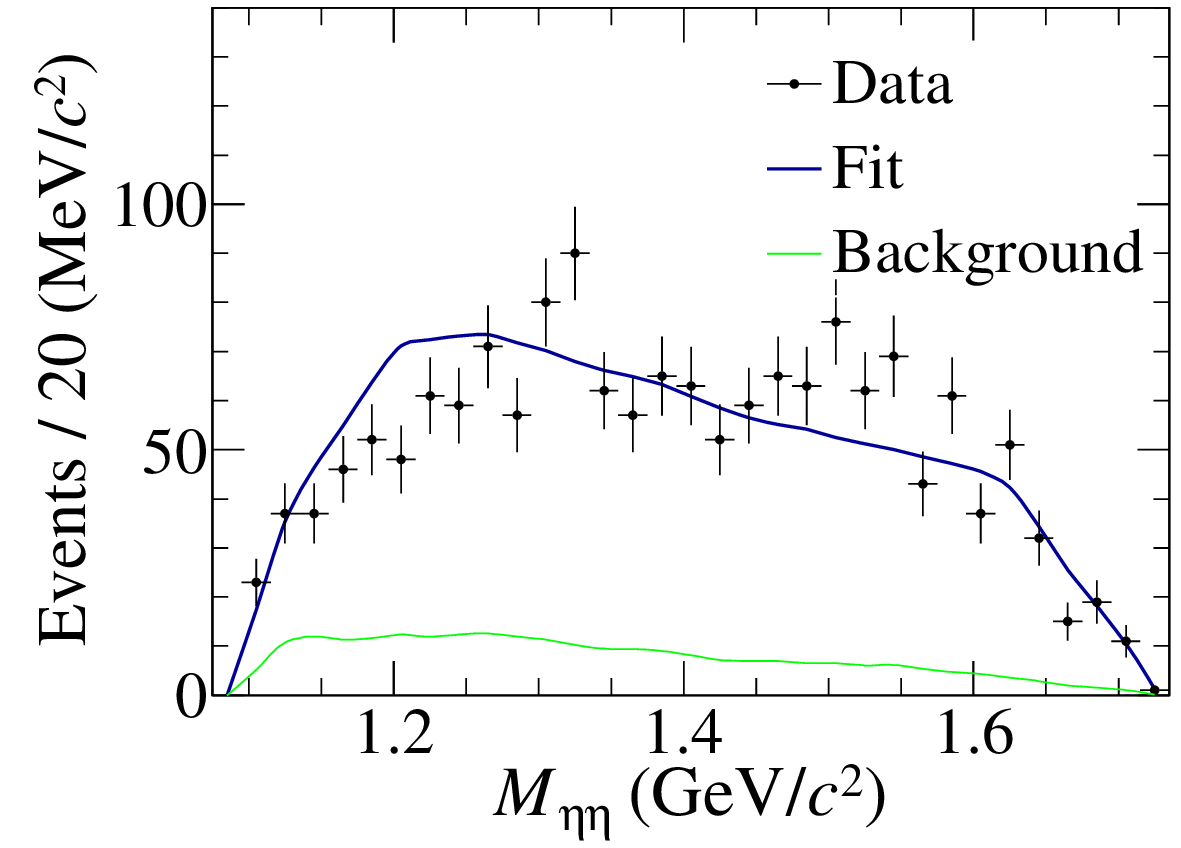}
%\put(-25,65){(d)}
\end{minipage}
\begin{minipage}[b]{0.23\textwidth}
\epsfig{width=0.98\textwidth,clip=true,file=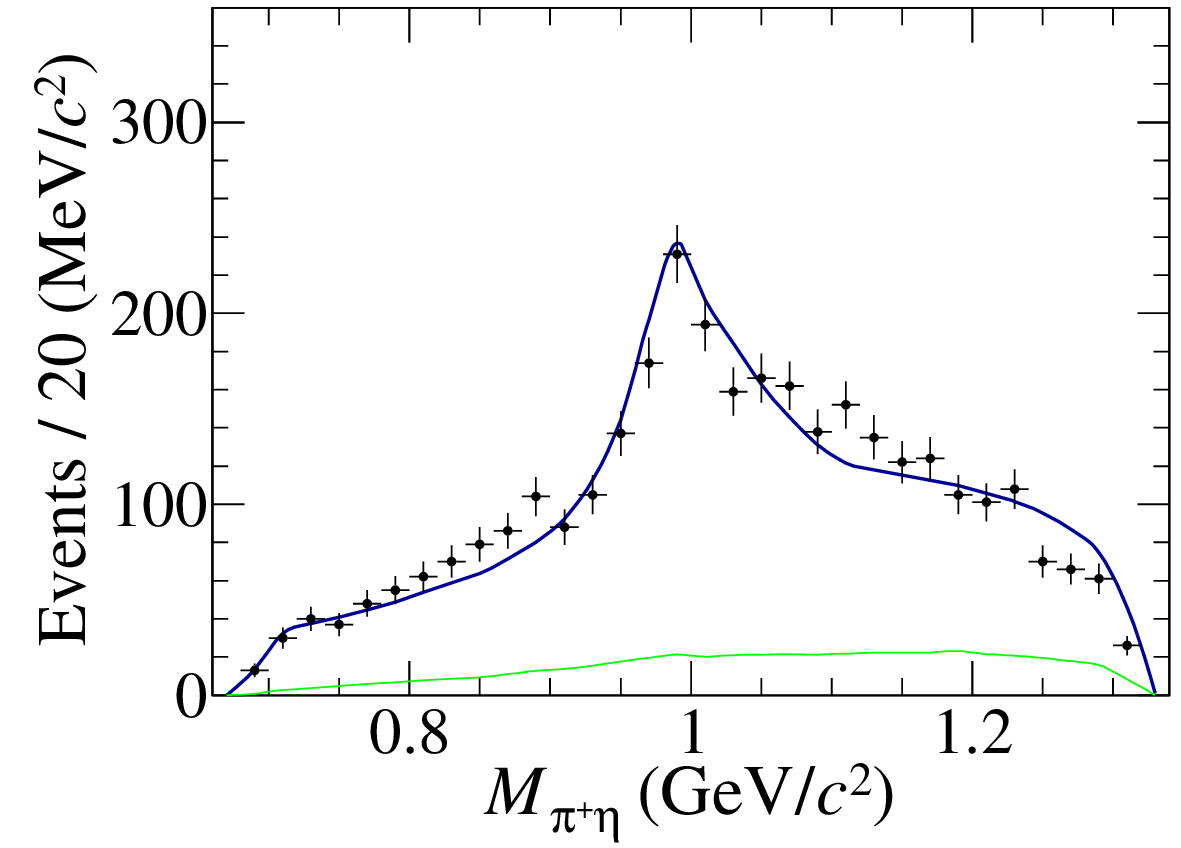}
\put(-75,65){s3}
%\put(-25,65){(e)}
\end{minipage}
\begin{minipage}[b]{0.23\textwidth}
\epsfig{width=0.98\textwidth,clip=true,file=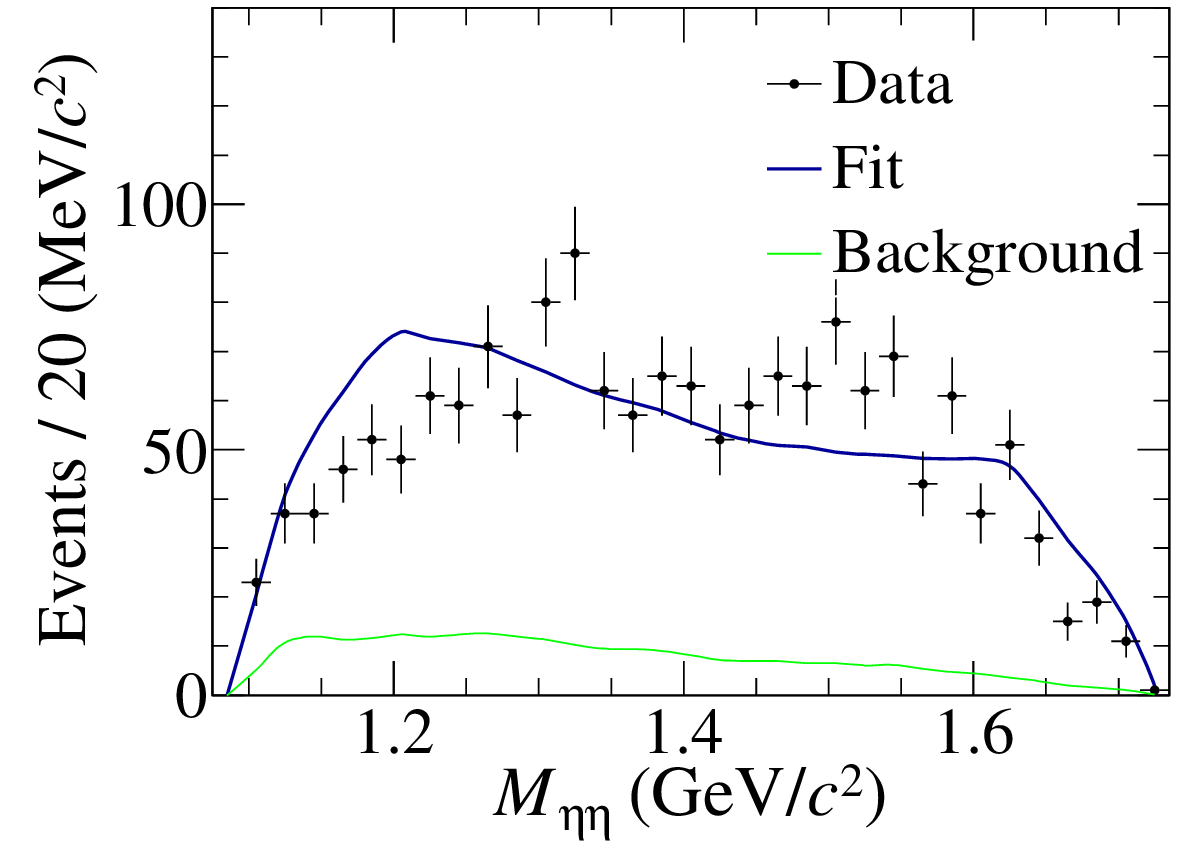}
%\put(-25,65){(f)}
\end{minipage}
\begin{minipage}[b]{0.23\textwidth}
\epsfig{width=0.98\textwidth,clip=true,file=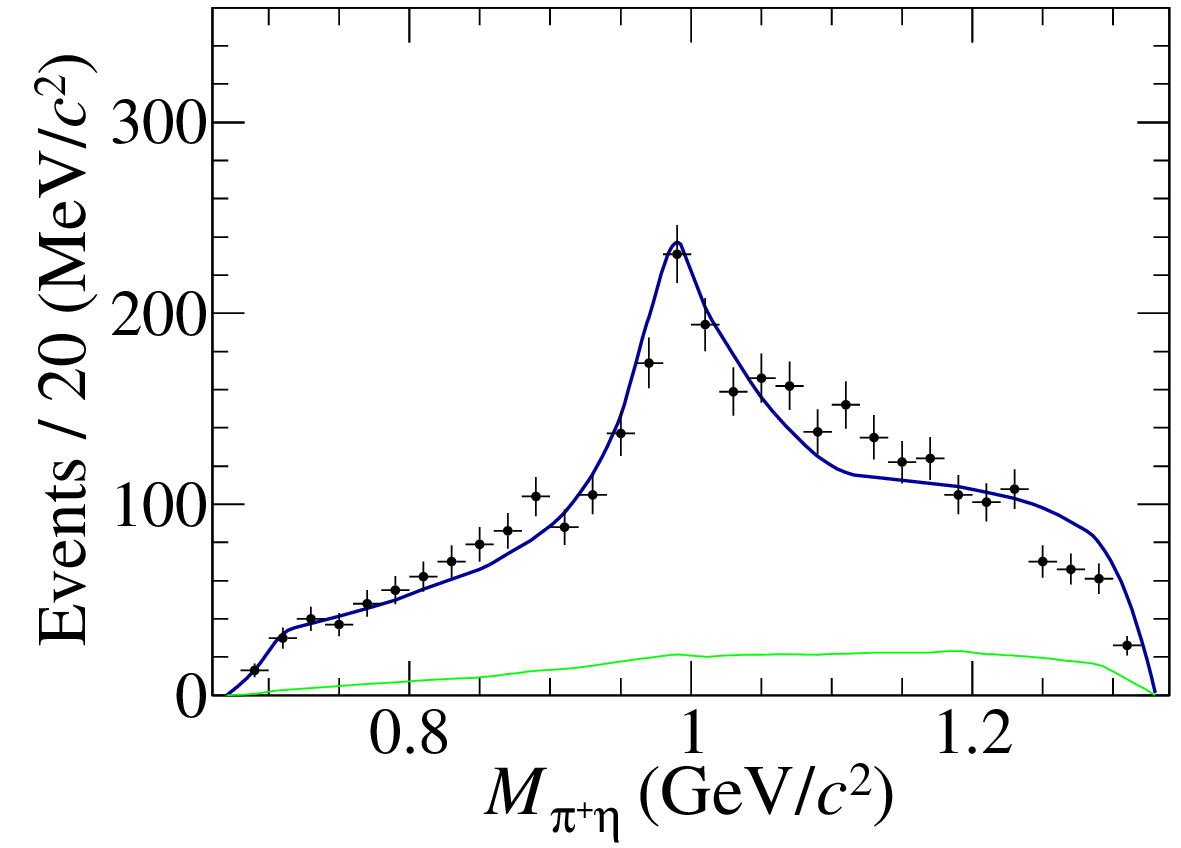}
\put(-75,65){s4}
%\put(-25,65){(g)}
\end{minipage}
\begin{minipage}[b]{0.23\textwidth}
\epsfig{width=0.98\textwidth,clip=true,file=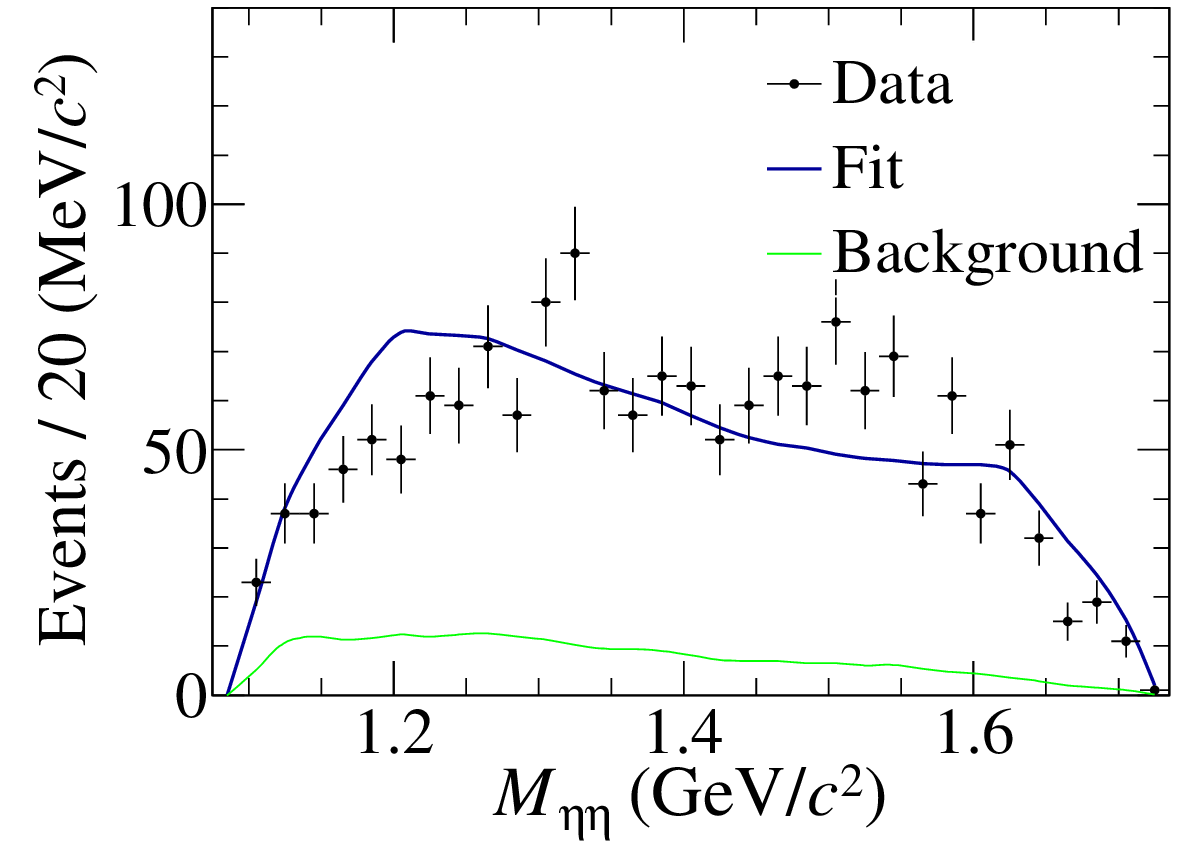}
%\put(-25,65){(h)}
\end{minipage}
\caption{The same as Fig.~\ref{fig:Flattebaseline}, but for the fit model using the $K$-matrix form to parameterize the $a_{0}(980)$ with the 
parameter sets (top) s1, (second row) s2, (third row) s3 and (bottom) s4. }
\label{fig:projectionKmatrixbaseline}
\end{center}
\end{figure}

The fitted production-vector parameters, together with the corresponding $\ln\mathcal{L}$ and $\chi^2/{\rm NDOF}$ values, are listed in Table~\ref{tab:Kmatrixbaseline}. 
The projections are shown in Fig.~\ref{fig:projectionKmatrixbaseline}. 

\begin{table}[htbp]
\begin{center}
\caption{Fit results for the $K$-matrix baseline models using the four parameter sets. 
The production-vector parameters $\beta_2$ and $f_1^{\rm prod}$ are determined from the fit.}
\resizebox{0.40\textwidth}{!}{%
\begin{tabular}{c|cc|cc} \hline 
Parameter set  & $\beta_{2}$   & $f_{1}^{\mathrm{prod}}$ & $\ln\mathcal{L}$  & $\chi^{2}/\mathrm{NDOF}$ \\ \hline
s1             &$-7.1\pm1.0$ & $0.63\pm0.13$         & 84.8     & $159.1/89$                 \\  
s2             &$-9.5\pm1.3$ & $0.47\pm0.10$         & 92.9     & $158.3/89$                 \\
s3             &$-9.9\pm1.3$ & $1.06\pm0.18$         & 66.6     & $185.2/89$                 \\
s4             &$-16.0\pm1.9$& $1.07\pm0.16$         & 75.0     & $180.8/89$                 \\
\hline
\end{tabular}
}
\label{tab:Kmatrixbaseline}
\end{center}
\end{table}

\begin{figure}[htbp]
\begin{center}
\begin{minipage}[b]{0.23\textwidth}
\epsfig{width=0.98\textwidth,clip=true,file=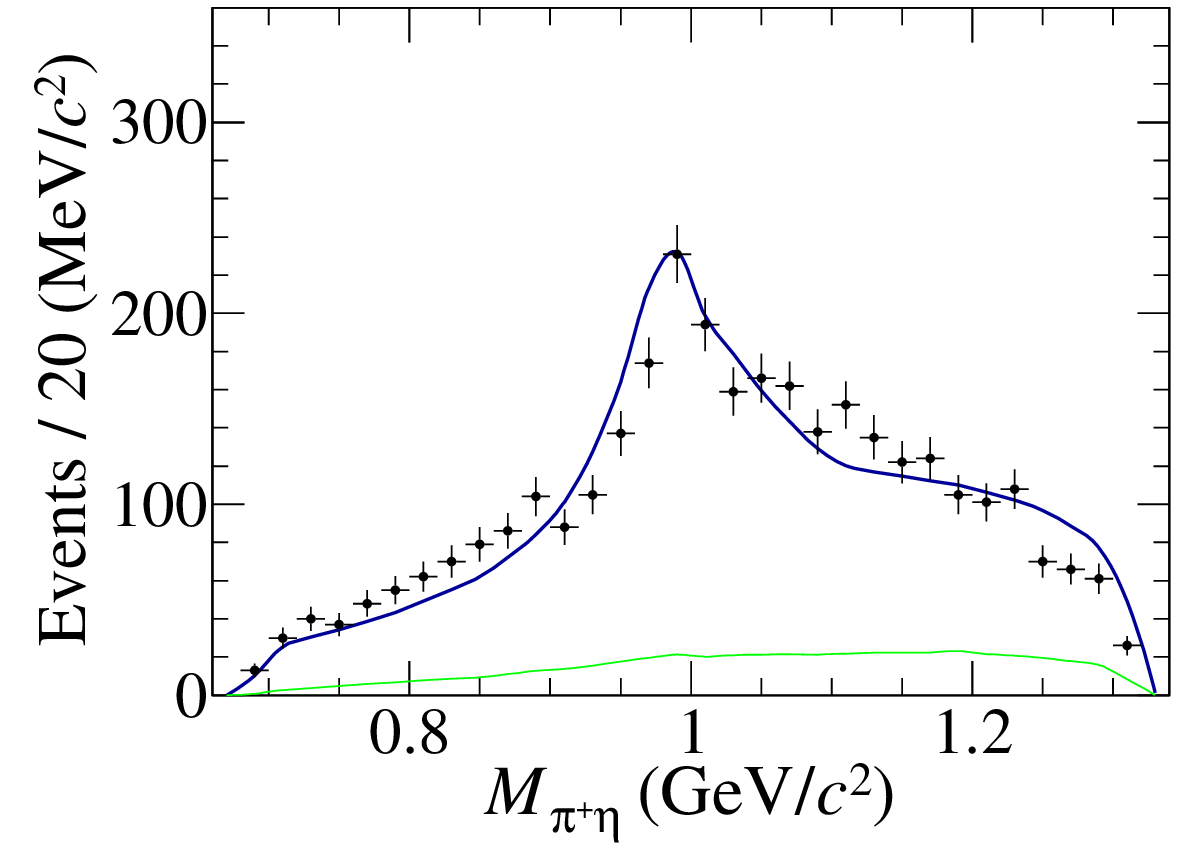}
\put(-75,65){as1}
%\put(-25,65){(a)}
\end{minipage}
\begin{minipage}[b]{0.23\textwidth}
\epsfig{width=0.98\textwidth,clip=true,file=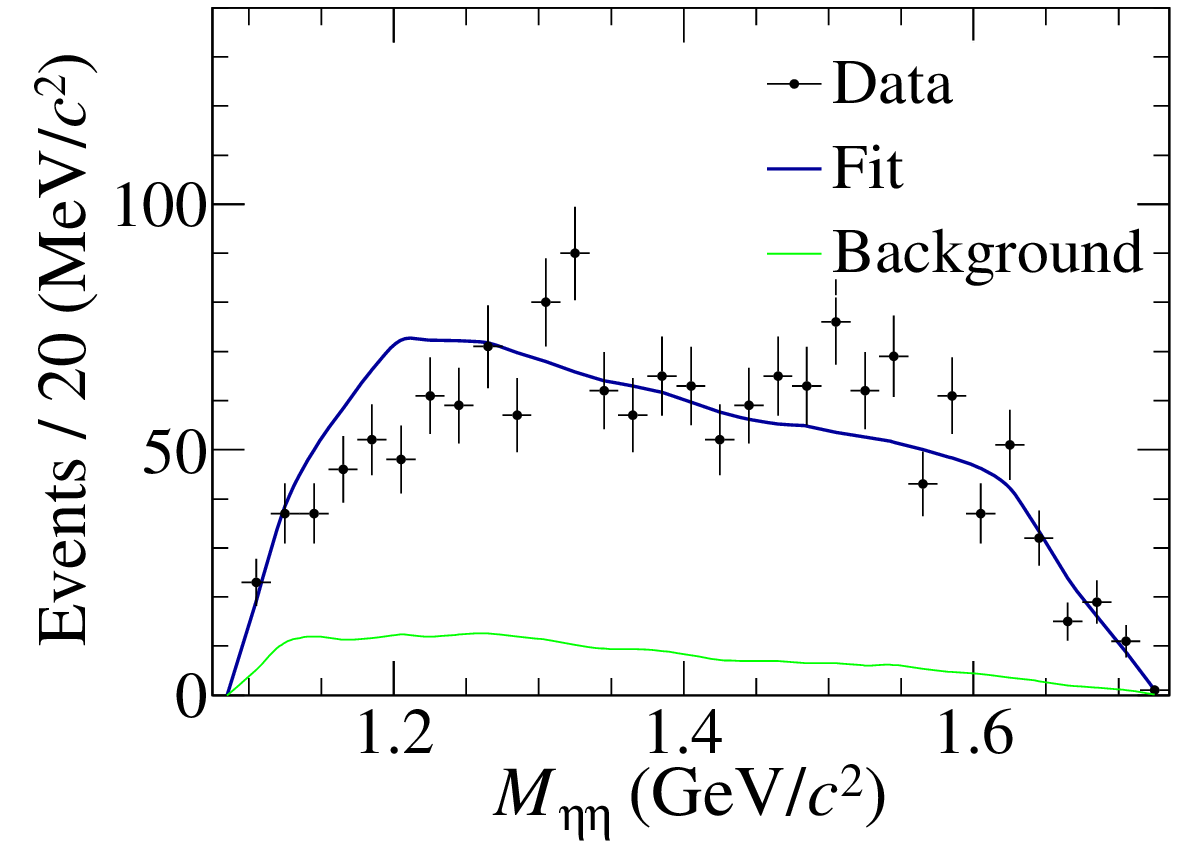}
%\put(-25,65){(b)}
\end{minipage}
\begin{minipage}[b]{0.23\textwidth}
\epsfig{width=0.98\textwidth,clip=true,file=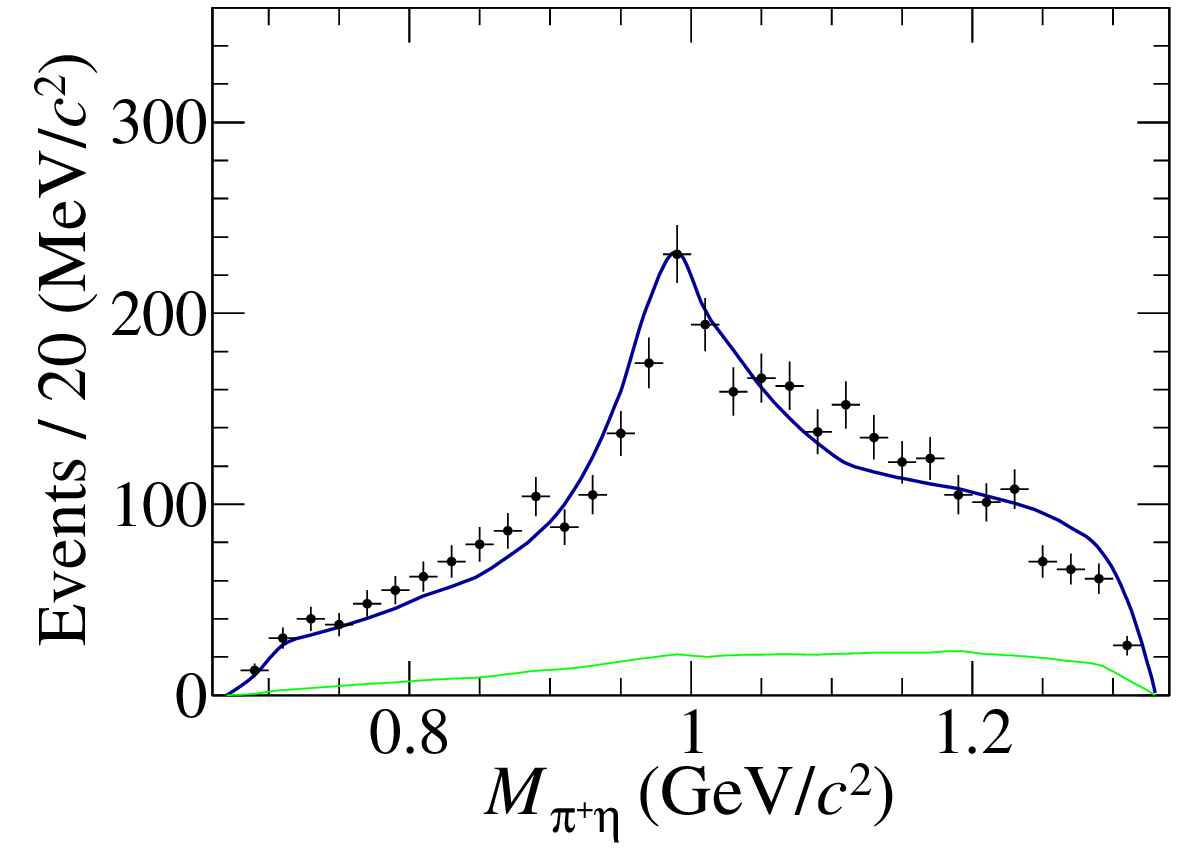}
\put(-75,65){as2}
%\put(-25,65){(c)}
\end{minipage}
\begin{minipage}[b]{0.23\textwidth}
\epsfig{width=0.98\textwidth,clip=true,file=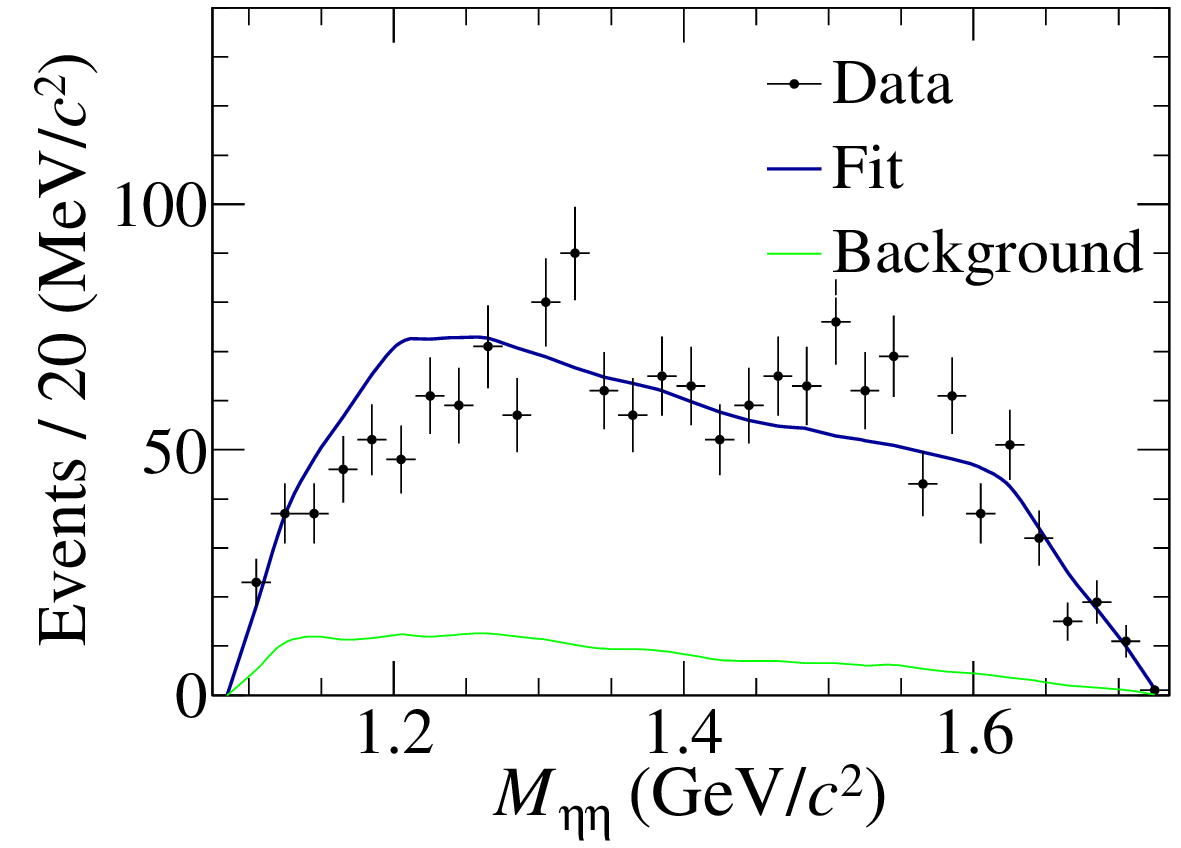}
%\put(-25,65){(d)}
\end{minipage}
\begin{minipage}[b]{0.23\textwidth}
\epsfig{width=0.98\textwidth,clip=true,file=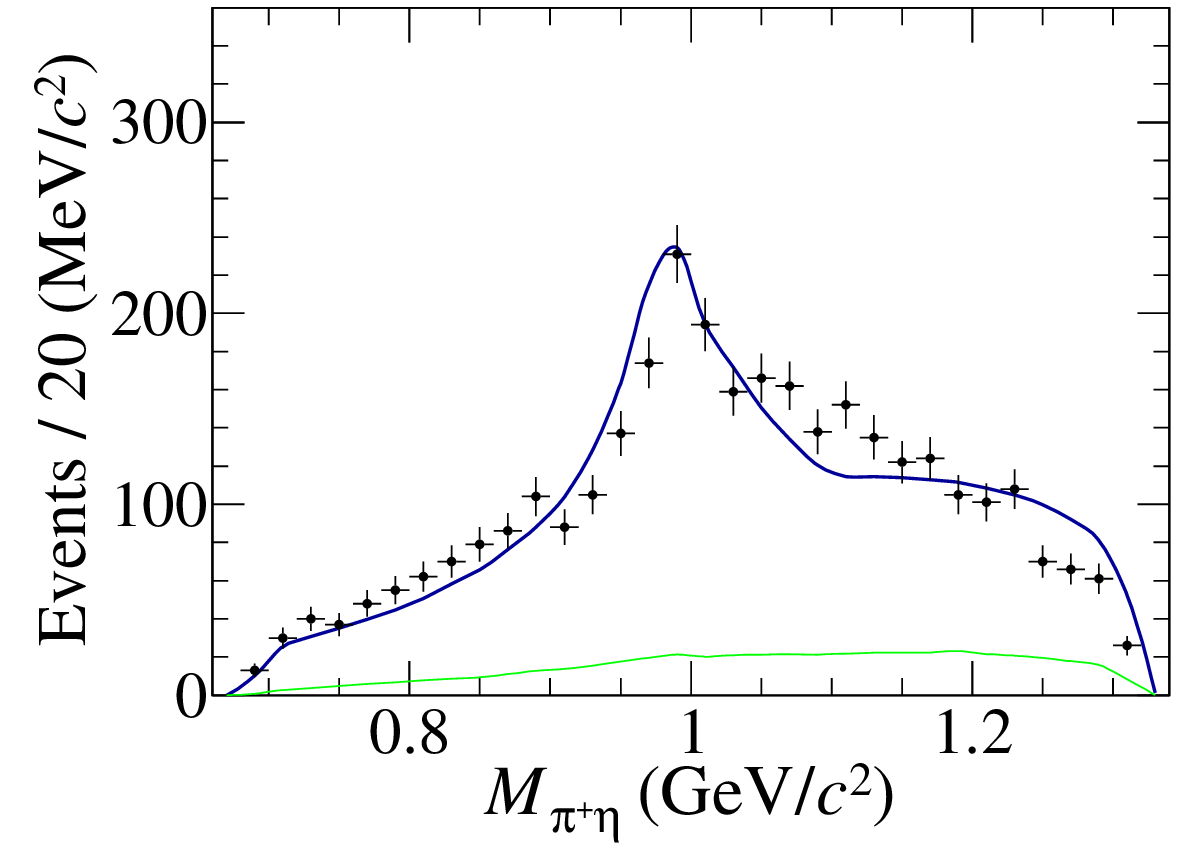}
\put(-75,65){as3}
%\put(-25,65){(e)}
\end{minipage}
\begin{minipage}[b]{0.23\textwidth}
\epsfig{width=0.98\textwidth,clip=true,file=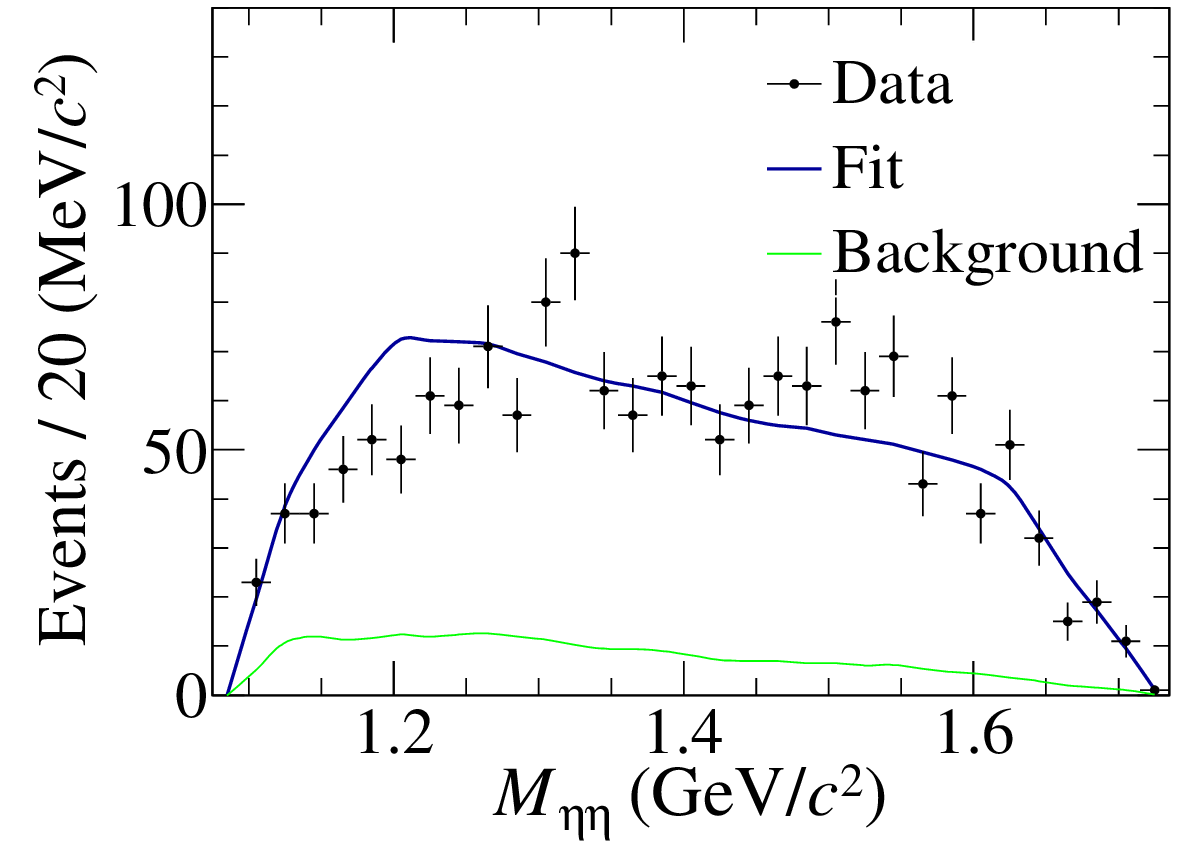}
%\put(-25,65){(f)}
\end{minipage}
\begin{minipage}[b]{0.23\textwidth}
\epsfig{width=0.98\textwidth,clip=true,file=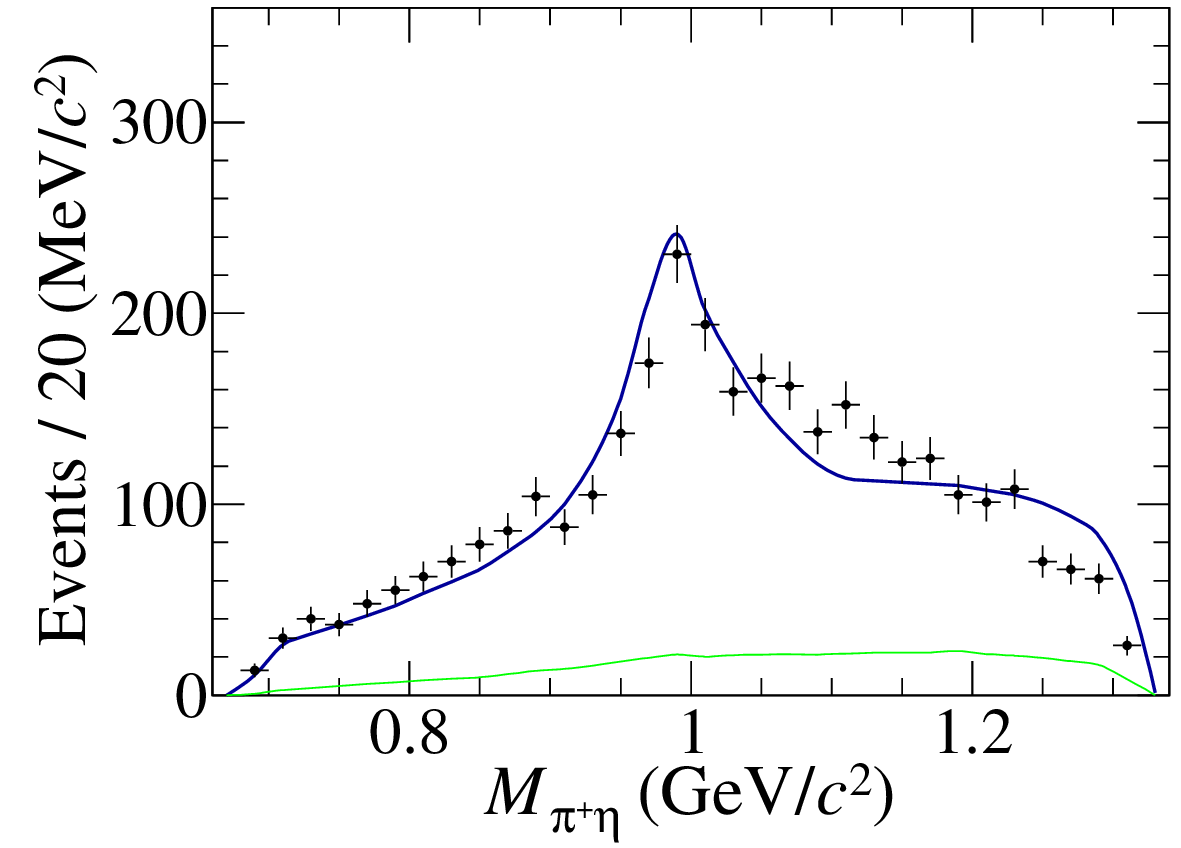}
\put(-75,65){as4}
%\put(-25,65){(g)}
\end{minipage}
\begin{minipage}[b]{0.23\textwidth}
\epsfig{width=0.98\textwidth,clip=true,file=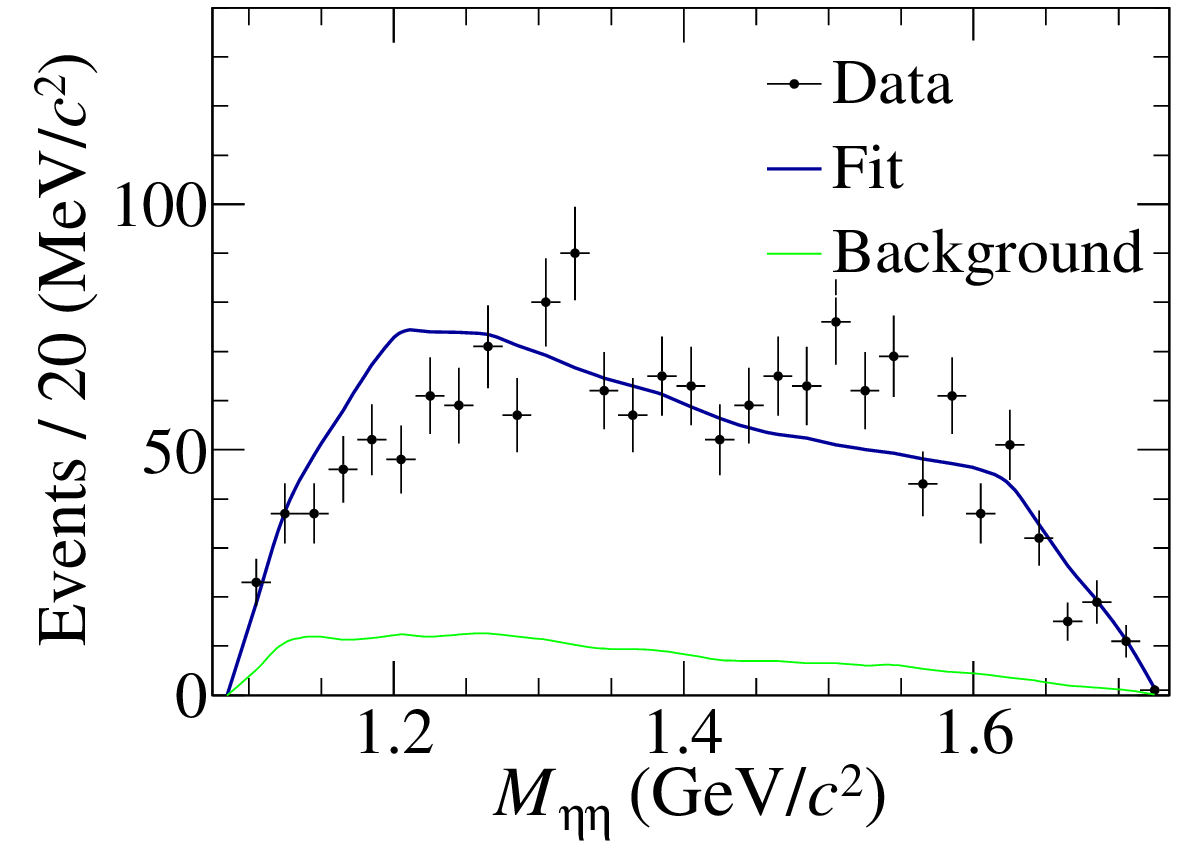}
\put(-25,65){(h)}
\end{minipage}
\caption{The same as Fig.~\ref{fig:projectionKmatrixbaseline}, but for the fit model with an altered $P$-vector background term. }
\label{fig:projectionalteredKmatrixbaseline}
\end{center}
\end{figure}

As an alternative description of the smooth production term, the form in Eq.~\ref{eq:Pvector_smooth} is replaced by the first-order polynomial form in Eq.~\ref{eq:polyPvector}. 
The fitted values of $\beta_2$, $c_{1,0}$, and $c_{1,1}$, together with the corresponding $\ln\mathcal{L}$ and $\chi^2/{\rm NDOF}$ values, are listed in Table~\ref{tab:alteredKmatrixbaseline}. 
The projections are shown in Fig.~\ref{fig:projectionalteredKmatrixbaseline}. The results given with the altered $P$-vector parameterization are denoted as 
``as1''$\sim$``as4''. 
This alternative form does not lead to a satisfactory improvement in the fit quality.

%%===============
\begin{table}[htbp]
\begin{center}
\caption{Fit results for the $K$-matrix baseline models with the smooth production term in the $P$ vector replaced by a first-order polynomial.}
\resizebox{0.5\textwidth}{!}{%
\begin{tabular}{c|ccc|cc} \hline 
Parameter set  & $\beta_{2}$    & $c_{1,0}$      & $c_{1,1}$      & $\ln\mathcal{L}$  & $\chi^{2}/\mathrm{NDOF}$ \\ \hline
s1             &$-10.4\pm2.5$ & $2.04\pm0.44$  &$-0.28\pm0.37$  & 83.5     & $164.3/88$               \\  
s2             &$-6.6\pm1.6$  & $1.44\pm0.32$  &$-1.05\pm0.27$  & 91.0     & $154.2/88$               \\
s3             &$-25.1\pm5.2$ & $4.38\pm0.85$  & $0.66\pm0.65$  & 77.9     & $179.4/88$               \\
s4             &$-18.5\pm2.8$ & $3.09\pm0.51$  &$-0.95\pm0.39$  & 74.3     & $183.1/88$               \\
\hline
\end{tabular}
}
\label{tab:alteredKmatrixbaseline}
\end{center}
\end{table}

\begin{figure}[htbp]
\begin{center}
\begin{minipage}[b]{0.23\textwidth}
\epsfig{width=0.98\textwidth,clip=true,file=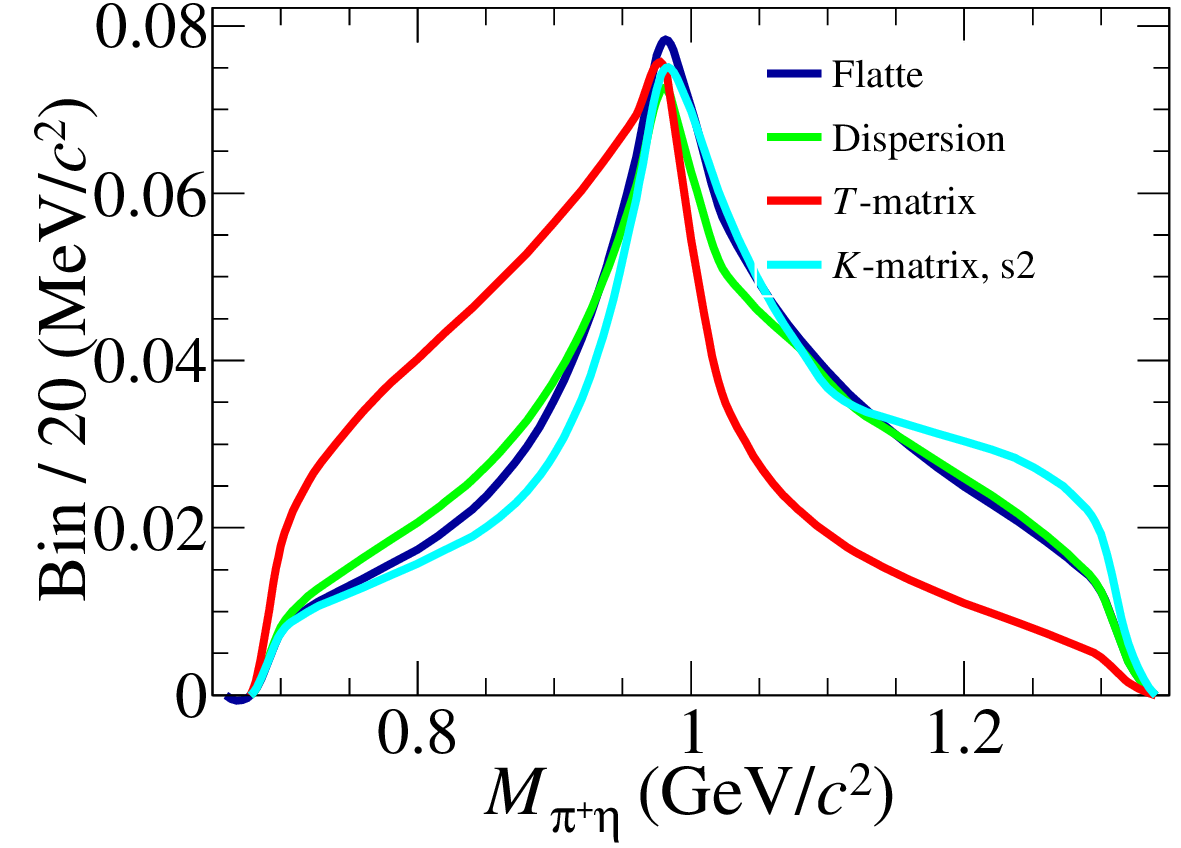}
\put(-85,65){(s2)}
\end{minipage}
\begin{minipage}[b]{0.23\textwidth}
\epsfig{width=0.98\textwidth,clip=true,file=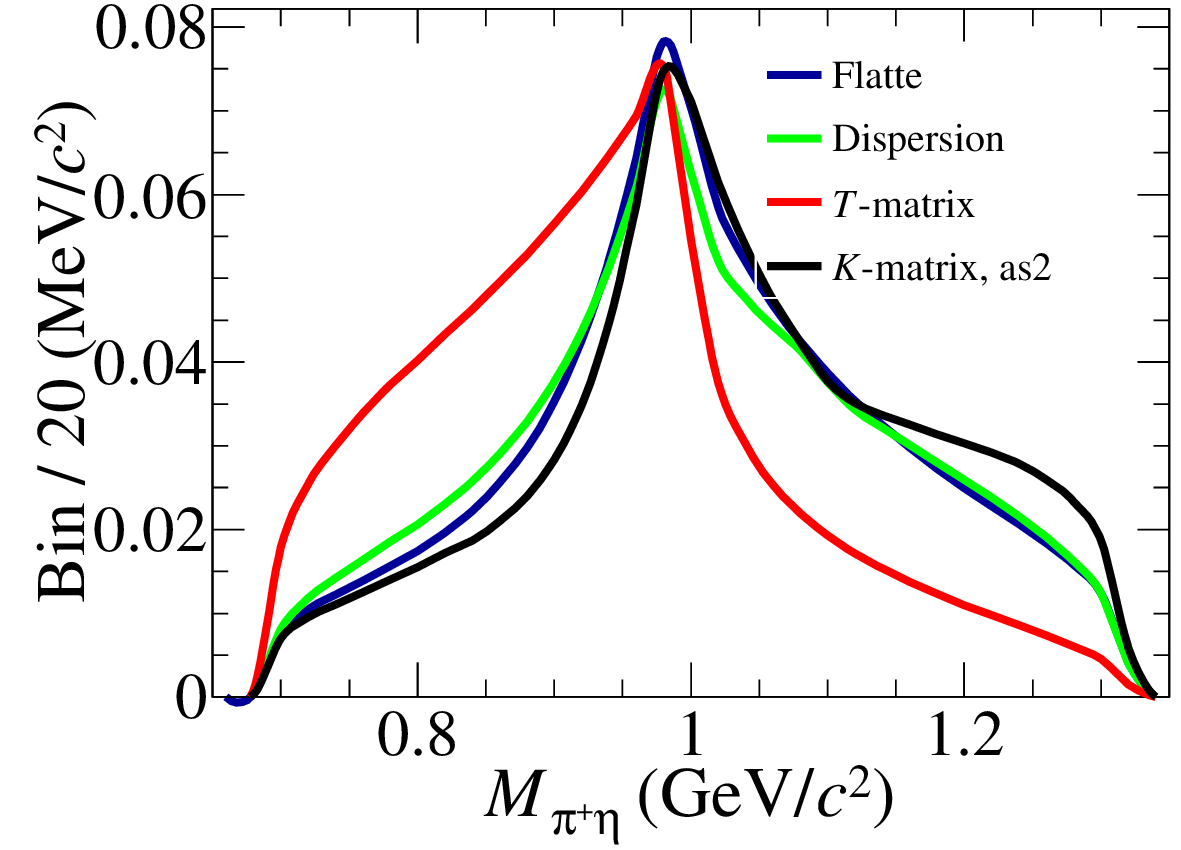}
\put(-85,65){(as2)}
\end{minipage}
\caption{Similar with Fig.~\ref{fig:a0lineshapegen}, normalized $\pi^+\eta$ mass distributions obtained with the $K$-matrix formula are added with (s2) cyan 
and (as2) black lines. }
\label{fig:a0lineshapegen2}
\end{center}
\end{figure}

% As the line-shapes displayed in Fig.~\ref{fig:a0lineshapegen}, the generated level $a_{0}(980)$ line-shape for s2 and as2 are 
% added and shown with cyan and black lines in Fig.~\ref{fig:a0lineshapegen2}, respectively. 
% In which, the $K$-matrix line shapes are found to be increased at the region of $M(\pi^{+}\eta)>1.1~\rm{GeV}/c^2$. For region near the peak position, 
% the two lines are similar with the Flatt\'e and dispersively modified Flatt\'e lines. 
Figure~\ref{fig:a0lineshapegen2} compares the normalized $\pi^{+}\eta$ mass distributions generated using the fitted $K$-matrix solutions s2 and as2 with the distributions shown in Fig.~\ref{fig:a0lineshapegen}.
The s2 and as2 distributions are shown by the cyan and black lines, respectively.
Compared with the Flatt\'e and dispersively modified Flatt\'e descriptions, the $K$-matrix distributions show an enhancement above $M(\pi^{+}\eta)=1.1~\mathrm{GeV}/c^2$, while their shapes near the peak are similar.

Among the baseline fits, the parameter set s2 gives the best fit quality and is therefore used for the subsequent tests with additional amplitudes. 
The additional amplitudes listed in Table~\ref{tab:addamp} are added individually to the $K$-matrix baseline model. 
The corresponding fit results are summarized in Table~\ref{tab:KMatrixaddamp}, and the projections are shown in Fig.~\ref{fig:KMatrixaddamp} of Appendix~\ref{app:addamp_projections}. 

\begin{table}[htbp]
\footnotesize
\begin{center}
\caption{Fit results for models obtained by adding one additional amplitude to the $K$-matrix baseline model with the parameter set s2. The $\Delta\ln\mathcal{L}$ and significance are evaluated relative to this baseline model.}
\begin{tabular}{c|cc|ccc} \hline 
Amplitude   & $\beta_{2}$     & $f_{1}^{\mathrm{prod}}$ & $\Delta\ln\mathcal{L}$  & $\chi^{2}/\mathrm{NDOF}$ & Significance\\ \hline
I           &$-14.0\pm1.9$  & $0.53\pm0.11$           & $+19.5$  & $137.4/87$               & 5.9            \\  
II          &$-12.0\pm1.6$  & $0.59\pm0.10$           & $+27.3$  & $135.7/87$               & 7.1            \\
III         &$-7.5\pm1.5$   & $0.01\pm0.12$           & $+39.0$  & $114.6/87$               & 8.6            \\
IV          &$-9.9\pm1.5$   & $0.56\pm0.11$           & $+11.5$  & $140.1/87$               & 4.4            \\
V           &$-7.7\pm1.5$   & $0.37\pm0.11$           & $+9.8$   & $146.5/87$               & 4.0            \\
VI          &$-8.6\pm1.8$   & $0.47\pm0.14$           & $+11.8$  & $145.3/87$               & 4.5            \\
VII         &$-8.6\pm2.0$   & $0.46\pm0.17$           & $+9.0$   & $155.2/87$               & 3.8            \\
VIII        &$-14.3\pm2.4$  & $0.98\pm0.19$           & $+13.7$  & $136.6/87$               & 4.9            \\
IX          &$-33.3\pm12.4$ & $0.30\pm0.24$           & $+49.8$  & $90.9/87$                & 9.6            \\
X           &$-26.2\pm4.1$  & $0.87\pm0.24$           & $+46.8$  & $98.1/87$                & 9.4            \\
XI          &$-9.0\pm1.8$   & $0.41\pm0.14$           & $+9.3$   & $141.2/87$               & 3.9            \\
XII         &$-16.3\pm4.1$  & $0.98\pm0.32$           & $+4.3$   & $151.7/87$               & 2.4            \\
\hline
\end{tabular}
\label{tab:KMatrixaddamp}
\end{center}
\end{table}

Similar to the behavior observed in the test set C, the models ``baseline+III'', ``baseline+IX'', and ``baseline+X'', corresponding to the additional $f_0(1710)\pi^+$ amplitude, the constant term, and the $(\pi^+\eta)_V\eta$ amplitude, respectively, exhibit large interference with the dominant $\pi\eta$ $S$-wave amplitude. 
These solutions are therefore not considered stable independent descriptions of the data. 
For the remaining amplitudes with large apparent significances, no corresponding clear structures are observed in the relevant invariant-mass projections. 
As in the previous test sets, these apparent improvements are mainly driven by correlations with the dominant $\pi\eta$ $S$-wave amplitude rather than by well-resolved additional components.

Therefore, the $K$-matrix formalism with the reference scattering parameters, even with alternative production-vector forms or additional conventional amplitudes, does not provide a satisfactory description of the data.

\subsection{Model-independent Amplitude Analysis for $(\eta\eta)_{S-{\rm wave}}$}
\label{sec:MIPWAres}

% Due to the data statistical limit, we construct the magnitude and phase interpolation curves using five complex coefficients 
% $c^{\rm MI}_{i}=\rho_{i}^{\rm MI}e^{i\phi_{i}^{\rm MI}}$, $i=0\sim 4$. 
% Here, all the magnitudes and phases are respected to the reference channel $D^{+} \to a_{0}(980)^+\eta$, $a_{0}(980)^+ \to \pi^+ \eta$. 
% The method is the same with the model-independent amplitude analysis for $(\pi^{+}\pi^{-})_{S-{\rm wave}}$ in $D_{s}^{+} \to \pi^{+}\pi^{+}\pi^{-}$~\cite{BESIII:2021jnf}. 
Due to the limited data sample, the physical $M(\eta\eta)$ range is divided into four equal-width intervals, and five complex coefficients are defined at the interval boundaries. The corresponding mass nodes are
$M_i=2m_{\eta}+i[(m_{D^+}-m_{\pi^+})-2m_{\eta}]/4$, $i=0,\ldots,4$, or approximately $M_i=(1.096,\,1.254,\,1.413,\,1.571,\,1.730)~{\rm GeV}/c^2$. 
At each node, the complex coefficient is parameterized as $c^{\rm MI}_{i}=\rho_{i}^{\rm MI}e^{i\phi_{i}^{\rm MI}}$, where the phases $\phi_i^{\rm MI}$ are expressed in radians. 
All magnitudes and phases are defined relative to the reference channel $D^{+} \to a_{0}(980)^+\eta$, $a_{0}(980)^+ \to \pi^+ \eta$. 
The interpolation procedure and endpoint treatment follow the model-independent amplitude-analysis method used for $(\pi^{+}\pi^{-})_{S-{\rm wave}}$ in $D_{s}^{+} \to \pi^{+}\pi^{+}\pi^{-}$~\cite{BESIII:2021jnf}.
Because the $T$-matrix fits are substantially worse than those obtained using the Flatt\'e and dispersively modified Flatt\'e parameterizations of $P_{a_{0}(980)}$, and because the $K$-matrix formalism contains undetermined $P$-vector parameters, the model-independent amplitude-analysis tests for $(\eta\eta)_{S-{\rm wave}}$ are performed only with the Flatt\'e and dispersively modified Flatt\'e descriptions of $P_{a_{0}(980)}$. 

\begin{figure}[htbp]
\begin{center}
\begin{minipage}[b]{0.23\textwidth}
\epsfig{width=0.98\textwidth,clip=true,file=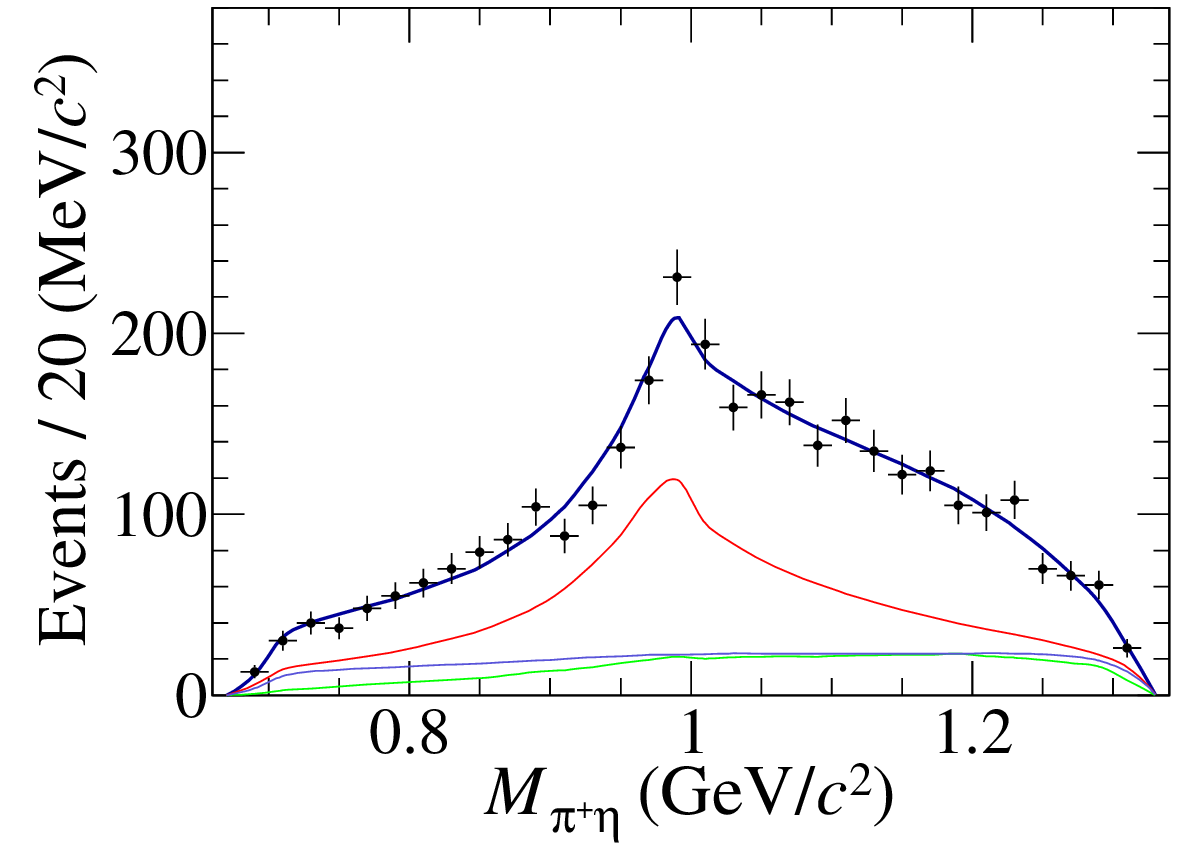}
\put(-25,65){(${\rm S}_{1}^{\rm F}$)}
\end{minipage}
\begin{minipage}[b]{0.23\textwidth}
\epsfig{width=0.98\textwidth,clip=true,file=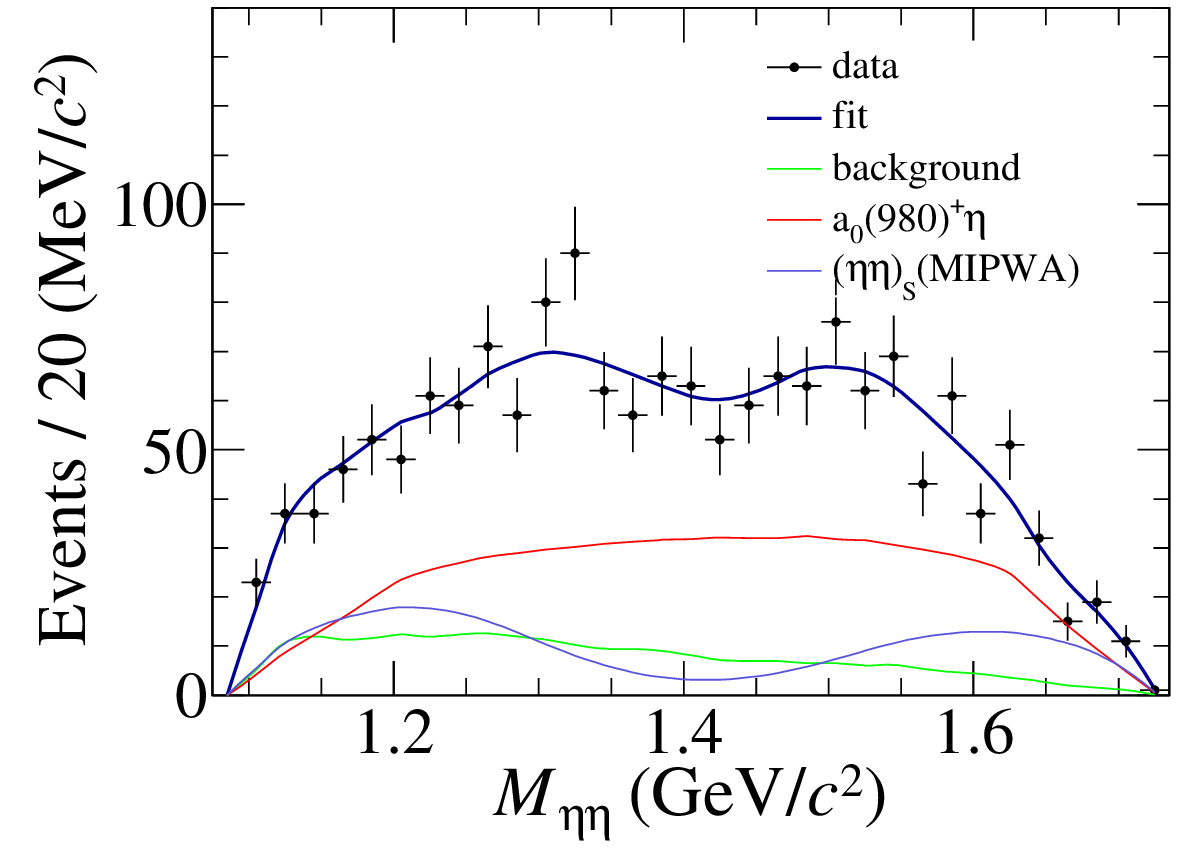}
%\put(-25,65){(b)}
\end{minipage}
%=================
\begin{minipage}[b]{0.23\textwidth}
\epsfig{width=0.98\textwidth,clip=true,file=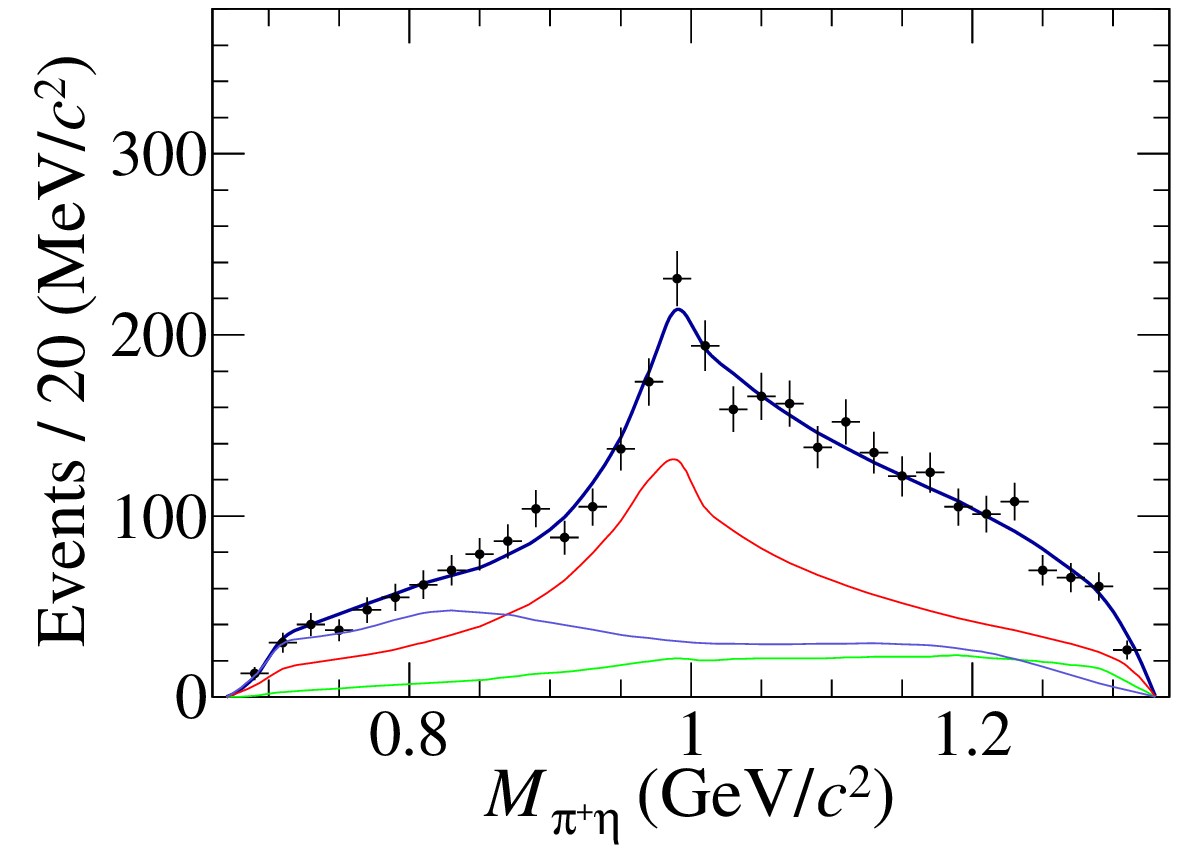}
\put(-25,65){(${\rm S}_{2}^{\rm F}$)}
\end{minipage}
\begin{minipage}[b]{0.23\textwidth}
\epsfig{width=0.98\textwidth,clip=true,file=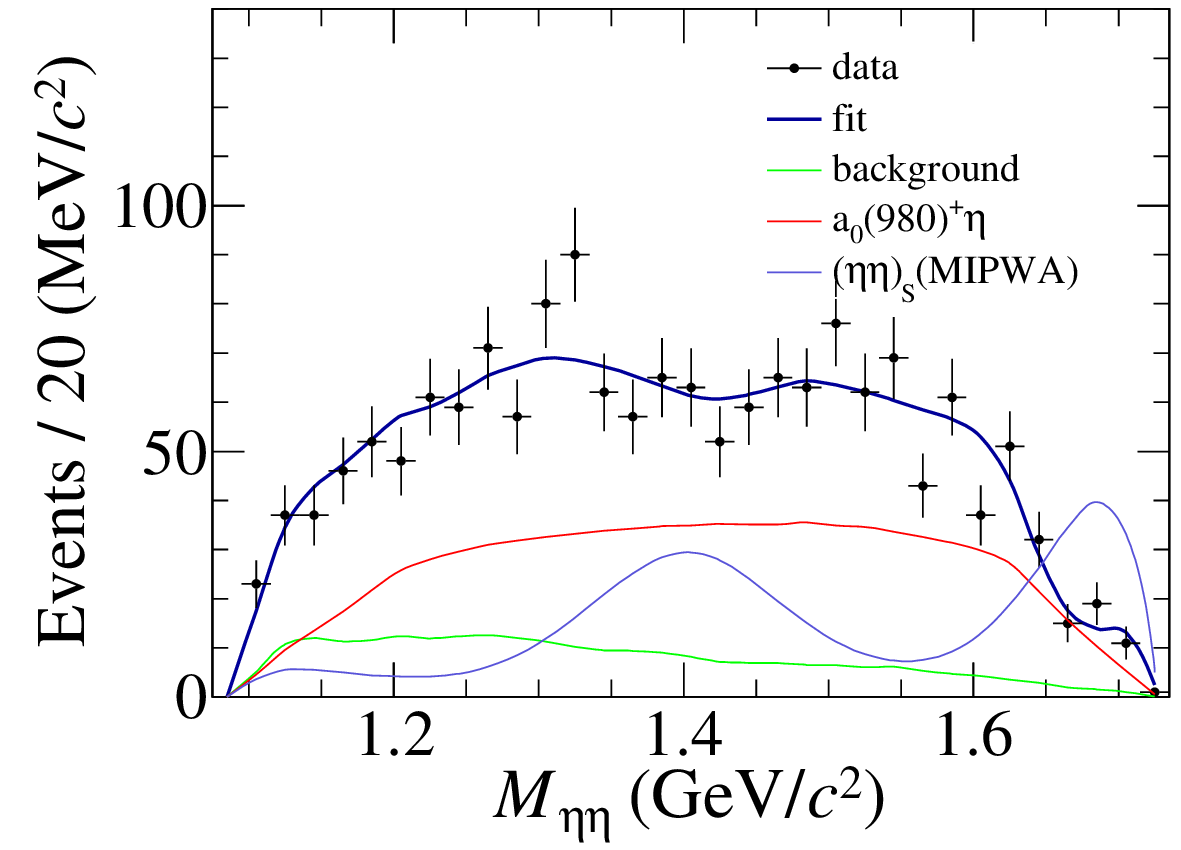}
%\put(-25,65){(b)}
\end{minipage}
%=================
\caption{The projections of (left) $M(\pi^{+}\eta)$ and (right) $M(\eta\eta)$ for the fits to the model that using 
a model-independent $(\eta\eta)_{S-{\rm wave}}$ instead of the constant term 
with (top) ${\rm S}^{\rm F}_1$ and (bottom) ${\rm S}^{\rm F}_2$. The Flatt\'e formula is used to parameterize the $P_{a_{0}(980)}$.
The black dots with error bars are the data; the blue, red and purple lines are the total fit, the intermediate state $a_{0}(980)$ and model-independent $(\eta\eta)_{S-{\rm wave}}$, respectively; and the green lines are the background. }
\label{fig:MIPWAFlatte}
\end{center}
\end{figure}

\begin{figure}[htbp]
\begin{center}
%=================
\begin{minipage}[b]{0.23\textwidth}
\epsfig{width=0.98\textwidth,clip=true,file=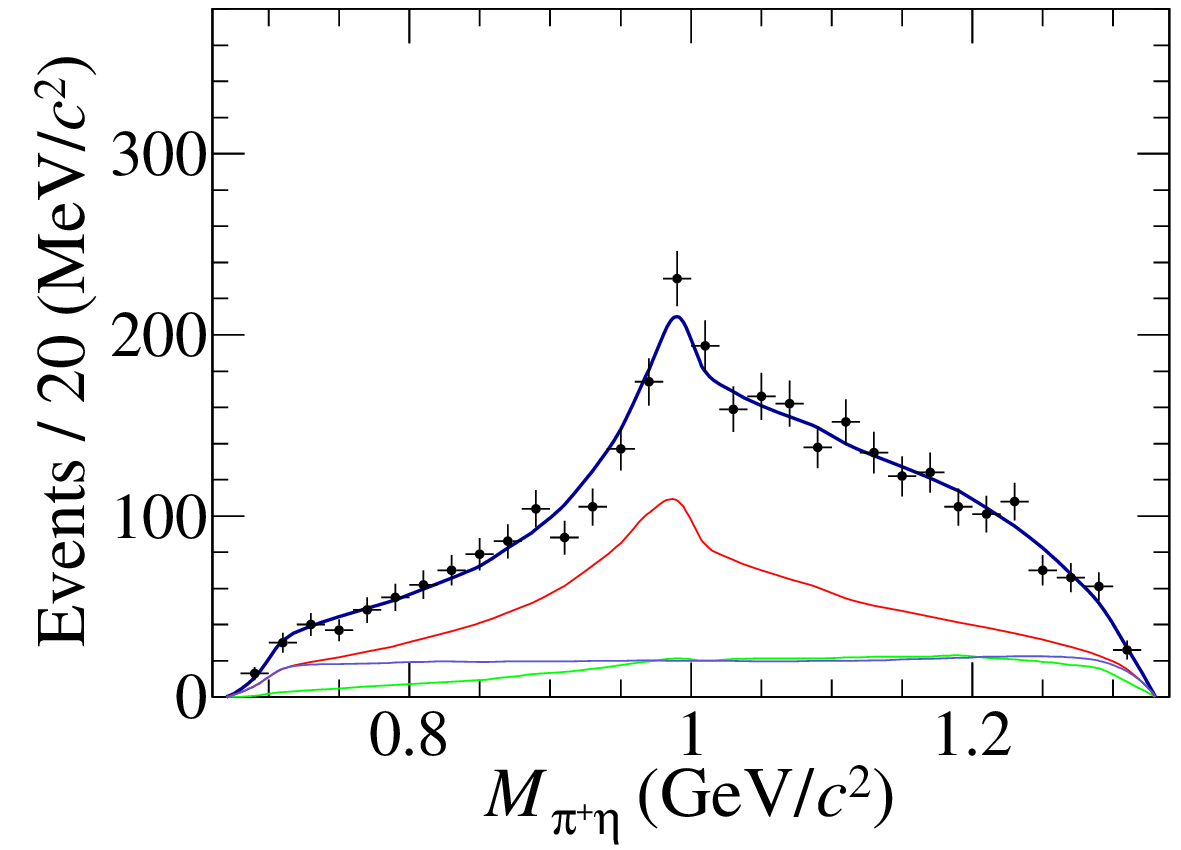}
\put(-25,65){(${\rm S}_{1}^{\rm D}$)}
\end{minipage}
\begin{minipage}[b]{0.23\textwidth}
\epsfig{width=0.98\textwidth,clip=true,file=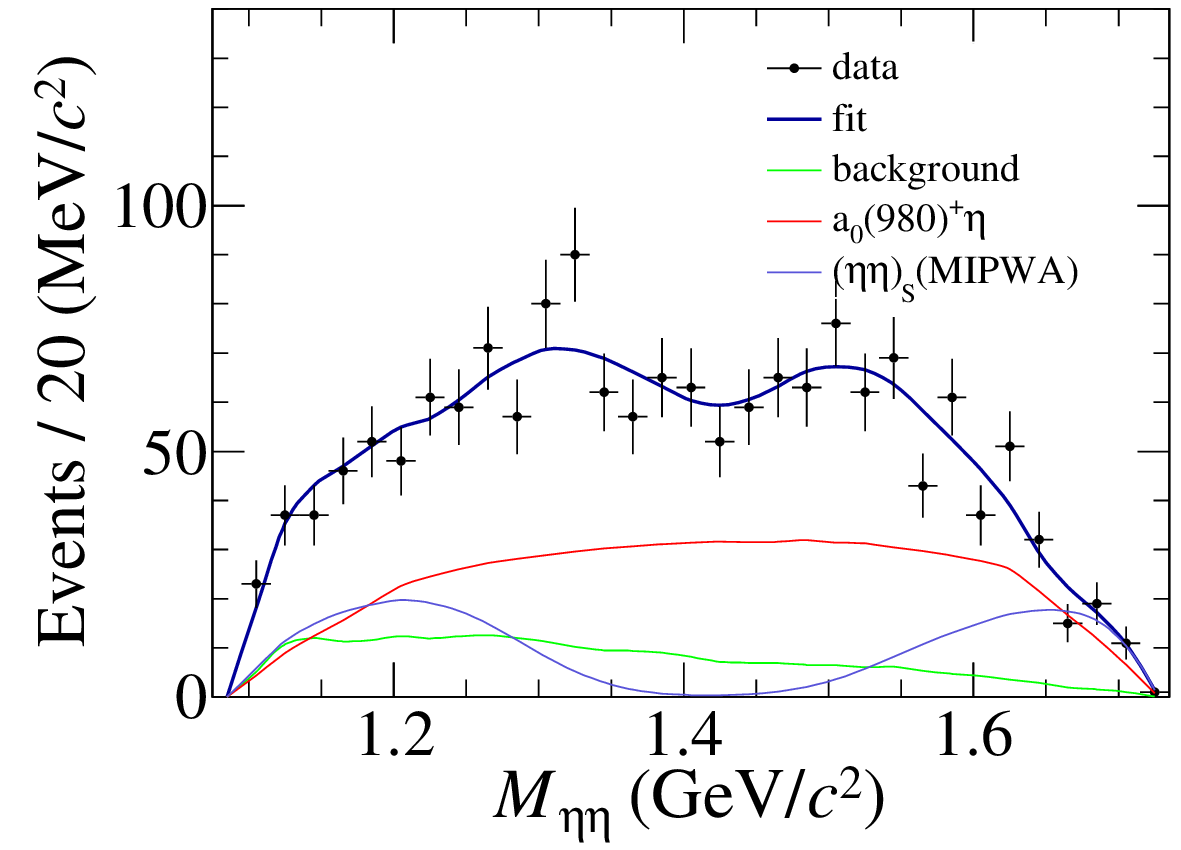}
%\put(-25,65){(b)}
\end{minipage}
\begin{minipage}[b]{0.23\textwidth}
\epsfig{width=0.98\textwidth,clip=true,file=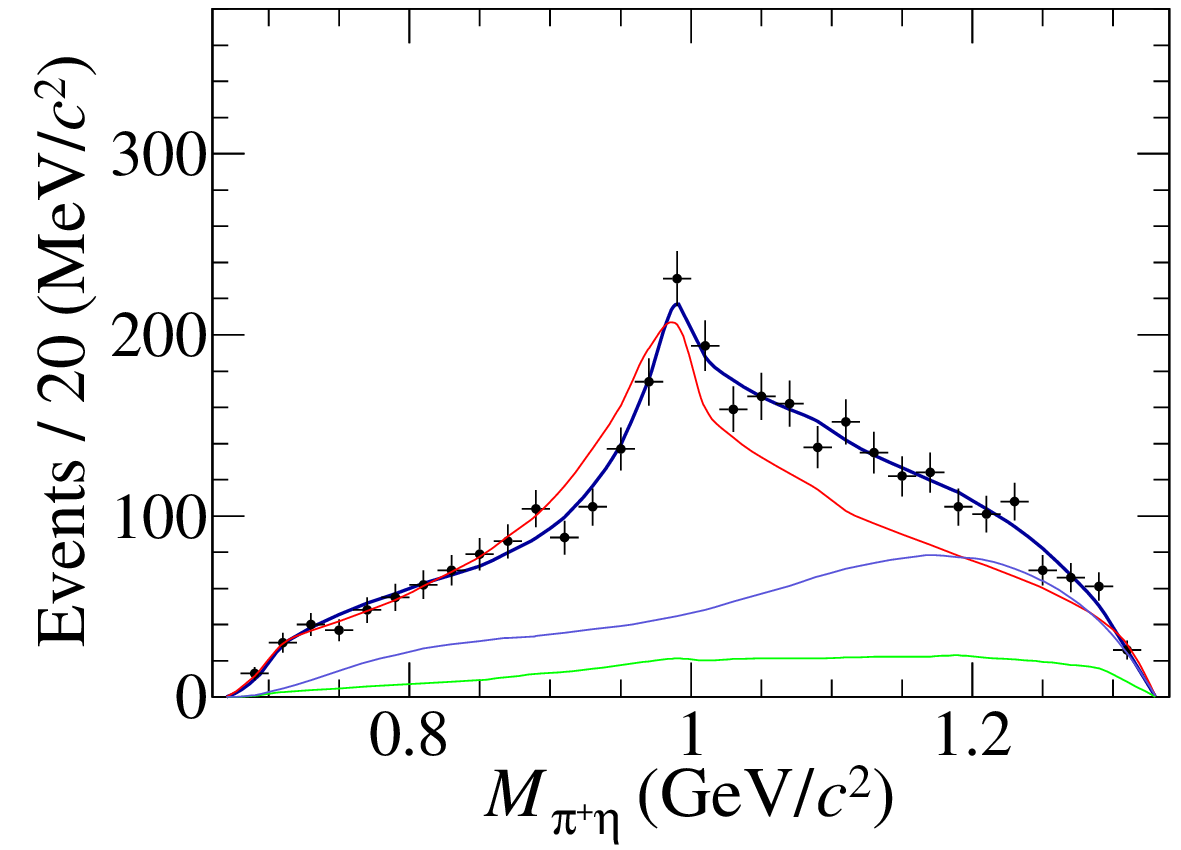}
\put(-25,65){(${\rm S}_{2}^{\rm D}$)}
\end{minipage}
\begin{minipage}[b]{0.23\textwidth}
\epsfig{width=0.98\textwidth,clip=true,file=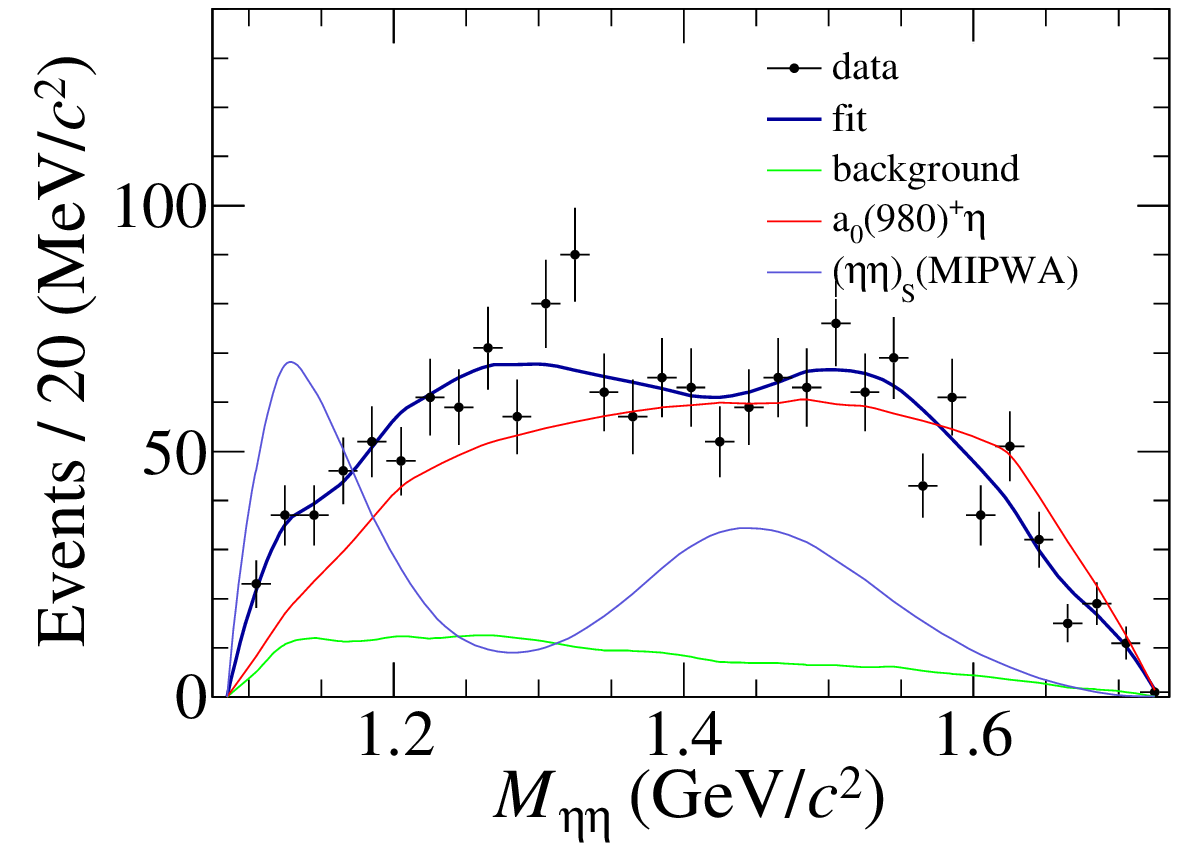}
%\put(-25,65){(b)}
\end{minipage}
%=================
\caption{The same as Fig.~\ref{fig:MIPWAFlatte}, but for the models using the dispersively modified Flatt\'e for the $P_{a_{0}(980)}$. }
\label{fig:MIPWAdispersive}
\end{center}
\end{figure}

Detailed results for the five complex coefficients, $\Delta\ln\mathcal{L}$, likelihood-based statistical significances, $\chi^{2}/{\rm NDOF}$, and 
the FFs for $D^+ \to (\eta\eta)_{S-{\rm wave}}\pi^{+}$ and $D^{+} \to a_{0}(980)^{+}\eta$ 
are listed in Tables~\ref{tab:resMIPWAFlatte} and~\ref{tab:resMIPWAdispersive}. 
The projections onto $M(\pi^{+}\eta)$ and $M(\eta\eta)$ are shown in Figs.~\ref{fig:MIPWAFlatte} and~\ref{fig:MIPWAdispersive}. 
Two solutions are found for each parameterization and are denoted by ${\rm S}_{1,2}^{\rm F}$ and ${\rm S}_{1,2}^{\rm D}$. 
The superscripts ${\rm F}$ and ${\rm D}$ indicate the Flatt\'e and dispersively modified Flatt\'e parameterizations of $P_{a_{0}(980)}$, respectively. 
% Since the fits for $T$-matrix formalism are much worse than the Flatt\'e and dispersively modified Flatt\'e $P_{a_{0}(980)}$, 
% and there are unknown $P$-vector parameters for $K$-matrix formalism, the tests for model-independent amplitude analysis for $(\eta\eta)_{S-{\rm wave}}$
% are performed for Flatt\'e and dispersively modified Flatt\'e $P_{a_{0}(980)}$. 

% The details results for the five complex coefficients, $\Delta\ln\mathcal{L}$, likelihood based statistical significance, $\chi^{2}/{\rm NDOF}$, and 
% the FFs for $D^+ \to (\eta\eta)_{S-{\rm wave}}\pi^{+}$ and $D^{+} \to a_{0}(980)^{+}\eta$ 
% are listed in Tab.~\ref{tab:resMIPWAFlatte} and~\ref{tab:resMIPWAdispersive}. 
% The projections on $M(\pi^{+}\eta)$ and $M(\eta\eta)$ are shown in Fig.~\ref{fig:MIPWAFlatte} and~\ref{fig:MIPWAdispersive}. 
% Here, two solutions are found for both models, which are denote as ${\rm S}_{1(2)}^{\rm F(D)}$. 
% The superscript represents the $P_{a_{0}(980)}$ is parameterized with Flatt\'e or dispersively modified Flatt\'e formalism. 
%, while for two of them, 
%the resulted $(\eta\eta)_{S-{\rm wave}}$ line shapes are not consistent and therefore rejected. 

\begin{table}[htbp]
%\footnotesize
\begin{center}
\caption{Fit results for the model-independent amplitude analysis for $(\eta\eta)_{S-{\rm wave}}$. 
The $P_{a_{0}(980)}$ is parameterized with the Flatt\'e formula. 
The uncertainties for the five magnitudes and phases, and the fit fractions are statistical only.
% The five complex coefficients correspond to ten free real parameters.
The five complex coefficients correspond to ten free real parameters. The $\Delta\ln\mathcal{L}$ and significance are evaluated relative to the Flatt\'e baseline model.}
\begin{tabular}{c|cc} \hline 
$~$                      & ${\rm S}^{\rm F}_1$&  ${\rm S}^{\rm F}_2$ \\ \hline
$\rho_{0}^{\rm MI}$      & $3.64\pm1.85$      &  $3.00\pm1.20$                  \\
$\rho_{1}^{\rm MI}$      & $3.08\pm0.70$      &  $1.66\pm0.55$                  \\
$\rho_{2}^{\rm MI}$      & $1.28\pm0.48$      &  $3.76\pm0.50$                  \\
$\rho_{3}^{\rm MI}$      & $2.85\pm0.57$      &  $2.16\pm0.53$                  \\
$\rho_{4}^{\rm MI}$      & $3.14\pm3.03$      &  $9.05\pm0.83$                  \\
$\phi_{0}^{\rm MI}$      & $3.57\pm0.68$      &  $3.46\pm0.80$                  \\
$\phi_{1}^{\rm MI}$      & $3.19\pm0.18$      &  $2.62\pm0.29$                  \\
$\phi_{2}^{\rm MI}$      & $1.57\pm0.46$      &  $3.21\pm0.17$                  \\
$\phi_{3}^{\rm MI}$      & $2.26\pm0.19$      &  $2.10\pm0.24$                  \\
$\phi_{4}^{\rm MI}$      & $1.76\pm0.81$      &  $3.72\pm0.39$                  \\
$\Delta\ln\mathcal{L}$   & $+41.6$            &  $+43.8$                        \\
Sig.~($\sigma$)          & $7.4$              &  $7.7$                          \\
$\chi^{2}/{\rm NDOF}$    & $89.6/81$          &  $86.6/81$                      \\
$FF(a_{0}(980)^+)$ (\%)  & $55.0\pm7.9$       &  $60.2\pm8.4$                   \\
$FF[(\eta\eta)_{S-{\rm wave}}]$ (\%) & $22.4\pm7.7$ & $35.9\pm4.9$               \\
\hline
\end{tabular}
\label{tab:resMIPWAFlatte}
\end{center}
\end{table}

\begin{table}[htbp]
\footnotesize
\begin{center}
\caption{Fit results for the model-independent amplitude analysis for $(\eta\eta)_{S-{\rm wave}}$. 
The $P_{a_{0}(980)}$ is parameterized with the dispersively modified Flatt\'e formula. 
The uncertainties for the five magnitudes and phases, and the fit fractions are statistical only.
% The five complex coefficients correspond to ten free real parameters.
The five complex coefficients correspond to ten free real parameters. The $\Delta\ln\mathcal{L}$ and significance are evaluated relative to the dispersive baseline model.}
\begin{tabular}{c|cc} \hline 
$~$                      & ${\rm S}^{\rm D}_1$&  ${\rm S}^{\rm D}_2$   \\ \hline
$\rho_{0}^{\rm MI}$      & $4.53\pm3.07$      & $10.78\pm1.38$         \\
$\rho_{1}^{\rm MI}$      & $3.81\pm0.92$      & $2.38\pm0.94$          \\
$\rho_{2}^{\rm MI}$      & $0.51\pm0.66$      & $3.87\pm0.67$          \\
$\rho_{3}^{\rm MI}$      & $3.34\pm0.72$      & $2.99\pm0.59$          \\
$\rho_{4}^{\rm MI}$      & $5.83\pm4.71$      & $0.32\pm0.71$          \\
$\phi_{0}^{\rm MI}$      & $3.79\pm0.62$      & $5.25\pm0.35$          \\
$\phi_{1}^{\rm MI}$      & $3.48\pm0.15$      & $3.48\pm0.15$          \\
$\phi_{2}^{\rm MI}$      & $2.09\pm0.39$      & $3.46\pm0.14$          \\
$\phi_{3}^{\rm MI}$      & $2.64\pm0.15$      & $2.89\pm0.17$          \\
$\phi_{4}^{\rm MI}$      & $2.45\pm0.63$      & $3.22\pm0.57$          \\
$\Delta\ln\mathcal{L}$   & $+35.1$            & $+37.4$                \\
Sig.~($\sigma$)          & $6.6$              & $6.9$                  \\
$\chi^{2}/{\rm NDOF}$    & $91.0/81$          & $81.6/81$              \\
$FF(a_{0}(980)^+)$ (\%)  & $79.2\pm9.7$       & $103.6\pm14.3$         \\
$FF[(\eta\eta)_{S-{\rm wave}}](\%)$ & $28.5\pm11.2$& $51.2\pm9.3$           \\
\hline
\end{tabular}
\label{tab:resMIPWAdispersive}
\end{center}
\end{table}

%From $M(\eta\eta)$ projections, the structure given by the model-independent $(\eta\eta)_{S-{\rm wave}}$ line shapes 
%from solutions ${\rm S}^{\rm F}_2$ and ${\rm S}^{\rm D}_2$ 
%are not consistent with other solutions and the $|S|^{2}$ shape shown in Fig.~\ref{fig:metaetapw}. 
%Therefore, they are firstly rejected.

\begin{figure}[htbp]
\begin{center}
\begin{minipage}[b]{0.23\textwidth}
\epsfig{width=0.98\textwidth,clip=true,file=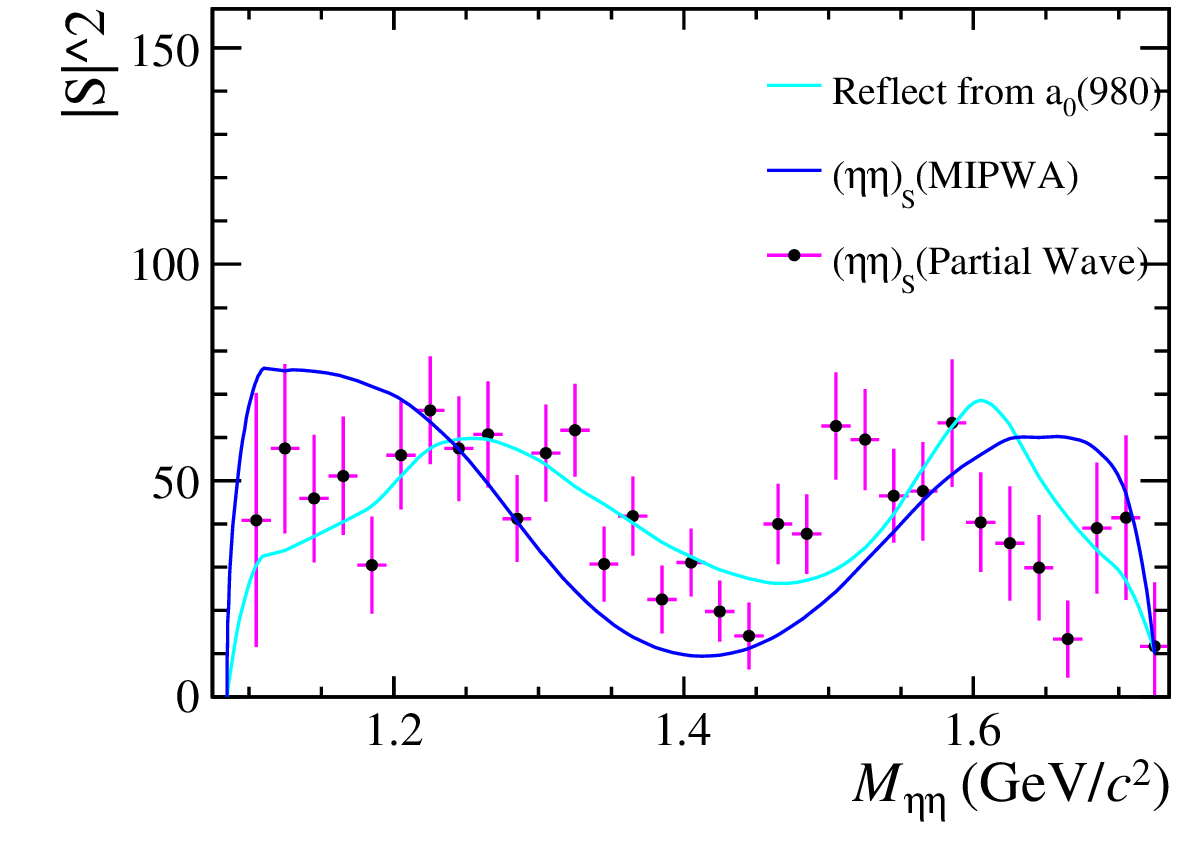}
\put(-85,65){(${\rm S}_{1}^{\rm F}$)}
\end{minipage}
\begin{minipage}[b]{0.23\textwidth}
\epsfig{width=0.98\textwidth,clip=true,file=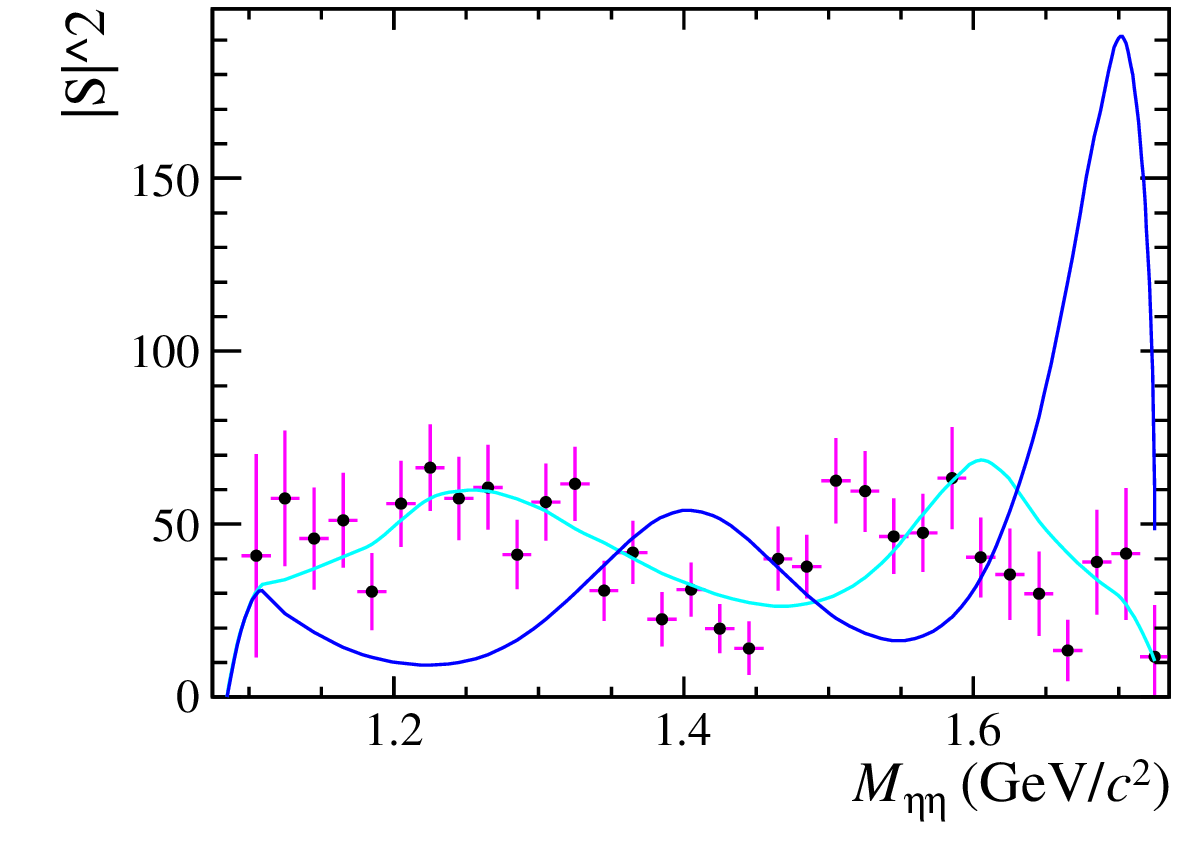}
\put(-85,65){(${\rm S}_{2}^{\rm F}$)}
\end{minipage}
\begin{minipage}[b]{0.23\textwidth}
\epsfig{width=0.98\textwidth,clip=true,file=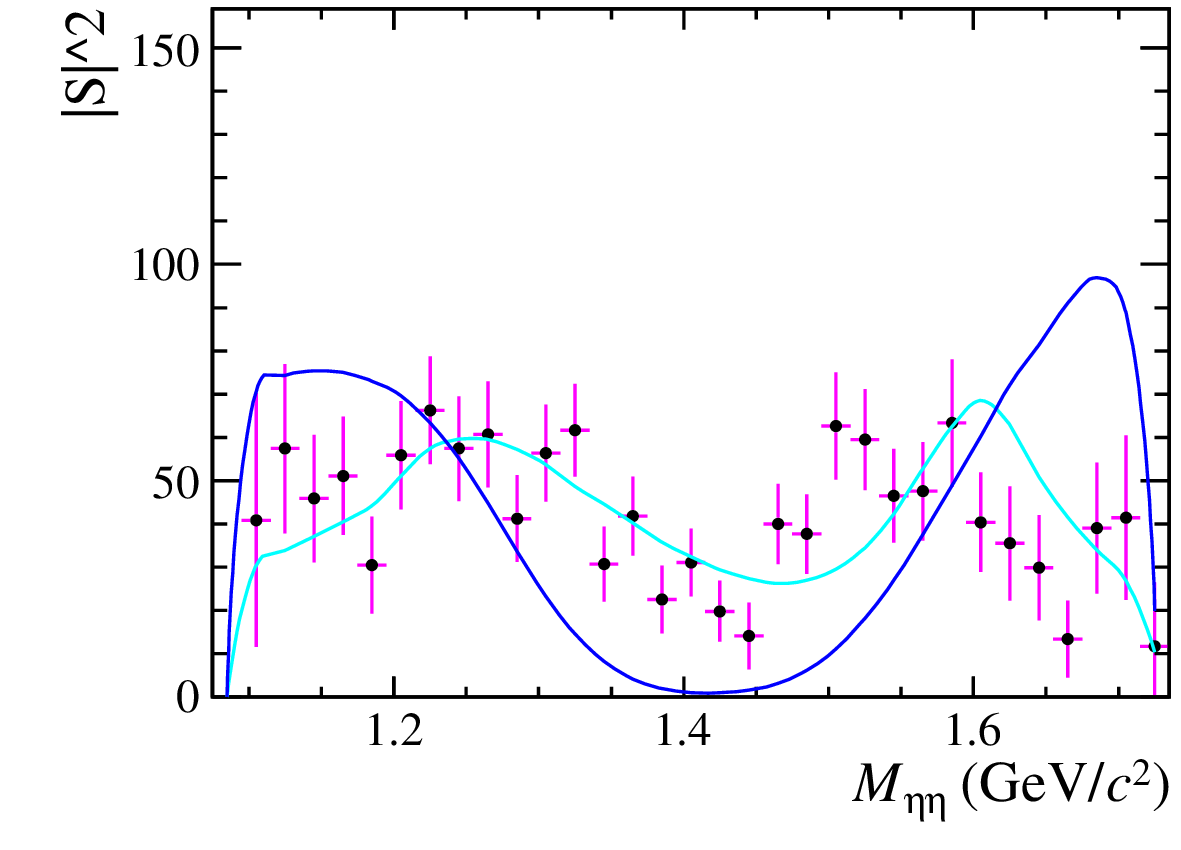}
\put(-85,65){(${\rm S}_{1}^{\rm D}$)}
\end{minipage}
\begin{minipage}[b]{0.23\textwidth}
\epsfig{width=0.98\textwidth,clip=true,file=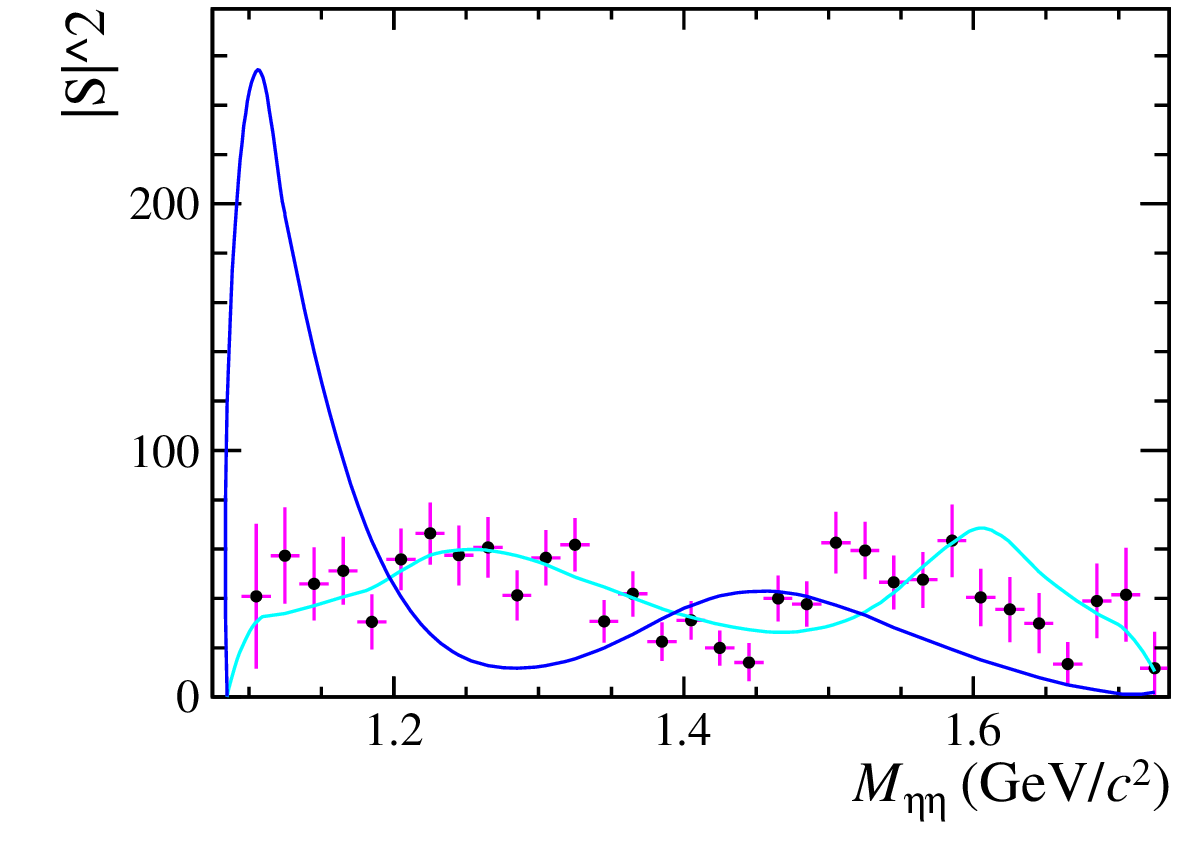}
\put(-85,65){(${\rm S}_{2}^{\rm D}$)}
\end{minipage}
\caption{The comparisons between the (blue solid line) model-independent $(\eta\eta)_{S-{\rm wave}}$,  (cyan solid line) the expected 
line shape reflected from the $a_{0}(980)$, and (dots with errors) the Legendre-moment-weighted $(\eta\eta)_{S-{\rm wave}}$ 
for solutions ${\rm S}_{1}^{\rm F}$, ${\rm S}_{2}^{\rm F}$, ${\rm S}_{1}^{\rm D}$, and ${\rm S}_{2}^{\rm D}$.}
\label{fig:MIPWA_vs_pw}
\end{center}
\end{figure}

% Here, we also compare the line-shape of $(\eta\eta)_{S-{\rm wave}}$ obtained from model-independent amplitude analysis 
% ${\rm S}^{\rm F}_1$ (and ${\rm S}^{\rm D}_1$), the expected line shape reflected from a normal $a_{0}(980)$ and that from Legendre-moment-weighted projections on $M(\eta\eta)$. The results are shown in Fig.~\ref{fig:MIPWA_vs_pw}. 
To examine the stability of the model-independent
$(\eta\eta)_{S-{\rm wave}}$ solutions, for each of the four
solutions we compare the fitted $(\eta\eta)_{S-{\rm wave}}$ line
shape with the kinematic reflection of the corresponding
$a_0(980)^+\eta$ amplitude and with the Legendre-moment-weighted
$M(\eta\eta)$ distribution.
The comparisons are shown in Fig.~\ref{fig:MIPWA_vs_pw}.
%From which, the model-independent $(\eta\eta)_{S-{\rm wave}}$ is quite similar with the expected line shape reflected from the $a_{0}(980)$, which may result large interference.  
%the fits with model-independent $(\eta\eta)_{S-{\rm wave}}$ are also rejected. 

%Furthermore, the highly enhanced structures also indicate great interference even the total interference is acceptable for ${\rm S}^{\rm F}_2$. 
% For ${\rm S}^{\rm F}_1$ and ${\rm S}^{\rm D}_1$, the model independent $(\eta\eta)_{S-{\rm wave}}$ are similar, and
% the $a_{0}(980)$ line shapes shown as in Fig.~\ref{fig:a0lineshapegen} are also similar, but the fit fractions of the dominant channel 
% $D^{+} \to a_{0}(980)^{+}\eta$, $a_{0}(980)^+\to \pi^+ \eta$ is quite different. 
% %This fact indicates that the fits with model-independent $(\eta\eta)_{S-{\rm wave}}$ are not stable.    
% In addition, shown as in Fig.~\ref{fig:MIPWA_vs_pw}, different with the line-shape given by the data, the structure around the lower and upper boundary 
% also indicates great interference even the total interference is acceptable for ${\rm S}^{\rm D}_1$. 
Although the model-independent $(\eta\eta)_{S-{\rm wave}}$ line shapes obtained for ${\rm S}^{\rm F}_1$ and ${\rm S}^{\rm D}_1$ are similar, the corresponding fit fractions of the dominant $D^{+} \to a_{0}(980)^{+}\eta$, $a_{0}(980)^+\to \pi^+ \eta$ amplitude differ substantially, even though the generated $a_{0}(980)$ line shapes shown in Fig.~\ref{fig:a0lineshapegen} are themselves similar. 
This behavior indicates that the separation of the two contributions is strongly affected by correlations between them and is not uniquely determined by the data. 
Moreover, the deviations between the fitted model-independent line shapes and the Legendre-moment-weighted distribution near the kinematic boundaries, where the sensitivity is limited, further show that the additional $(\eta\eta)_{S-{\rm wave}}$ contribution is poorly constrained in these regions.

The $(\eta\eta)_{S-{\rm wave}}$ line shapes obtained for ${\rm S}^{\rm F}_2$ and ${\rm S}^{\rm D}_2$ differ substantially, particularly near the kinematic boundaries. 
As shown in Fig.~\ref{fig:meteetapwa0eta_noloop_nofpar}, the reflection from the $a_{0}(980)^+$ amplitude already accounts for much of the observed behavior in the lower $M(\eta\eta)$ region, leaving any additional $(\eta\eta)_{S-{\rm wave}}$ contribution weakly constrained. 
Although ${\rm S}^{\rm F}_2$ gives a relatively smooth line shape in the lower-mass region, it exhibits a pronounced enhancement at higher masses. 
Together with the sizable interference with the $a_{0}(980)^+\eta$ amplitude, these differences illustrate the strong solution dependence of the extracted $(\eta\eta)_{S-{\rm wave}}$ contribution.

In summary, including a model-independent $(\eta\eta)_{S-{\rm wave}}$ amplitude substantially improves the likelihood and provides a good description of the data. 
However, multiple solutions with comparable fit qualities yield substantially different fit fractions and line shapes, and the extracted $(\eta\eta)_{S-{\rm wave}}$ contribution exhibits sizable interference with the dominant $a_{0}(980)^+\eta$ amplitude. 
The results also depend strongly on the parameterization of $P_{a_{0}(980)}$. 
With the current data sample, the two contributions therefore cannot be separated in a stable and unique manner. 
For this reason, the model-independent $(\eta\eta)_{S-{\rm wave}}$ solutions are not adopted in the nominal signal model.

\subsection{Signal model conclusion}

The studies in Secs.~\ref{sec:testsetA}--\ref{sec:testsetD} show that the discrepancy between the data and the baseline descriptions cannot be resolved by adding the conventional resonant or non-resonant amplitudes listed in Table~\ref{tab:addamp}. 
% The studies on the model-independent $(\eta\eta)_{S-{\rm wave}}$ presented in Sec.~\ref{sec:MIPWAres} show similar case 
% with the results that further involving the ``IX. constant term'', which introduces high correlation and makes the fit
The model-independent $(\eta\eta)_{S-{\rm wave}}$ study in Sec.~\ref{sec:MIPWAres} exhibits behavior similar to that observed when the ``IX. constant term'' is included: in both cases, strong correlations with the dominant $a_0(980)^+\eta$ amplitude render the decomposition unstable.

Although some additional amplitudes lead to sizable improvements in the likelihood, their fitted contributions are not stable. 
In particular, their significances change substantially when more than one additional amplitude is included, and several of the apparent improvements are accompanied by large interference with the dominant $\pi\eta$ $S$-wave amplitude. 
Moreover, no clear corresponding structures are observed in the relevant invariant-mass projections.

This conclusion is consistent across the Flatt\'e, dispersively modified Flatt\'e, $T$-matrix, and $K$-matrix descriptions tested above. 
The large significances obtained for some amplitudes in the test sets C and D mainly reflect the poor description of the dominant $\pi\eta$ line shape in the corresponding baseline models. 
They are therefore not interpreted as evidence for genuine additional intermediate states. 
Based on these studies, no additional amplitude is included as an independent component in the nominal signal model. 
The nominal signal model contains only the decay chain $D^{+}\to a_{0}(980)^{+}\eta$, $a_{0}(980)^{+}\to\pi^{+}\eta$.

\section{Further amplitude-analysis tests}

To improve the description of the data, further tests are performed by allowing the $a_0(980)$ parameters to float in the amplitude fit. 
Two cases are considered:
\begin{itemize}
    \item floating the $a_0(980)$ parameters in the Flatt\'e parameterization;
    \item floating the $a_0(980)$ parameters in the dispersively modified Flatt\'e parameterization.
\end{itemize}
In the following tests, model IX denotes the constant-amplitude model, while the two model-independent $(\eta\eta)_{S-{\rm wave}}$ solutions are denoted by ${\rm IX}^{\rm MI_i}$, with $i=1,2$. For the Flatt\'e parameterization, ${\rm IX}^{\rm MI_i}$ corresponds to solution ${\rm S}^{\rm F}_i$, whereas for the dispersively modified Flatt\'e parameterization it corresponds to solution ${\rm S}^{\rm D}_i$. The results for models IX, ${\rm IX}^{\rm MI_1}$, and ${\rm IX}^{\rm MI_2}$ are presented for completeness. Because of their strong solution dependence and sizable interference with the dominant $a_{0}(980)^+\eta$ amplitude, these models are not adopted in the nominal signal model and are not included in the evaluation of the systematic uncertainties.
% Here, for the fits with model-independent $(\eta\eta)_{S-{\rm wave}}$, they are denoted with ``${\rm IX^{\rm MI_{i}}}$'' for the solutions 
% ${\rm S}^{\rm F}_i$ and ${\rm S}^{\rm D}_i$ ($i=1,2$). 
% The results of the fits to models IX, ${\rm IX^{\rm MI_{1}}}$ and ${\rm IX^{\rm MI_{2}}}$ are only presented and not further be included 
% in the discussions. 

Due to the limited statistics and the weak sensitivity to the $\pi\eta'$ channel, the coupling constant $g_{\pi\eta'}$ is fixed to the values reported in Refs.~\cite{CLEO:2011upl} and~\cite{BESIII:2016tqo} for the Flatt\'e and dispersively modified Flatt\'e parameterizations, respectively. 
We do not attempt to float the scattering parameters of the $K$ matrix, since they are constrained by coupled-channel scattering information and cannot be reliably determined from the present single decay channel alone.

For the Flatt\'e and dispersively modified Flatt\'e parameterizations, the corresponding pole positions, $\sqrt{s}_{\mathrm{pole}}$, are also determined. 
The pole position is a process-independent quantity that characterizes the resonance. 
It is written as $\sqrt{s}_{\mathrm{pole}}=M_{\mathrm{pole}}-i\Gamma_{\mathrm{pole}}/2$, where $M_{\mathrm{pole}}$ and $\Gamma_{\mathrm{pole}}$ are the pole mass and width, respectively. 
For the $a_0(980)$, the pole is expected to be located on an unphysical sheet close to the physical region and near the $K\bar K$ threshold~\cite{ParticleDataGroup:2024cfk}.

For each decay channel $j$, the right-hand branch cut starts at the threshold $s_j^+=(m_{j1}+m_{j2})^2$ and extends to $+\infty$ along the real axis. 
The left-hand cuts are not considered in the sheet classification. 
Two branches of the phase-space factor are defined as
\begin{equation}
\rho_j^{\lambda_j}(s)=\lambda_j\rho_j^{+}(s),\qquad \lambda_j=\pm ,
\end{equation}
where
\begin{equation}
\rho_j^{+}(s)=
\frac{1}{s}\sqrt{(s-s_j^+)(s-s_j^-)} ,
\qquad
s_j^\pm=(m_{j1}\pm m_{j2})^2 .
\end{equation}
Here, $m_{j1}$ and $m_{j2}$ are the masses of the two particles in channel $j$. 
The physical branch is chosen by approaching the physical axis from the upper half-plane. 

For the $a_0(980)$, three coupled channels are considered: $\pi\eta$, $K\bar K$, and $\pi\eta'$, giving $2^3=8$ Riemann sheets. 
A sheet is labelled by the signs $(\lambda_{\pi\eta},\lambda_{K\bar K},\lambda_{\pi\eta'})$, and the physical sheet is $(+++)$. 
The unphysical sheets closest to the physical region are $(-++)$, $(--+)$, and $(---)$. 
For the CLEO Flatt\'e parameters, $g_{\pi\eta'}=0$, so the $\pi\eta'$ channel is decoupled from the denominator. 
In this case, the third sheet label is immaterial, and the sheet $(-++)$ is equivalent to the two-channel sheet $(-+)$ in the $(\pi\eta,K\bar K)$ space. 
Following Ref.~\cite{Zhang:2024qkg}, the pole search is performed on the relevant nearby sheets.

Denoting the denominator of the $a_0(980)$ dynamical function by $D_{a_0}(s)$, poles are obtained by solving
\begin{equation}
D_{a_0}^{\lambda_{\pi\eta},\lambda_{K\bar K},\lambda_{\pi\eta'}}(s)=0
\end{equation}
on the corresponding Riemann sheets. 
The pole associated with the $a_0(980)$ is expected to be close to the $K\bar K$ threshold. 
In the present study, the search is focused on the $(-++)$ and $(--+)$ sheets. 
Poles found on the $(--+)$ sheet have much larger imaginary parts and are far from the near-threshold region relevant to the $a_0(980)$. 
Therefore, only poles on the $(-++)$ sheet are reported. 

With the nominal $a_0(980)$ parameters reported by CLEO~\cite{CLEO:2011upl} and the Flatt\'e parameterization, the pole position is found to be
\begin{equation}
\sqrt{s}_{\mathrm{pole}}=(1.022-i\,0.070)~{\mathrm{GeV}/c^{2}}.
\end{equation}
With the nominal $a_0(980)$ parameters reported by BESIII~\cite{BESIII:2016tqo} and the dispersively modified Flatt\'e parameterization, the pole position is
\begin{equation}
\sqrt{s}_{\mathrm{pole}}=(1.046-i\,0.064)~{\mathrm{GeV}/c^{2}}.
\end{equation}

\subsection{Floating the $a_0(980)$ parameters in the Flatt\'e parameterization}
\label{sec:floatparFlatte}

\begin{figure}[htbp]
\begin{center}
\begin{minipage}[b]{0.23\textwidth}
\epsfig{width=0.98\textwidth,clip=true,file=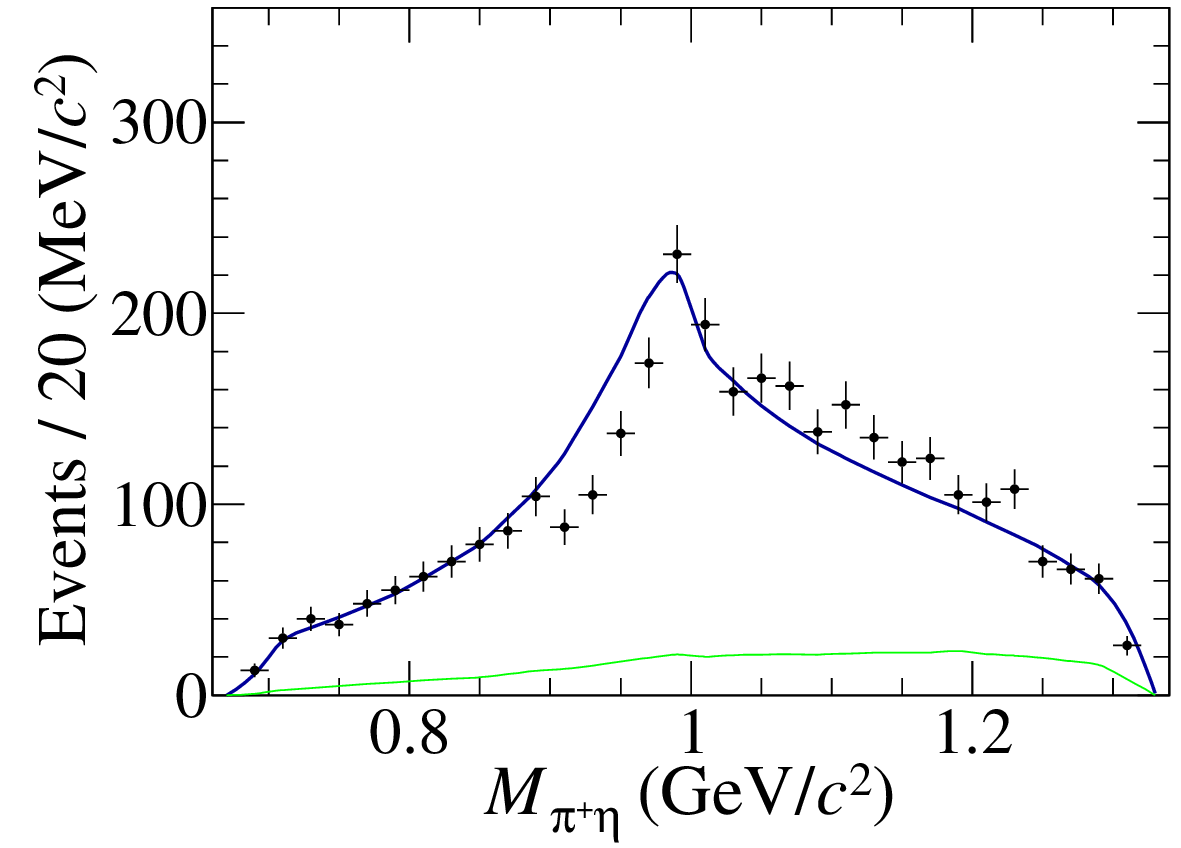}
%\put(-25,65){(a)}
\end{minipage}
\begin{minipage}[b]{0.23\textwidth}
\epsfig{width=0.98\textwidth,clip=true,file=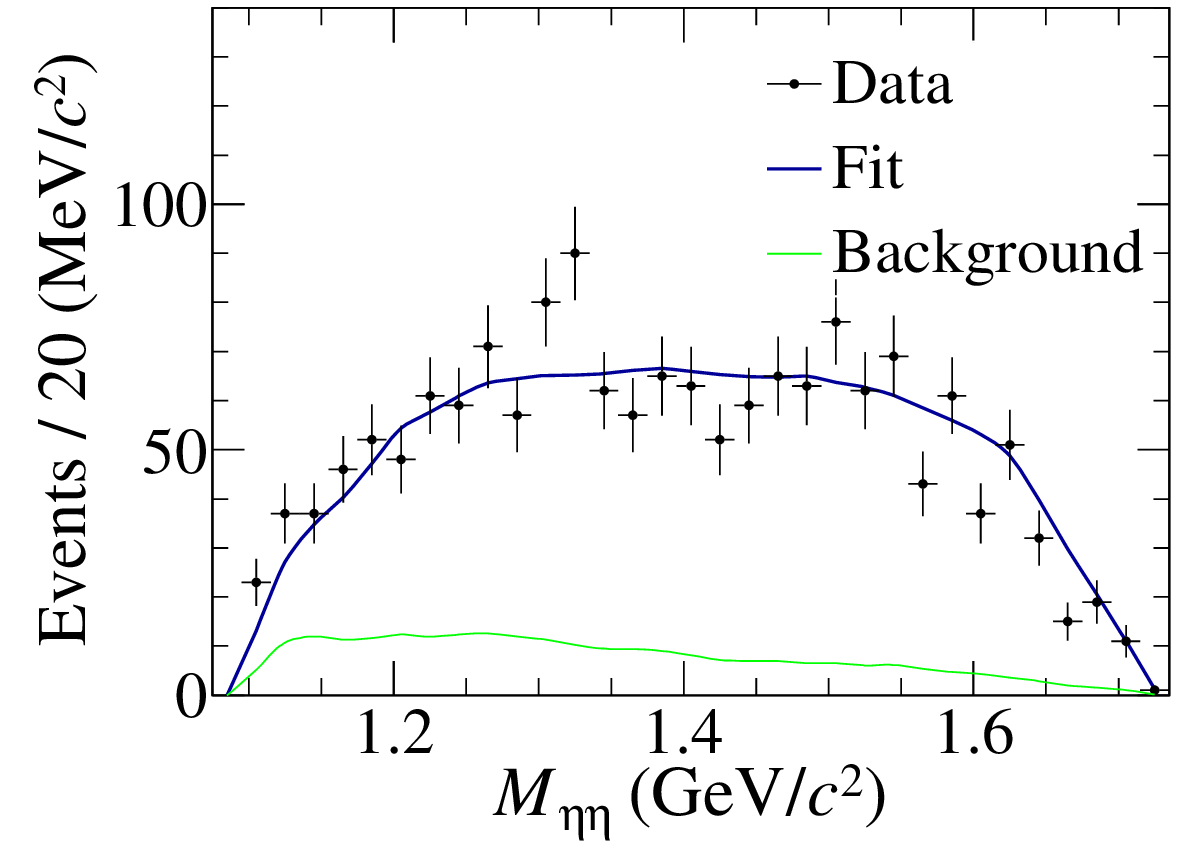}
%\put(-25,65){(b)}
\end{minipage}
\caption{The same as Fig.~\ref{fig:Flattebaseline}, but for baseline fit by floating the $a_{0}(980)$ coupling constants with using 
Flatt\'e parameterization for $P_{a_0(980)}$.}
\label{fig:flattefgsq}
\end{center}
\end{figure}

As discussed in Sec.~\ref{sec:a0onlyconfirmation}, the nominal signal model contains only the decay chain $D^+\to a_0(980)^+\eta$, $a_0(980)^+\to\pi^+\eta$. 
We first use this baseline model and leave the Flatt\'e parameters free. 
Since the value of $M_0$ reported in Ref.~\cite{CLEO:2011upl} is much more precisely determined than the coupling constants $g_{\pi\eta}^{2}$ and $g_{K\bar K}^{2}$, only the two coupling constants are floated in the first step, while $M_0$ is fixed to the CLEO value. 
The fit gives $\ln\mathcal{L}=109.3$, $g_{\pi\eta}^{2}=(0.449\pm0.038)~{\rm GeV}^2/c^4$, and $g_{K\bar K}^{2}=(0.371\pm0.098)~{\rm GeV}^2/c^4$, with $\chi^2/{\rm NDOF}=158.8/89$. 
The corresponding projections are shown in Fig.~\ref{fig:flattefgsq}. 
The data are still not well described.

%=====================================

\begin{figure}[htbp]
\begin{center}
\begin{minipage}[b]{0.23\textwidth}
\epsfig{width=0.98\textwidth,clip=true,file=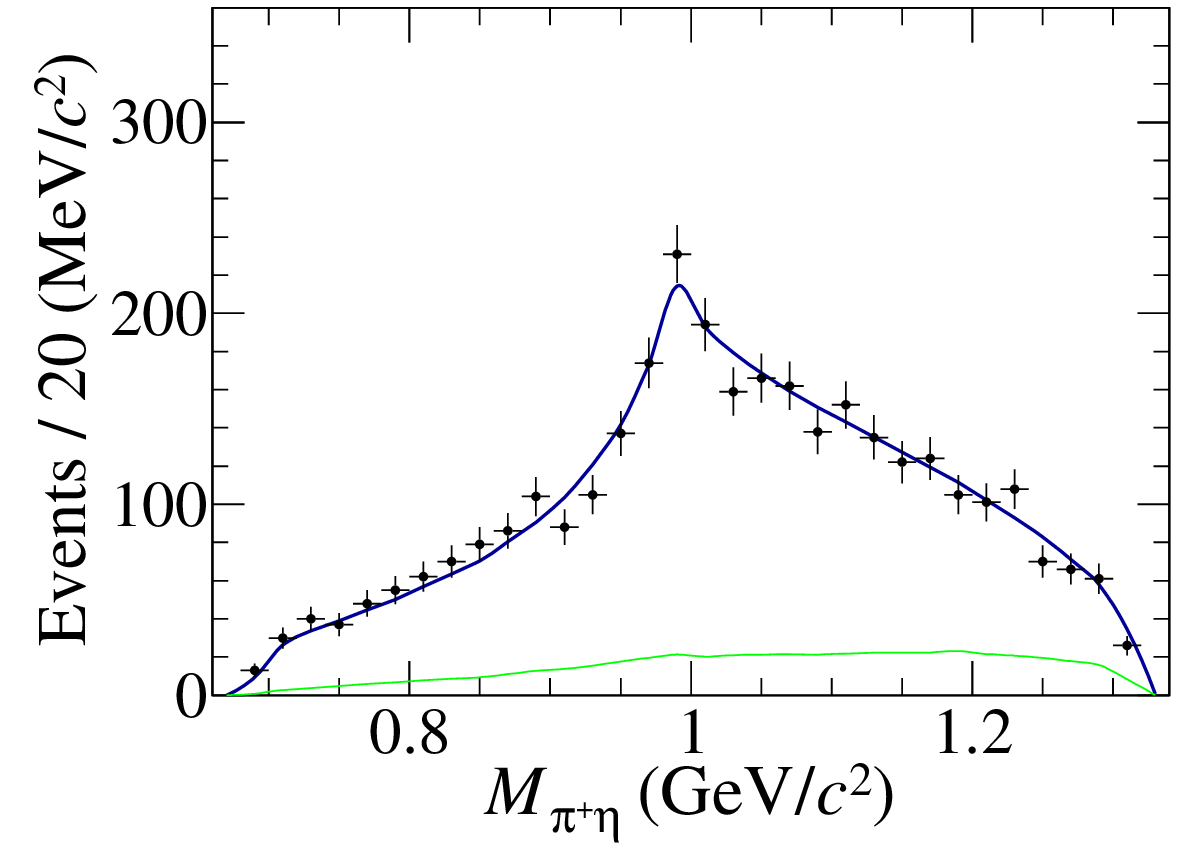}
%\put(-25,65){(a)}
\end{minipage}
\begin{minipage}[b]{0.23\textwidth}
\epsfig{width=0.98\textwidth,clip=true,file=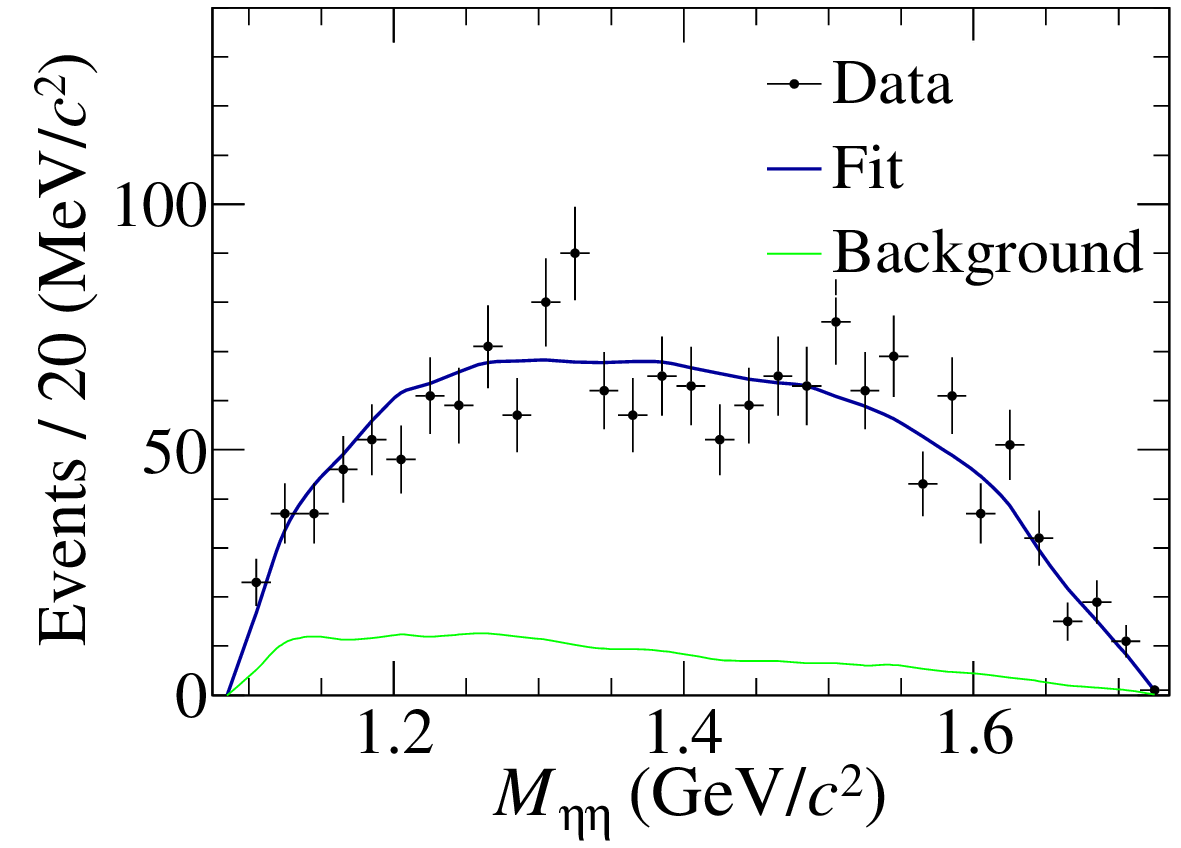}
%\put(-25,65){(b)}
\end{minipage}
\caption{The same as Fig.~\ref{fig:Flattebaseline}, but for baseline fit by floating all the $a_{0}(980)$ parameters 
with using Flatt\'e parameterization for $P_{a_0(980)}$.}
\label{fig:fparfit1}
\end{center}
\end{figure}

A further fit is performed by also floating $M_0$. 
The fit returns $M_0=(1.082\pm0.017_{\rm stat})~{\rm GeV}/c^2$, $g_{\pi\eta}^{2}=(0.469\pm0.047_{\rm stat})~{\rm GeV}^2/c^4$, and $g_{K\bar K}^{2}=(0.566\pm0.149_{\rm stat})~{\rm GeV}^2/c^4$. 
The fit gives $\ln\mathcal{L}=142.5$ and $\chi^2/{\rm NDOF}=85.4/88$. 
The projections are shown in Fig.~\ref{fig:fparfit1}.

Although the data are well described after floating all three Flatt\'e parameters, the resulting parameters differ significantly from the CLEO values. 
The corresponding pole position is $\sqrt{s}_{\rm pole} = [(1.096\pm0.015_{\rm stat})-i(0.042\pm0.024_{\rm stat})]~{\mathrm{GeV}/c^{2}}$.
The pole mass is therefore shifted well above the $K\bar K$ threshold, which is unexpected for the near-threshold $a_0(980)$ state.

To examine whether this pole shift can be attributed to systematic effects, systematic uncertainties on the floated parameters and pole position are evaluated. 
The following sources are considered. 
The detailed results are summarized in Table~\ref{tab:ampsysflatte}.
\begin{enumerate}
\item[(I)] The fixed value of $g_{\pi\eta'}$ is varied within the uncertainty reported in Ref.~\cite{CLEO:2011upl}.
\item[(II)] The signal purity $f_s$ is varied within its uncertainty to estimate the effect of the background fraction.
\item[(III)] The uncertainty due to the background shape includes three contributions: the choice of the $\Delta E$ window used to select the background sample, the kernel bandwidth used in extracting the background shape, and the statistical fluctuation of the background sample. 
The last contribution is estimated with the bootstrap method~\cite{Efron1979,EfronTibshirani1993}.
\item[(IV)] The uncertainty due to the fit bias is obtained from the input-output (IO) check described in Sec.~\ref{sec:pull}.
\item[(V)] The uncertainty associated with the reconstruction efficiency includes tracking, PID, and $\eta$ reconstruction.
\item[(VI)] The uncertainty due to possible additional amplitudes is evaluated by repeating the fit with each additional amplitude considered in Table~\ref{tab:addamp}. For each parameter, despite of the fits with an unstable result or with the pole position can not be found at $(-++)$ sheet, 
the largest deviation from the nominal result is assigned as the systematic uncertainty from this source. 
\item[(VII)] The underlying non-interfering $a_{0}(980)$ peaking-background contribution, which is estimated with using the sideband sample instead of the 
inclusive MC in the background PDF determination. 
\end{enumerate}

%%======================================
\begin{table}[htbp]
\begin{center}
\caption{Systematic uncertainties for the $a_0(980)$ parameters and pole position obtained with the Flatt\'e parameterization. 
The values are given in units of $10^{-2}$, with mass-related quantities in ${\rm GeV}/c^2$ and coupling constants in ${\rm GeV}^2/c^4$.}
\begin{tabular}{c|ccccccc|c} \hline 
Parameter                  &  I  &  II & III & IV  &  V   & VI   & VII  & Total\\ \hline
$M_{0}$                    & 0.1 & 0.3 & 0.7 & 0.2 & 0.1  & 2.5  & 1.1  & 2.8  \\  
$g_{\pi\eta}^{2}$          & 0.4 & 0.6 & 2.2 & 0.5 & 0.0  & 6.0  & 2.4  & 6.8  \\
$g_{K\bar{K}}^{2}$         & 0.9 & 1.2 & 6.4 & 0.9 & 0.1  & 14.2 & 8.4  & 17.7 \\
$M_{\mathrm{pole}}$        & 0.0 & 0.2 & 0.6 & 0.1 & 0.1  & 1.7  & 0.8  & 1.9  \\
$\Gamma_{\mathrm{pole}}/2$ & 0.2 & 0.1 & 0.9 & 0.1 & 0.0  & 3.7  & 1.1  & 3.9  \\
\hline
\end{tabular}
\label{tab:ampsysflatte}
\end{center}
\end{table}

Based on the systematic study, the final pole mass is 
\begin{equation}
    M_{\mathrm{pole}} = (1.096\pm0.015_{\rm stat}\pm0.019_{\rm syst})~{\rm GeV}/c^{2}.
\end{equation}
The difference between this $M_{\mathrm{pole}}$ and the pole mass calculated from the CLEO parameters cannot be covered by the systematic uncertainty.
Thus, although floating the Flatt\'e parameters improves the fit quality, it drives the pole mass well above the near-threshold region expected for the $a_0(980)$.
%%======================================

\subsubsection{IO check}
\label{sec:pull}

The IO check is performed using 300 data-sized MC samples. 
In these samples, the signal process is generated according to the amplitude model $D^{+}\to a_{0}(980)^{+}\eta$, $a_{0}(980)^{+}\to\pi^{+}\eta$, with the $a_{0}(980)$ parameters set to the values obtained from data. 
The same event-selection and amplitude-analysis procedures as those used for data are applied to these MC samples.

For each parameter, the pull is defined as
\begin{equation}
{\rm pull} = \frac{V_{\rm out}-V_{\rm in}}{\sigma_{\rm out}},
\end{equation}
where $V_{\rm out}\pm\sigma_{\rm out}$ is the fitted result from a pseudo-experiment, and $V_{\rm in}$ is the generated input value. 
The pull distributions obtained from the 300 samples are fitted with Gaussian functions. 
The fitted mean, $\mu_p$, is used to evaluate possible fit bias, while the fitted width, $\sigma_p$, is used to check whether the statistical uncertainty is properly estimated.

If $\mu_p$ differs significantly from zero, the fitted value in data is corrected according to
\begin{equation}
    V_c = V - \mu_p\sigma_{\rm stat},
\end{equation}
where $V\pm\sigma_{\rm stat}$ is the result obtained from data and $V_c$ is the corrected value. 
The uncertainty associated with this correction is assigned as
\begin{equation}
    \sigma_{\rm bias} = \sigma_{\mu_p}\,\sigma_{\rm stat},
\end{equation}
where $\sigma_{\mu_p}$ is the uncertainty of the fitted pull mean. 
If no significant bias is observed, no correction is applied, and the possible residual bias is assigned as a systematic uncertainty,
\begin{equation}
    \sigma_{\rm bias} = \sqrt{\mu_p^2+\sigma_{\mu_p}^2}\,\sigma_{\rm stat}.
\end{equation}

If $\sigma_p$ is larger than unity by more than three standard deviations, an additional systematic uncertainty is assigned to account for the possible underestimation of the statistical uncertainty:
\begin{equation}
    \sigma_c = \sqrt{\sigma_p^2-1}\,\sigma_{\rm stat}.
\end{equation}
The results of the IO checks are listed in Table~\ref{tab:pullflatte}.

\begin{table}[htbp]
\begin{center}
\caption{IO check results for the $a_0(980)$ parameters and pole quantities obtained with the Flatt\'e parameterization. }
\begin{tabular}{ccc} \hline 
Parameter            & $\mu_{p}$     & $\sigma_{p}$   \\ \hline
$M_{0}$              &$-0.11\pm0.06$ & $1.02\pm0.04$  \\  
$g_{\pi\eta}^{2}$    & $0.10\pm0.06$ & $0.97\pm0.04$  \\
$g_{K\bar{K}}^{2}$   &$-0.30\pm0.06$ & $1.12\pm0.05$  \\
$M_{\mathrm{pole}}$  &$-0.01\pm0.06$ & $0.97\pm0.04$  \\
$\Gamma_{\mathrm{pole}}/2$ & $0.37\pm0.06$ & $0.95\pm0.04$  \\
\hline
\end{tabular}
\label{tab:pullflatte}
\end{center}
\end{table}

For $M_0$, $g_{\pi\eta}^{2}$, and $M_{\rm pole}$, no significant fit bias is observed from the fitted values of $\mu_p$. 
Corrections are applied to $g_{K\bar K}^{2}$ and $\Gamma_{\rm pole}/2$, for which the fitted pull means deviate significantly from zero. 
The corrected results are $g_{K\bar K,c}^{2} = (0.610\pm0.149_{\rm stat}\pm0.177_{\rm syst})~{\rm GeV}^{2}/c^{4}$ and $\Gamma_{{\rm pole},c}/2 = (0.033\pm0.024_{\rm stat}\pm0.039_{\rm syst})~{\rm GeV}/c^{2}$.

\subsubsection{Tests with additional amplitudes in the floating Flatt\'e fit}
\label{sec:addampflattefloat}

The same additional amplitudes as those considered in Sec.~\ref{sec:testsetA} are tested in the floating-parameter Flatt\'e fit. 
%The constant term is excluded from these tests, following the discussion in Sec.~\ref{sec:testsetA}. 
For each additional amplitude, the fit is repeated with $M_0$, $g_{\pi\eta}^2$, and $g_{K\bar K}^2$ allowed to float. 
% The resulting $\ln\mathcal{L}$ increments, fit qualities, significances of the additional amplitudes, and shifts in the fitted $a_0(980)$ parameters and pole position are summarized in Table~\ref{tab:flatteaddampfloat}. 
The resulting $\Delta\ln\mathcal{L}$ values, fit qualities, significances of the additional amplitudes, and shifts in the fitted $a_0(980)$ parameters and pole position are summarized in Table~\ref{tab:flatteaddampfloat}.
Here, ``NA'' indicates that no pole associated with the $a_0(980)$ is found on the $(-++)$ sheet; such fits are not used in the evaluation of the systematic uncertainty from additional amplitudes. 
%For the ``IX constant term'' and alternative approach of model-independent $(\eta\eta)_{S-{\rm wave}}$, both are not considered due to the large interference. 
%The corresponding details can be found at App.~\ref{sec:MIPWAfpars}.

\begin{table}[htbp]
\begin{center}
\caption{Fit results obtained by adding one additional amplitude to the floating-parameter Flatt\'e baseline model. 
% The IX and ${\rm IX^{\rm MI}}$ are also presented but not considered into the systematic uncertainty due to the large interference and the 
% highly uncertain fit results. 
% All shifts are relative to the baseline fit with all Flatt\'e parameters floated. 
Results for model IX and the two model-independent solutions are presented for completeness, but they are not included in the evaluation of the systematic uncertainty because of the unstable fits. 
The $\Delta\ln\mathcal{L}$, significance, and all parameter and pole shifts are evaluated relative to the baseline fit with all Flatt\'e parameters floated.
The shifts are given in units of $10^{-2}$, with ${\rm GeV}/c^2$ for 
$\Delta M_0$, $\Delta M_{\rm pole}$, and $\Delta(\Gamma_{\rm pole}/2)$, 
and ${\rm GeV}^2/c^4$ for $\Delta g_{\pi\eta}^2$ and $\Delta g_{K\bar K}^2$. 
Here, ``NA'' indicates that no pole associated with the $a_0(980)$ is found on the $(-++)$ sheet.}
\resizebox{0.5\textwidth}{!}{%
\begin{tabular}{c|ccc|ccc|cc} \hline 
Amplitude 
& $\Delta\ln\mathcal{L}$ 
& $\chi^2/{\rm NDOF}$ 
& Sig. ($\sigma$) 
& $\Delta M_0$ 
& $\Delta g_{\pi\eta}^{2}$ 
& $\Delta g_{K\bar K}^{2}$ 
& $\Delta M_{\rm pole}$ 
& $\Delta(\Gamma_{\rm pole}/2)$ \\ \hline
I    & $+1.4$ & $86.0/86$ & 1.1 & $+2.1$ & $+1.1$ & $+13.4$ & $+1.3$ & $+2.6$ \\
II   & $+3.7$ & $83.6/86$ & 2.2 & $+2.5$ & $+1.8$ & $+14.2$ & $+1.6$ & $+2.7$ \\
III  & $+1.7$ & $85.1/86$ & 1.3 & $+2.2$ & $-3.0$ & $+13.1$ & $+0.9$ & $+3.7$ \\
IV   & $+0.2$ & $85.6/86$ & 0.2 & $-0.1$ & $+0.5$ & $+2.7$  & $-0.1$ & $+0.3$ \\
V    & $+1.6$ & $83.2/86$ & 1.3 & $-1.5$ & $-5.7$ & $-8.5$  & $-1.4$ & $-0.2$ \\
VI   & $+0.8$ & $83.9/86$ & 0.8 & $-1.4$ & $-5.4$ & $-7.6$  & $-1.4$ & $-0.2$ \\
VII  & $+1.2$ & $84.4/86$ & 1.0 & $-1.8$ & $-5.1$ & $-9.9$  & $-1.7$ & $-0.7$ \\
VIII & $+0.7$ & $84.3/86$ & 0.7 & $-1.1$ & $-6.0$ & $-11.1$ & $-1.1$ & $-0.5$ \\
IX   & $+2.1$ & $85.8/86$ & 1.6 & $+4.8$ & $-5.0$ & $+19.2$ & NA     & NA     \\
${\rm IX^{\rm MI_1}}$ & $+9.4$ & $79.2/78$ & 2.0 & $+0.3$ & $-2.6$ & $+10.6$ & $-0.4$ & $-2.5$ \\
${\rm IX^{\rm MI_2}}$ &$+11.0$ & $80.9/78$ & 2.4 & $+4.3$ & $+1.9$ & $+21.6$ & NA   & NA     \\
X    & $+2.8$ & $85.5/86$ & 1.9 & $+4.2$ & $+1.8$ & $+27.8$ & NA     & NA     \\
XI   & $+1.8$ & $83.5/86$ & 1.4 & $-1.1$ & $-3.3$ & $-3.4$  & $-1.1$ & $+0.1$ \\
XII  & $+0.8$ & $84.2/86$ & 0.7 & $-0.9$ & $-3.3$ & $-6.1$  & $-0.9$ & $-0.4$ \\
\hline
\end{tabular}
}
\label{tab:flatteaddampfloat}
\end{center}
\end{table}
                    
As shown in Table~\ref{tab:flatteaddampfloat}, once all Flatt\'e parameters are allowed to float, none of the additional amplitudes has a significance above $3\sigma$. 
This is different from the fixed-parameter Flatt\'e tests in Sec.~\ref{sec:testsetA}, where some additional amplitudes showed sizable apparent significances. 
The result indicates that the improvements previously attributed to additional amplitudes can largely be absorbed by changes in the $a_0(980)$ line shape.

To test whether a missing additional amplitude could mimic the observed shift of the $a_0(980)$ parameters, further fits are performed in which the complex coefficients of the additional amplitudes are fixed to the values obtained in Sec.~\ref{sec:testsetA}, while the Flatt\'e parameters are refitted. 
% The resulting $\ln\mathcal{L}$ values, fit qualities, fitted $a_0(980)$ parameters, and pole positions are listed in Table~\ref{tab:flattefixcoeff} of Appendix~\ref{app:fixed_addamp_tests}. 
The resulting $\Delta\ln\mathcal{L}$ values, fit qualities, fitted $a_0(980)$ parameters, and pole positions are listed in Table~\ref{tab:flattefixcoeff} of Appendix~\ref{app:fixed_addamp_tests}.
Even with such artificially fixed additional components, 
the pole mass remains far above the near-threshold pole position calculated from the CLEO parameters.

Since the shift in $M_{\rm pole}$ is mainly driven by the fitted value of $M_0$, additional tests are performed by floating only $M_0$ while fixing the coupling constants to the CLEO values~\cite{CLEO:2011upl}. 
For each additional amplitude, three strategies are tested: fixing the full complex coefficient, fixing only its magnitude, or fixing only its phase to the value obtained in Sec.~\ref{sec:testsetA}. 
The corresponding Flatt\'e results are summarized in the Flatt\'e columns of Tables~\ref{tab:finaltests0_combined}, \ref{tab:finaltests1_combined}, and \ref{tab:finaltests2_combined} in Appendix~\ref{app:fixed_addamp_tests}, respectively.

In all these tests, the pole mass remains well above the near-threshold value calculated from the CLEO parameters. 
Moreover, for the same additional amplitude, the fitted $M_{\rm pole}$ can vary substantially depending on which part of the additional-amplitude coefficient is fixed. 
This instability indicates that these additional amplitudes do not provide a robust physical solution to the observed line shape discrepancy. 
In the fits listed in Table~\ref{tab:finaltests2_combined}, where the phase of the additional amplitude is fixed and the magnitude is allowed to float, the fitted magnitudes for the models with amplitudes I, III, IV, and X are compatible with zero within statistical uncertainties. 
This further indicates that these components are not required by the data once the $a_0(980)$ line shape is allowed to vary.

These studies show that the Flatt\'e parameterization can describe the data only by shifting the $a_0(980)$ pole mass well above the near-threshold region. 
Possible additional amplitudes, even when introduced with coefficients fixed from the fixed-parameter studies, cannot bring the pole mass back to the value calculated from the CLEO parameters. 
This demonstrates a tension between fit quality and the physical pole position in the direct-production Flatt\'e description.

%=====================================================================================
\subsection{Floating the $a_0(980)$ parameters in the dispersively modified Flatt\'e parameterization}
\label{sec:floatpardispersive}

\begin{figure}[htbp]
\begin{center}
\begin{minipage}[b]{0.23\textwidth}
\epsfig{width=0.98\textwidth,clip=true,file=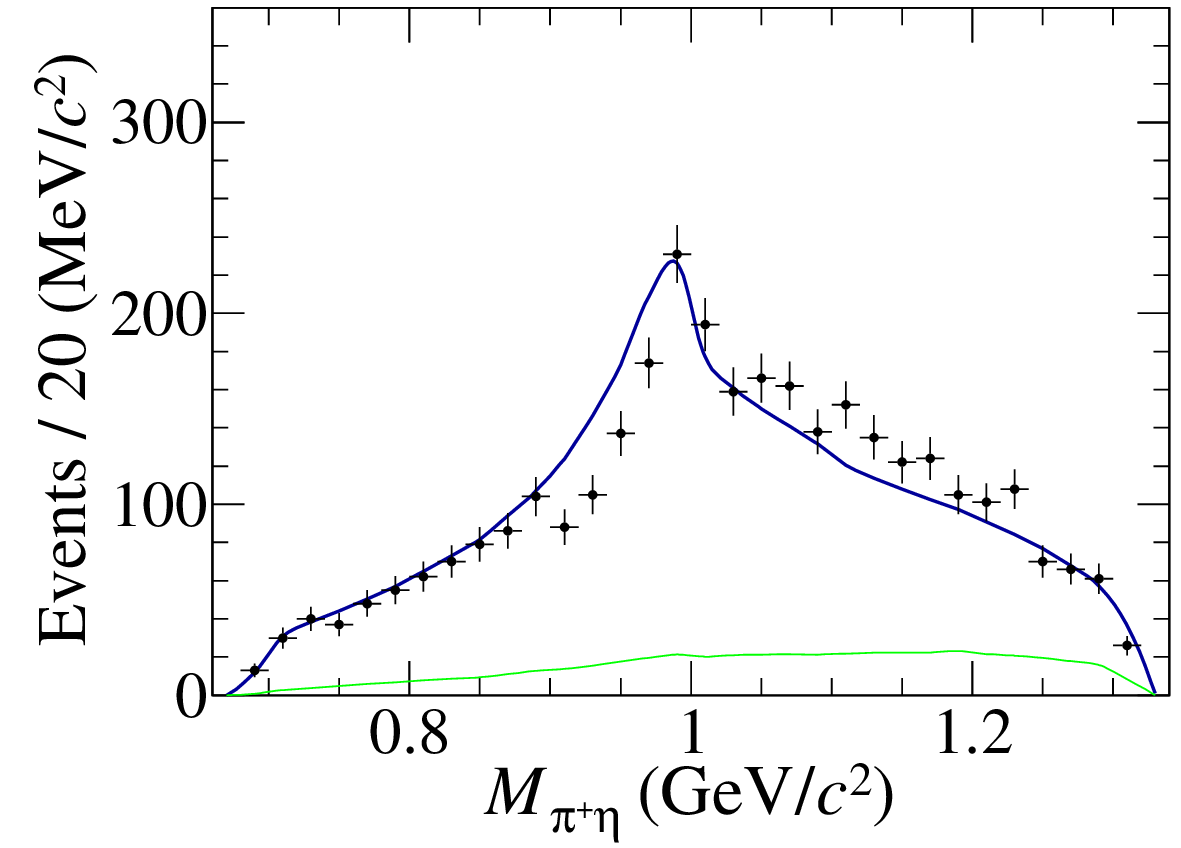}
%\put(-25,65){(a)}
\end{minipage}
\begin{minipage}[b]{0.23\textwidth}
\epsfig{width=0.98\textwidth,clip=true,file=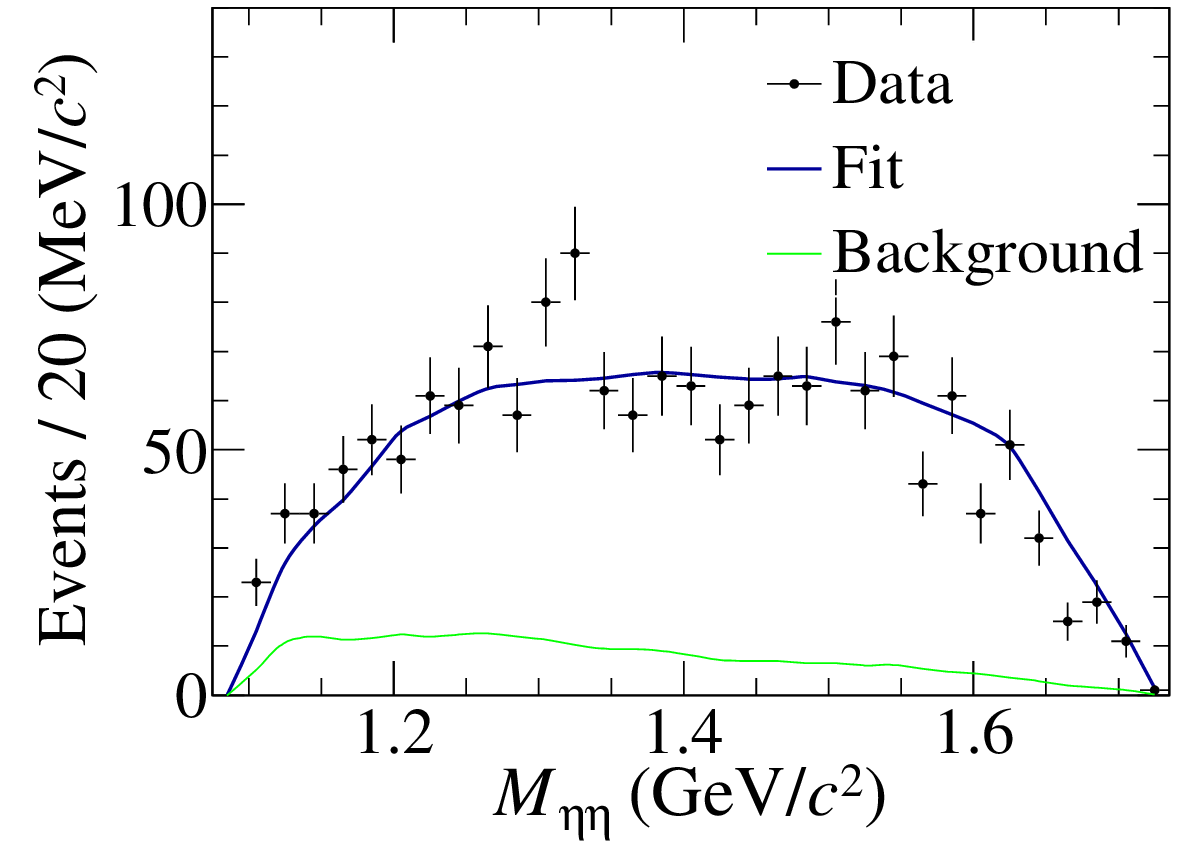}
%\put(-25,65){(b)}
\end{minipage}
\caption{The same as Fig.~\ref{fig:Flattebaseline}, but for the baseline fit by floating the $a_0(980)$ coupling constants with 
using dispersively modified Flatt\'e parameterization for $P_{a_0(980)}$.}
\label{fig:dispersivefgsq}
\end{center}
\end{figure}

As in the Flatt\'e case, we first perform a fit in which only the coupling constants $g_{\pi\eta}^{2}$ and $g_{K\bar K}^{2}$ are floated, while $M_0$ is fixed to the value reported in Ref.~\cite{BESIII:2016tqo}. 
The fit gives $\ln\mathcal{L}=112.5$, $g_{\pi\eta}^{2}=(0.345\pm0.021)~{\rm GeV}^2/c^4$, and $g_{K\bar K}^{2}=(0.341\pm0.057)~{\rm GeV}^2/c^4$, with $\chi^2/{\rm NDOF}=159.6/89$. 
The corresponding projections are shown in Fig.~\ref{fig:dispersivefgsq}. 
The data are still not well described when only the two coupling constants are floated.

\begin{figure}[htbp]
\begin{center}
\begin{minipage}[b]{0.23\textwidth}
\epsfig{width=0.98\textwidth,clip=true,file=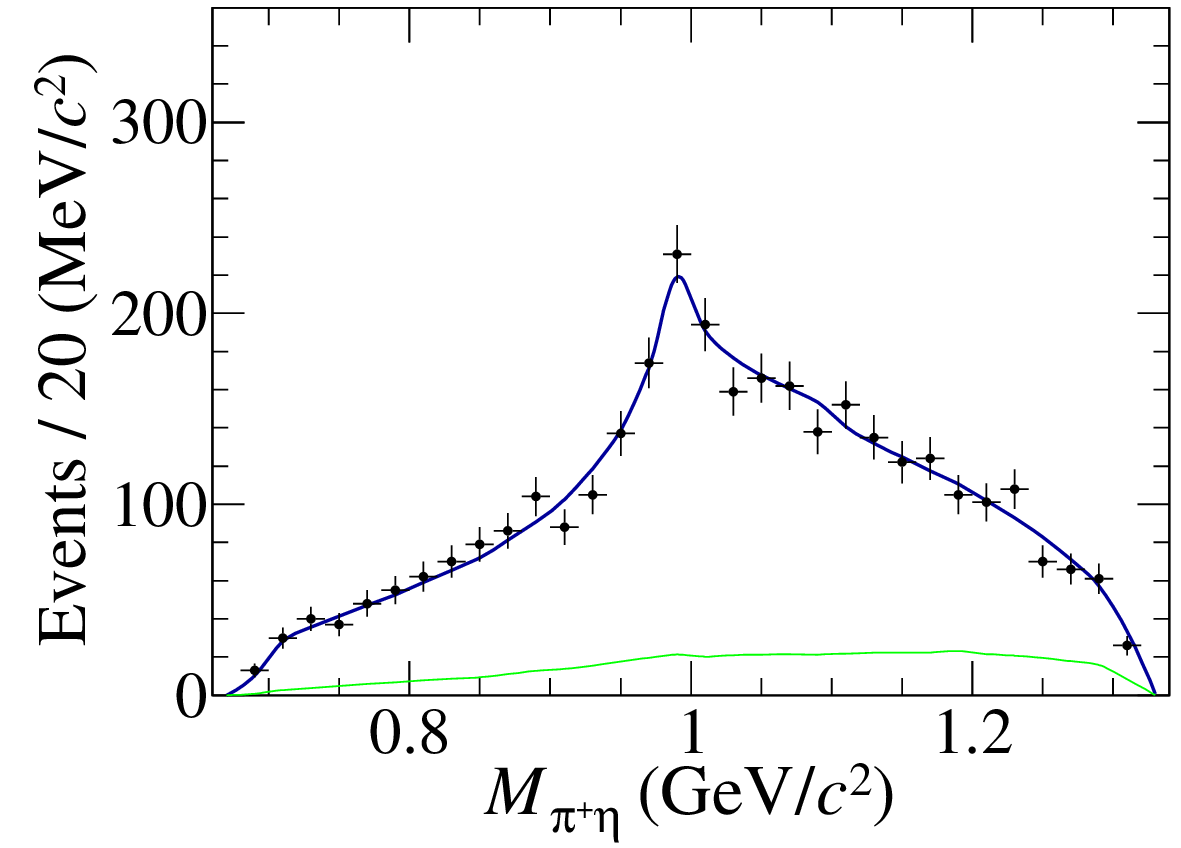}
%\put(-25,65){(a)}
\end{minipage}
\begin{minipage}[b]{0.23\textwidth}
\epsfig{width=0.98\textwidth,clip=true,file=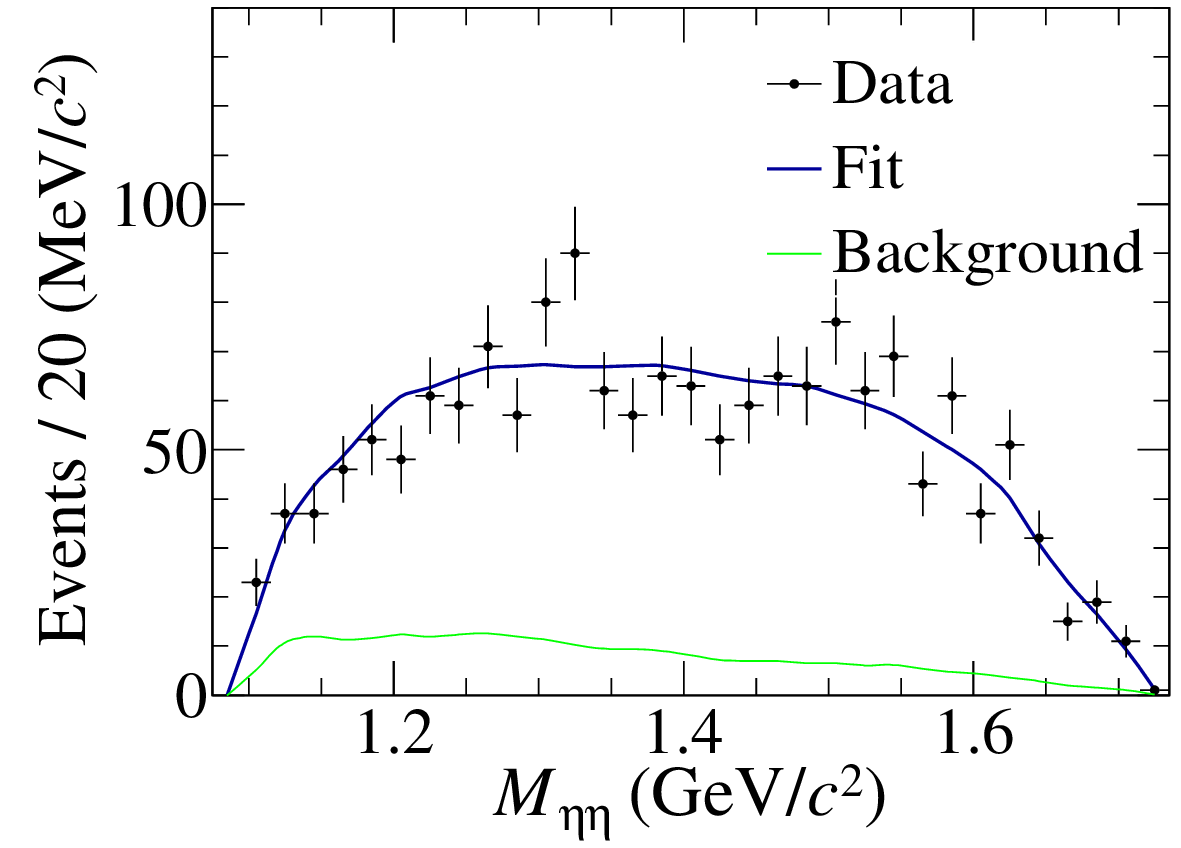}
%\put(-25,65){(b)}
\end{minipage}
\caption{The same as Fig.~\ref{fig:Flattebaseline}, but for the baseline fit by floating all $a_0(980)$ parameters 
with using dispersively modified Flatt\'e parameterization for $P_{a_0(980)}$.}
\label{fig:fpardispersive}
\end{center}
\end{figure}

A further fit is performed by also floating $M_0$, while $g_{\pi\eta'}^{2}$ is fixed to the value reported in Ref.~\cite{BESIII:2016tqo} due to the limited sensitivity to the $\pi\eta'$ channel. 
The fit returns $M_0=(1.045\pm0.007_{\rm stat})~{\rm GeV}/c^2$, $g_{\pi\eta}^{2}=(0.319\pm0.026_{\rm stat})~{\rm GeV}^{2}/c^{4}$, and $g_{K\bar K}^{2}=(0.416\pm0.064_{\rm stat})~{\rm GeV}^{2}/c^{4}$. 
The fit gives $\ln\mathcal{L}=144.6$ and $\chi^2/{\rm NDOF}=84.5/88$. 
The projections are shown in Fig.~\ref{fig:fpardispersive}. 
The corresponding pole position is $\sqrt{s}_{\rm pole} = [(1.087\pm0.009_{\rm stat})-i(0.008^{+0.014}_{-0.008})]~{\rm GeV}/c^{2}$. 
Both $M_{\rm pole}$ and $\Gamma_{\rm pole}/2$ differ significantly from the pole position calculated using the BESIII parameters reported in Ref.~\cite{BESIII:2016tqo}. 
The pole mass is shifted well above the $K\bar K$ threshold, and $\Gamma_{\rm pole}/2$ becomes very small. 
Therefore, as in the Flatt\'e case, although the data are well described after floating the $a_0(980)$ parameters, the resulting pole position is difficult to reconcile with the near-threshold nature of the $a_0(980)$.

The systematic uncertainties of the floated $a_0(980)$ parameters and pole quantities are also evaluated. 
The considered sources and evaluation procedures are the same as those used in the Flatt\'e case discussed in Sec.~\ref{sec:floatparFlatte}. 
Only the IO check results and the effects from possible additional amplitudes are presented below, in Secs.~\ref{sec:pulldispersive} and~\ref{sec:addampdispersivefloat}, respectively. 
The resulting systematic uncertainties are summarized in Table~\ref{tab:ampsysdispersive}.

%==============================================================================
\begin{table}[htbp]
\begin{center}
\caption{Systematic uncertainties for the $a_0(980)$ parameters and pole quantities obtained with the dispersively modified Flatt\'e parameterization. 
The values are given in units of $10^{-2}$; the units are ${\rm GeV}/c^2$ for $M_0$, $M_{\rm pole}$, and $\Gamma_{\rm pole}/2$, and ${\rm GeV}^2/c^4$ for the coupling constants. 
The fit-bias uncertainty, source IV, is not assigned for $\Gamma_{\rm pole}/2$ because, after retaining only poles on the $(-++)$ sheet in the IO check, its pull distribution cannot be described by a Gaussian function. 
The dash indicates that the corresponding uncertainty is not assigned and is not included in the total uncertainty.}
\begin{tabular}{c|ccccccc|c} \hline 
Parameter                & I   &  II & III & IV  &  V  & VI  & VII & Total\\ \hline
$M_{0}$                  & 0.3 & 0.1 & 0.2 & 0.0 & 0.0 & 0.4 & 0.2 & 0.5  \\  
$g_{\pi\eta}^{2}$        & 0.3 & 0.1 & 1.0 & 0.2 & 0.1 & 1.7 & 0.6 & 2.1  \\
$g_{K\bar{K}}^{2}$       & 0.7 & 0.5 & 4.1 & 0.4 & 0.3 & 3.0 & 5.7 & 7.6  \\
$M_{\mathrm{pole}}$      & 0.0 & 0.2 & 0.4 & 0.1 & 0.0 & 0.7 & 0.3 & 0.9  \\
$\Gamma_{\mathrm{pole}}/2$ & 0.1 & 0.1 & 0.9 & - & 0.0 & 0.3 & 0.1 & 1.0  \\
\hline
\end{tabular}
\label{tab:ampsysdispersive}
\end{center}
\end{table}

%==============================================================================
\subsubsection{The IO Check}
\label{sec:pulldispersive}

For $\Gamma_{\rm pole}/2$, the nominal value is very close to zero, suggesting that the pole is located close to the real axis and near the boundary between neighboring Riemann sheets. 
In the IO check, the pole associated with the $a_0(980)$ is found on a neighboring Riemann sheet in some pseudoexperiments. 
Following the pole-selection criterion used in the nominal analysis, only poles found on the $(-++)$ sheet are retained. 
For the retained pseudo-experiments, the pull distribution of $M_{\rm pole}$ can still be described by a Gaussian function, and the corresponding fit-bias uncertainty is evaluated. 
However, the pull distribution of $\Gamma_{\rm pole}/2$ cannot be described by a Gaussian function. 
Therefore, no fit-bias uncertainty is assigned for $\Gamma_{\rm pole}/2$.
The IO check results for the remaining parameters are listed in Table~\ref{tab:IOdispersive}.

\begin{table}[htbp]
\begin{center}
\caption{IO check results for the $a_0(980)$ parameters and pole mass obtained with the dispersively modified Flatt\'e parameterization.}
\begin{tabular}{ccc} \hline 
Parameter            & $\mu_{p}$     & $\sigma_{p}$   \\ \hline
$M_{0}$              & $0.02\pm0.06$ & $1.02\pm0.04$  \\  
$g_{\pi\eta}^{2}$    & $0.38\pm0.06$ & $1.03\pm0.04$  \\
$g_{K\bar{K}}^{2}$   &$-0.29\pm0.06$ & $1.05\pm0.04$  \\
$M_{\mathrm{pole}}$  & $0.18\pm0.07$ & $0.98\pm0.05$  \\
\hline
\end{tabular}
\label{tab:IOdispersive}
\end{center}
\end{table}

Following the same correction procedure as that used in Sec.~\ref{sec:pull}, corrections are applied to 
$g_{\pi\eta}^{2}$, $g_{K\bar K}^{2}$, and $M_{\rm pole}$ according to the fitted pull means. 
The corrected results are $g_{\pi\eta,c}^{2} = (0.309\pm0.026_{\rm stat}\pm0.021_{\rm syst})~{\rm GeV}^{2}/c^{4}$, $g_{K\bar K,c}^{2} = (0.435\pm0.064_{\rm stat}\pm0.076_{\rm syst})~{\rm GeV}^{2}/c^{4}$ and $M_{{\rm pole},c} = (1.085\pm0.009_{\rm stat}\pm0.009_{\rm syst})~{\rm GeV}/c^{2}$.
Even after considering the systematic uncertainties, the corrected pole mass remains well above the pole mass calculated from the BESIII reference parameters.

%==============================================================================
%==============================================================================

\subsubsection{Tests with additional amplitudes in the floating dispersive fit}
\label{sec:addampdispersivefloat}

The same additional amplitudes as those considered in Sec.~\ref{sec:testsetB} are tested in the floating-parameter dispersively modified Flatt\'e fit. 
%The constant term is excluded from these tests, following the discussion in Sec.~\ref{sec:testsetB}. 
For each additional amplitude, the fit is repeated with $M_0$, $g_{\pi\eta}^2$, and $g_{K\bar K}^2$ allowed to float, while $g_{\pi\eta'}^2$ is fixed to the value reported in Ref.~\cite{BESIII:2016tqo}. 
The resulting $\Delta\ln\mathcal{L}$ values, fit qualities, significances of the additional amplitudes, and shifts in the fitted $a_0(980)$ parameters and pole quantities are summarized in Table~\ref{tab:dispersiveFit1b}.
Here, ``NA'' indicates that no pole associated with the $a_0(980)$ is found on the $(-++)$ sheet; such fits are not used in the evaluation of the systematic uncertainty from additional amplitudes. 
For the ``IX constant term'' and alternative approach of model-independent $(\eta\eta)_{S-{\rm wave}}$, both are not considered due to the unstable fits. 
%The corresponding details can be found at App.~\ref{sec:MIPWAfpars}. 

\begin{table}[htbp]
\begin{center}
\caption{Fit results obtained by adding one additional amplitude to the floating-parameter dispersively modified Flatt\'e baseline model. 
% The constant term is not included. 
Results for model IX and the two model-independent solutions are presented for completeness, but they are not included in the evaluation of the systematic uncertainty because of the unstable fits.
The $\Delta\ln\mathcal{L}$, significance, and all parameter and pole shifts are evaluated relative to the baseline fit with the dispersive parameters floated.
The shifts are given in units of $10^{-2}$, with ${\rm GeV}/c^2$ for $\Delta M_0$, $\Delta M_{\rm pole}$, and $\Delta(\Gamma_{\rm pole}/2)$, and ${\rm GeV}^2/c^4$ for $\Delta g_{\pi\eta}^2$ and $\Delta g_{K\bar K}^2$. 
Here, ``NA'' indicates that no pole associated with the $a_0(980)$ is found on the $(-++)$ sheet.}
\resizebox{0.5\textwidth}{!}{%
\begin{tabular}{c|ccc|ccc|cc} \hline 
Amplitude 
& $\Delta\ln\mathcal{L}$ 
& $\chi^2/{\rm NDOF}$ 
& Sig. ($\sigma$) 
& $\Delta M_0$ 
& $\Delta g_{\pi\eta}^{2}$ 
& $\Delta g_{K\bar K}^{2}$ 
& $\Delta M_{\rm pole}$ 
& $\Delta(\Gamma_{\rm pole}/2)$ \\ \hline
I    & $+0.4$ & $85.4/86$ & 0.5 & $+0.5$ & $-1.0$ & $+3.0$ & NA     & NA     \\
II   & $+2.4$ & $84.1/86$ & 1.6 & $+0.8$ & $-1.1$ & $+4.6$ & NA     & NA     \\
III  & $+0.4$ & $85.0/86$ & 0.4 & $+0.4$ & $-1.6$ & $+2.4$ & NA     & NA     \\
IV   & $+0.2$ & $84.8/86$ & 0.2 & $-0.1$ & $-0.0$ & $+1.0$ & $-0.1$ & $-0.3$ \\
V    & $+1.1$ & $83.0/86$ & 1.0 & $-0.4$ & $-1.7$ & $-3.0$ & $-0.7$ & $+0.3$ \\
VI   & $+0.3$ & $83.6/86$ & 0.3 & $-0.2$ & $-1.3$ & $-1.9$ & $-0.4$ & $+0.1$ \\
VII  & $+0.3$ & $84.4/86$ & 0.3 & $-0.4$ & $-0.9$ & $-2.0$ & $-0.5$ & $+0.3$ \\
VIII & $+0.2$ & $84.2/86$ & 0.2 & $+0.1$ & $-0.8$ & $-1.5$ & $-0.1$ & $+0.0$ \\
IX   & $+0.7$ & $85.4/86$ & 0.7 & $+1.5$ & $-3.2$ & $+5.6$ & NA     & NA     \\
${\rm IX^{\rm MI_1}}$&$+6.7$&$81.2/78$&1.3&$+0.4$&$+1.1$&$+4.2$&$+0.7$&$-0.8$\\ 
${\rm IX^{\rm MI_2}}$&$+12.9$&$79.1/78$&2.9&$+0.8$&$-2.2$&$+7.1$& NA& NA     \\
X    & $+1.1$ & $85.8/86$ & 1.0 & $-1.0$ & $-1.9$ & $+6.7$ & NA     & NA     \\
XI   & $+0.6$ & $84.1/86$ & 0.6 & $-0.3$ & $-0.9$ & $-1.0$ & $-0.4$ & $+0.0$ \\
XII  & $+0.2$ & $84.0/86$ & 0.2 & $-0.1$ & $-0.4$ & $-1.4$ & $-0.2$ & $+0.2$ \\
\hline
\end{tabular}
}
\label{tab:dispersiveFit1b}
\end{center}
\end{table}

As shown in Table~\ref{tab:dispersiveFit1b}, once the dispersive parameters are allowed to float, none of the additional amplitudes has a significance above $3\sigma$. 
This is different from the fixed-parameter dispersive tests in Sec.~\ref{sec:testsetB}, where some additional amplitudes showed sizable apparent significances. 
The result indicates that the improvements previously attributed to additional amplitudes can largely be absorbed by changes in the $a_0(980)$ line shape.

To test whether a missing additional amplitude could mimic the observed shift of the $a_0(980)$ parameters, further fits are performed in which the complex coefficients of the additional amplitudes are fixed to the values obtained in Sec.~\ref{sec:testsetB}, while the dispersive parameters are refitted. 
% The resulting $\ln\mathcal{L}$ values, fit qualities, fitted $a_0(980)$ parameters, and pole positions are listed in Table~\ref{tab:dispersivefixcoeff} of Appendix~\ref{app:fixed_addamp_tests}.
The resulting $\Delta\ln\mathcal{L}$ values, fit qualities, fitted $a_0(980)$ parameters, and pole positions are listed in Table~\ref{tab:dispersivefixcoeff} of Appendix~\ref{app:fixed_addamp_tests}.
Even with such artificially fixed additional components, the pole mass cannot be brought back to the near-threshold pole position calculated from the BESIII reference parameters.

% In the fits to the models with additional amplitudes, after floating all the parameters, the contributions from the additional amplitudes are negligible, which 
% are different with the case observed in Sec.~\ref{sec:testsetB}. 
% %
% So new tests with fixing the coefficients of additional amplitudes at the results presented in Sec.~\ref{sec:testsetB} are performed.
% %
% The $\ln\mathcal{L}$, fit quality, values of $a_{0}(980)$ parameters and the pole positions are listed in Tab.~\ref{tab:dispersivefixcoeff}. 
% %
% Similar with the case that $a_{0}(980)$ line shape parameterized with Flatt\'e formula, the pole position still can not be restored with artificially considering an additional state with the $a_{0}(980)$ line shape parameterized with dispersive formula. 
%===================================================================

%===================================================================

As in Sec.~\ref{sec:addampflattefloat}, additional tests are performed by floating only $M_0$ while fixing the coupling constants to the BESIII values~\cite{BESIII:2016tqo}. 
For each additional amplitude, three strategies are tested: fixing the full complex coefficient, fixing only its magnitude, or fixing only its phase to the value obtained in Sec.~\ref{sec:testsetB}. 
The corresponding dispersively modified Flatt\'e results are summarized in the dispersive columns of Tables~\ref{tab:finaltests0_combined}, \ref{tab:finaltests1_combined}, and \ref{tab:finaltests2_combined} in Appendix~\ref{app:fixed_addamp_tests}, respectively.
% The resulting $\ln\mathcal{L}$ values, fit qualities, $M_0$, and $M_{\rm pole}$ values are listed in Tables~\ref{tab:dispersivefinaltests0}, \ref{tab:dispersivefinaltests1}, and \ref{tab:dispersivefinaltests2}, respectively.

In all these tests, the pole mass remains well above the near-threshold value calculated from the BESIII reference parameters. 
This indicates that possible additional amplitudes, even when introduced with coefficients fixed from the fixed-parameter studies, cannot provide a robust solution to the observed pole-position shift.

These studies show that, as in the Flatt\'e case, the dispersively modified Flatt\'e parameterization can describe the data only by shifting the pole mass away from the near-threshold region. 
The additional-amplitude tests do not bring the pole mass back to the value calculated from the BESIII reference parameters, demonstrating the same tension between fit quality and the physical pole position.

\section{Summary of the tests}

The studies presented above show that the observed $a_0(980)$ line shape in $D^+\to a_0(980)^+\eta$, $a_0(980)^+\to\pi^+\eta$, cannot be satisfactorily described by the conventional direct-production amplitude models tested in this work. 
With the reference $a_0(980)$ parameters, the Flatt\'e, dispersively modified Flatt\'e, $T$-matrix, and $K$-matrix descriptions do not reproduce the data satisfactorily, even when additional conventional resonant or nonresonant amplitudes are included. 

When the $a_0(980)$ parameters are allowed to float in the Flatt\'e or dispersively modified Flatt\'e parameterizations, the fit quality is significantly improved. 
However, the resulting pole mass is driven well above the $K\bar K$ threshold and cannot be brought back to the near-threshold region by systematic variations or by introducing additional amplitudes. 
Thus, the improved fit quality is obtained at the cost of a pole position that is difficult to reconcile with the usual near-threshold character of the $a_0(980)$.

These tests indicate that the distortion observed in the $a_0(980)$ line shape cannot be interpreted as evidence for a stable additional intermediate resonance within the conventional amplitude models considered here. 
Instead, the results point to a limitation of simple direct-production descriptions and suggest that mechanisms beyond the conventional tree-level amplitude picture may be relevant.

\section{Branching fraction measurement}

To measure the branching fraction of $D^+\to\pi^+\eta\eta$, a selection optimized for the branching-fraction measurement is applied. 
The requirements $0.505<M(\gamma\gamma)_\eta<0.570~{\rm GeV}/c^2$ and $\chi^2(\eta)<50$ are imposed instead of the MVA selection used in the amplitude analysis.

The branching fraction is measured with the DT method~\cite{MARK-III:1985hbd}:
\begin{equation}
    \mathcal{B}(D^+\to\pi^+\eta\eta) = \frac{Y_{\rm DT}}{Y_{\rm ST}\,\epsilon_{\rm sig}\,\mathcal{B}_{\rm sub}^{2}},
\end{equation}
where $Y_{\rm DT}$ is the DT yield, $Y_{\rm ST}$ is the total single tag (ST) yield, $\epsilon_{\rm sig}$ is the weighted signal efficiency, and $\mathcal{B}_{\rm sub}$ denotes the branching fraction of $\eta\to\gamma\gamma$. 
The total ST yield is $Y_{\rm ST}=(10646.9\pm3.8)\times10^3$. 
The weighted signal efficiency is defined as
\begin{equation}
    \epsilon_{\rm sig} = \frac{\sum_i Y_{\rm ST}^{(i)}\,\epsilon_{\rm DT}^{(i)}/\epsilon_{\rm ST}^{(i)}}{Y_{\rm ST}},
\end{equation}
where $Y_{\rm ST}^{(i)}$, $\epsilon_{\rm ST}^{(i)}$, and $\epsilon_{\rm DT}^{(i)}$ are the ST yield, ST efficiency, and DT efficiency for the $i^{\rm th}$ tag mode, respectively.

\begin{figure}[htbp]
\begin{center}
\begin{minipage}[b]{0.35\textwidth}
\epsfig{width=0.98\textwidth,clip=true,file=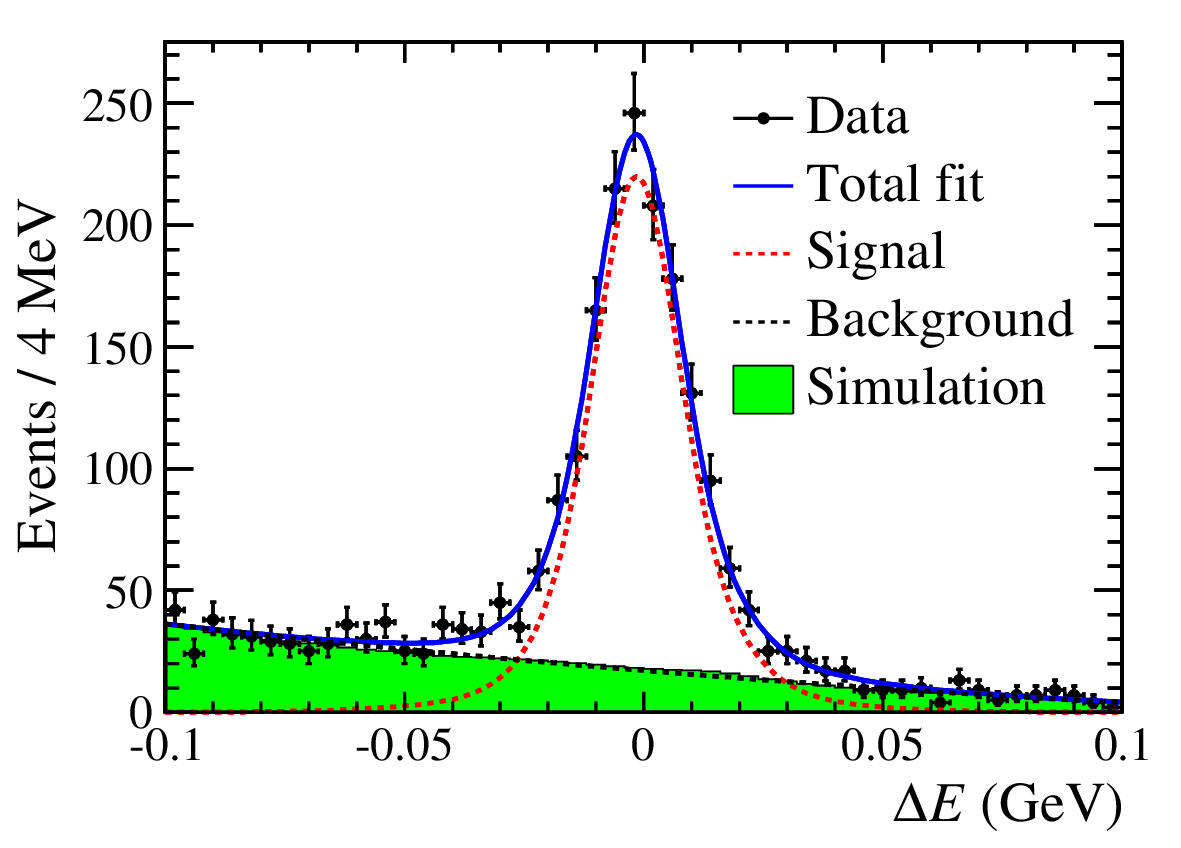}
\end{minipage}
\caption{Fit to the $\Delta E$ distribution. 
The black points with error bars represent data, the solid curve shows the total fit, the red dashed curve shows the signal component, and the black dashed curve shows the background component. 
The green histogram shows the background estimated from the inclusive MC sample.}
\label{fig:DT}
\end{center}
\end{figure}

The DT yield is extracted by fitting the $\Delta E$ distribution before applying the $\Delta E$ signal-window requirement, as shown in Fig.~\ref{fig:DT}. 
The signal shape is described by the sum of a bifurcated Gaussian function and a double-Gaussian function. 
The bifurcated Gaussian function is a modified Gaussian function with different widths on the left and right sides of the peak. 
The two signal components share a common mean value. 
All signal-shape parameters except the common mean are fixed to the values obtained from the signal MC sample. 
The background shape is described by a second-order Chebyshev polynomial, which is validated with the inclusive MC sample. 
The fit gives $Y_{\rm DT}=1506\pm49$.

The DT efficiency $\epsilon_{\rm DT}^{(i)}$ is determined with the signal MC sample. 
The signal MC sample is generated according to the amplitude model obtained from the fit to data using the Flatt\'e parameterization with floated $a_0(980)$ parameters. 
This model reproduces the observed kinematic distributions and is used for the efficiency determination. 
The uncertainty associated with the generator model is evaluated by varying the model parameters according to the covariance matrix from the fit to data. The branching fraction is measured to be
\begin{equation}
    \mathcal{B}(D^+\to\pi^+\eta\eta) = (3.67\pm0.12_{\rm stat}\pm0.06_{\rm syst})\times10^{-3}.
\end{equation}

The systematic uncertainty is estimated to be 1.7\%. 
The contributions are as follows. 
The uncertainties associated with PID, tracking, and $\eta$ reconstructions are 0.5\%, 0.5\%, and 0.6\%, respectively, and are determined from hadronic DT $D\bar D$ control samples~\cite{BESIII:2024njj}. 
The uncertainty from the signal shape is 0.6\%, estimated by varying the fixed signal-shape parameters within their uncertainties and by replacing the nominal signal shape with the sum of a double-sided Crystal Ball function and a Gaussian function. 
The uncertainty from the background shape is 0.7\%, estimated by fixing the polynomial parameters to those obtained from a fit to the simulated background sample from inclusive MC. 
The uncertainty associated with the $M_{\rm BC}$ signal window is found to be negligible. 
The uncertainty due to ST yield is 0.3\%, taken from Ref.~\cite{BESIII:2024njj}. 
The uncertainty from the fit procedure is 0.3\%, estimated with the inclusive MC sample using the same fitting procedure as applied to data. 
The MC-generator uncertainty is found to be negligible by varying the parameters in the generator model according to the covariance matrix obtained from the fit to data. 
The uncertainty from $\mathcal{B}(\eta\to\gamma\gamma)$ is 0.9\% for two $\eta$ mesons, taken from the PDG~\cite{ParticleDataGroup:2024cfk}.

\section{Conclusion}

Using $20.3~{\rm fb}^{-1}$ of $e^+e^-$ collision data collected with the BESIII detector at $\sqrt{s}=3.773~{\rm GeV}$, we perform the first amplitude analysis of the decay $D^+\to\pi^+\eta\eta$. 
A prominent contribution from the decay chain $D^+\to a_0(980)^+\eta$, $a_0(980)^+\to\pi^+\eta$, is observed. 
No additional conventional resonant or non-resonant amplitude is found to provide a stable independent contribution to the signal model. 
The branching fraction of $D^+\to\pi^+\eta\eta$ is measured to be $(3.67\pm0.12_{\rm stat}\pm0.06_{\rm syst})\times10^{-3}$ using the DT method. 
Within the nominal amplitude model, in which only the $D^+\to a_0(980)^+\eta$, $a_0(980)^+\to\pi^+\eta$ decay chain is included, this value corresponds to the product branching fraction of the intermediate process.

The observed $\pi^+\eta$ mass spectrum exhibits an $a_0(980)$ line shape that differs from those observed in other processes. 
A systematic study is performed using several conventional descriptions of the $a_0(980)$ amplitude, including the Flatt\'e parameterization, the dispersively modified Flatt\'e parameterization, the $T$-matrix formalism, and the $K$-matrix formalism. 
With reference $a_0(980)$ parameters, these descriptions do not reproduce the observed line shape satisfactorily, even when additional conventional cascade decay amplitudes are included.

When the $a_0(980)$ parameters are allowed to float in the Flatt\'e and dispersively modified Flatt\'e parameterizations, the fit quality is significantly improved. 
However, the resulting pole mass is driven well above the $K\bar K$ threshold. 
This pole position is difficult to reconcile with the usual near-threshold character of the $a_0(980)$. 
Systematic variations and tests with additional amplitudes do not bring the pole mass back to the near-threshold region.

These results demonstrate a tension between the fit quality and the physical pole position in conventional direct-production amplitude models. 
The observed line shape distortion therefore cannot be explained solely by the choice of the $a_0(980)$ line shape parameterization or by interference with small additional conventional amplitudes. 
This suggests that additional dynamical effects beyond the conventional direct-production amplitude descriptions may be relevant for describing the $a_0(980)$ line shape in $D^+\to\pi^+\eta\eta$.

In addition, the model-independent $(\eta\eta)_{S-{\rm wave}}$ study does not establish a stable and unique separation of this contribution from the dominant $a_0(980)^+\eta$ amplitude with the present data sample.
Replacement of the $\pi\eta$ or $\eta\eta$ S-waves by a model-independent binned amplitude may also shed further light on the $a_{0}(980)$ shape distortion in future analyses with more data. 
% In addition, the model-independent $(\eta\eta)_{S-{\rm wave}}$ study does not establish a stable and unique separation of this contribution from the dominant $a_0(980)^+\eta$ amplitude with the present data sample.
% Future analyses with larger data samples, in which the $\pi^{+}\eta$ or $\eta\eta$ $S$ wave is described by a model-independent binned amplitude, may shed further light on the origin of the observed $a_0(980)$ line-shape distortion.

The BESIII Collaboration thanks the staff of BEPCII (https://cstr.cn/31109.02.BEPC) and the IHEP computing center for their strong support. This work is supported in part by National Key R\&D Program of China under Contracts Nos. 2023YFA1606000, 2023YFA1606704; National Natural Science Foundation of China (NSFC) under Contracts Nos. 12205384, 12175239, 12221005, 11635010, 11735014, 11935015, 11935016, 11935018, 12025502, 12035009, 12035013, 12061131003, 12192260, 12192261, 12192262, 12192263, 12192264, 12192265, 12221005, 12225509, 12235017, 12361141819; the Chinese Academy of Sciences (CAS) Large-Scale Scientific Facility Program; the CAS Center for Excellence in Particle Physics (CCEPP); Joint Large-Scale Scientific Facility Funds of the NSFC and CAS under Contract No. U1832207; CAS under Contract No. YSBR-101; 100 Talents Program of CAS; the Excellent Youth Foundation of Henan Scientific Commitee under Contract No.~242300421044; The Institute of Nuclear and Particle Physics (INPAC) and Shanghai Key Laboratory for Particle Physics and Cosmology; German Research Foundation DFG under Contract No. FOR5327; Istituto Nazionale di Fisica Nucleare, Italy; Knut and Alice Wallenberg Foundation under Contracts Nos. 2021.0174, 2021.0299; Ministry of Development of Turkey under Contract No. DPT2006K-120470; National Research Foundation of Korea under Contract No. NRF-2022R1A2C1092335; National Science and Technology fund of Mongolia; National Science Research and Innovation Fund (NSRF) via the Program Management Unit for Human Resources \& Institutional Development, Research and Innovation of Thailand under Contract No. B50G670107; Polish National Science Centre under Contract No. 2019/35/O/ST2/02907; Swedish Research Council under Contract No. 2019.04595; The Swedish Foundation for International Cooperation in Research and Higher Education under Contract No. CH2018-7756; U. S. Department of Energy under Contract No. DE-FG02-05ER41374.

%\newpage

\clearpage
\newpage
\appendix

\onecolumngrid
\section{The Legendre-moment-weight Formalism}
\label{sec:Legendre}
Assuming $N$ is the number of events for a given mass interval of $m(\eta\eta)$, 
the angular distribution $\frac{dN}{d{\rm cos}\Theta}$ can be expanded in terms of spherical harmonic functions:
\begin{equation}
    \frac{dN}{d{\rm cos}\Theta} = 2\pi \sum^{2L_{\rm max}}_{k=0} \langle Y^0_k \rangle Y^0_k ({\rm cos} \Theta),
    \label{harmonicfunc}
\end{equation}
where, $Y^0_k({\rm cos}\Theta) = \sqrt{(2k+1)/4\pi}P_k({\rm cos}\Theta)$, and $P_k({\rm cos}\Theta)$ is the $k^{\rm th}$ order Legendre polynomial.
In general $D$ decays, only the $S$-, $P$-, and $D$-wave are considered. 
Therefore, 
\begin{equation}
    \begin{aligned}
        & \sqrt{4\pi} \left \langle Y^0_0 \right \rangle = |S|^2 + |P|^2 + |D|^2,\\
        & \sqrt{4\pi} \left \langle Y^0_1 \right \rangle = 2|S||P|{\rm cos}\phi_{SP} + \frac{4}{\sqrt{5}}|P||D|{\rm cos}\phi_{PD},\\
        & \sqrt{4\pi} \left \langle Y^0_2 \right \rangle = \frac{2}{\sqrt{5}}|P|^2 + \frac{2\sqrt{5}}{7}|D|^{2} + 2|S||D|{\rm cos}\phi_{SD},\\
        & \sqrt{4\pi} \left \langle Y^0_3 \right \rangle = \frac{6\sqrt{105}}{35}|P||D|{\rm cos}\phi_{PD},\\
        & \sqrt{4\pi} \left \langle Y^0_4 \right \rangle = \frac{6}{7}|D|^2.
    \end{aligned}   
    \label{eq:generalformalism}
\end{equation}

Here, it is worthwhile mentioning that in $\eta\eta$ combination, the orbital angular momentum can only be even due to the symmetry requirement. 
Therefore, only the $S$-wave and $D$-wave are considered and $|P|=0$, then Eqs.~\ref{eq:generalformalism} are simplified to 
\begin{equation}
    \begin{aligned}
        & \sqrt{4\pi} \left \langle Y^0_0 \right \rangle = |S|^2 + |D|^2,\\
        & \sqrt{4\pi} \left \langle Y^0_2 \right \rangle = \frac{2\sqrt{5}}{7}|D|^{2} + 2|S||D|{\rm cos}\phi_{SD},\\
        & \sqrt{4\pi} \left \langle Y^0_4 \right \rangle = \frac{6}{7}|D|^2.
    \end{aligned}   
    \label{eq:noP_formalism}
\end{equation}
Therefore, $|D|^{2} = \frac{7}{6}\sqrt{4\pi}\left \langle Y^0_4 \right \rangle$ and 
$|S|^{2} = \sqrt{4\pi}\left \langle Y^0_0 \right \rangle - \frac{7}{6}\sqrt{4\pi}\left \langle Y^0_4 \right \rangle$.

\section{Projections for fits with additional amplitudes}
\label{app:addamp_projections}
This appendix presents the complete projection plots for the fits in which one additional conventional amplitude is added to the baseline model. 
The results are shown for the Flatt\'e, dispersively modified Flatt\'e, $T$-matrix, and $K$-matrix parameterizations of the $a_0(980)$ contribution. 
For each tested model, the left panel shows the $M(\pi^+\eta)$ projection and the right panel shows the $M(\eta\eta)$ projection. 
The model labels I--XII follow the definitions given in Table~\ref{tab:addamp}.

% \onecolumngrid
% \subsection{Flatt\'e parameterization}
\begin{figure}[hbp]
\begin{center}
\begin{minipage}[b]{0.24\textwidth}
\epsfig{width=0.98\textwidth,clip=true,file=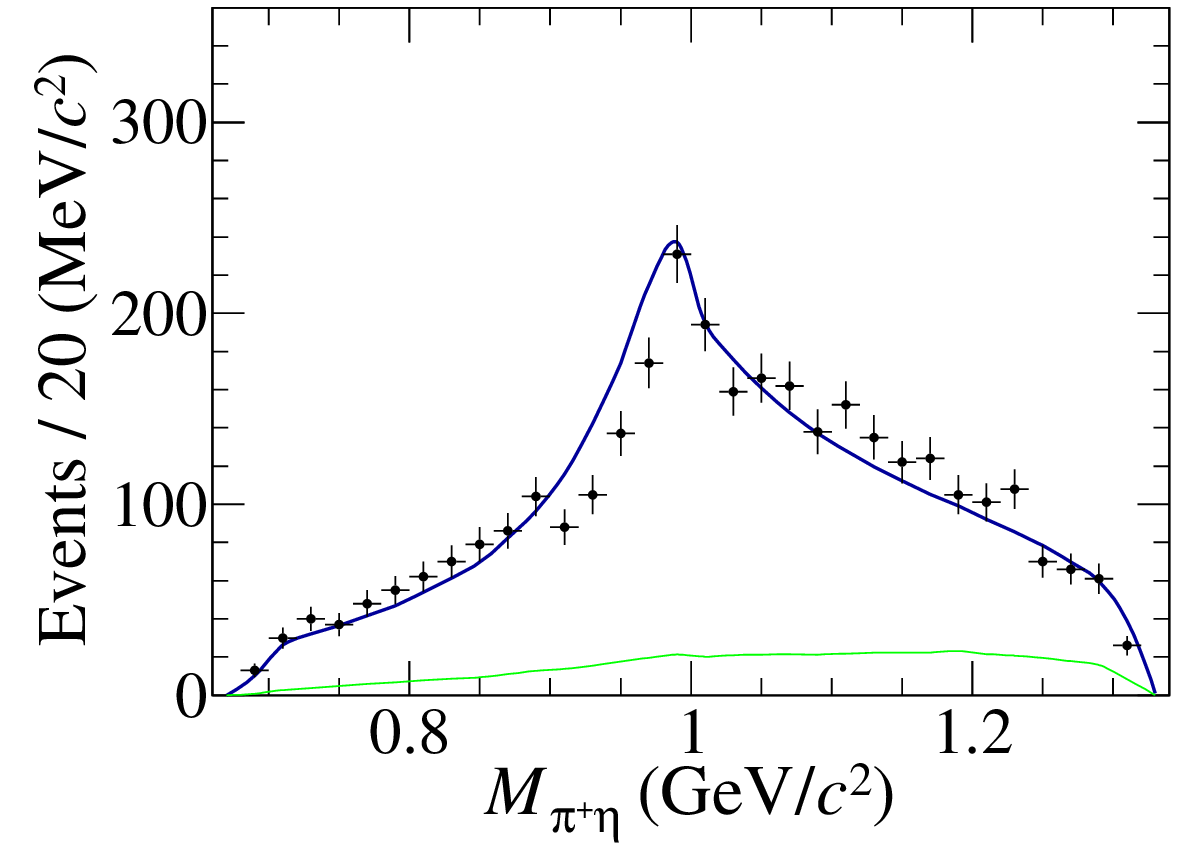}
%\put(-25,50){(a)}
\put(-85,70){I}
\end{minipage}
\begin{minipage}[b]{0.24\textwidth}
\epsfig{width=0.98\textwidth,clip=true,file=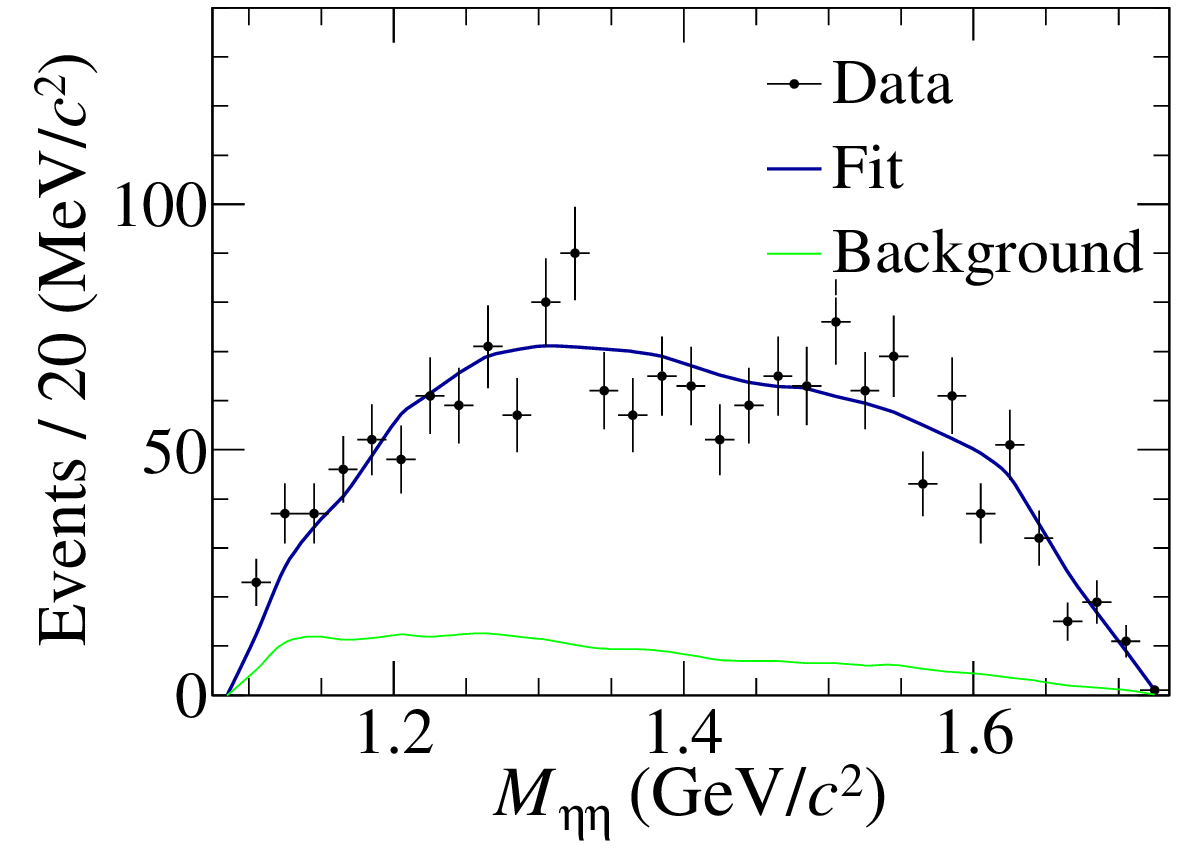}
%\put(-25,50){(b)}
\end{minipage}
\begin{minipage}[b]{0.24\textwidth}
\epsfig{width=0.98\textwidth,clip=true,file=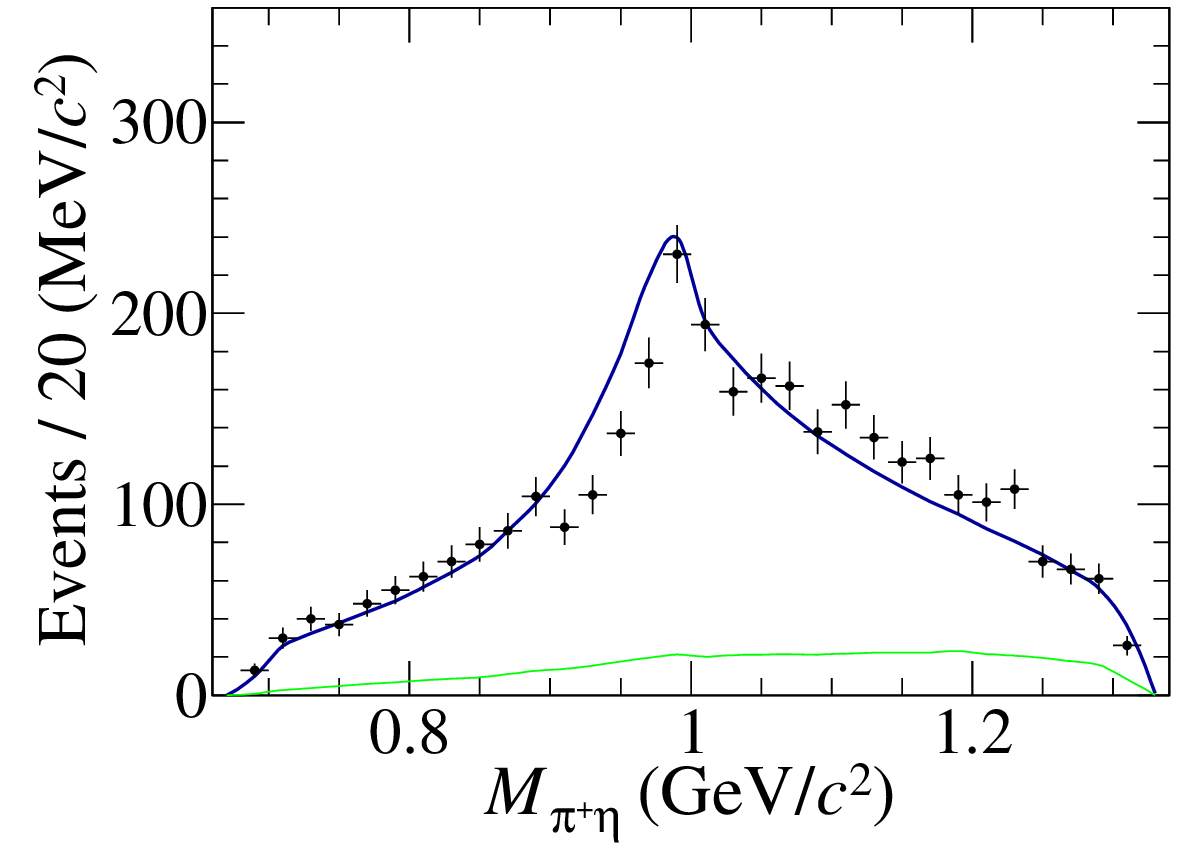}
%\put(-25,50){(c)}
\put(-85,70){II}
\end{minipage}
\begin{minipage}[b]{0.24\textwidth}
\epsfig{width=0.98\textwidth,clip=true,file=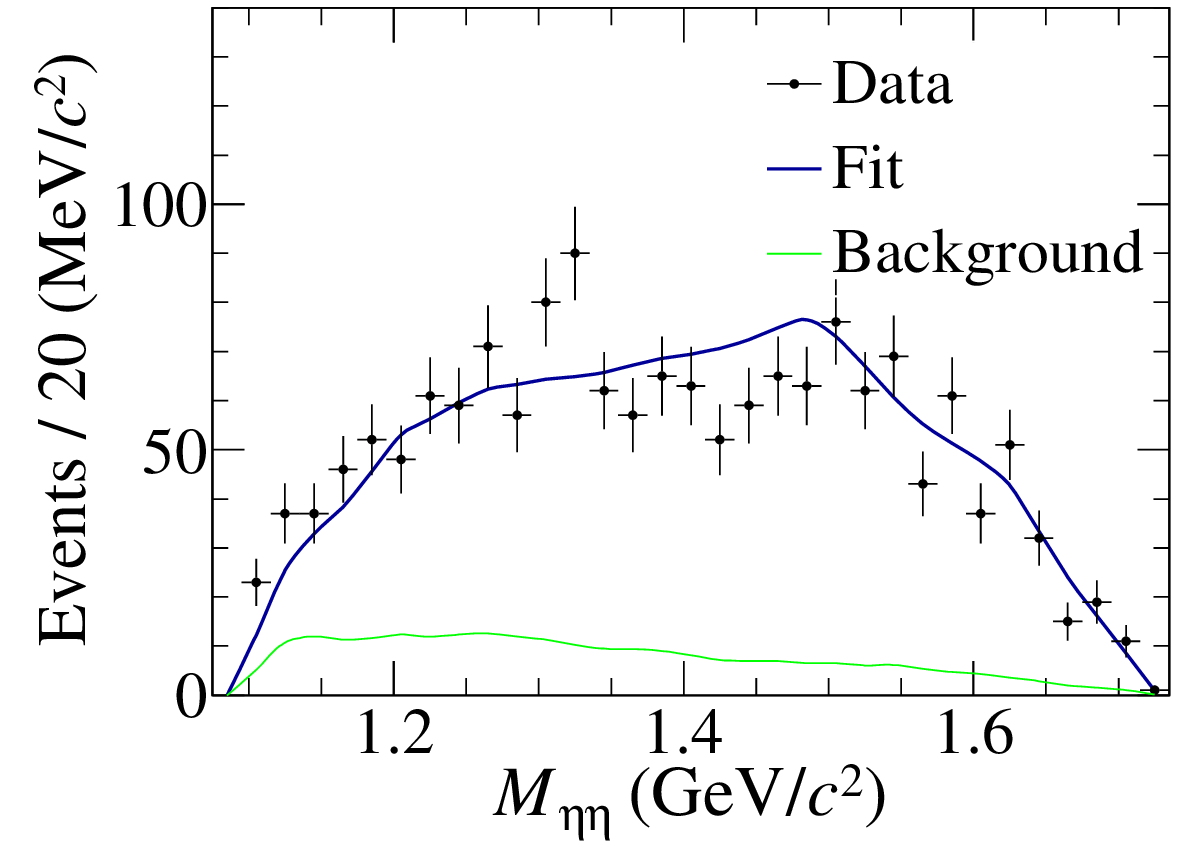}
%\put(-25,50){(d)}
\end{minipage}
\begin{minipage}[b]{0.24\textwidth}
\epsfig{width=0.98\textwidth,clip=true,file=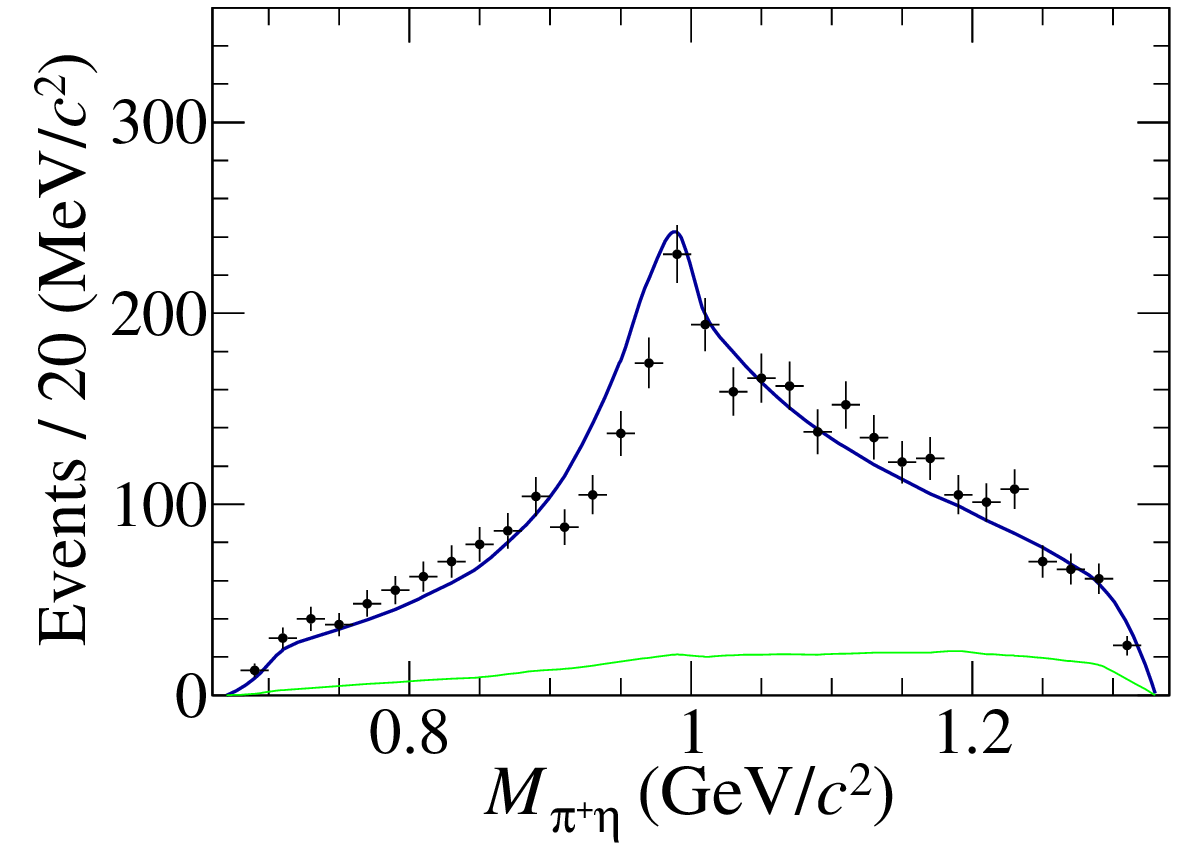}
%\put(-25,50){(e)}
\put(-85,70){III}
\end{minipage}
\begin{minipage}[b]{0.24\textwidth}
\epsfig{width=0.98\textwidth,clip=true,file=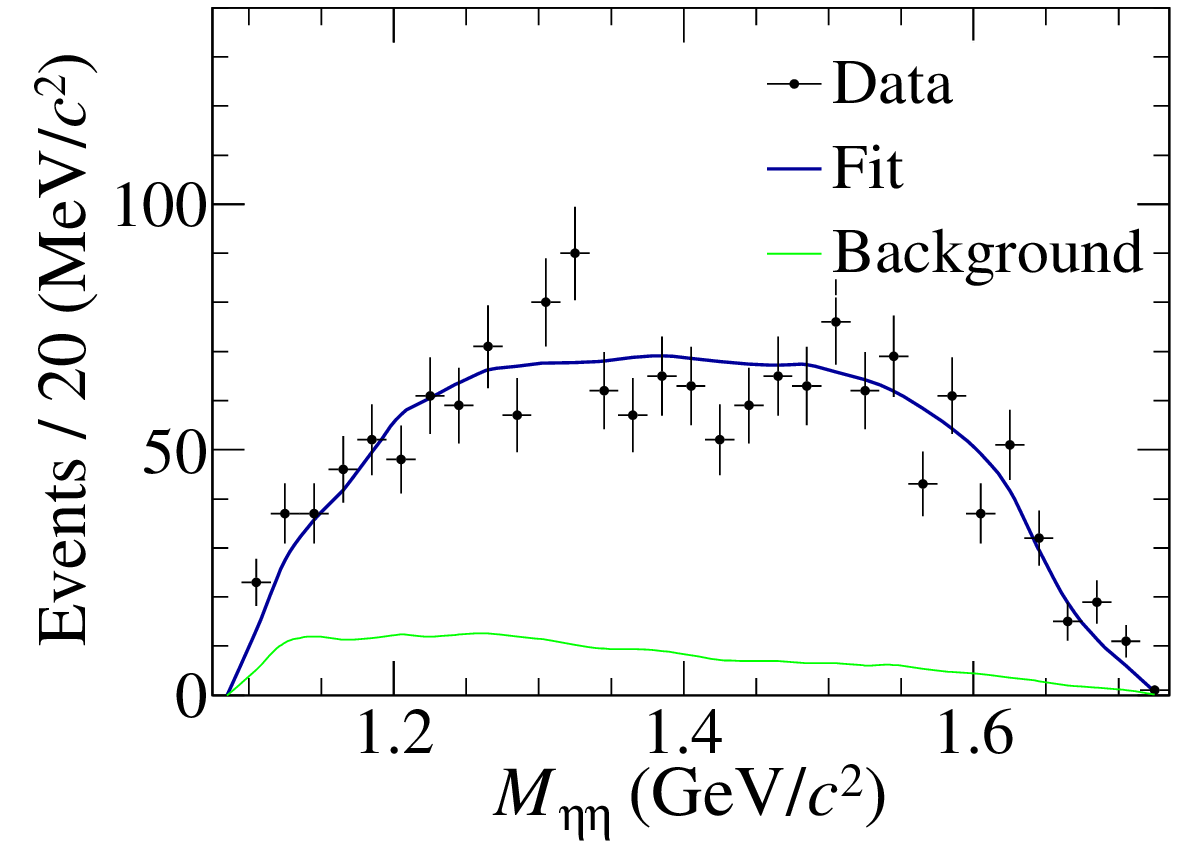}
%\put(-25,50){(f)}
\end{minipage}
\begin{minipage}[b]{0.24\textwidth}
\epsfig{width=0.98\textwidth,clip=true,file=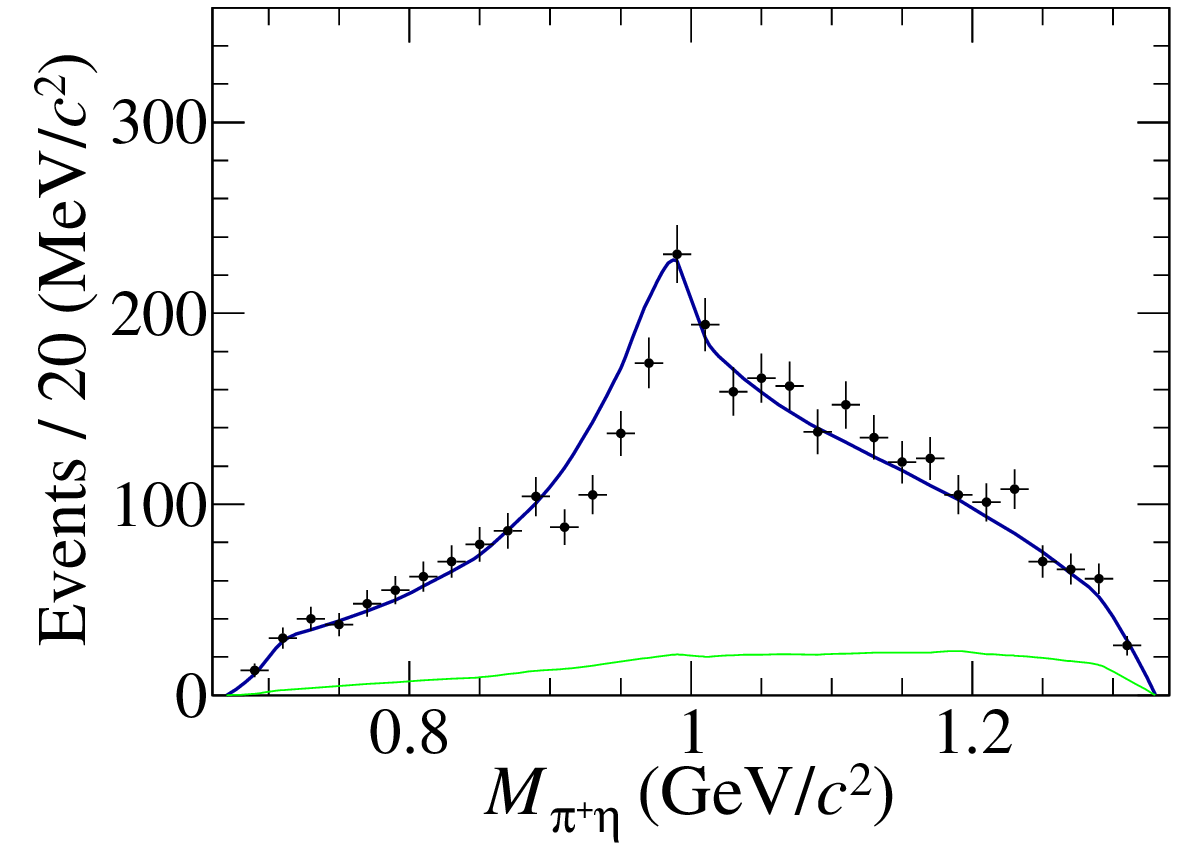}
%\put(-25,50){(g)}
\put(-85,70){IV}
\end{minipage}
\begin{minipage}[b]{0.24\textwidth}
\epsfig{width=0.98\textwidth,clip=true,file=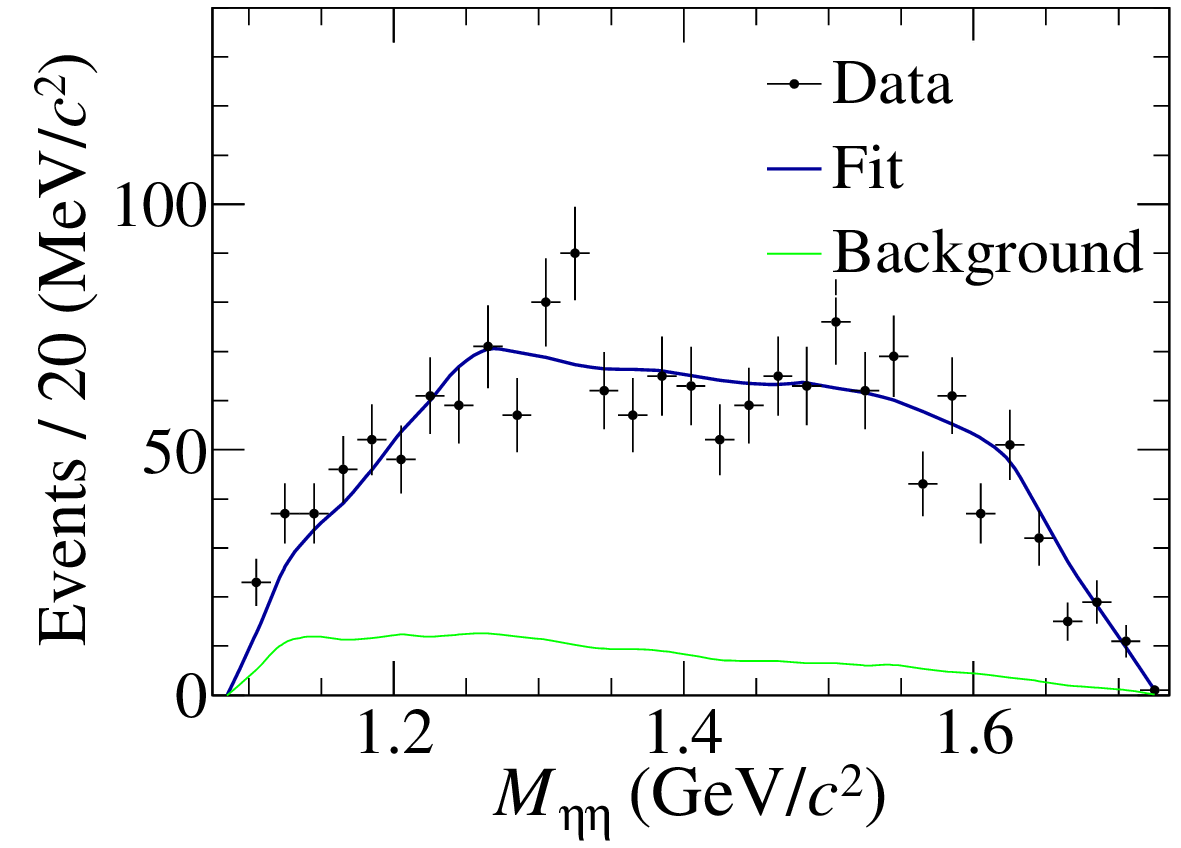}
%\put(-25,50){(h)}
\end{minipage}
\begin{minipage}[b]{0.24\textwidth}
\epsfig{width=0.98\textwidth,clip=true,file=./CLEOcpar/f2_1525/mpipeta_all.eps}
%\put(-25,50){(i)}
\put(-85,70){V}
\end{minipage}
\begin{minipage}[b]{0.24\textwidth}
\epsfig{width=0.98\textwidth,clip=true,file=./CLEOcpar/f2_1525/meta1eta2.eps}
%\put(-25,50){(j)}
\end{minipage}
\begin{minipage}[b]{0.24\textwidth}
\epsfig{width=0.98\textwidth,clip=true,file=./CLEOcpar/f2_1565/mpipeta_all.eps}
%\put(-25,50){(k)}
\put(-85,70){VI}
\end{minipage}
\begin{minipage}[b]{0.24\textwidth}
\epsfig{width=0.98\textwidth,clip=true,file=./CLEOcpar/f2_1565/meta1eta2.eps}
%\put(-25,50){(l)}
\end{minipage}
\begin{minipage}[b]{0.24\textwidth}
\epsfig{width=0.98\textwidth,clip=true,file=./CLEOcpar/f2_1640/mpipeta_all.eps}
%\put(-25,50){(m)}
\put(-85,70){VII}
\end{minipage}
\begin{minipage}[b]{0.24\textwidth}
\epsfig{width=0.98\textwidth,clip=true,file=./CLEOcpar/f2_1640/meta1eta2.eps}
%\put(-25,50){(n)}
\end{minipage}
\begin{minipage}[b]{0.24\textwidth}
\epsfig{width=0.98\textwidth,clip=true,file=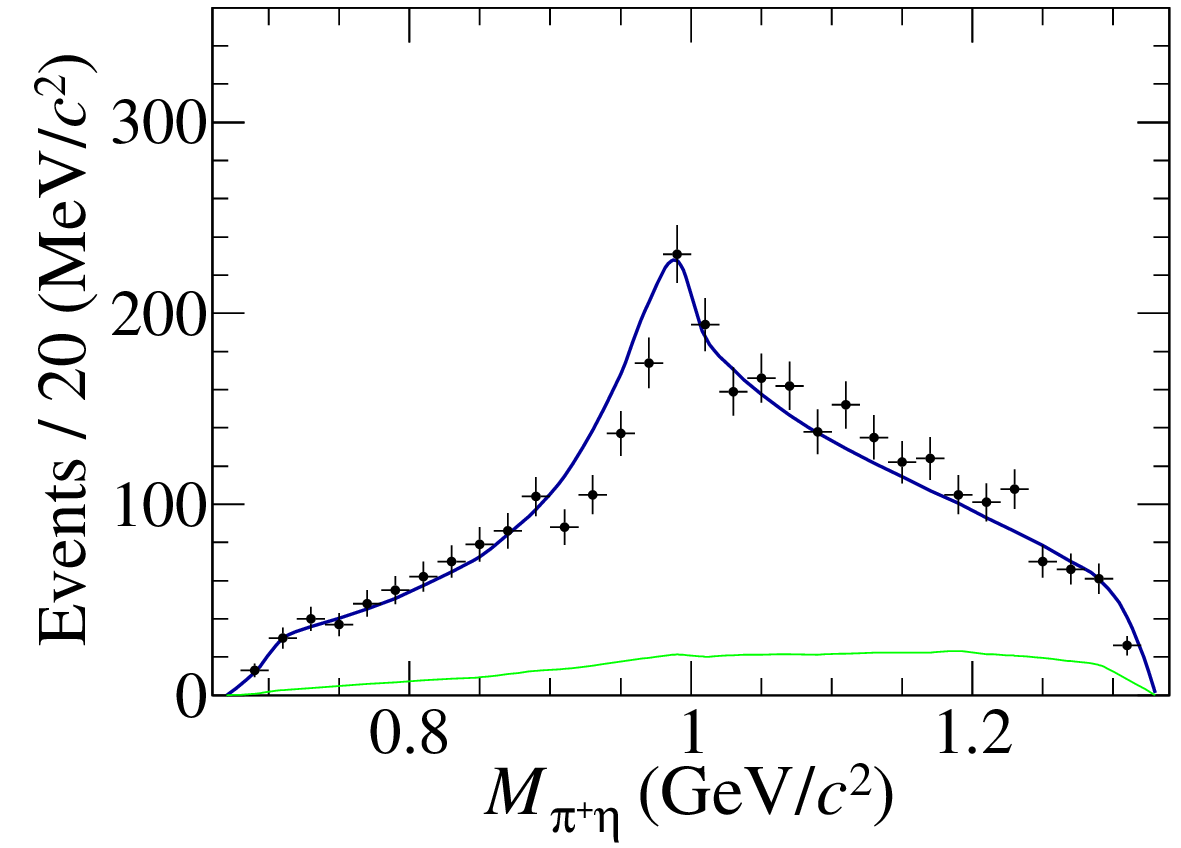}
%\put(-25,50){(o)}
\put(-85,70){VIII}
\end{minipage}
\begin{minipage}[b]{0.24\textwidth}
\epsfig{width=0.98\textwidth,clip=true,file=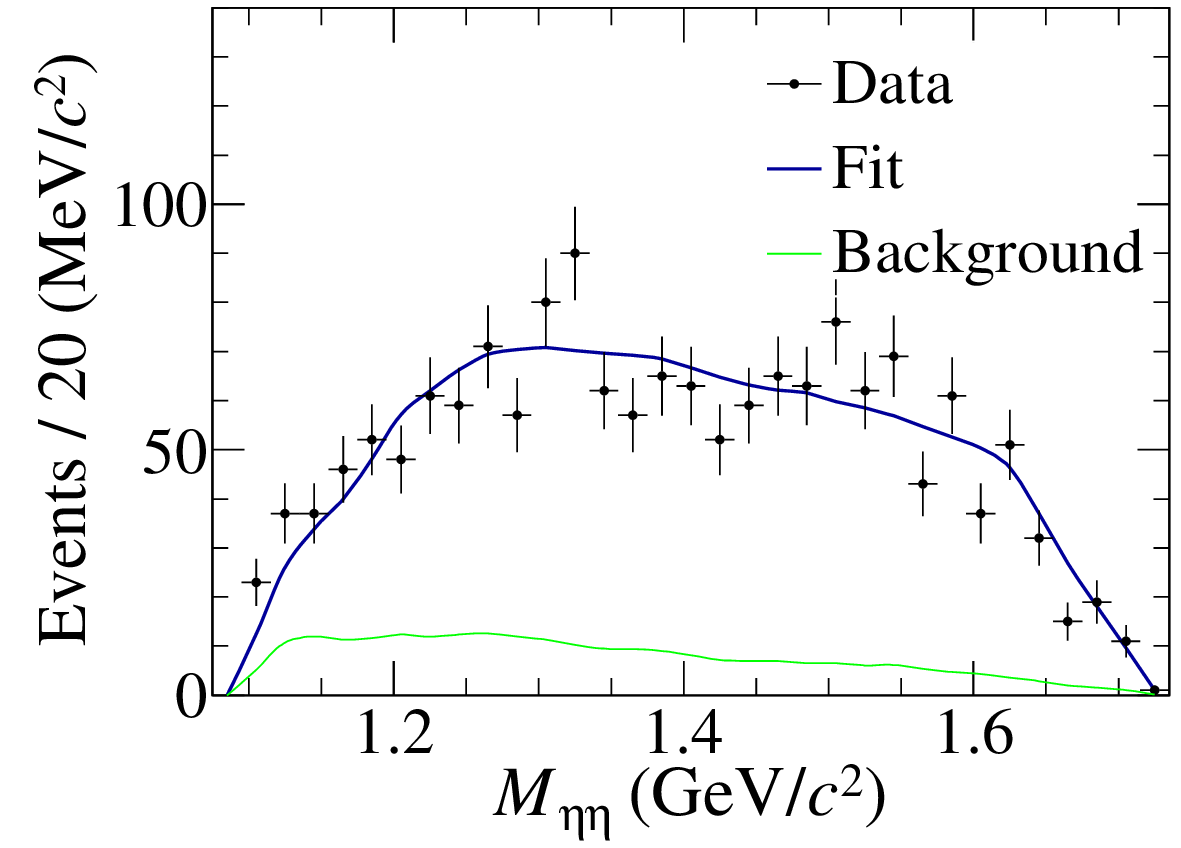}
%\put(-25,50){(p)}
\end{minipage}
\begin{minipage}[b]{0.24\textwidth}
\epsfig{width=0.98\textwidth,clip=true,file=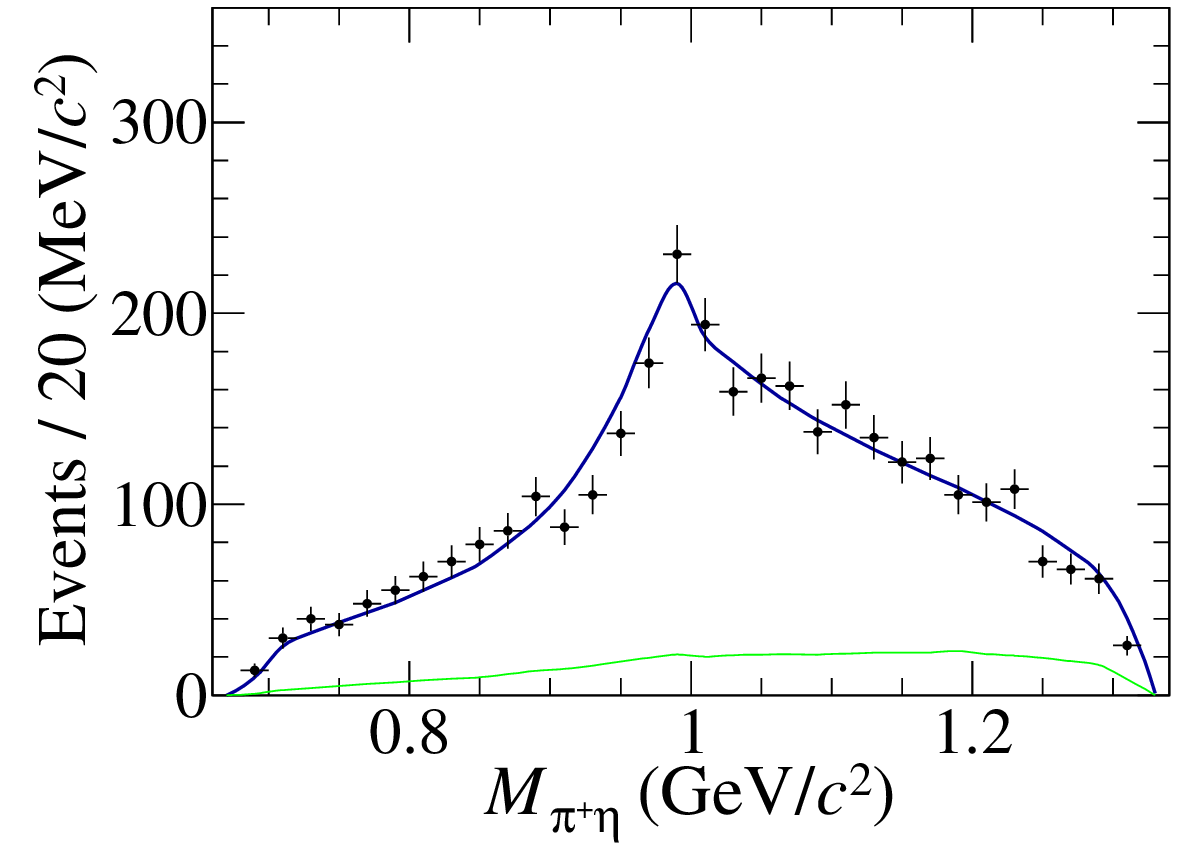}
%\put(-25,50){(q)}
\put(-85,70){IX}
\end{minipage}
\begin{minipage}[b]{0.24\textwidth}
\epsfig{width=0.98\textwidth,clip=true,file=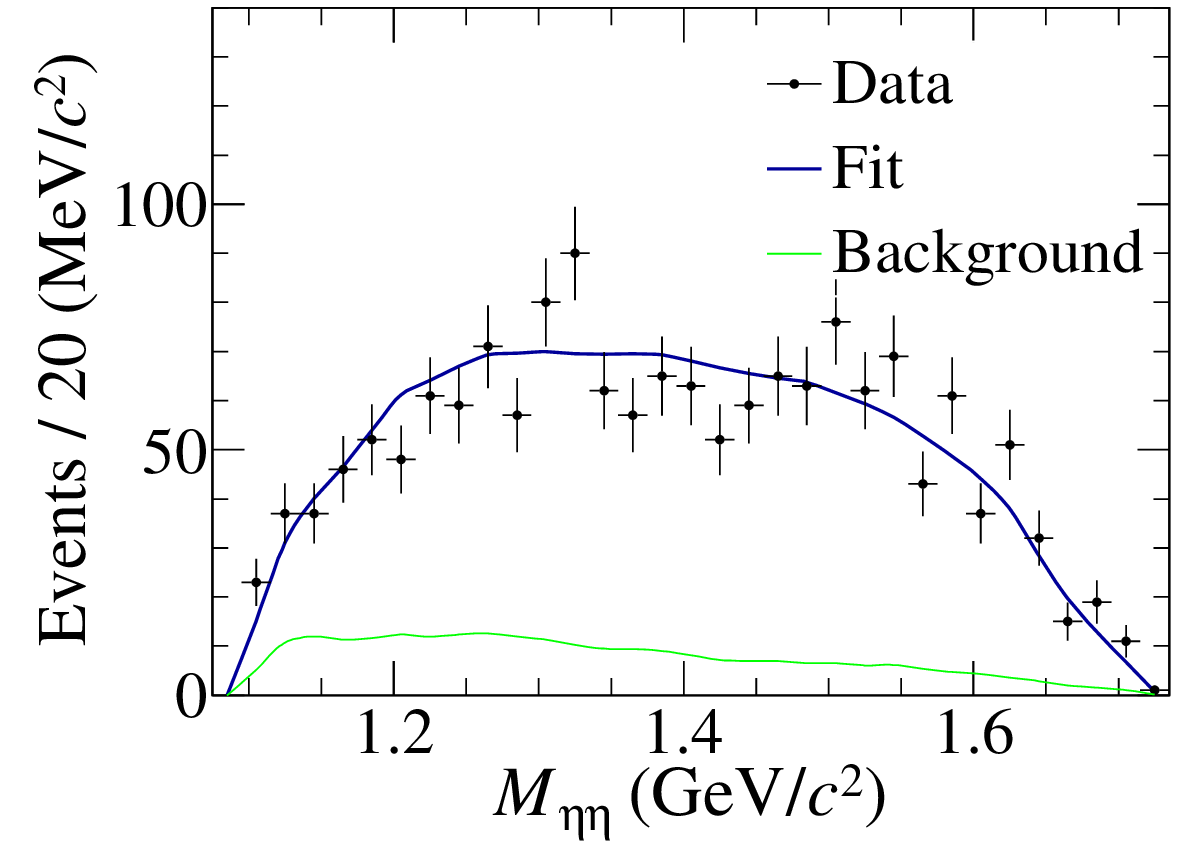}
%\put(-25,50){(r)}
\end{minipage}
\begin{minipage}[b]{0.24\textwidth}
\epsfig{width=0.98\textwidth,clip=true,file=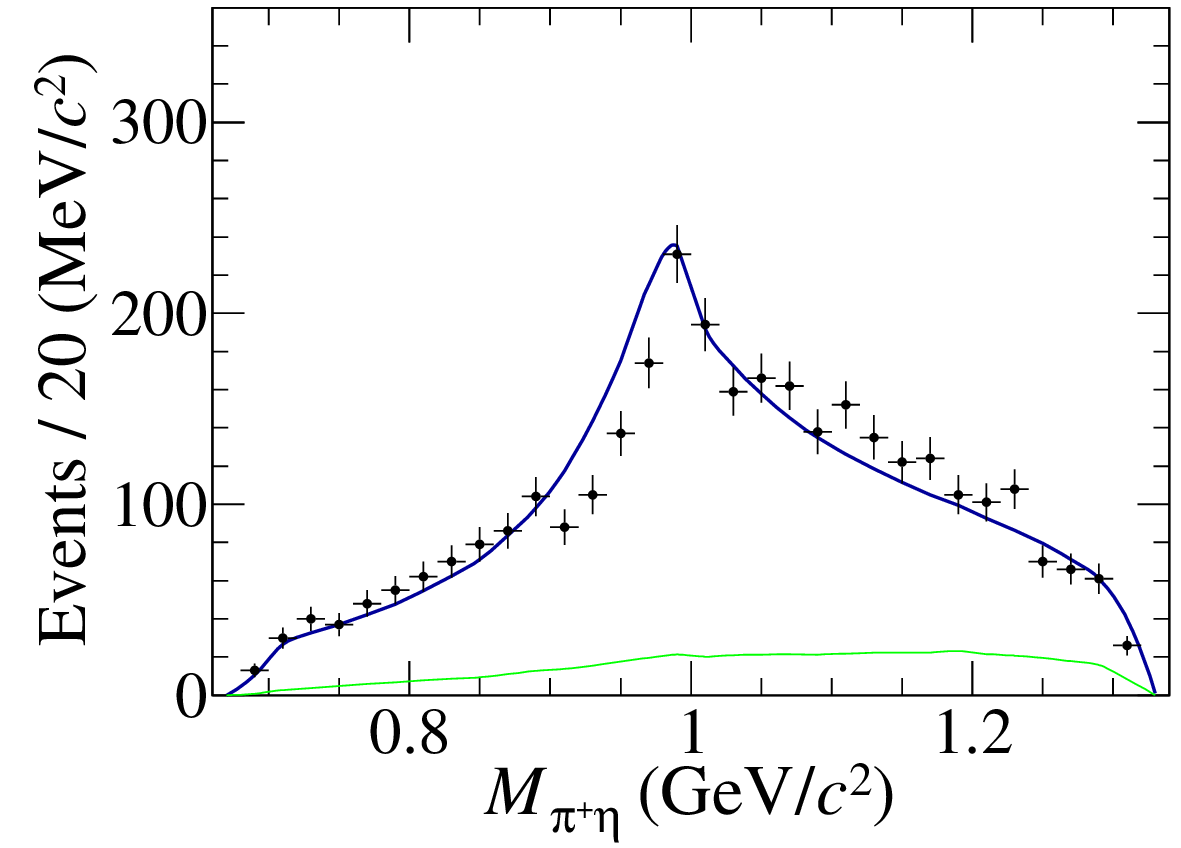}
%\put(-25,50){(s)}
\put(-85,70){X}
\end{minipage}
\begin{minipage}[b]{0.24\textwidth}
\epsfig{width=0.98\textwidth,clip=true,file=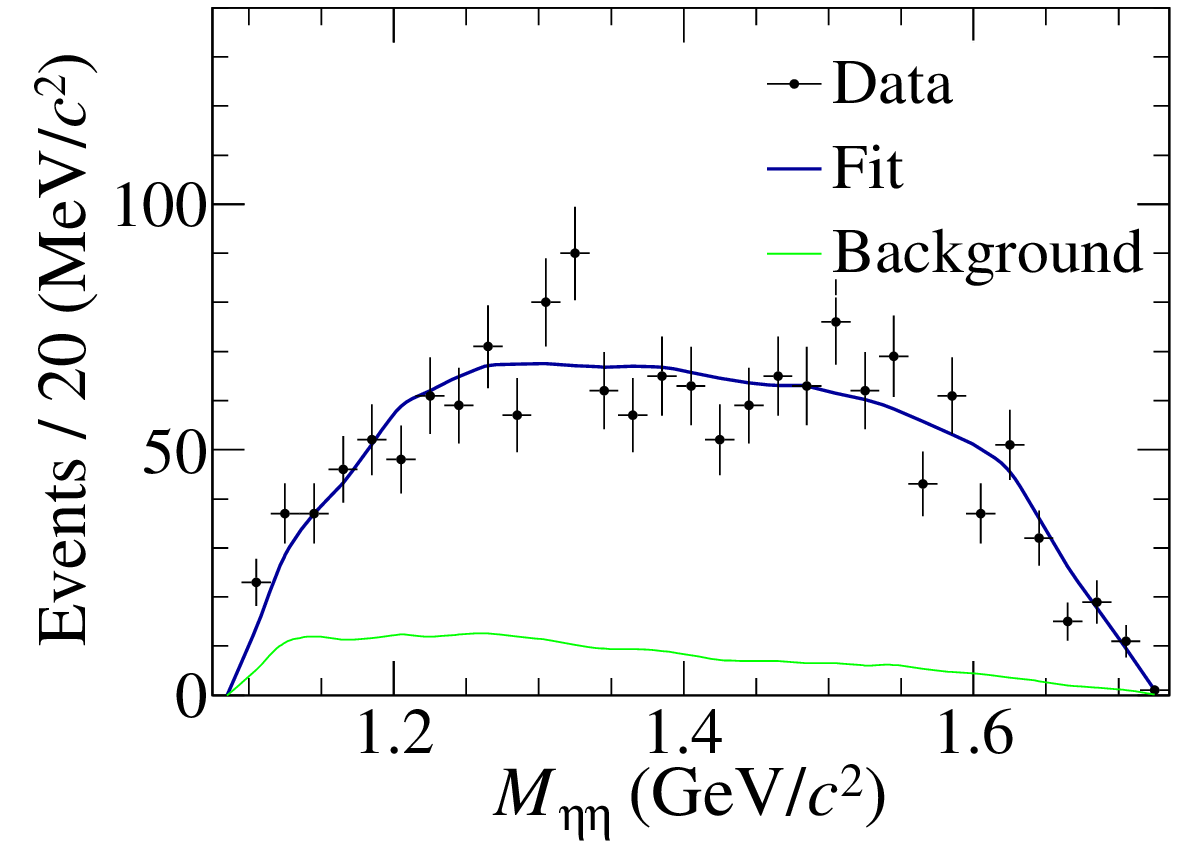}
%\put(-25,50){(t)}
\end{minipage}
\begin{minipage}[b]{0.24\textwidth}
\epsfig{width=0.98\textwidth,clip=true,file=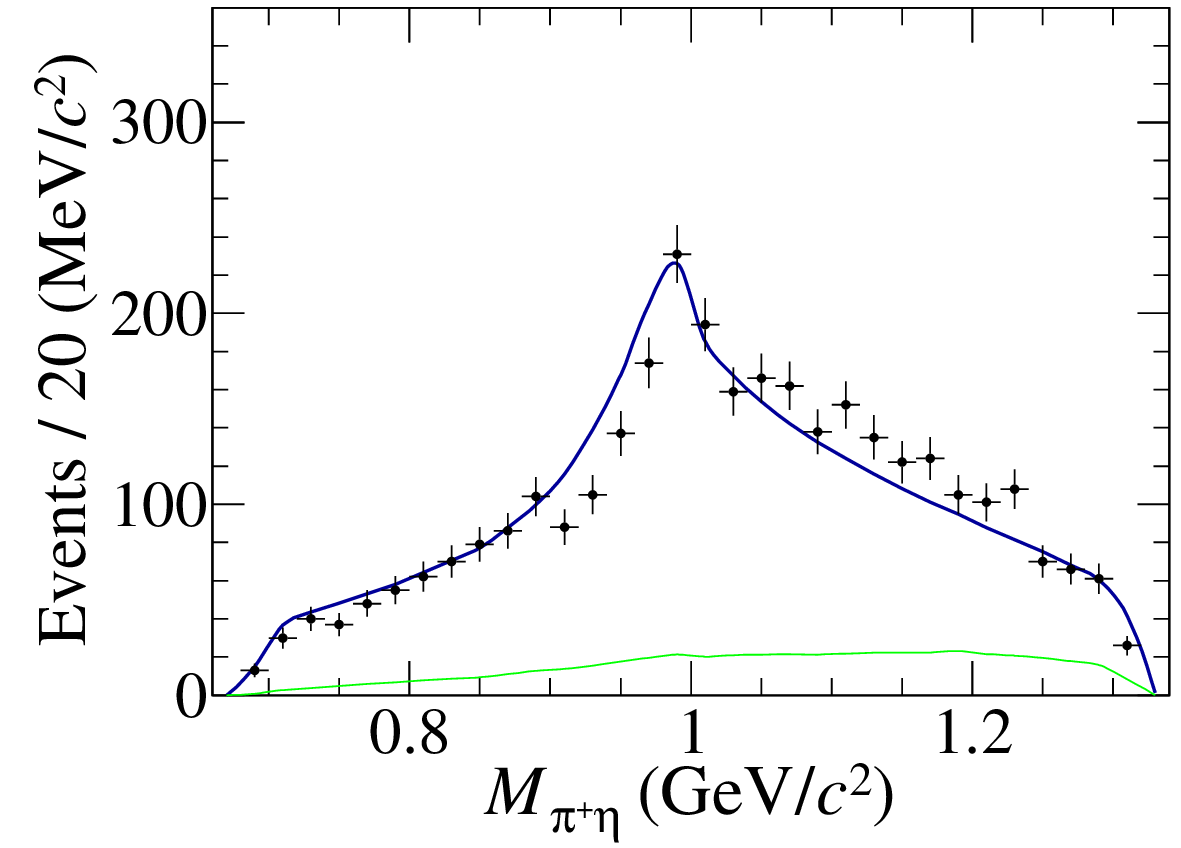}
%\put(-25,50){(u)}
\put(-85,70){XI}
\end{minipage}
\begin{minipage}[b]{0.24\textwidth}
\epsfig{width=0.98\textwidth,clip=true,file=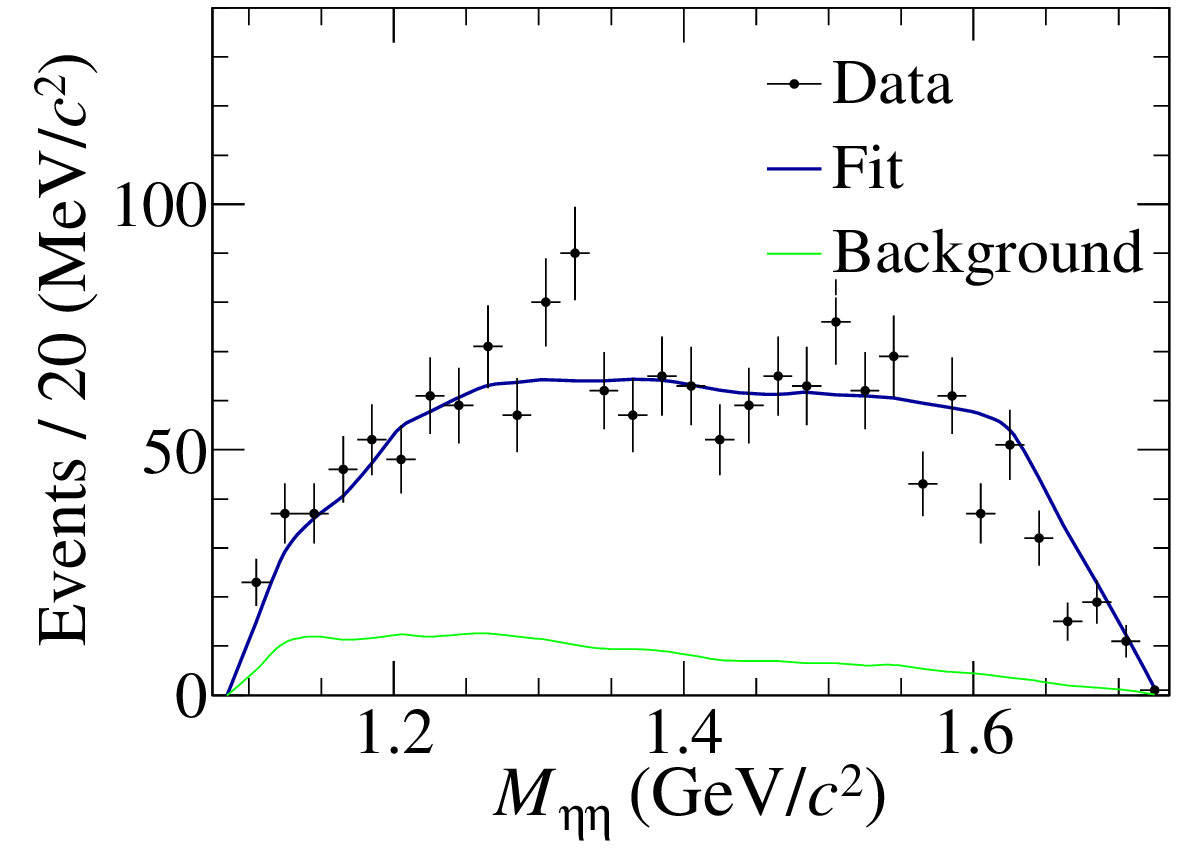}
%\put(-25,50){(v)}
\end{minipage}
\begin{minipage}[b]{0.24\textwidth}
\epsfig{width=0.98\textwidth,clip=true,file=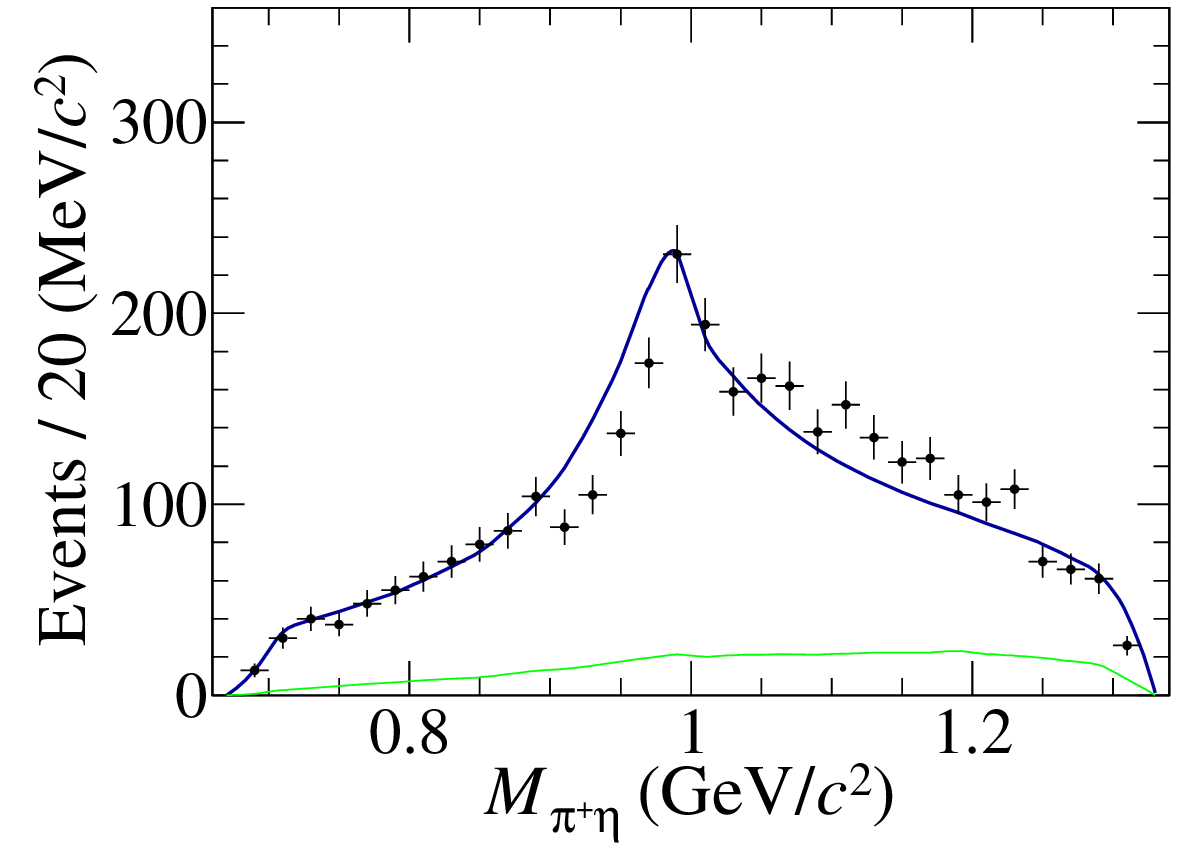}
%\put(-25,50){(w)}
\put(-85,70){XII}
\end{minipage}
\begin{minipage}[b]{0.24\textwidth}
\epsfig{width=0.98\textwidth,clip=true,file=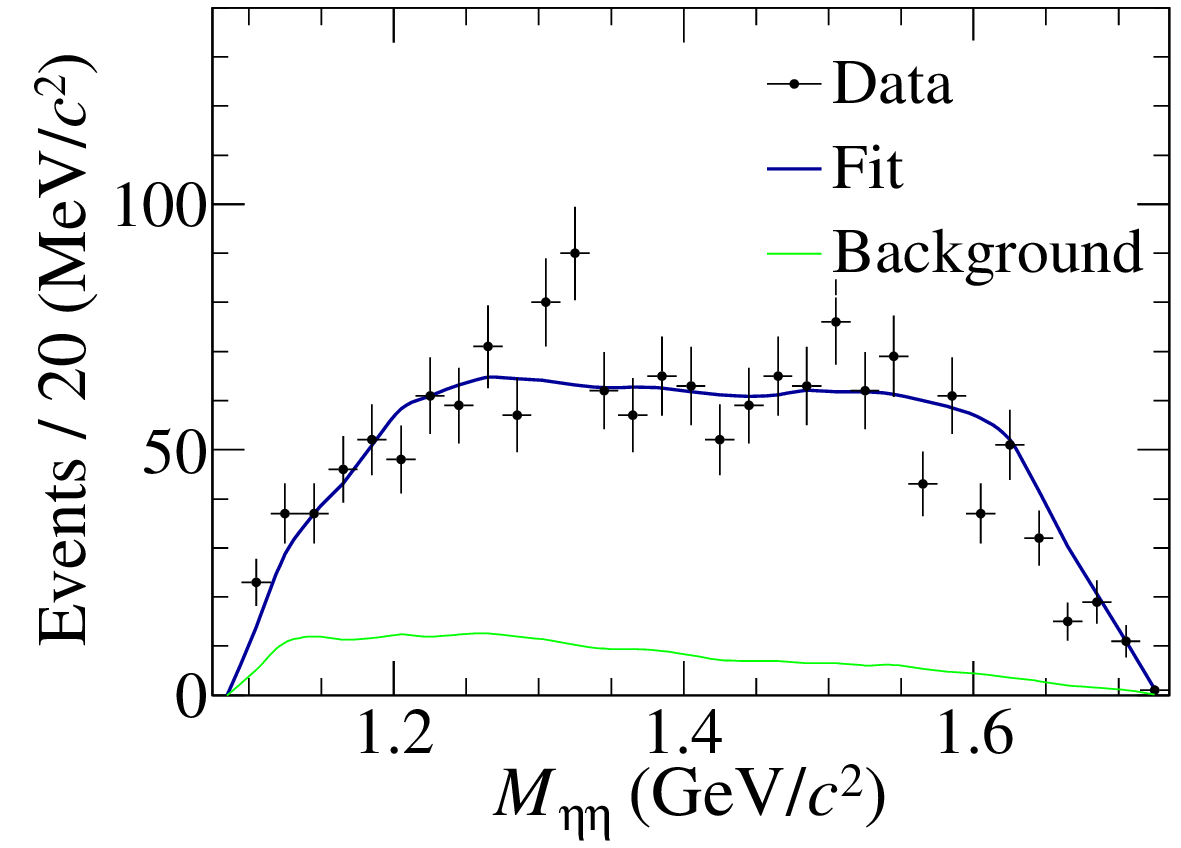}
%\put(-25,50){(x)}
\end{minipage}
\caption{The projections of (first and third columns) $M(\pi^{+}\eta)$ and (second and fourth columns) $M(\eta\eta)$ for model with further adding one amplitude using the Flatt\'e parameterization for $P_{a_{0}(980)}$. The black dots with error bars are the data; the blue lines are the total fit; the green lines are the background. The amplitude numbers are the same as those listed in Table~\ref{tab:addamp}.}
\label{fig:Flatteaddamp}
\end{center}
\end{figure}

% \subsection{Dispersive parameterization}
% \label{app:dispersive_addamp}

\begin{figure}[htbp]
\begin{center}
\begin{minipage}[b]{0.24\textwidth}
\epsfig{width=0.98\textwidth,clip=true,file=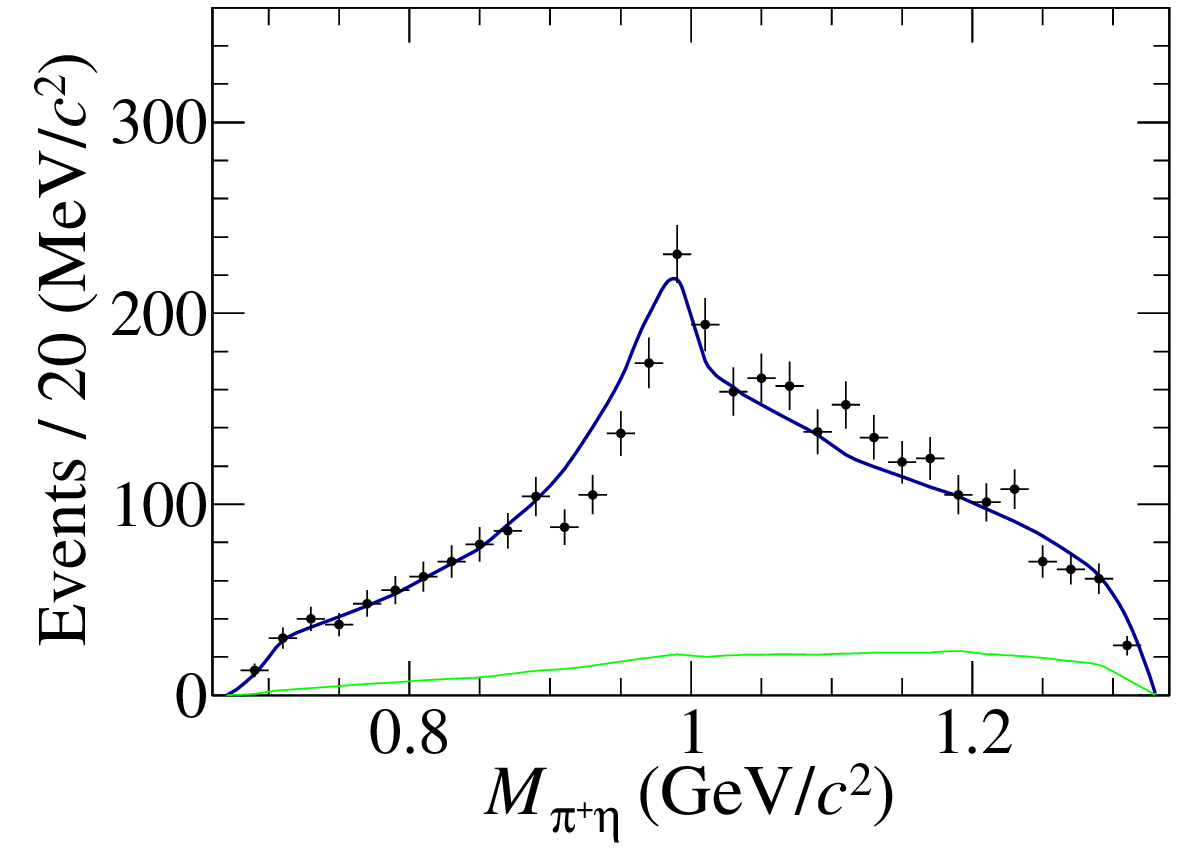}
%\put(-25,50){(a)}
\put(-85,70){I}
\end{minipage}
\begin{minipage}[b]{0.24\textwidth}
\epsfig{width=0.98\textwidth,clip=true,file=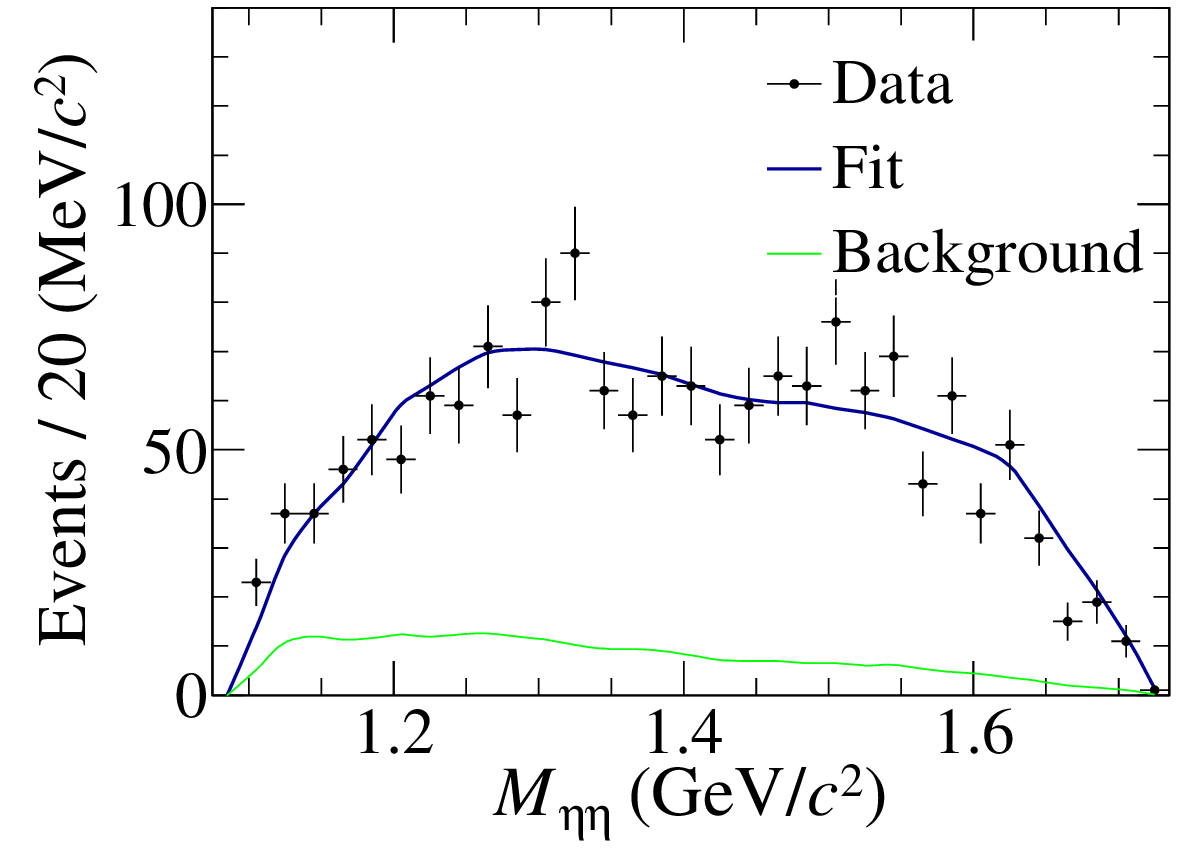}
%\put(-25,50){(b)}
\end{minipage}
\begin{minipage}[b]{0.24\textwidth}
\epsfig{width=0.98\textwidth,clip=true,file=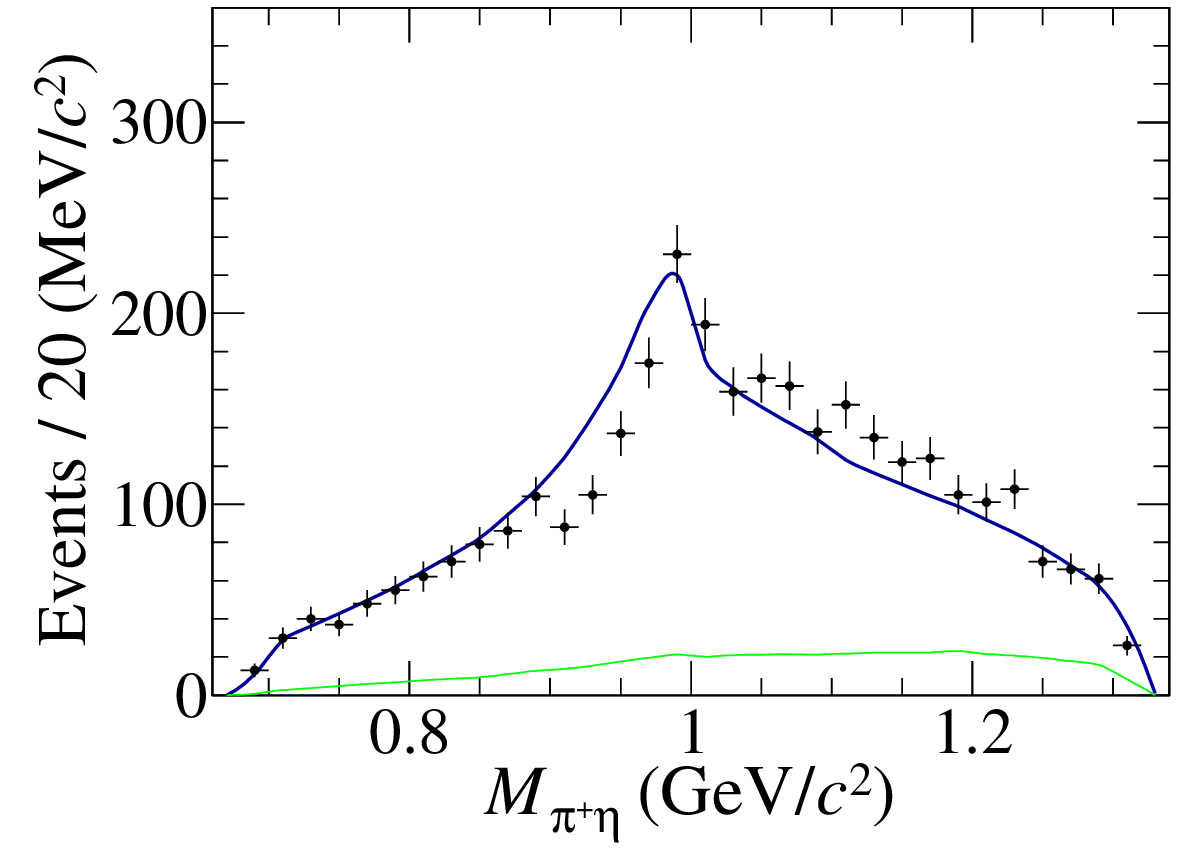}
%\put(-25,50){(c)}
\put(-85,70){II}
\end{minipage}
\begin{minipage}[b]{0.24\textwidth}
\epsfig{width=0.98\textwidth,clip=true,file=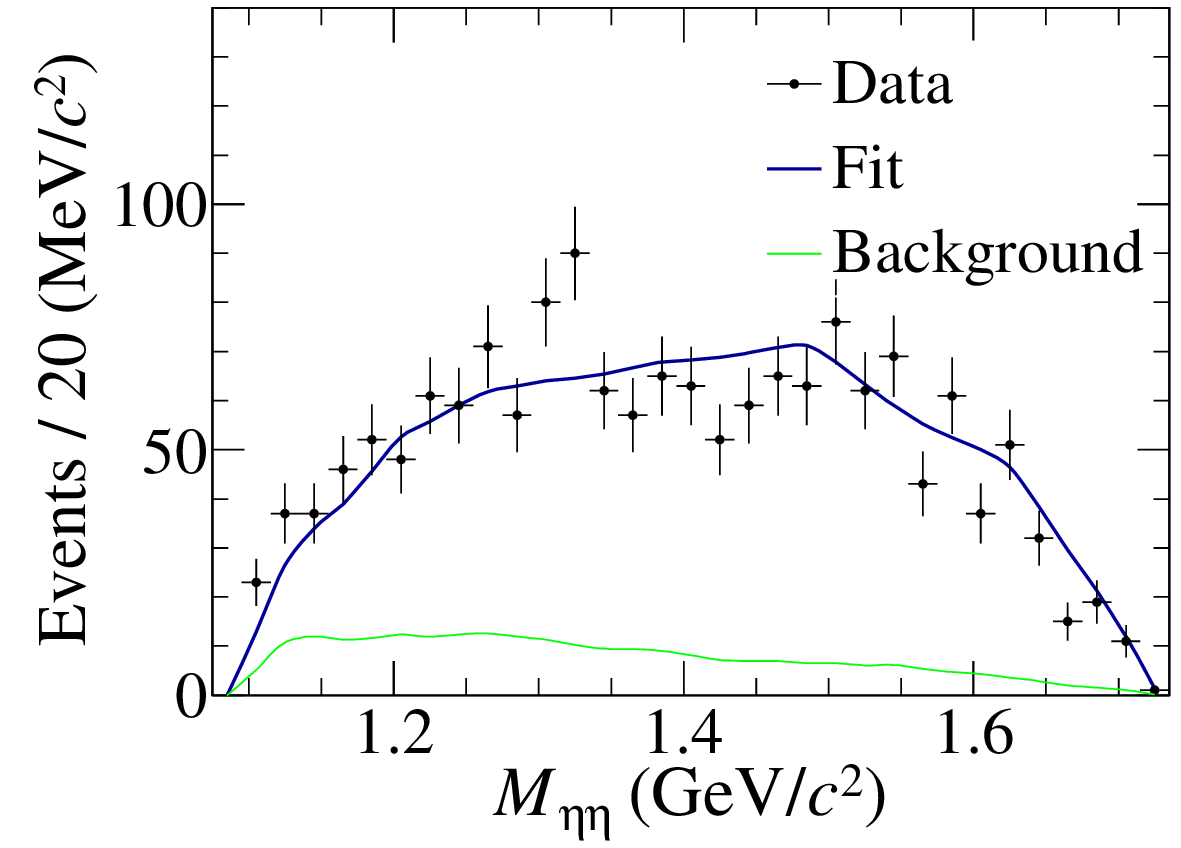}
%\put(-25,50){(d)}
\end{minipage}
\begin{minipage}[b]{0.24\textwidth}
\epsfig{width=0.98\textwidth,clip=true,file=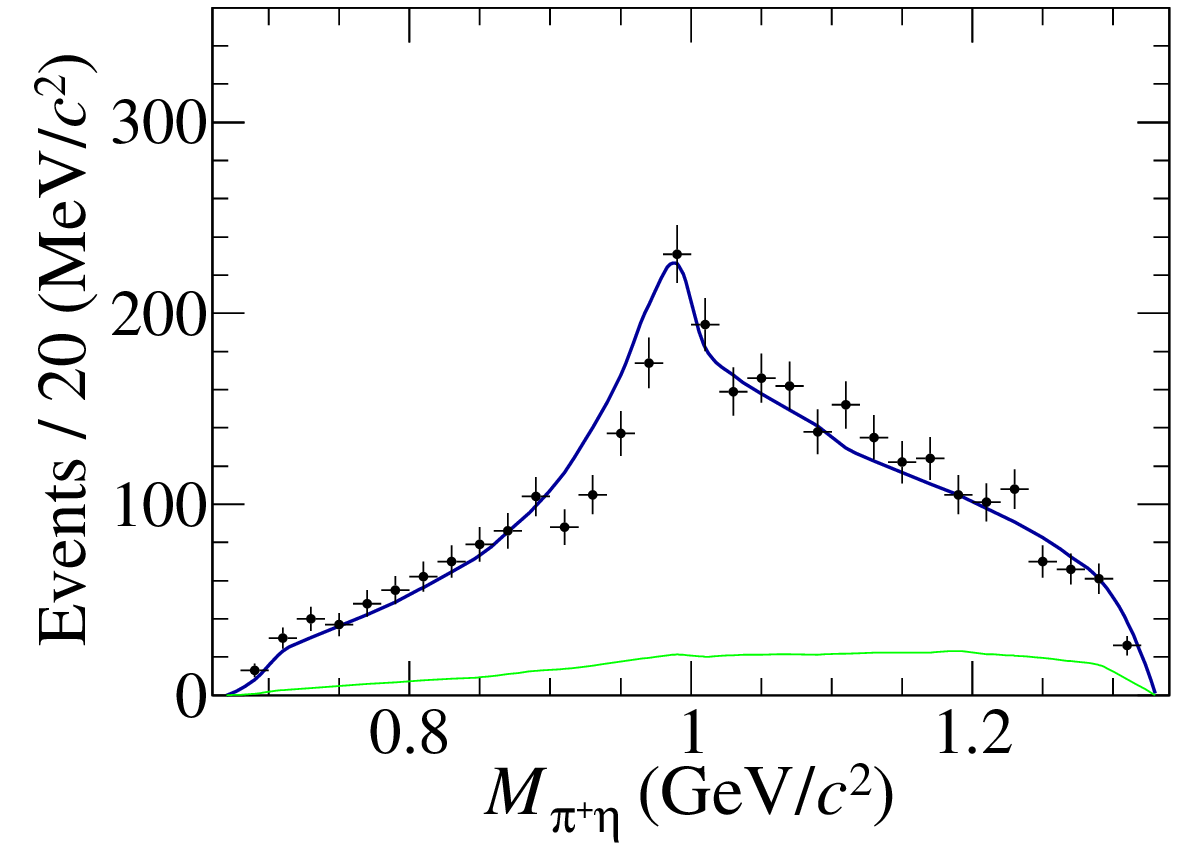}
%\put(-25,50){(e)}
\put(-85,70){III}
\end{minipage}
\begin{minipage}[b]{0.24\textwidth}
\epsfig{width=0.98\textwidth,clip=true,file=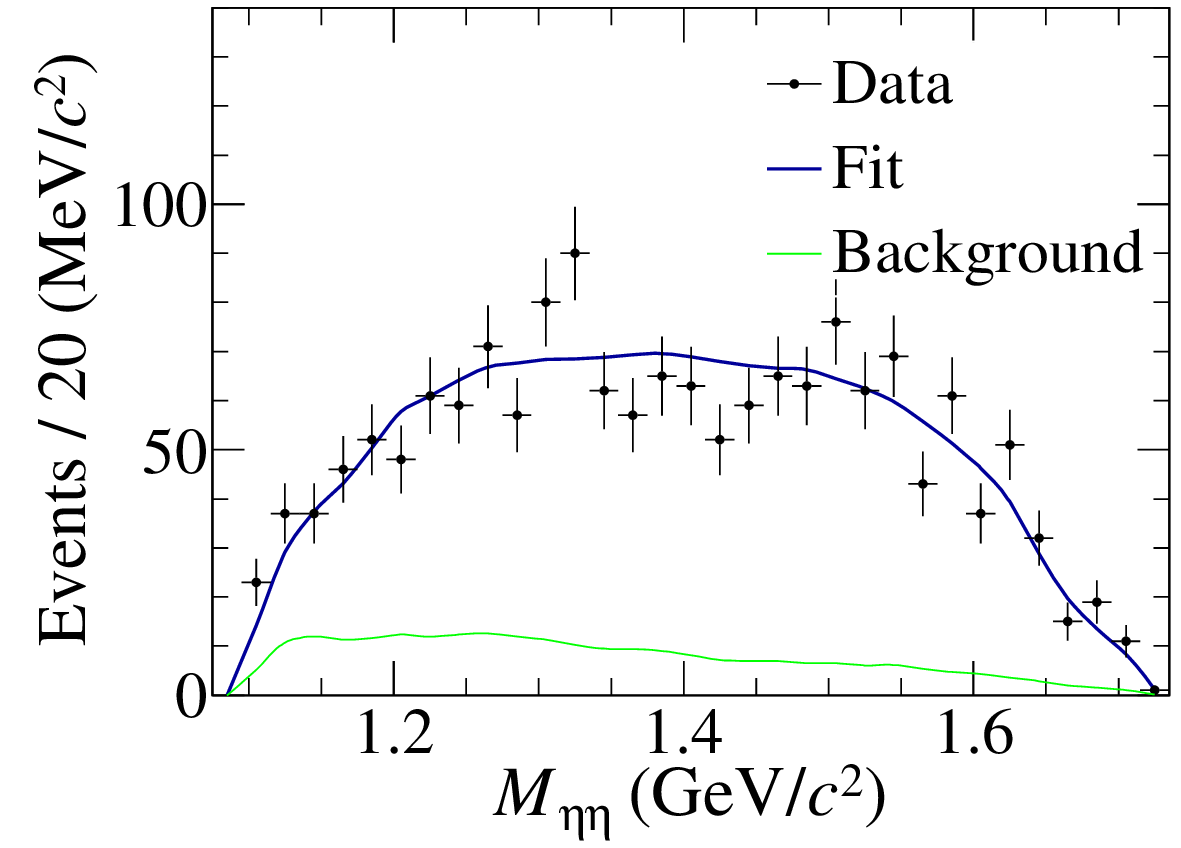}
%\put(-25,50){(f)}
\end{minipage}
\begin{minipage}[b]{0.24\textwidth}
\epsfig{width=0.98\textwidth,clip=true,file=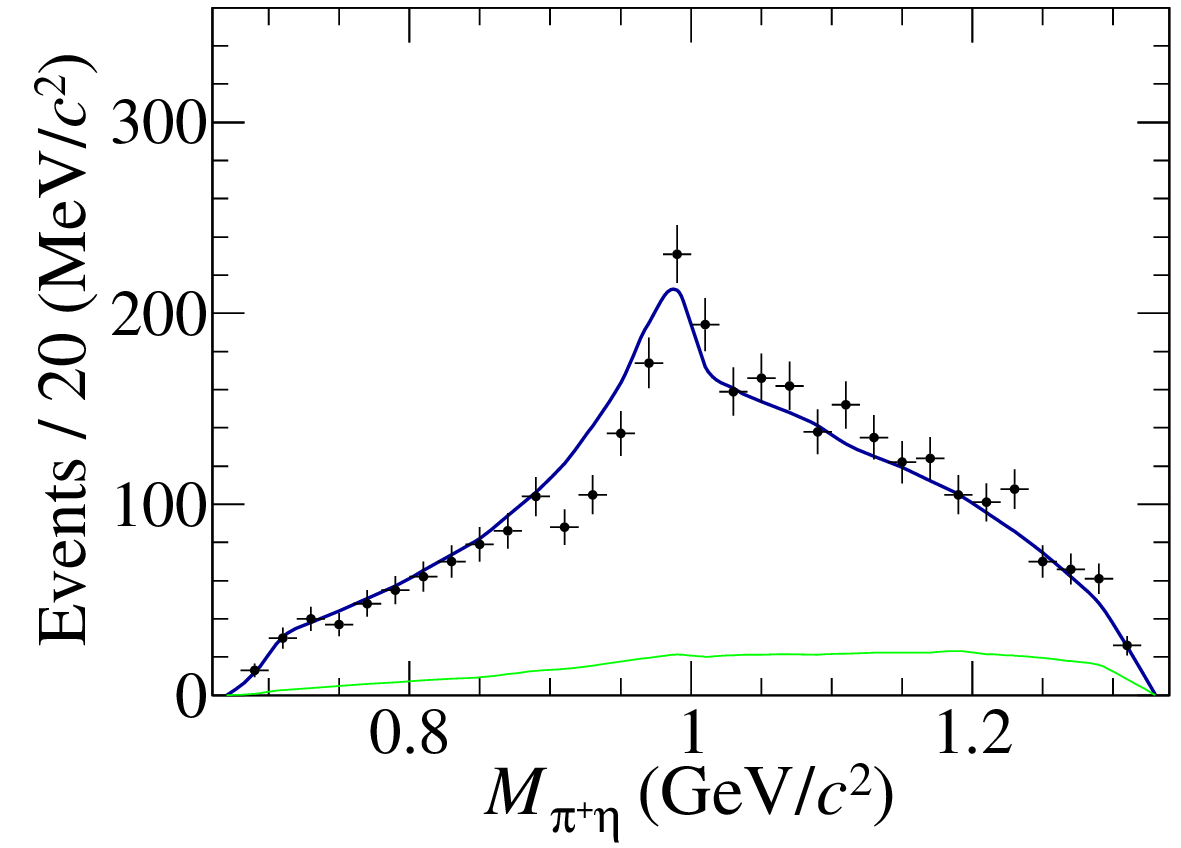}
%\put(-25,50){(g)}
\put(-85,70){IV}
\end{minipage}
\begin{minipage}[b]{0.24\textwidth}
\epsfig{width=0.98\textwidth,clip=true,file=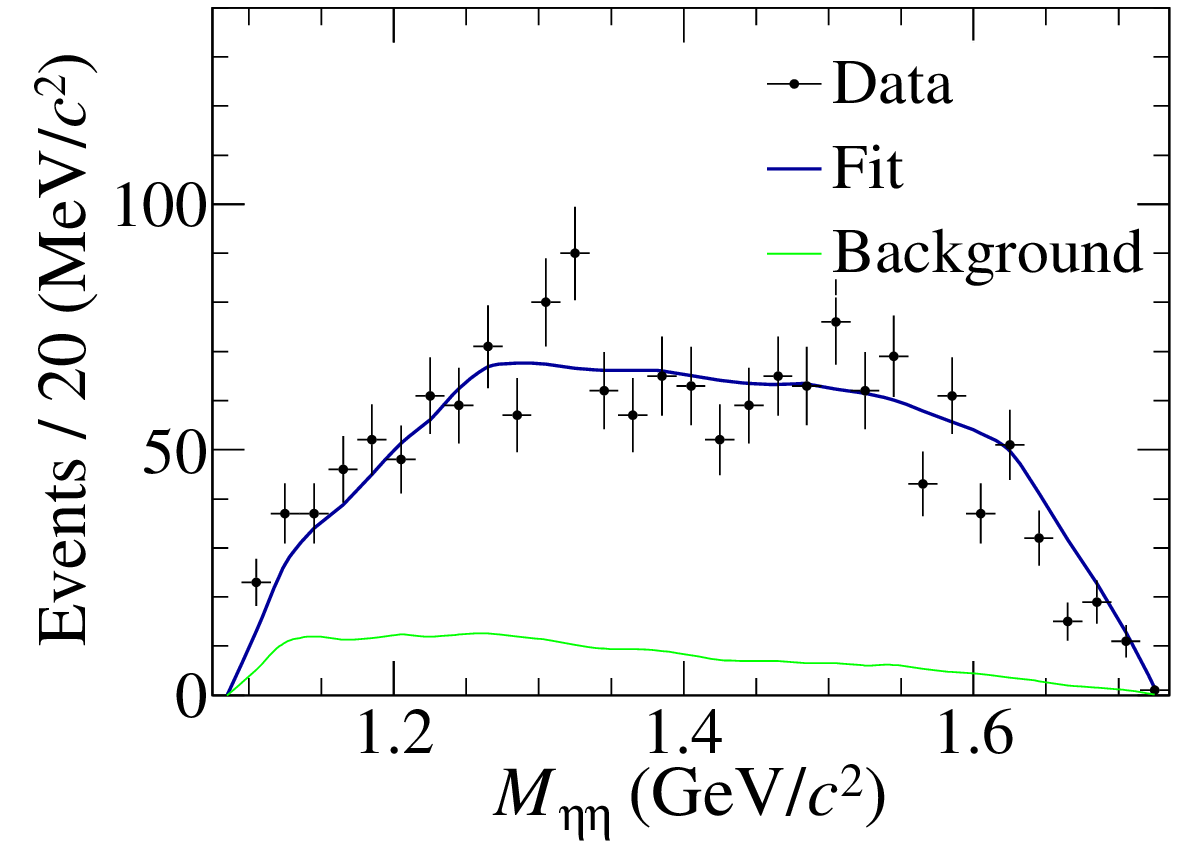}
%\put(-25,50){(h)}
\end{minipage}
\begin{minipage}[b]{0.24\textwidth}
\epsfig{width=0.98\textwidth,clip=true,file=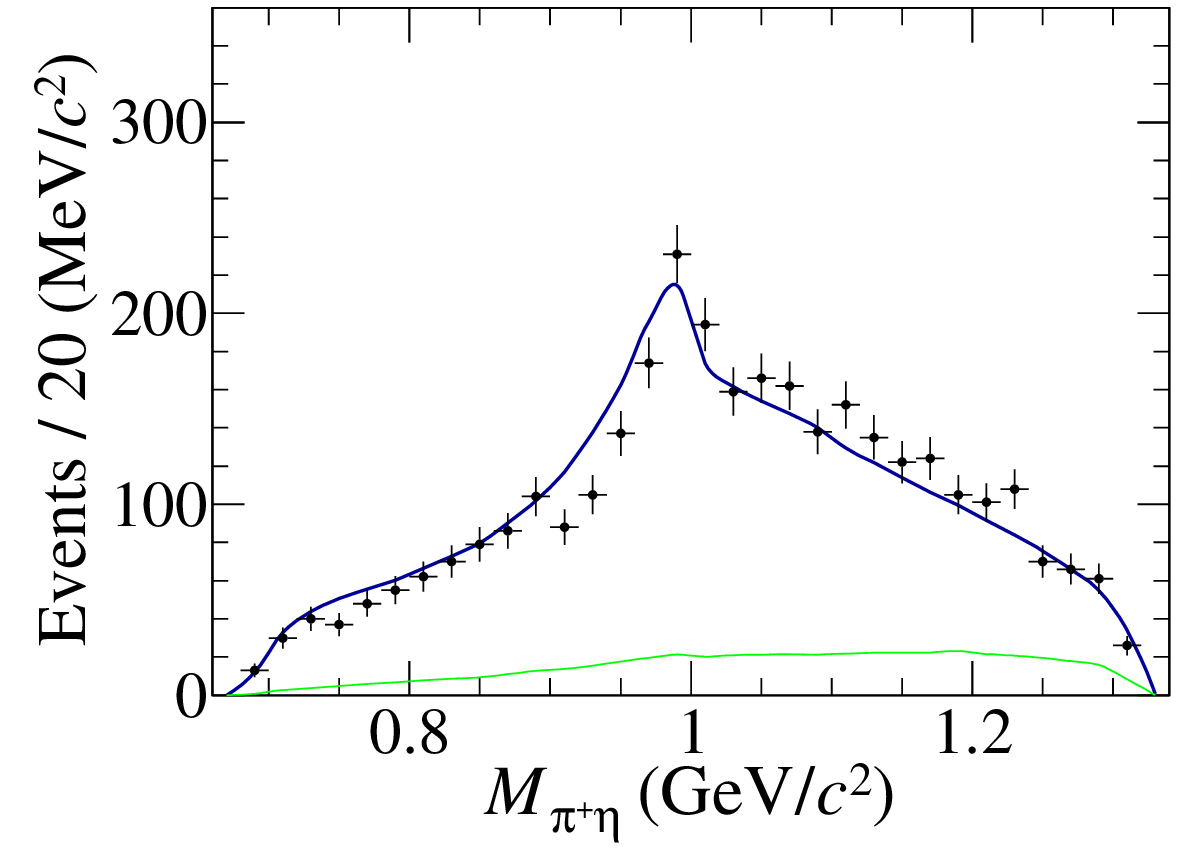}
%\put(-25,50){(i)}
\put(-85,70){V}
\end{minipage}
\begin{minipage}[b]{0.24\textwidth}
\epsfig{width=0.98\textwidth,clip=true,file=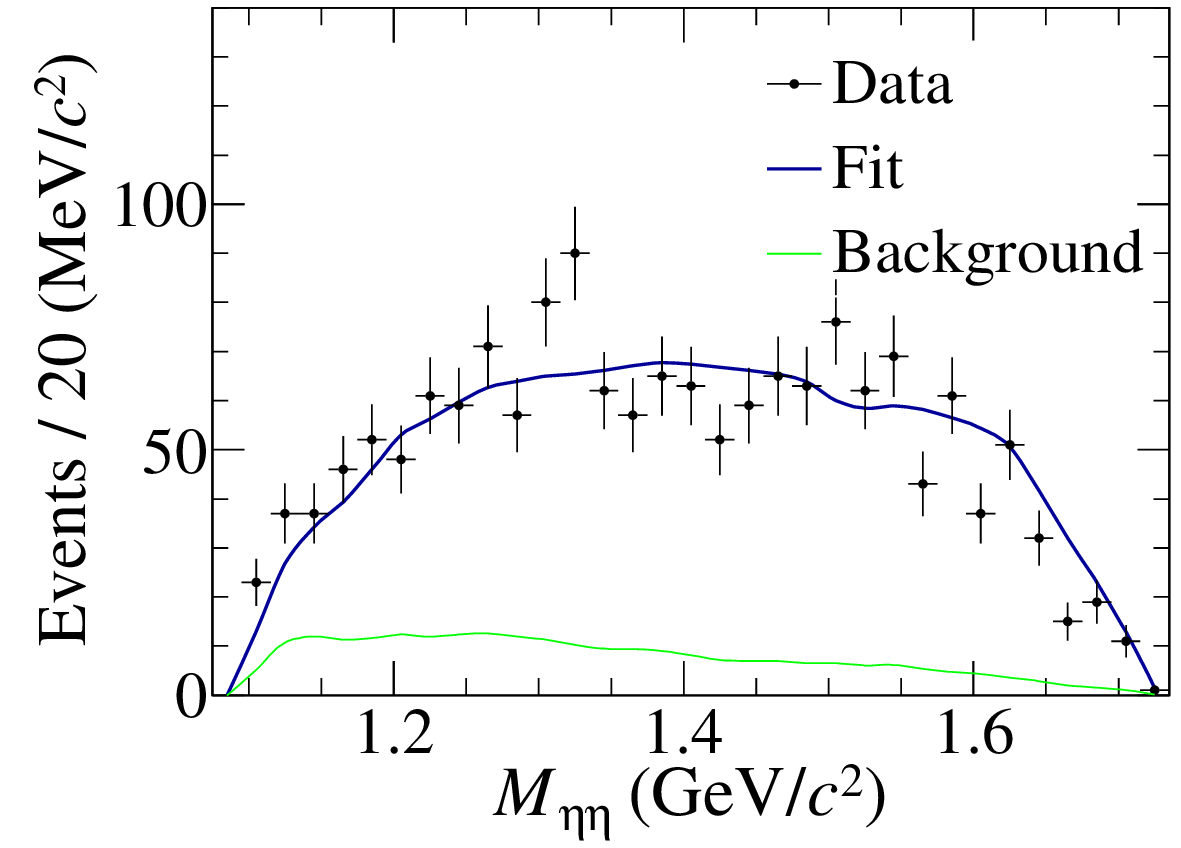}
%\put(-25,50){(j)}
\end{minipage}
\begin{minipage}[b]{0.24\textwidth}
\epsfig{width=0.98\textwidth,clip=true,file=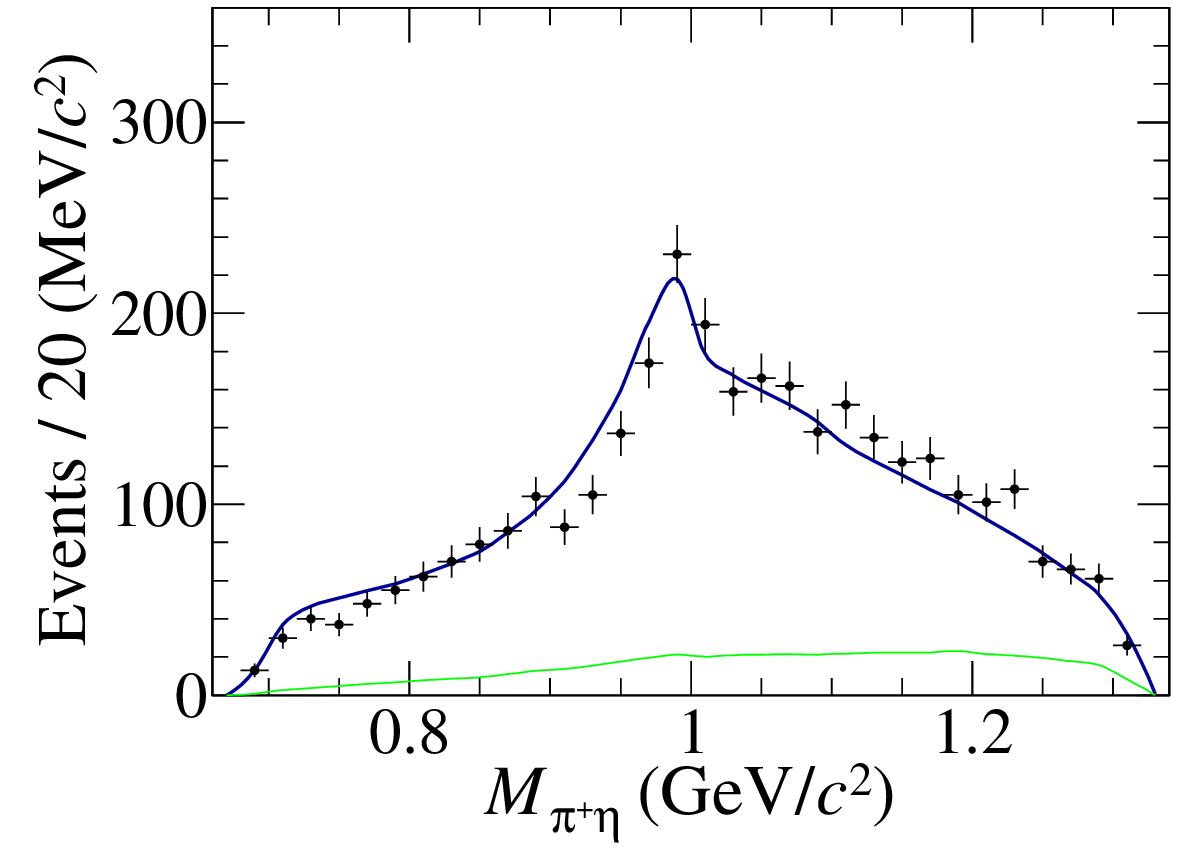}
%\put(-25,50){(k)}
\put(-85,70){VI}
\end{minipage}
\begin{minipage}[b]{0.24\textwidth}
\epsfig{width=0.98\textwidth,clip=true,file=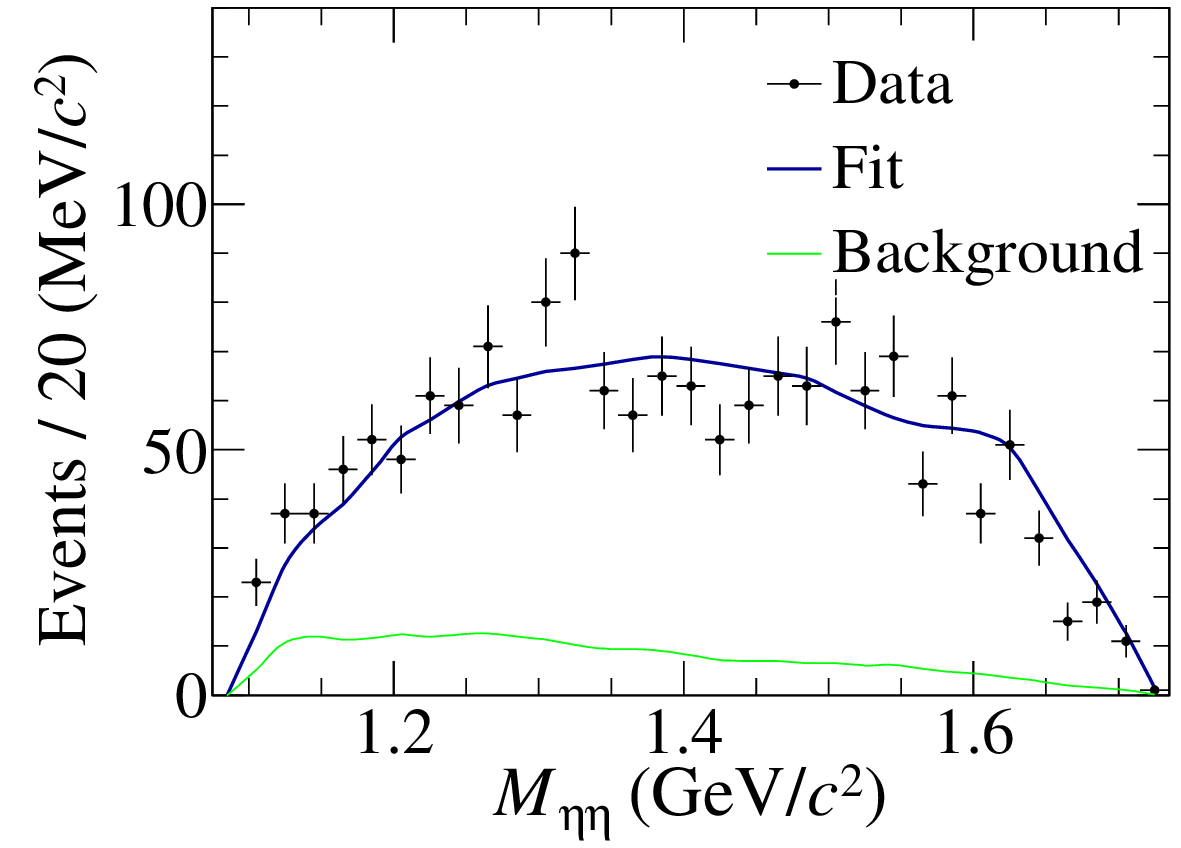}
%\put(-25,50){(l)}
\end{minipage}
\begin{minipage}[b]{0.24\textwidth}
\epsfig{width=0.98\textwidth,clip=true,file=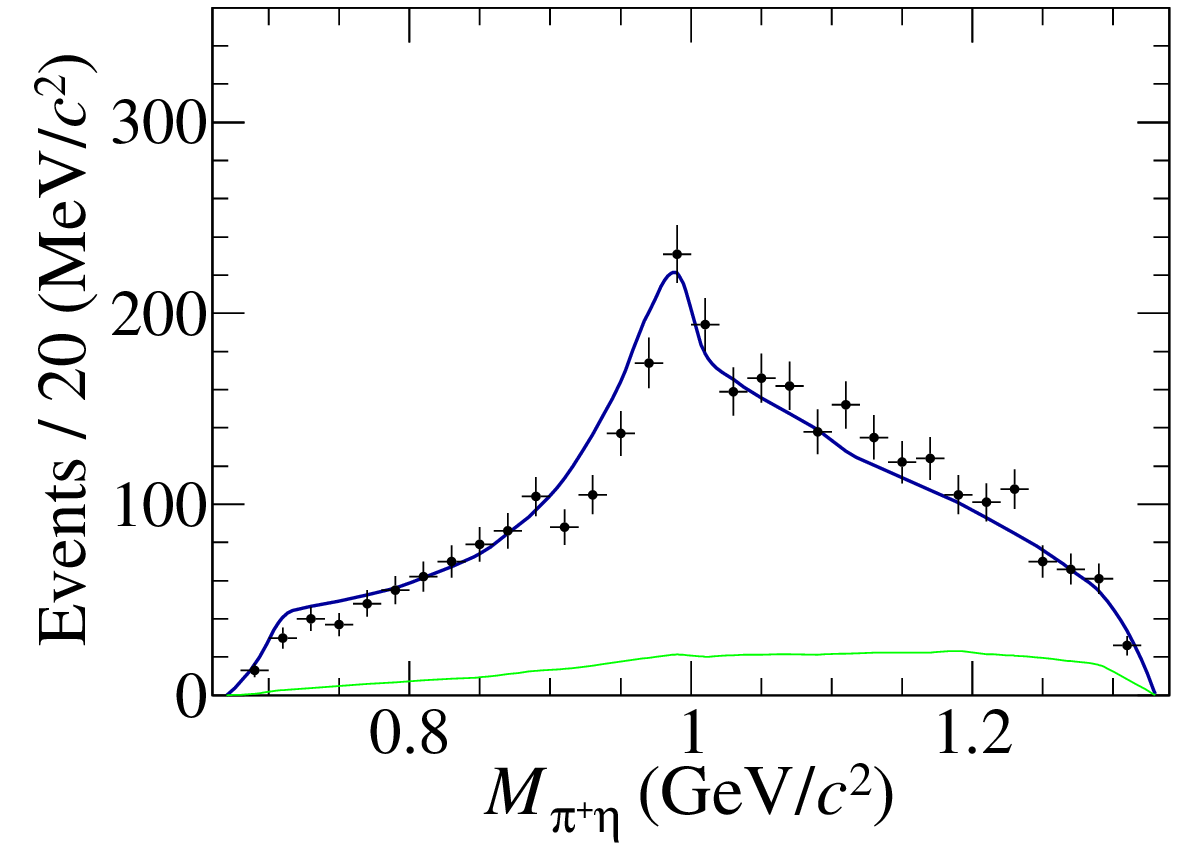}
%\put(-25,50){(m)}
\put(-85,70){VII}
\end{minipage}
\begin{minipage}[b]{0.24\textwidth}
\epsfig{width=0.98\textwidth,clip=true,file=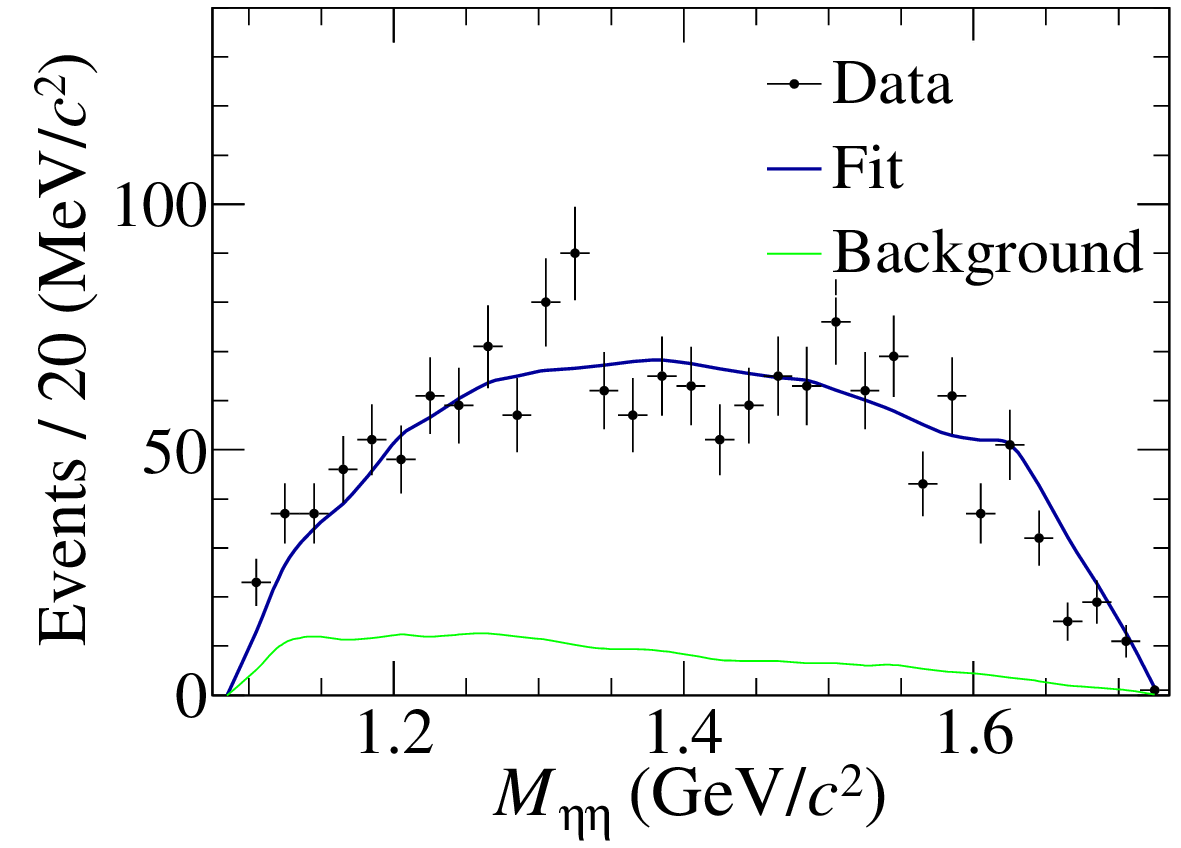}
%\put(-25,50){(n)}
\end{minipage}
\begin{minipage}[b]{0.24\textwidth}
\epsfig{width=0.98\textwidth,clip=true,file=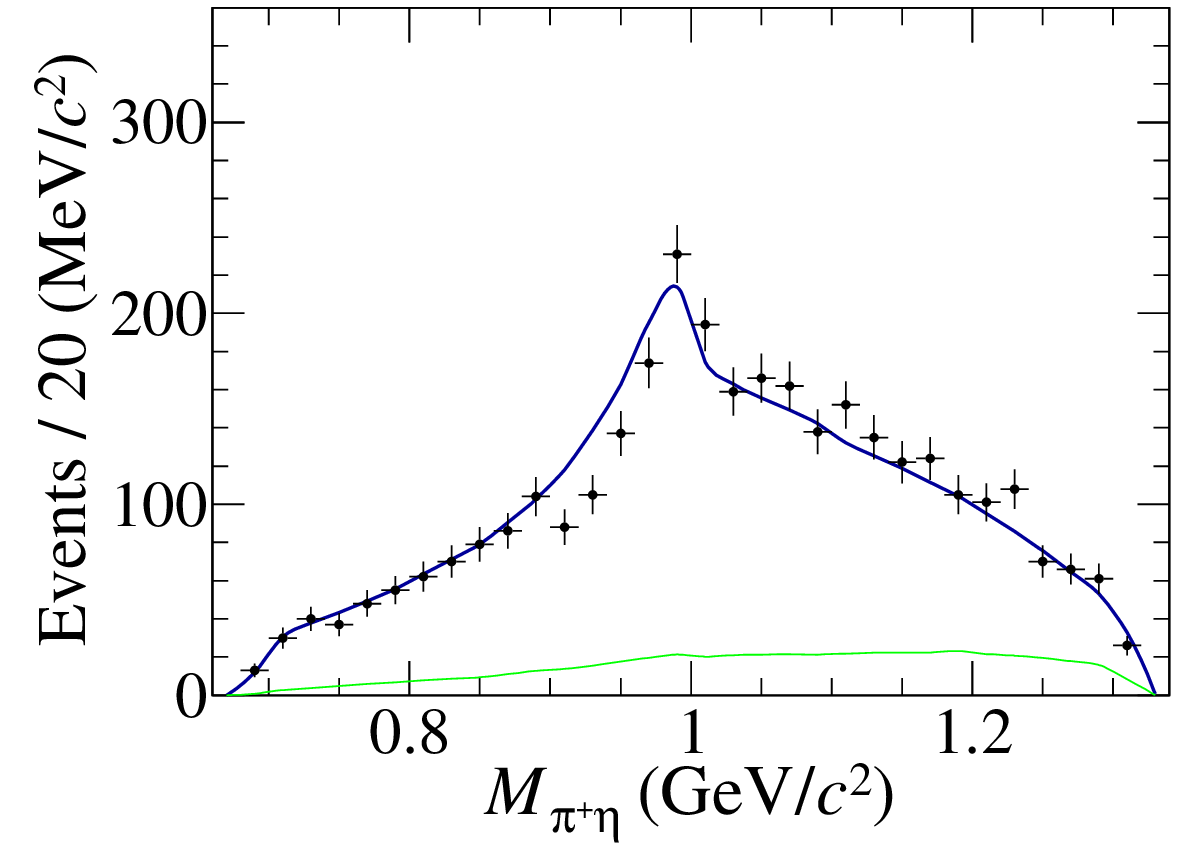}
%\put(-25,50){(o)}
\put(-85,70){VIII}
\end{minipage}
\begin{minipage}[b]{0.24\textwidth}
\epsfig{width=0.98\textwidth,clip=true,file=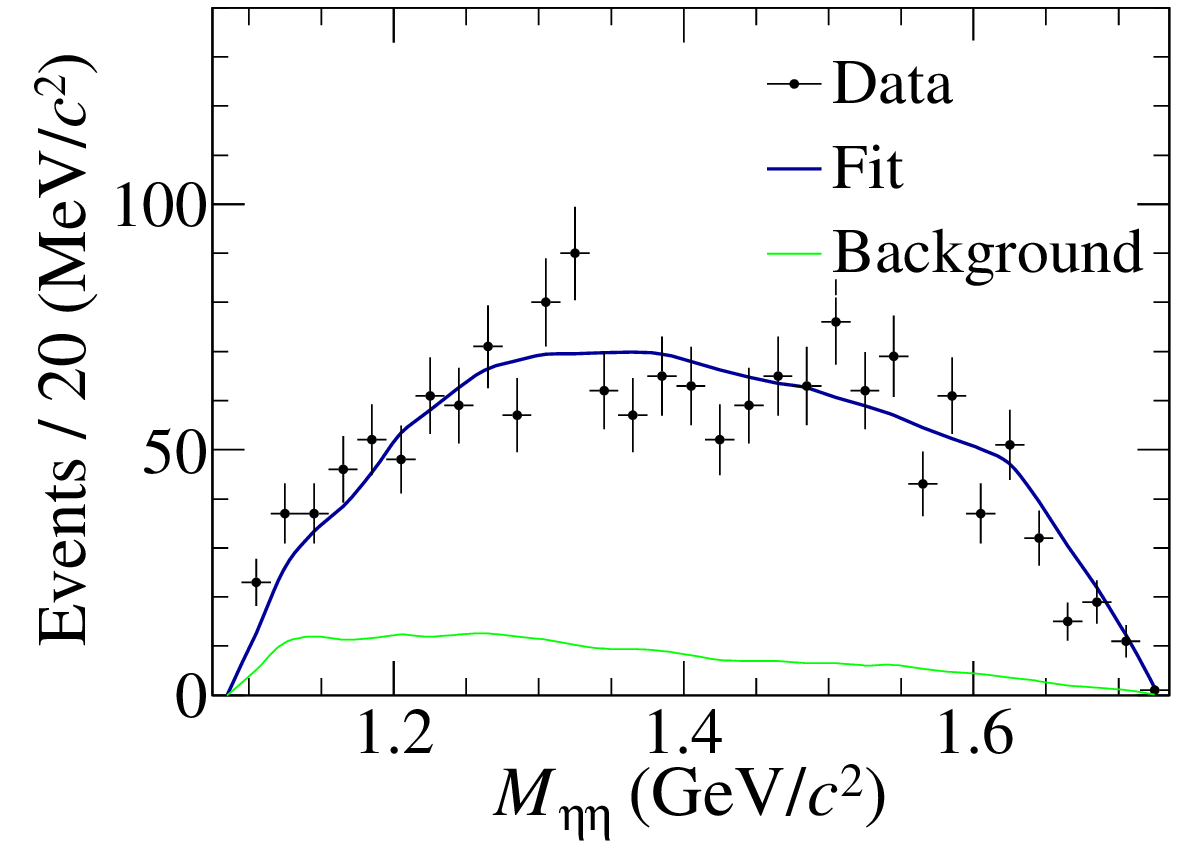}
%\put(-25,50){(p)}
\end{minipage}
\begin{minipage}[b]{0.24\textwidth}
\epsfig{width=0.98\textwidth,clip=true,file=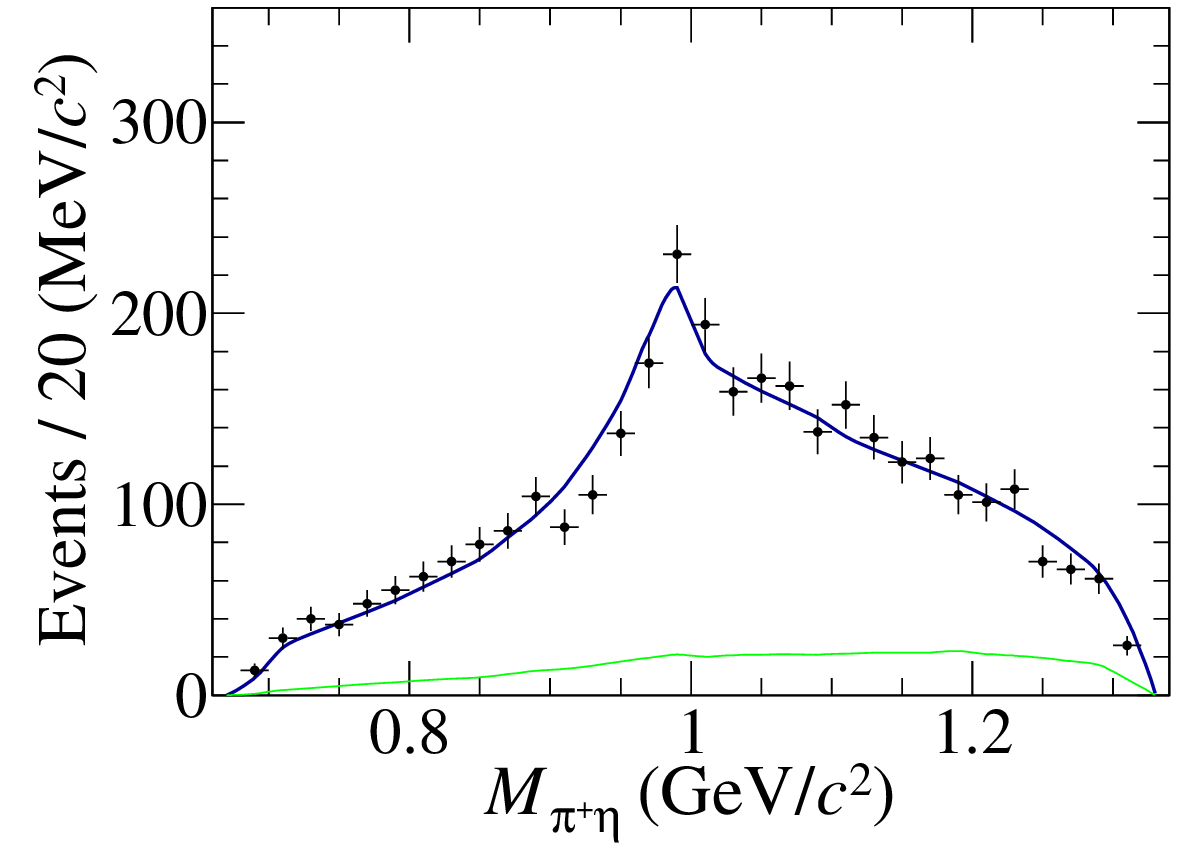}
%\put(-25,50){(q)}
\put(-85,70){IX}
\end{minipage}
\begin{minipage}[b]{0.24\textwidth}
\epsfig{width=0.98\textwidth,clip=true,file=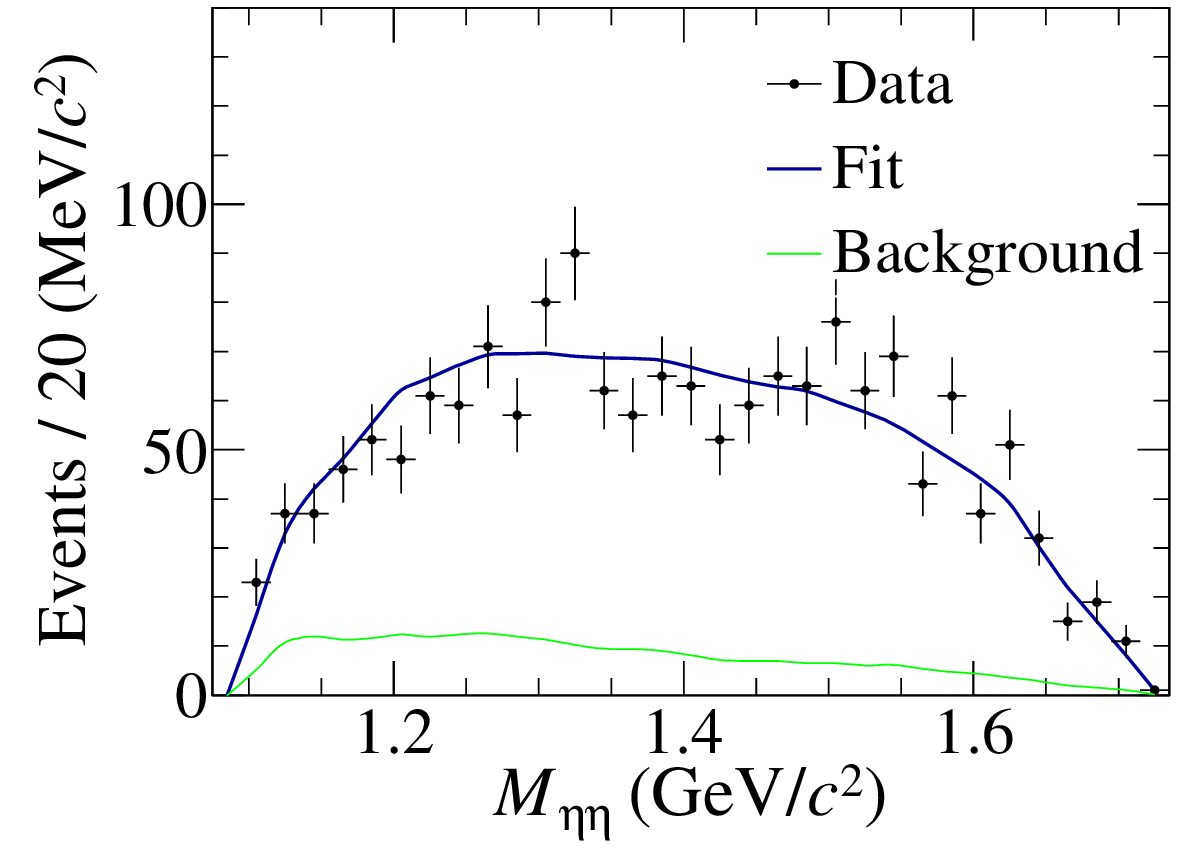}
%\put(-25,50){(r)}
\end{minipage}
\begin{minipage}[b]{0.24\textwidth}
\epsfig{width=0.98\textwidth,clip=true,file=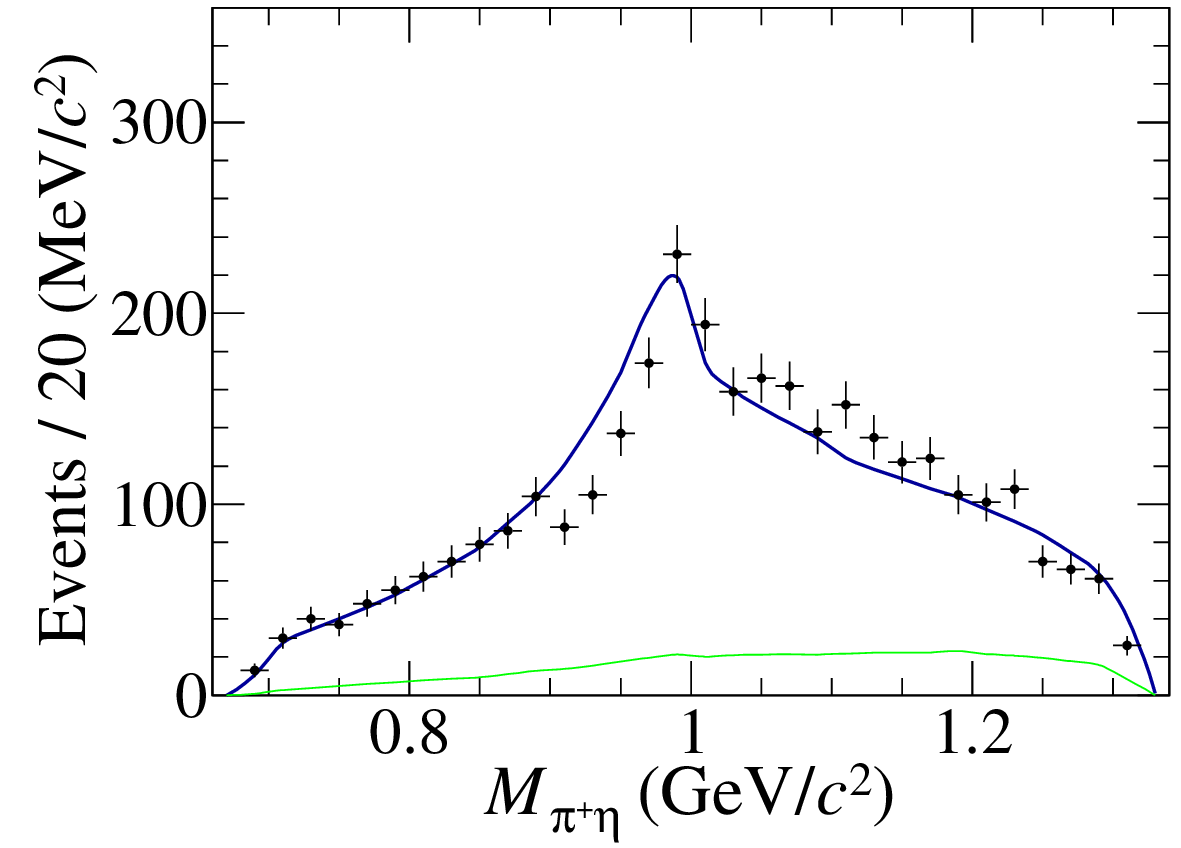}
%\put(-25,50){(s)}
\put(-85,70){X}
\end{minipage}
\begin{minipage}[b]{0.24\textwidth}
\epsfig{width=0.98\textwidth,clip=true,file=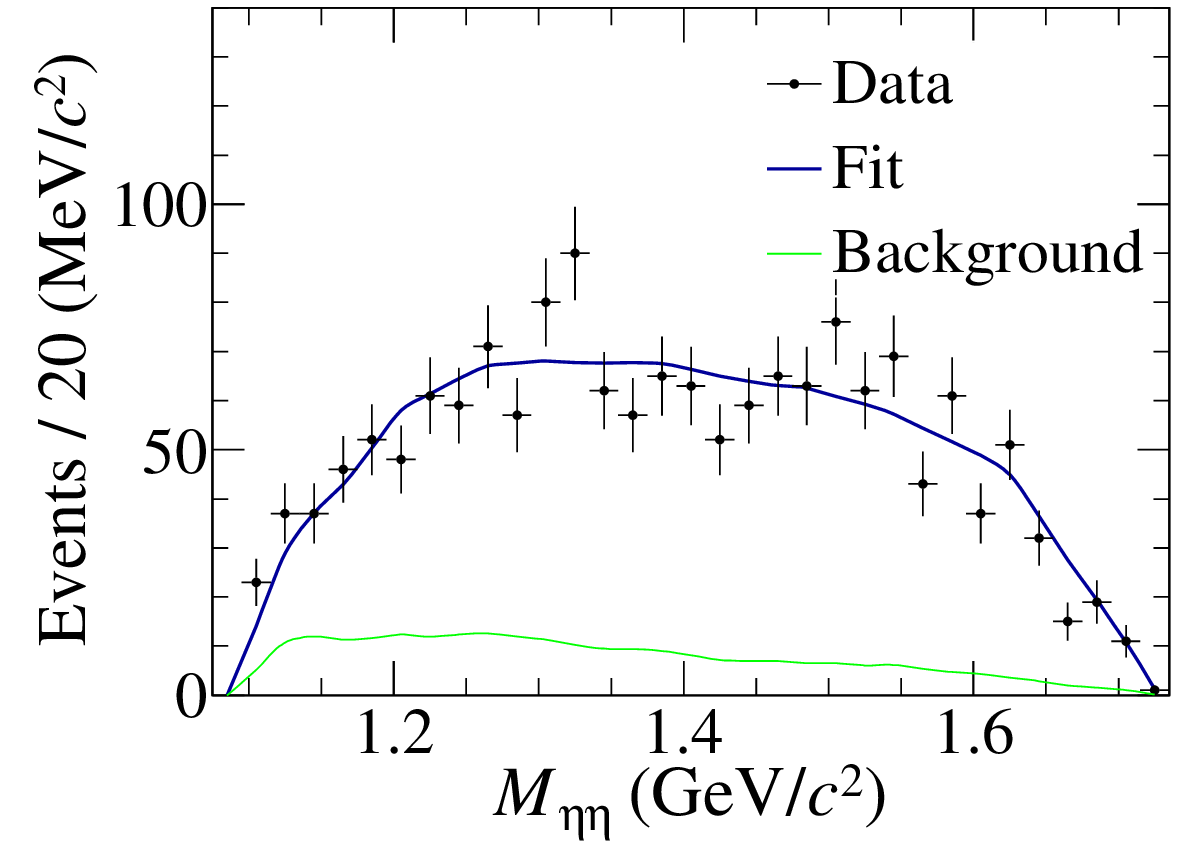}
%\put(-25,50){(t)}
\end{minipage}
\begin{minipage}[b]{0.24\textwidth}
\epsfig{width=0.98\textwidth,clip=true,file=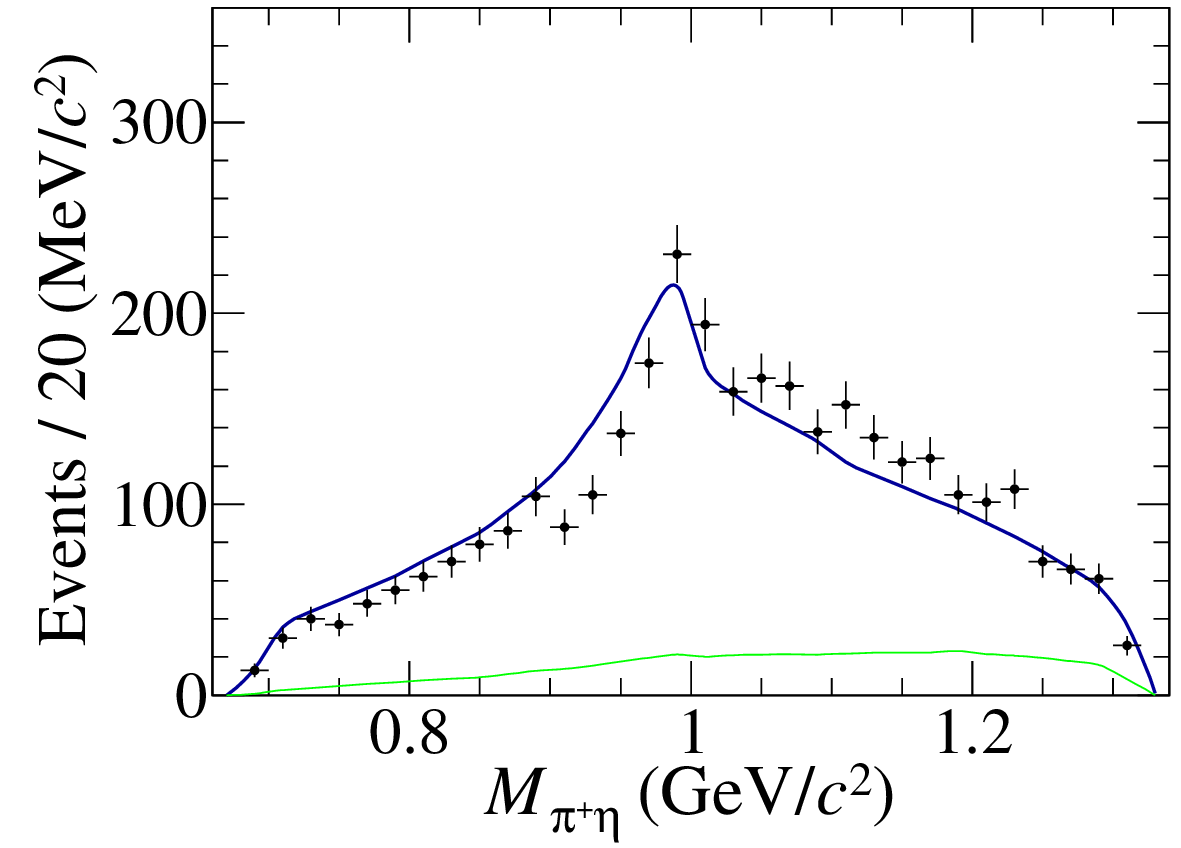}
%\put(-25,50){(u)}
\put(-85,70){XI}
\end{minipage}
\begin{minipage}[b]{0.24\textwidth}
\epsfig{width=0.98\textwidth,clip=true,file=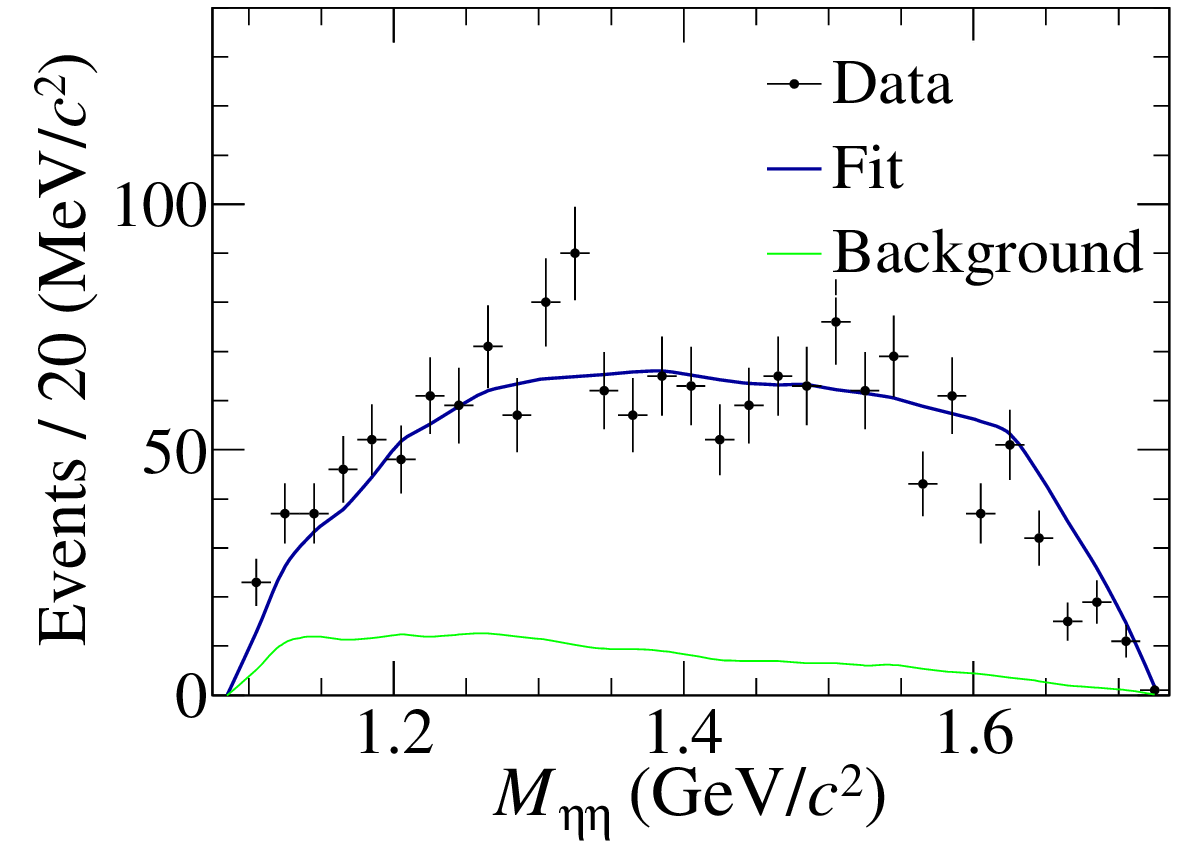}
%\put(-25,50){(v)}
\end{minipage}
\begin{minipage}[b]{0.24\textwidth}
\epsfig{width=0.98\textwidth,clip=true,file=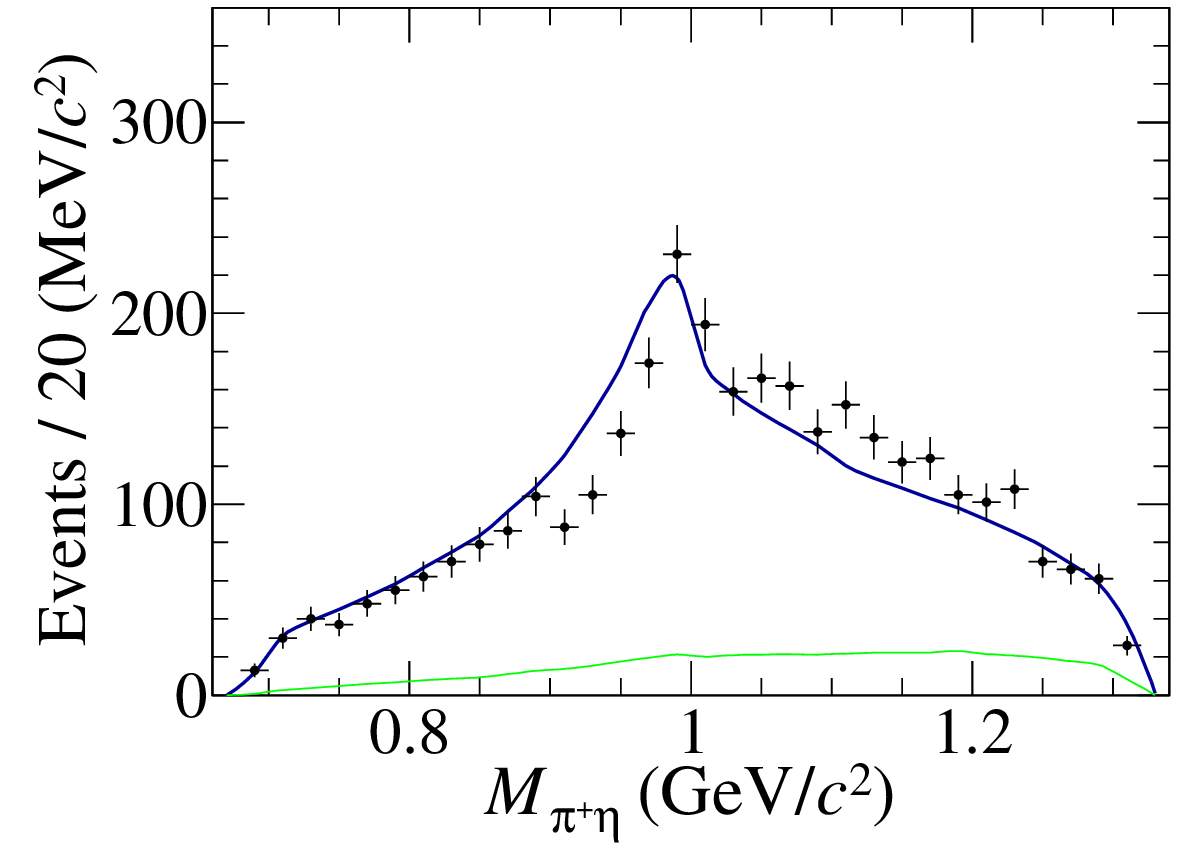}
%\put(-25,50){(w)}
\put(-85,70){XII}
\end{minipage}
\begin{minipage}[b]{0.24\textwidth}
\epsfig{width=0.98\textwidth,clip=true,file=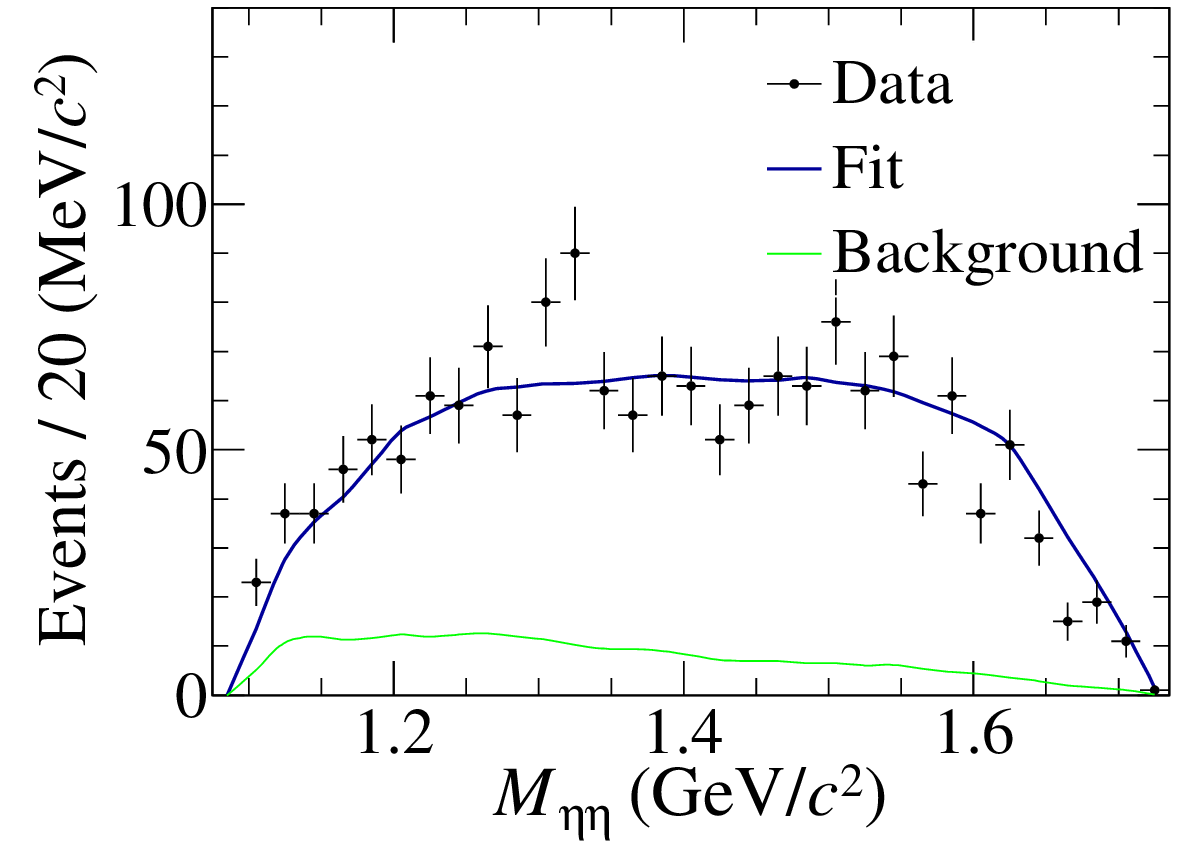}
%\put(-25,50){(x)}
\end{minipage}
\caption{The same as Fig.~\ref{fig:Flatteaddamp}, but for the models using the dispersive parameterization for $P_{a_{0}(980)}$.}
\label{fig:dispersive}
\end{center}
\end{figure}

\begin{figure}[htbp]
\begin{center}
\begin{minipage}[b]{0.24\textwidth}
\epsfig{width=0.98\textwidth,clip=true,file=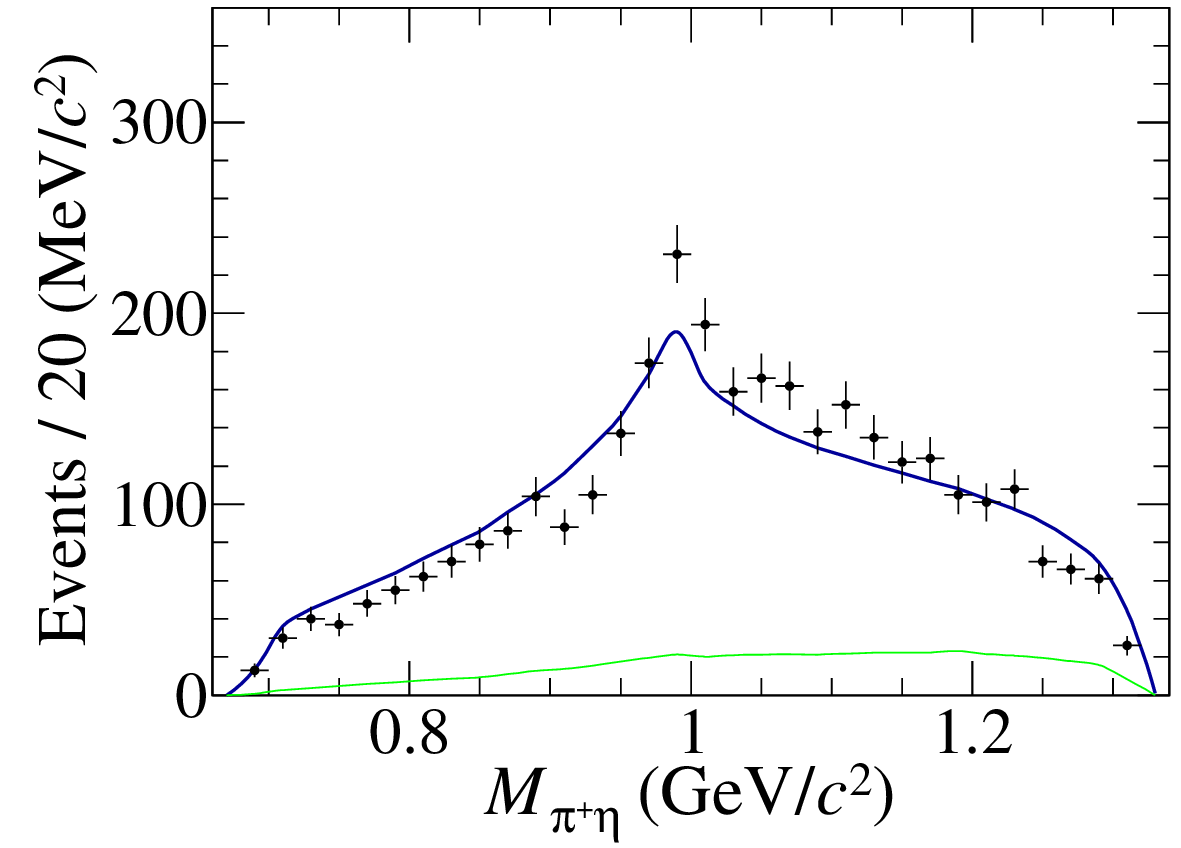}
%\put(-25,50){(a)}
\put(-85,70){I}
\end{minipage}
\begin{minipage}[b]{0.24\textwidth}
\epsfig{width=0.98\textwidth,clip=true,file=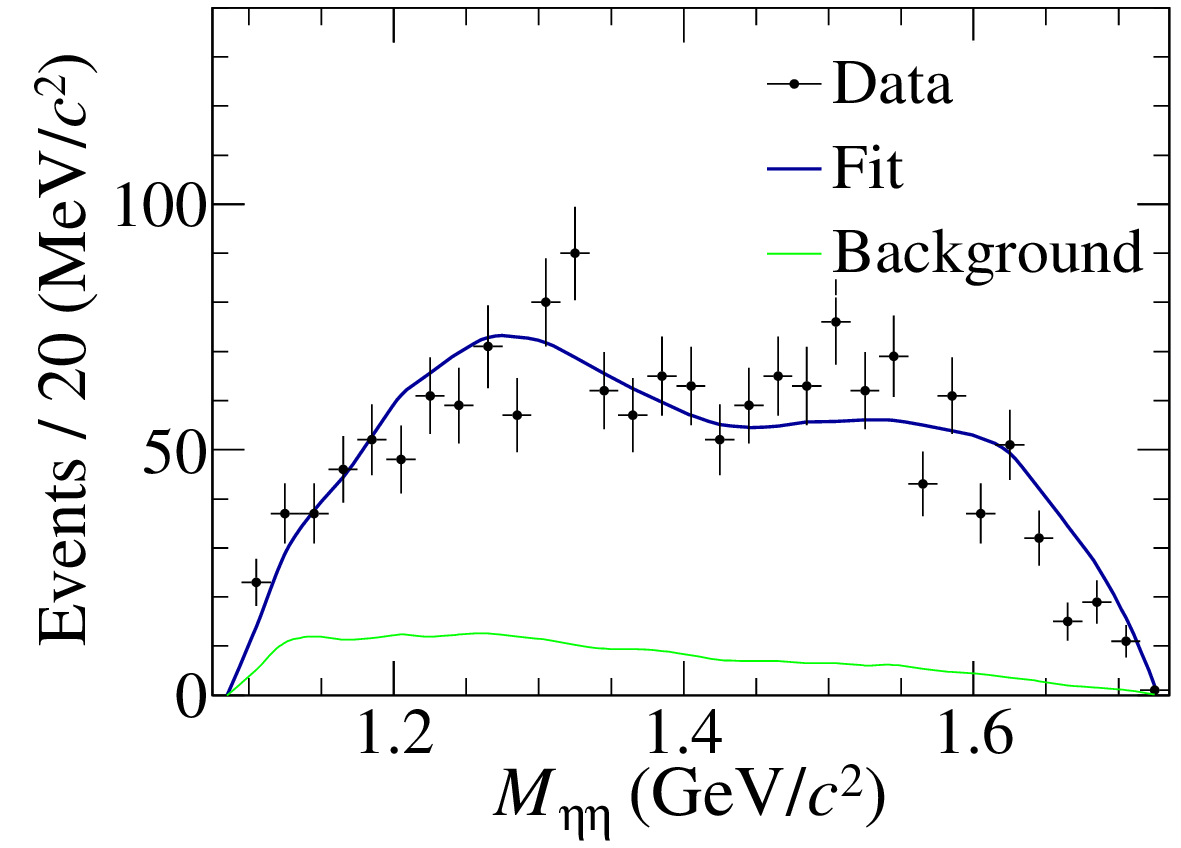}
%\put(-25,50){(b)}
\end{minipage}
\begin{minipage}[b]{0.24\textwidth}
\epsfig{width=0.98\textwidth,clip=true,file=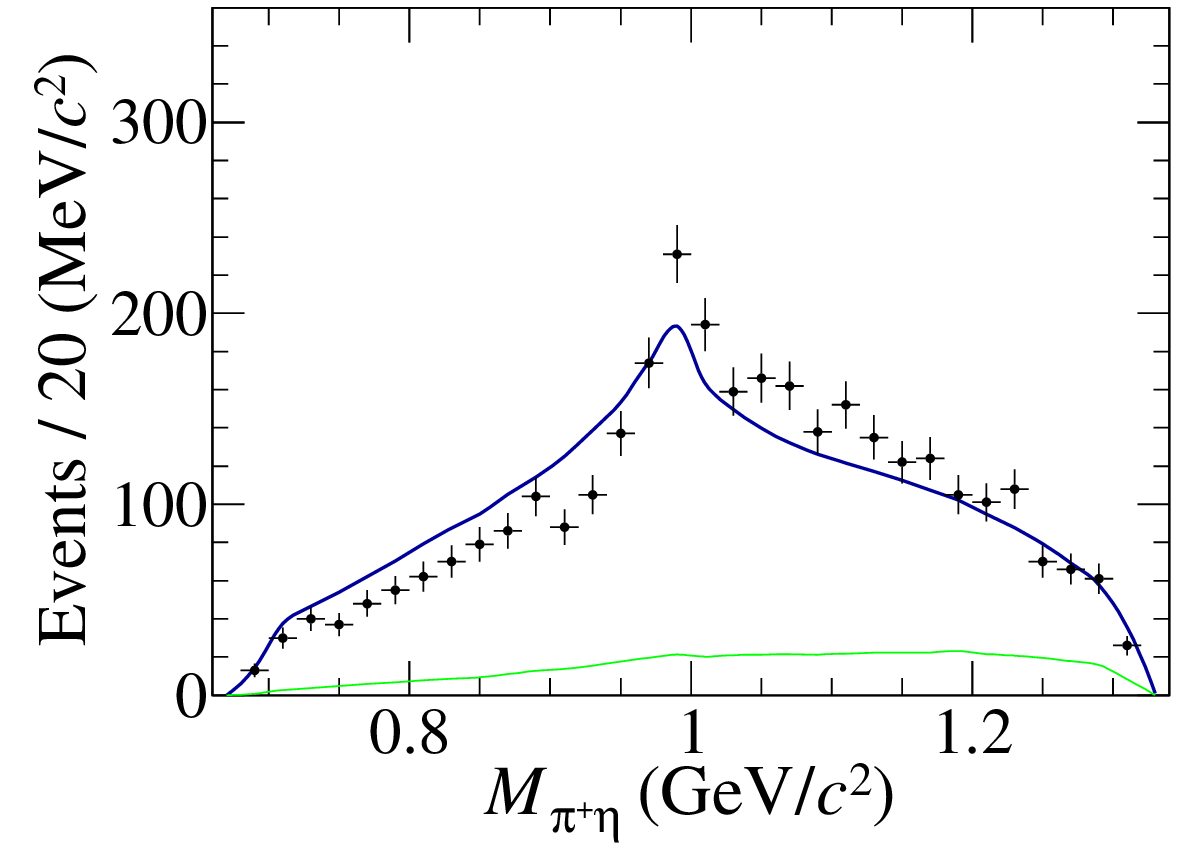}
%\put(-25,50){(c)}
\put(-85,70){II}
\end{minipage}
\begin{minipage}[b]{0.24\textwidth}
\epsfig{width=0.98\textwidth,clip=true,file=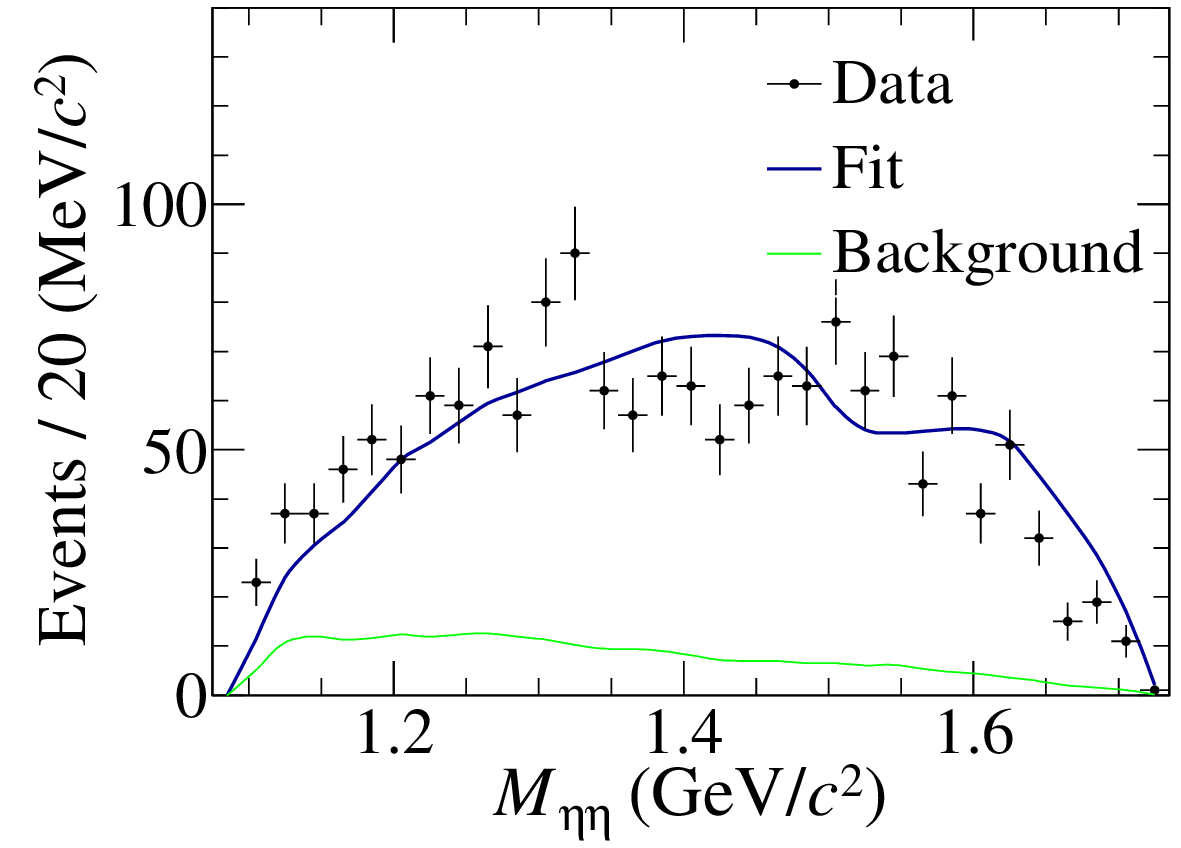}
%\put(-25,50){(d)}
\end{minipage}
\begin{minipage}[b]{0.24\textwidth}
\epsfig{width=0.98\textwidth,clip=true,file=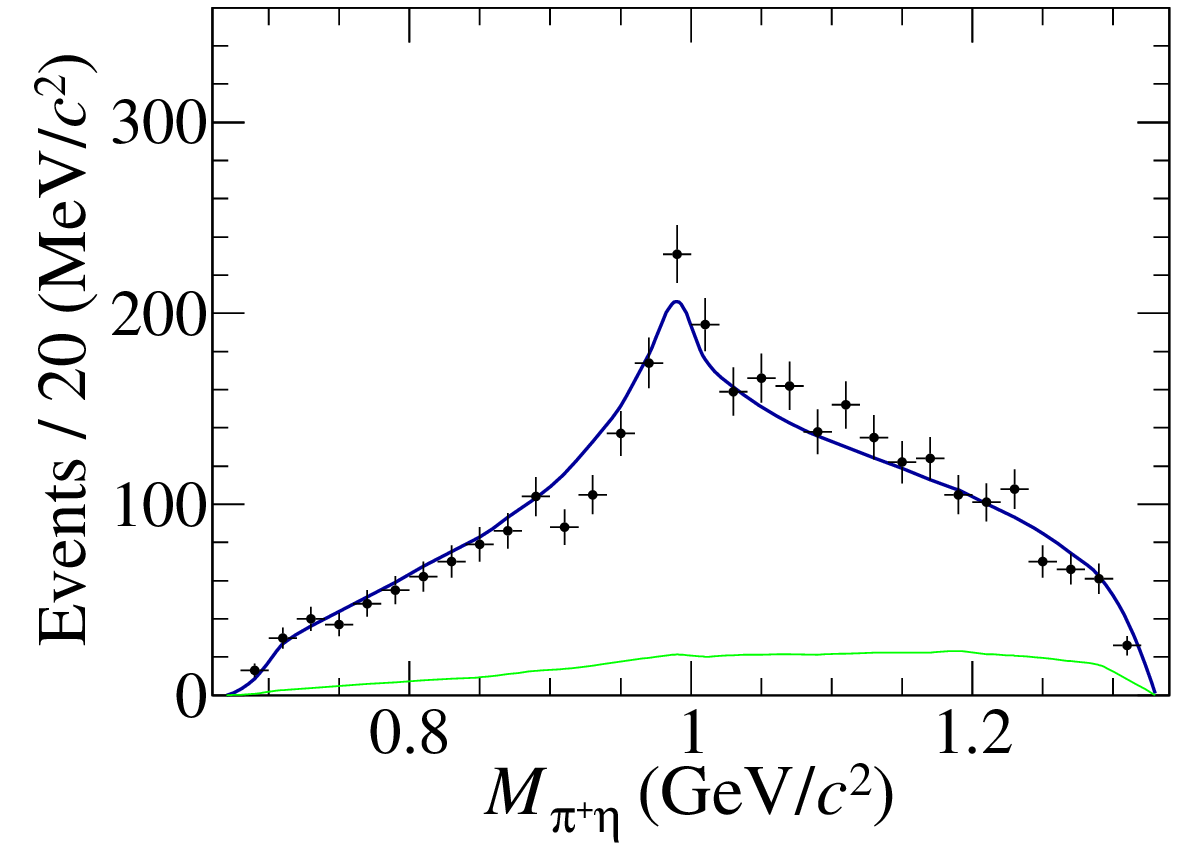}
%\put(-25,50){(e)}
\put(-85,70){III}
\end{minipage}
\begin{minipage}[b]{0.24\textwidth}
\epsfig{width=0.98\textwidth,clip=true,file=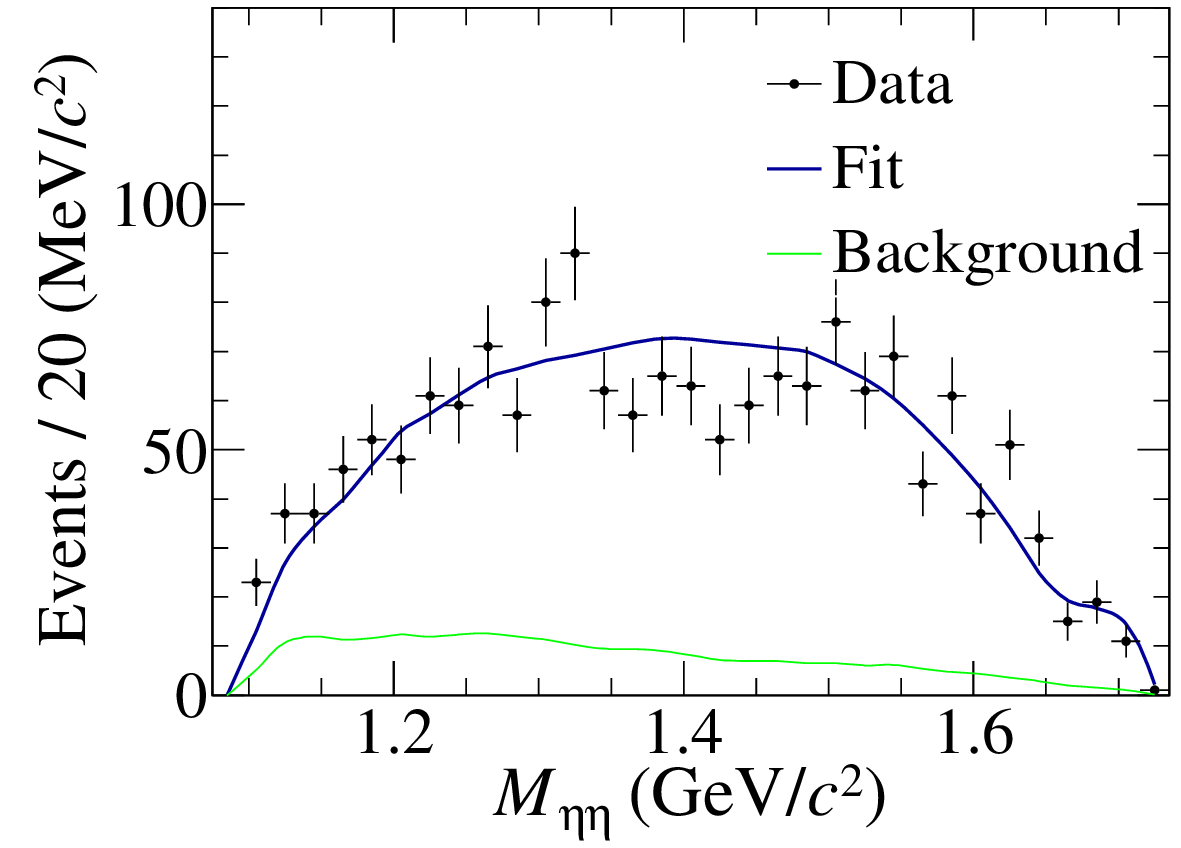}
%\put(-25,50){(f)}
\end{minipage}
\begin{minipage}[b]{0.24\textwidth}
\epsfig{width=0.98\textwidth,clip=true,file=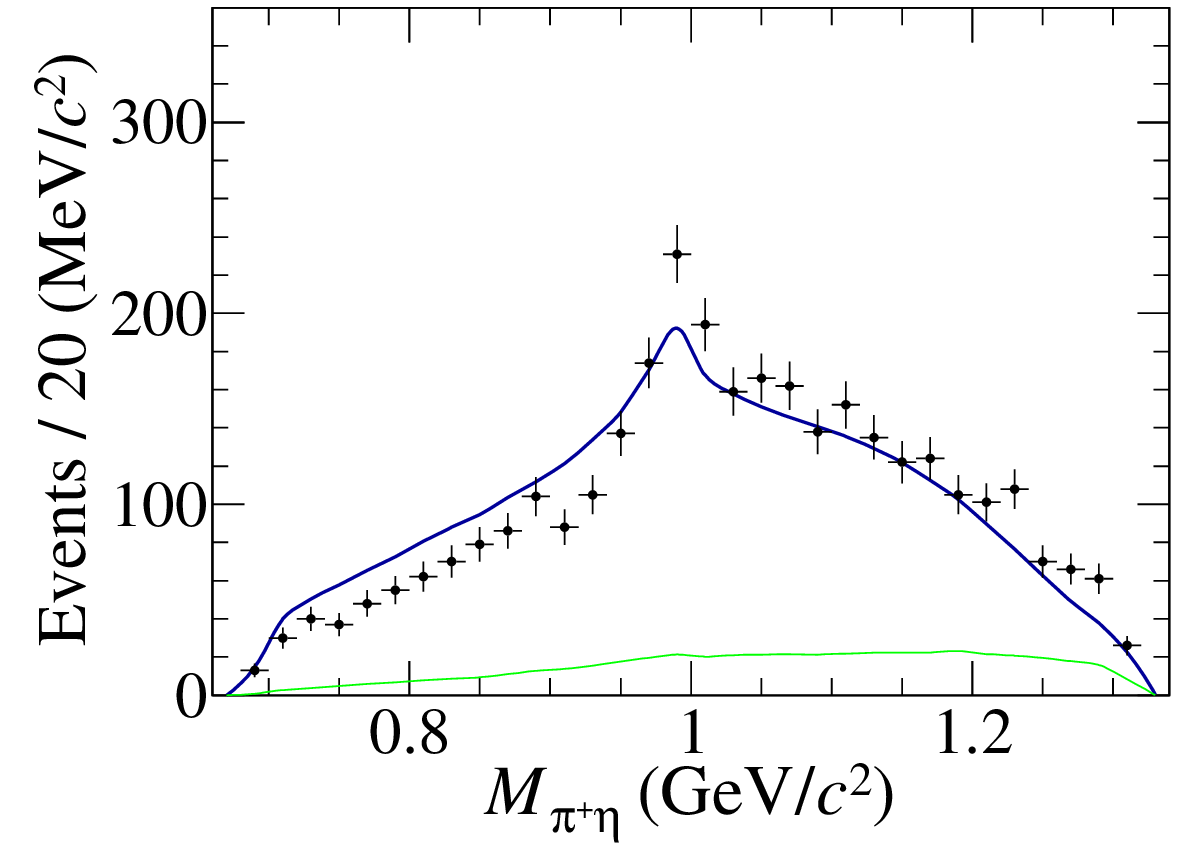}
%\put(-25,50){(g)}
\put(-85,70){IV}
\end{minipage}
\begin{minipage}[b]{0.24\textwidth}
\epsfig{width=0.98\textwidth,clip=true,file=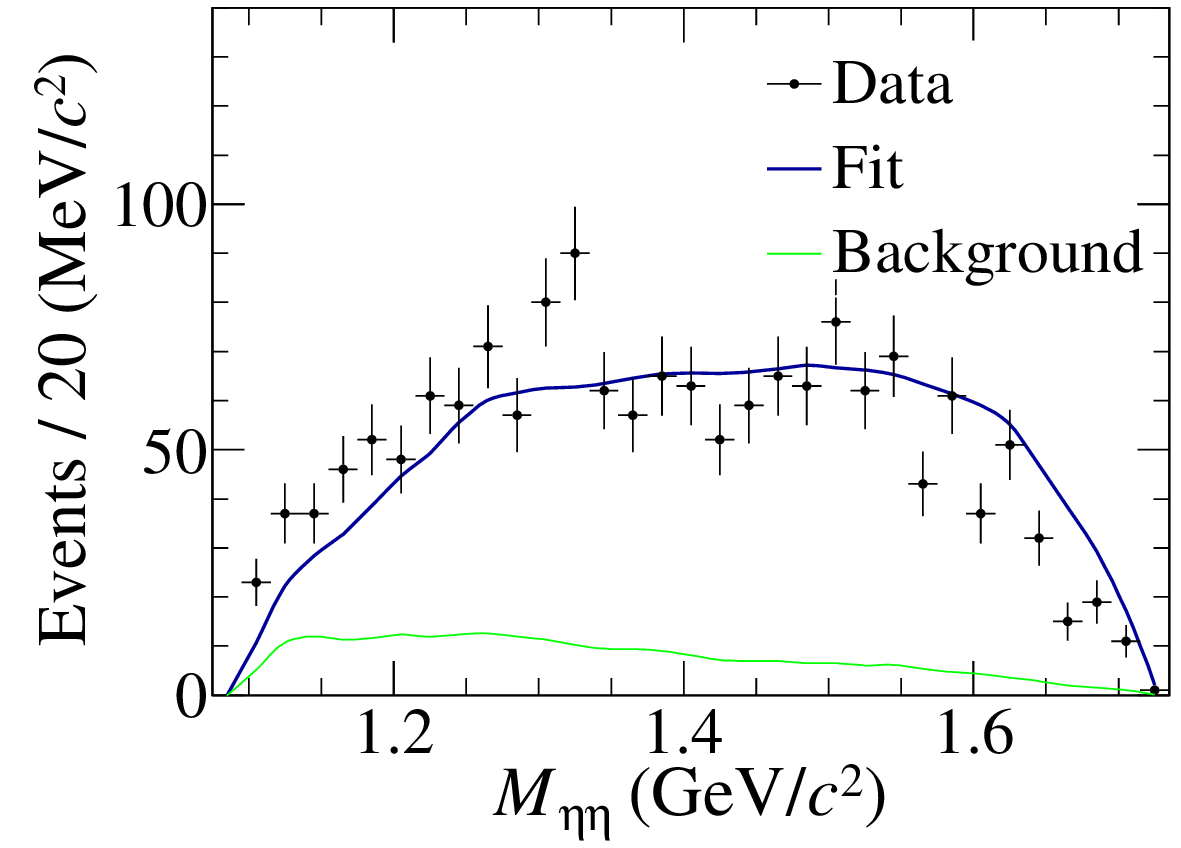}
%\put(-25,50){(h)}
\end{minipage}
\begin{minipage}[b]{0.24\textwidth}
\epsfig{width=0.98\textwidth,clip=true,file=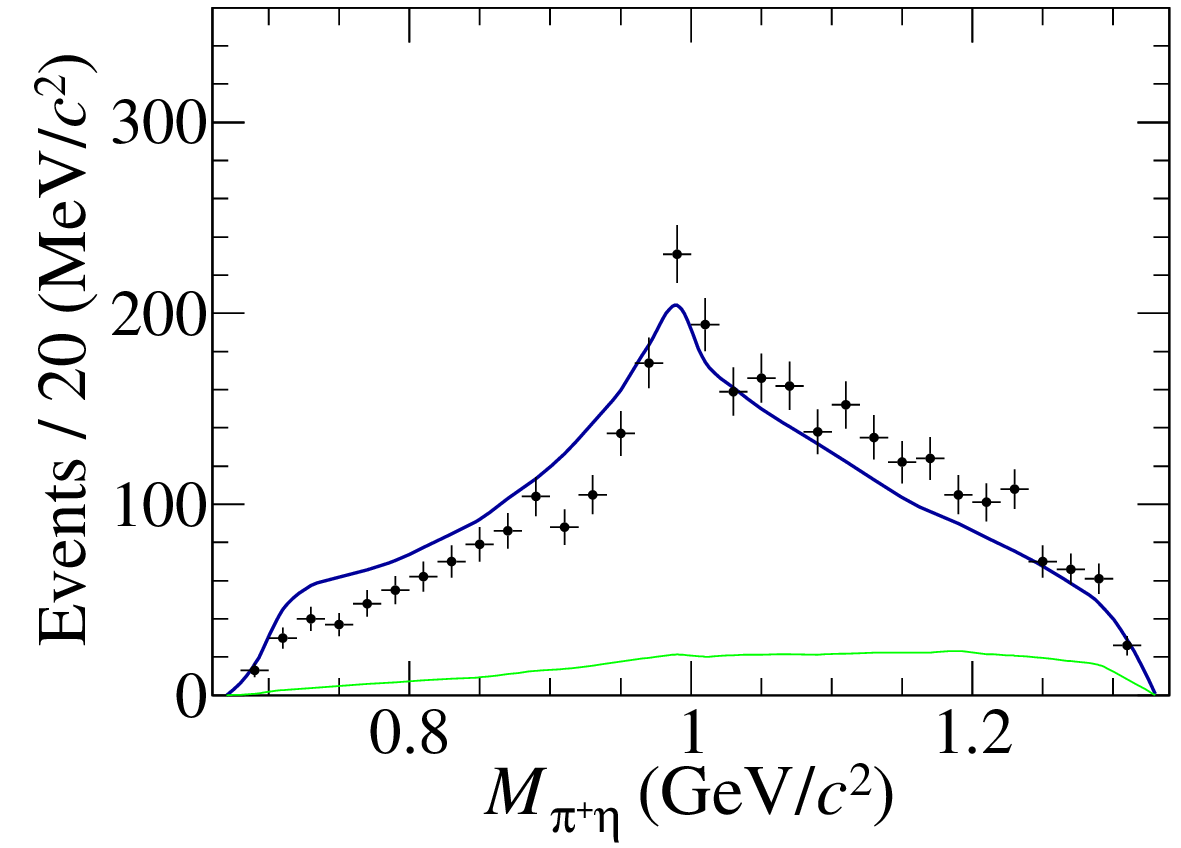}
%\put(-25,50){(i)}
\put(-85,70){V}
\end{minipage}
\begin{minipage}[b]{0.24\textwidth}
\epsfig{width=0.98\textwidth,clip=true,file=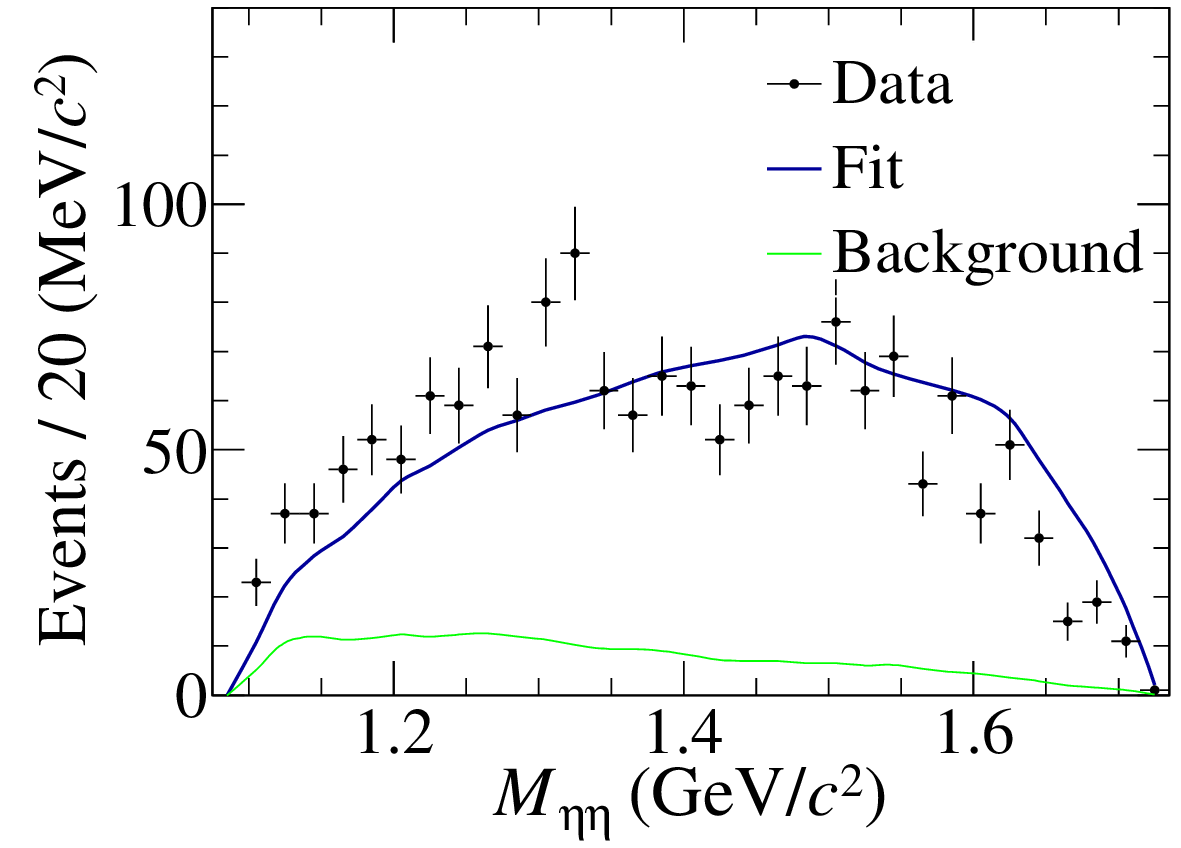}
%\put(-25,50){(j)}
\end{minipage}
\begin{minipage}[b]{0.24\textwidth}
\epsfig{width=0.98\textwidth,clip=true,file=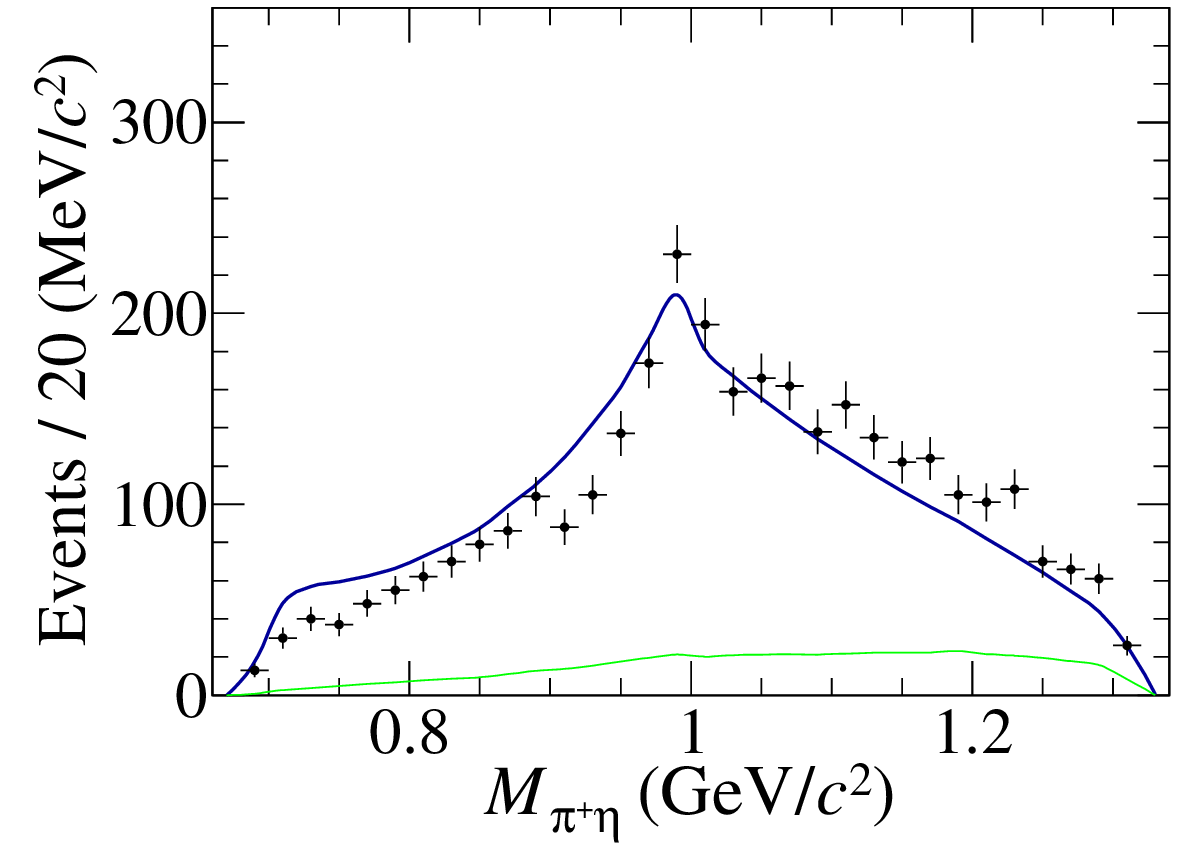}
%\put(-25,50){(k)}
\put(-85,70){VI}
\end{minipage}
\begin{minipage}[b]{0.24\textwidth}
\epsfig{width=0.98\textwidth,clip=true,file=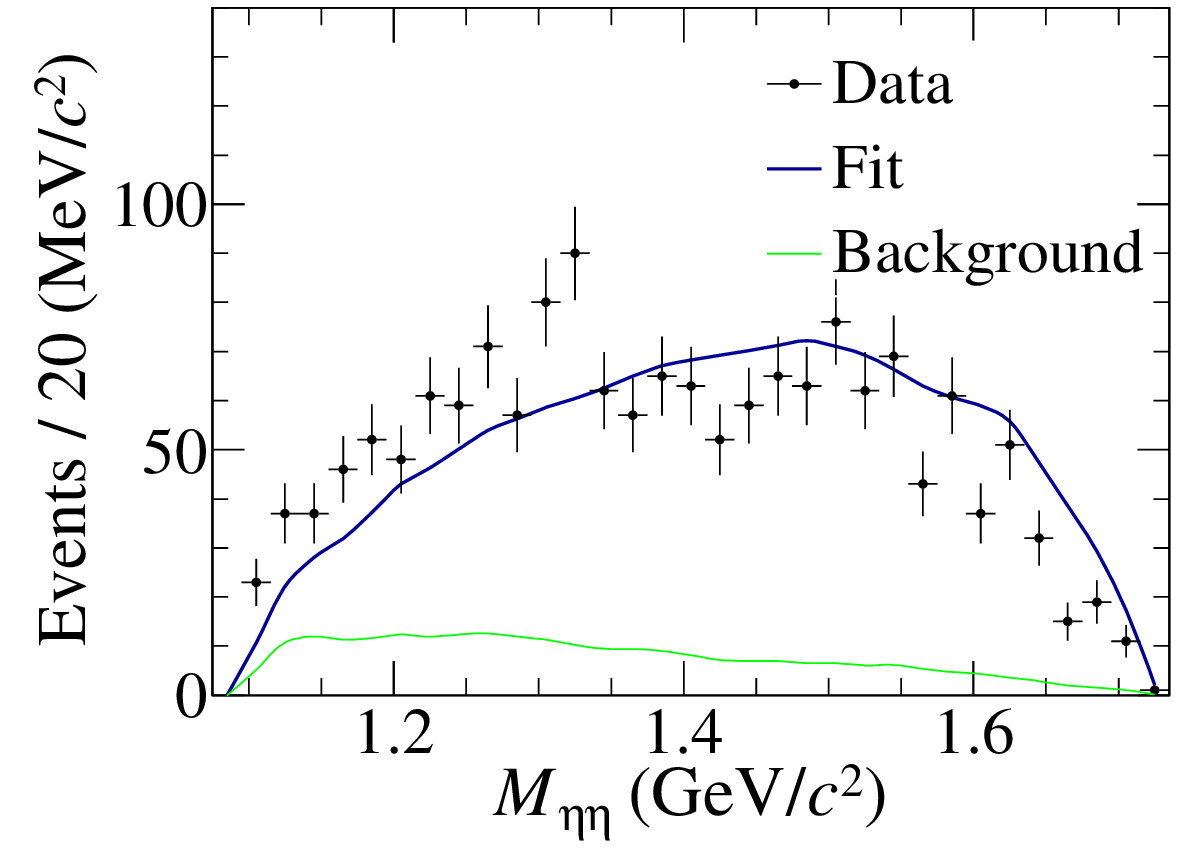}
%\put(-25,50){(l)}
\end{minipage}
\begin{minipage}[b]{0.24\textwidth}
\epsfig{width=0.98\textwidth,clip=true,file=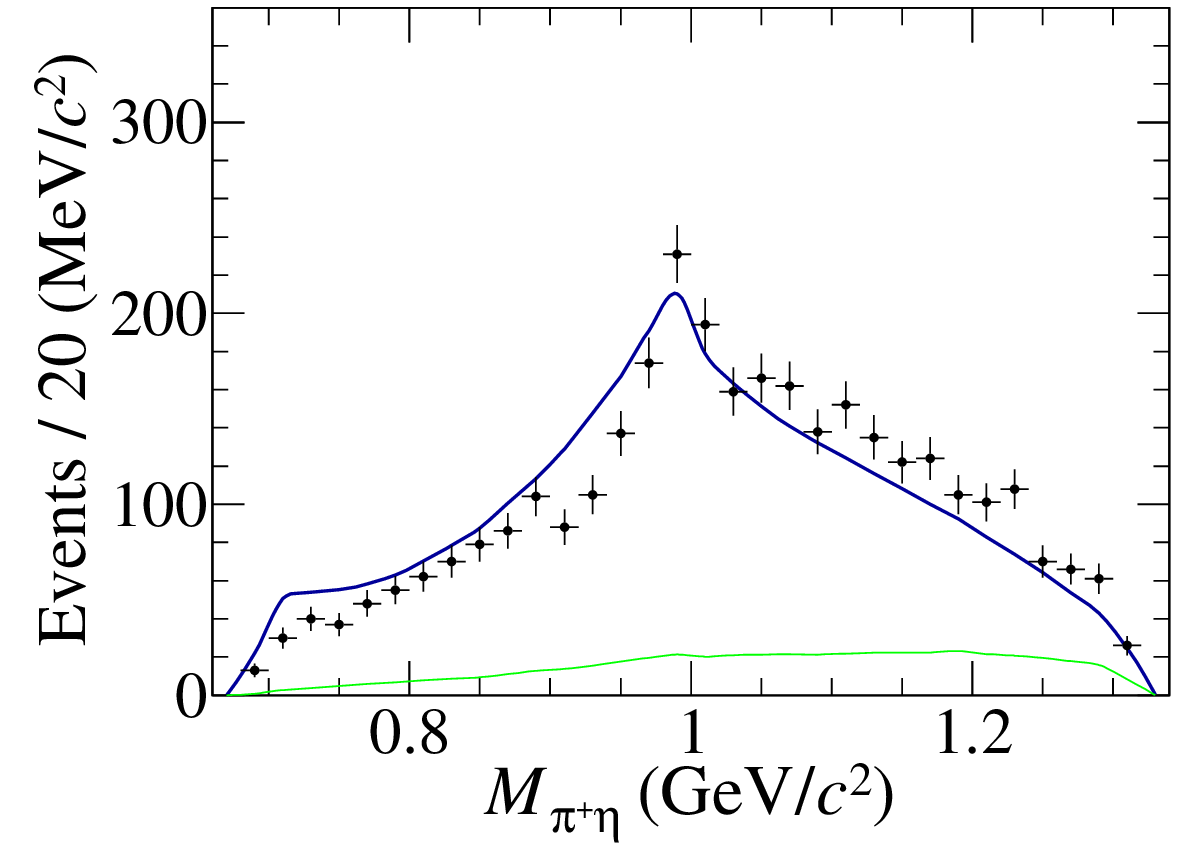}
%\put(-25,50){(m)}
\put(-85,70){VII}
\end{minipage}
\begin{minipage}[b]{0.24\textwidth}
\epsfig{width=0.98\textwidth,clip=true,file=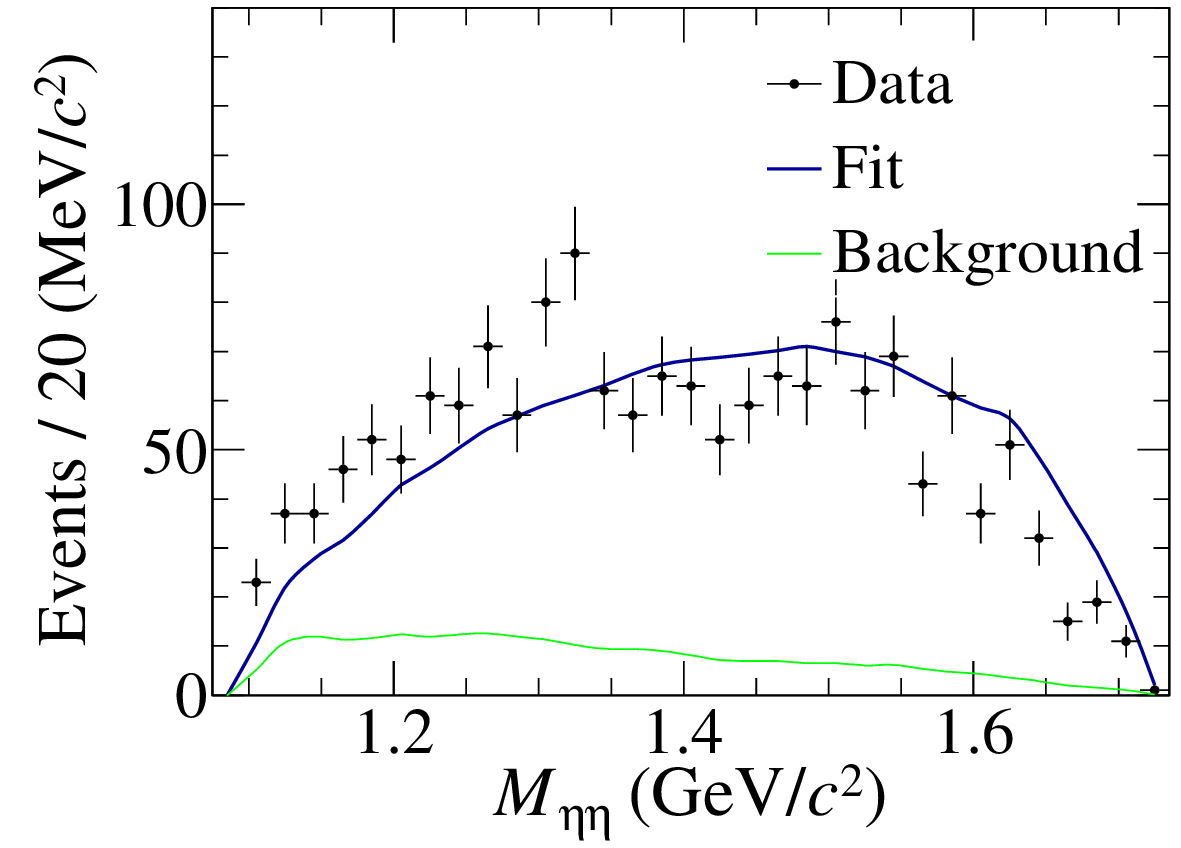}
%\put(-25,50){(n)}
\end{minipage}
\begin{minipage}[b]{0.24\textwidth}
\epsfig{width=0.98\textwidth,clip=true,file=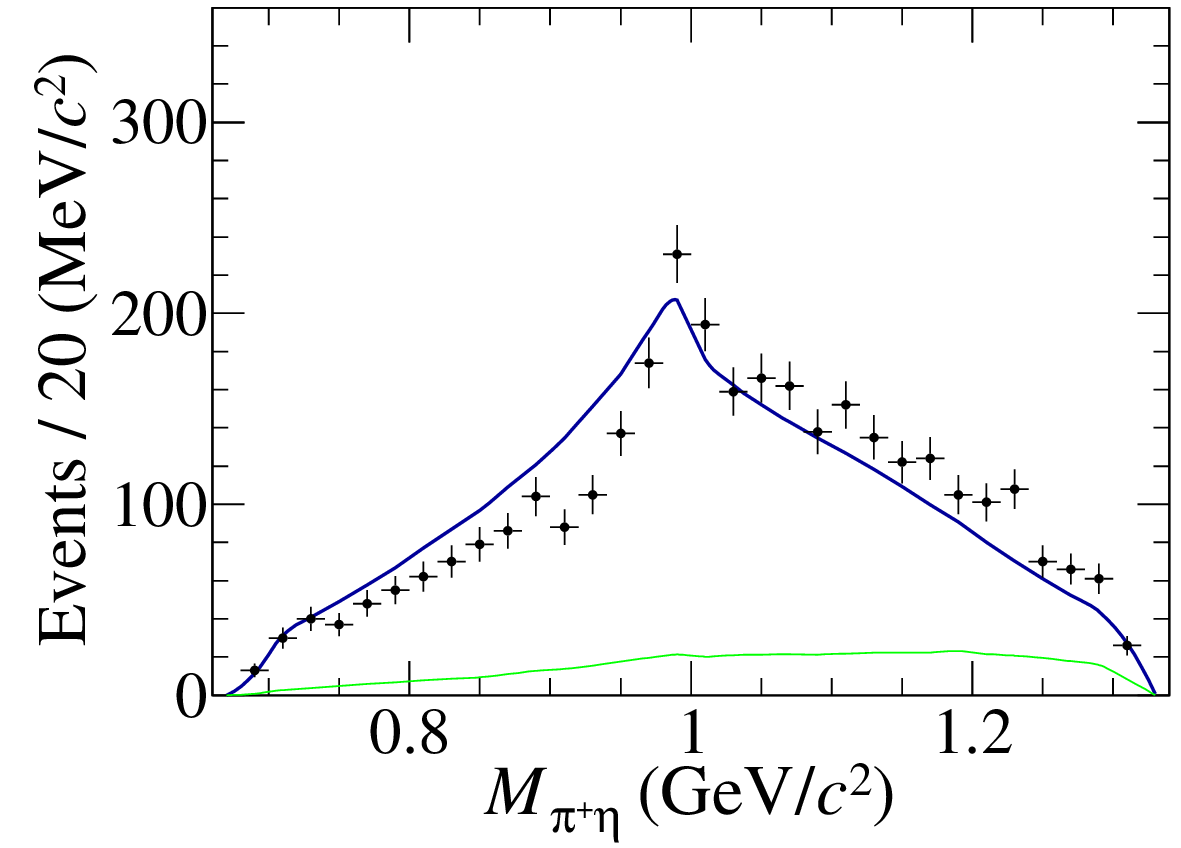}
%\put(-25,50){(o)}
\put(-85,70){VIII}
\end{minipage}
\begin{minipage}[b]{0.24\textwidth}
\epsfig{width=0.98\textwidth,clip=true,file=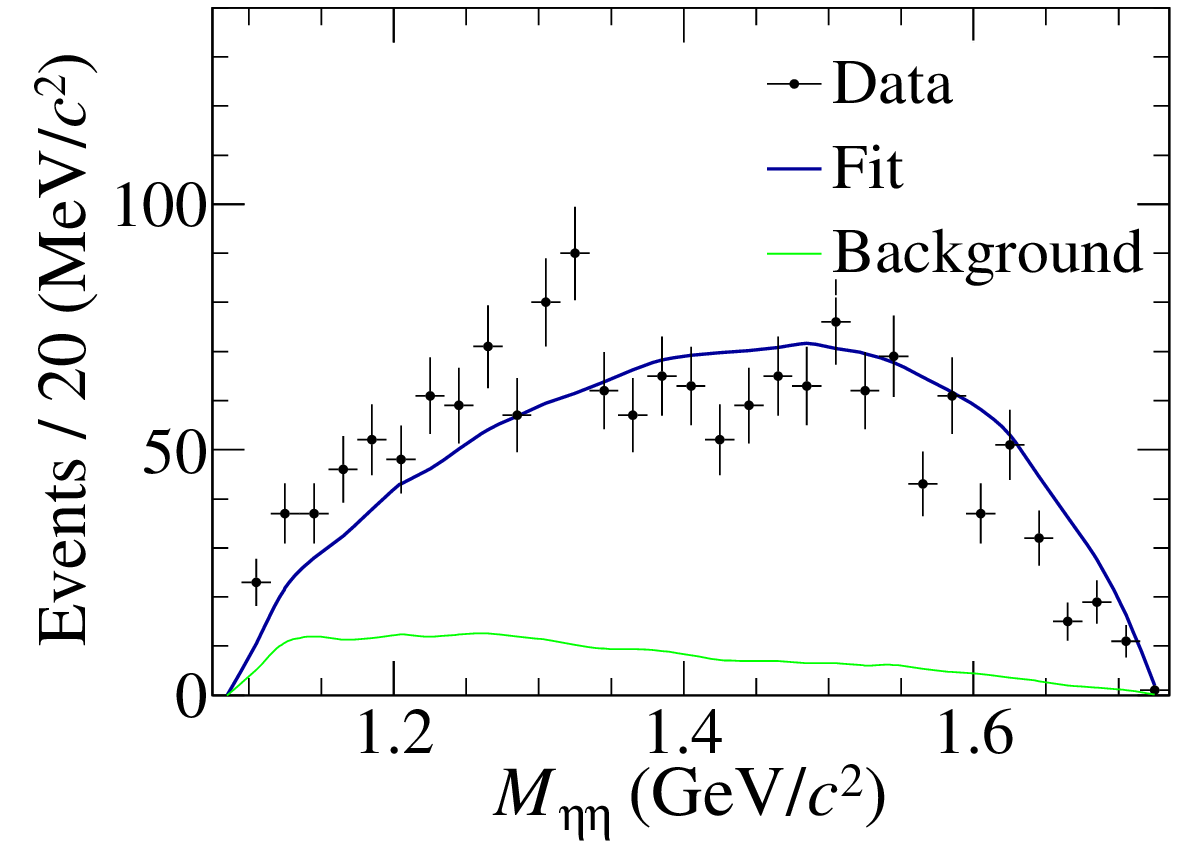}
%\put(-25,50){(p)}
\end{minipage}
\begin{minipage}[b]{0.24\textwidth}
\epsfig{width=0.98\textwidth,clip=true,file=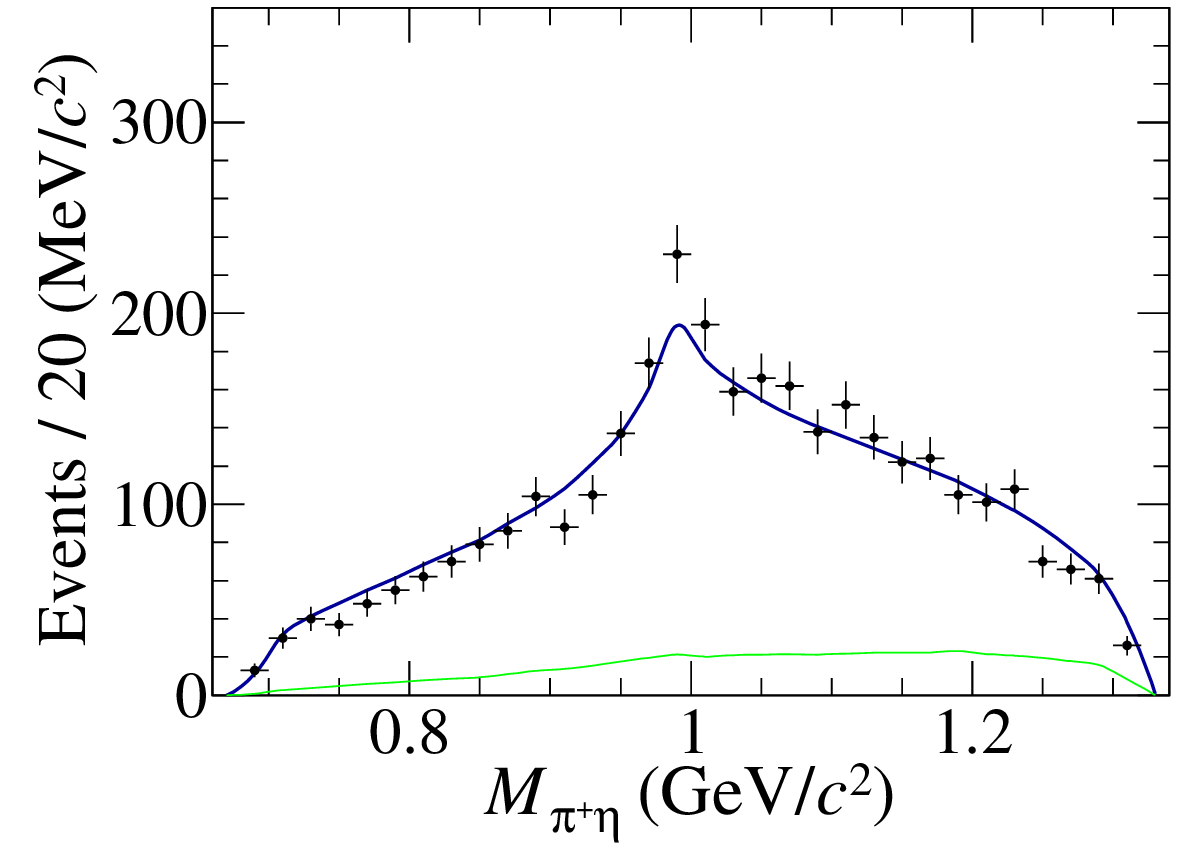}
%\put(-25,50){(q)}
\put(-85,70){IX}
\end{minipage}
\begin{minipage}[b]{0.24\textwidth}
\epsfig{width=0.98\textwidth,clip=true,file=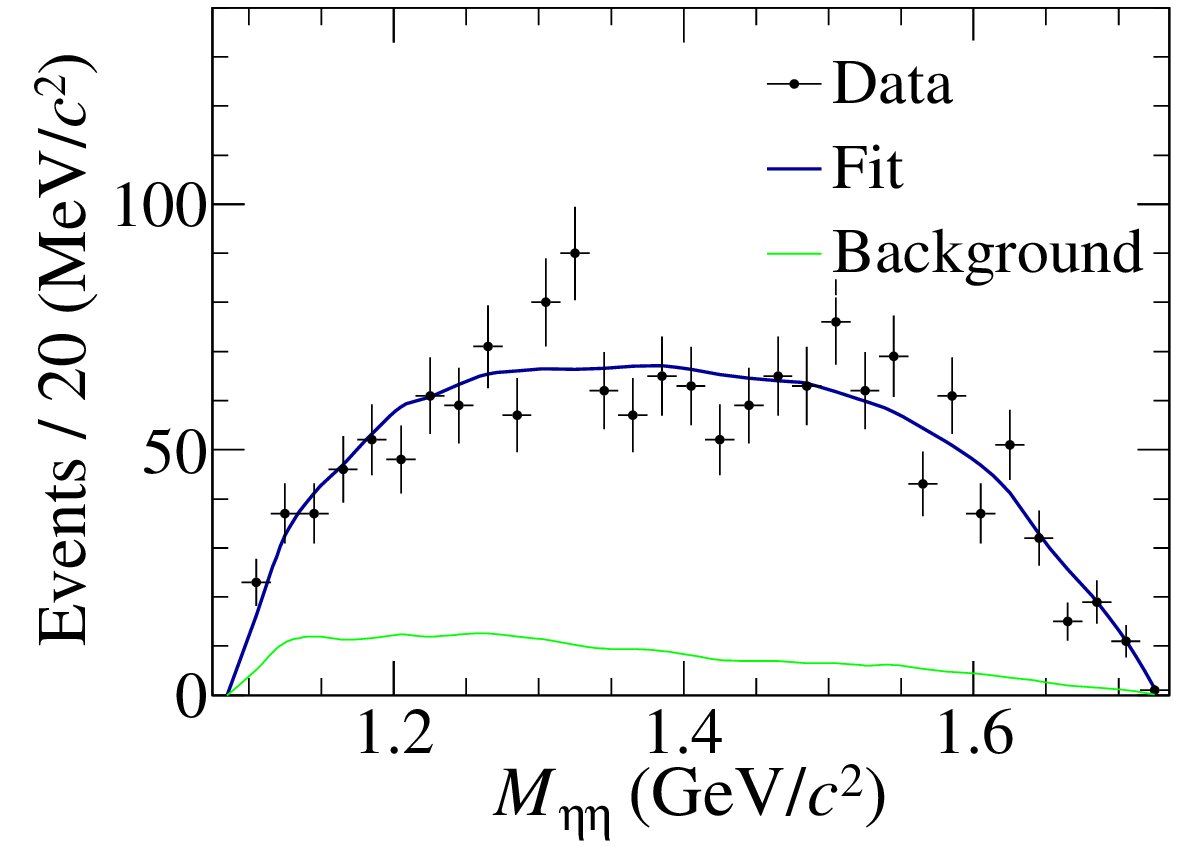}
%\put(-25,50){(r)}
\end{minipage}
\begin{minipage}[b]{0.24\textwidth}
\epsfig{width=0.98\textwidth,clip=true,file=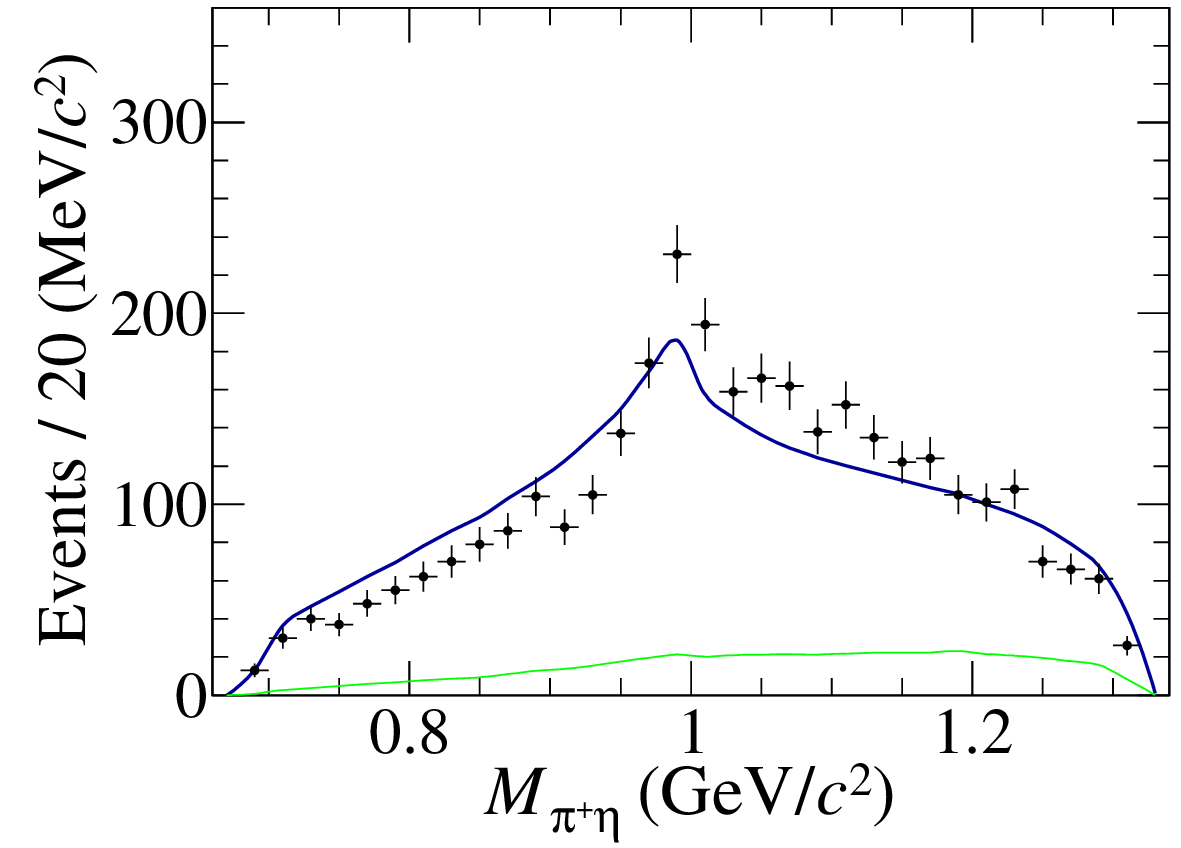}
%\put(-25,50){(s)}
\put(-85,70){X}
\end{minipage}
\begin{minipage}[b]{0.24\textwidth}
\epsfig{width=0.98\textwidth,clip=true,file=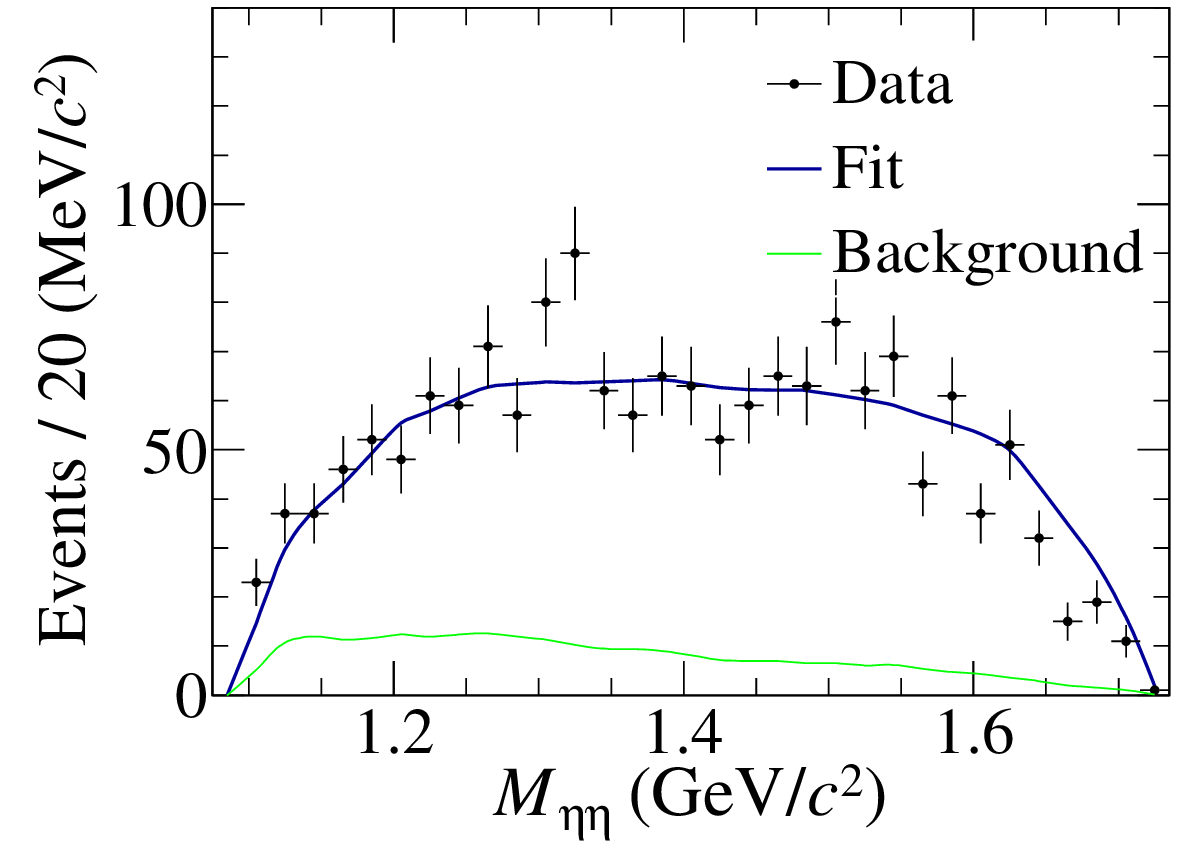}
%\put(-25,50){(t)}
\end{minipage}
\begin{minipage}[b]{0.24\textwidth}
\epsfig{width=0.98\textwidth,clip=true,file=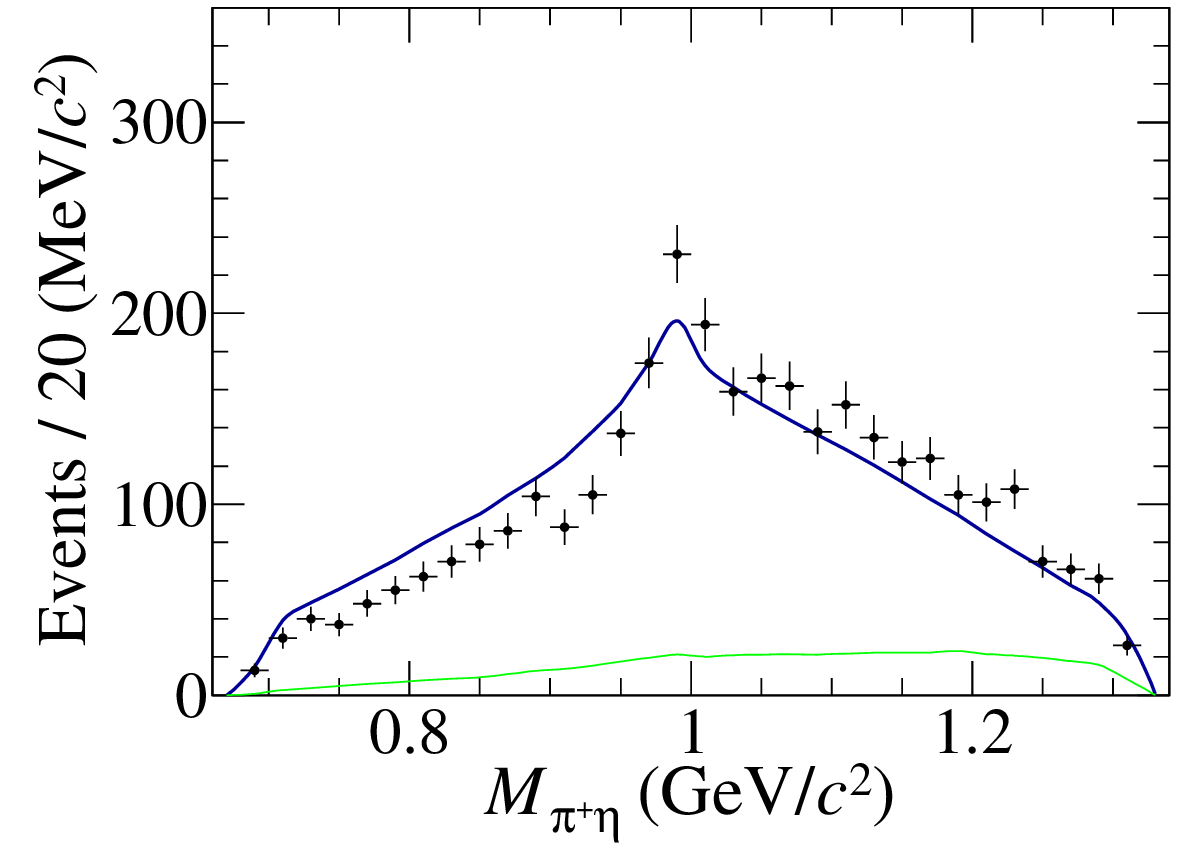}
%\put(-25,50){(u)}
\put(-85,70){XI}
\end{minipage}
\begin{minipage}[b]{0.24\textwidth}
\epsfig{width=0.98\textwidth,clip=true,file=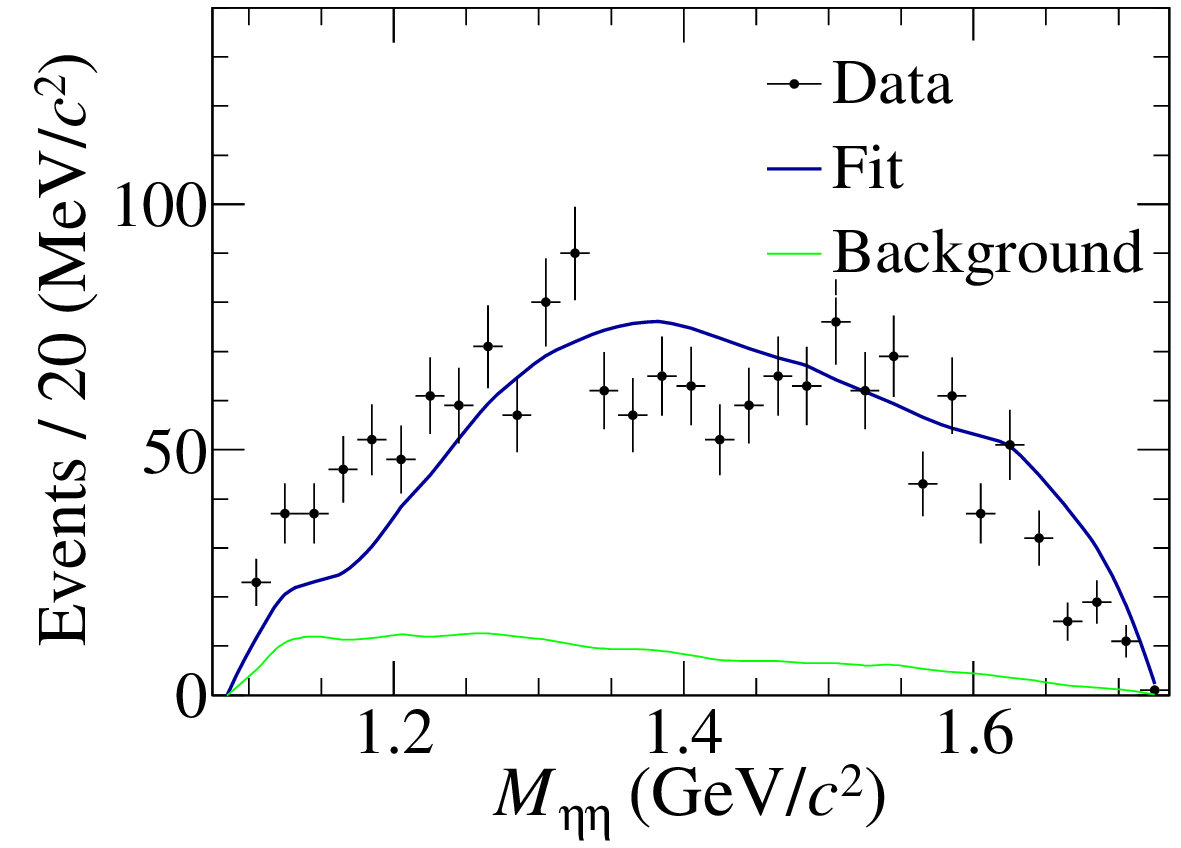}
%\put(-25,50){(v)}
\end{minipage}
\begin{minipage}[b]{0.24\textwidth}
\epsfig{width=0.98\textwidth,clip=true,file=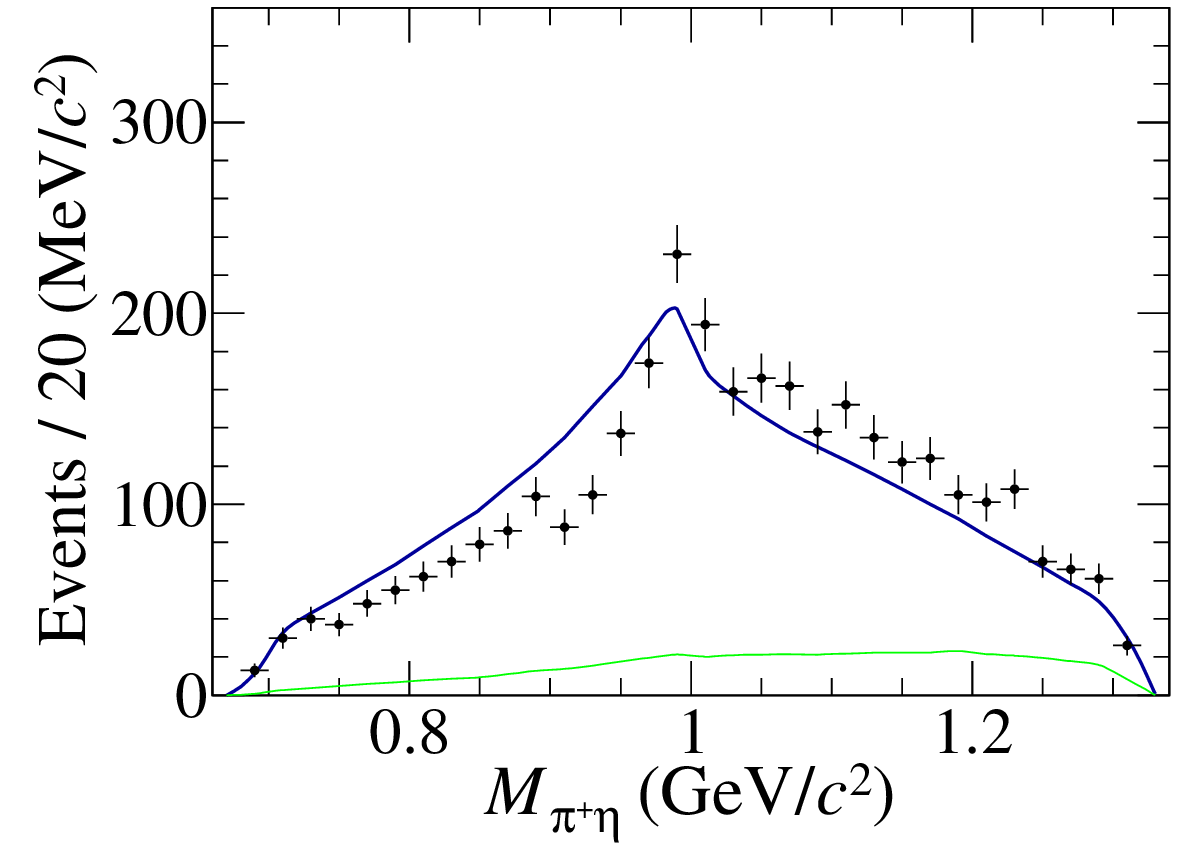}
%\put(-25,50){(w)}
\put(-85,70){XII}
\end{minipage}
\begin{minipage}[b]{0.24\textwidth}
\epsfig{width=0.98\textwidth,clip=true,file=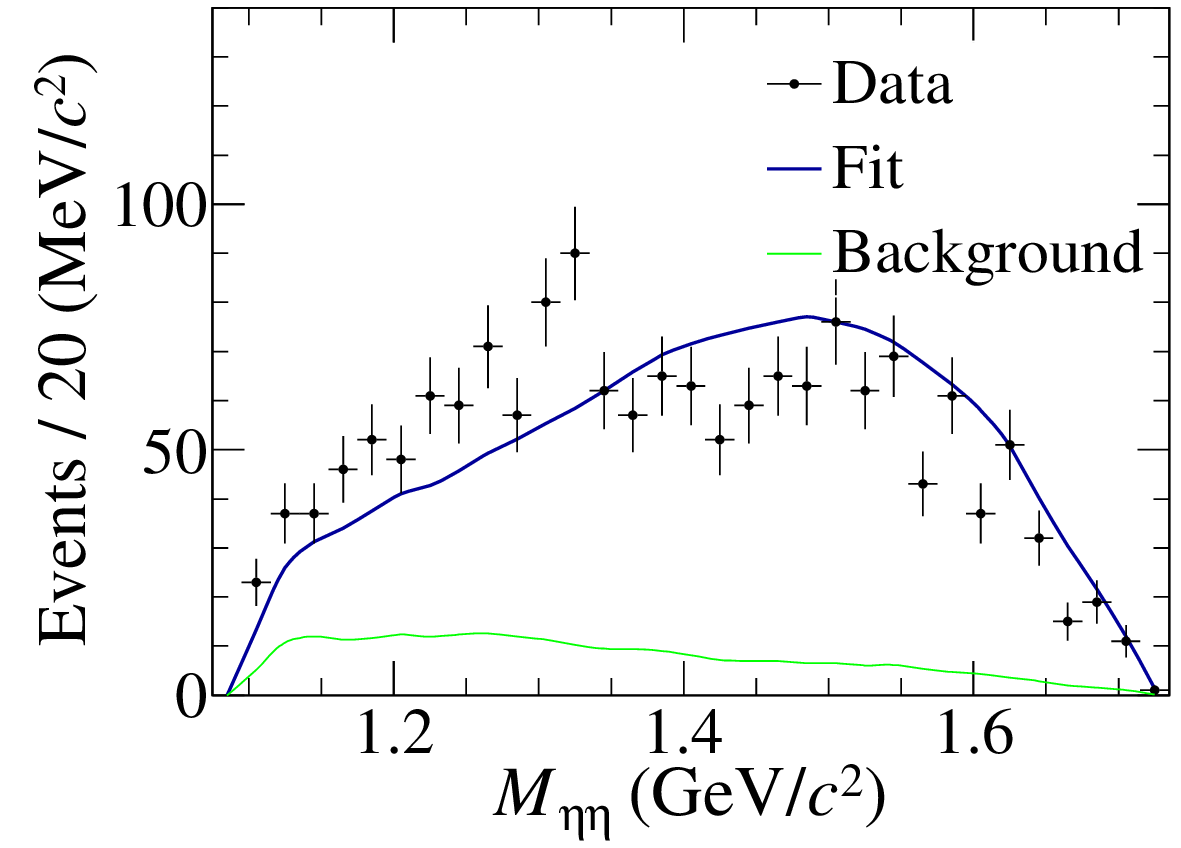}
%\put(-25,50){(x)}
\end{minipage}
\caption{The same as Fig.~\ref{fig:Flatteaddamp}, but for the models using the $T$-matrix form instead of the $a_{0}(980)$ amplitude.}
\label{fig:TMatrixadd}
\end{center}
\end{figure}

\begin{figure}[htbp]
\begin{center}
\begin{minipage}[b]{0.24\textwidth}
\epsfig{width=0.98\textwidth,clip=true,file=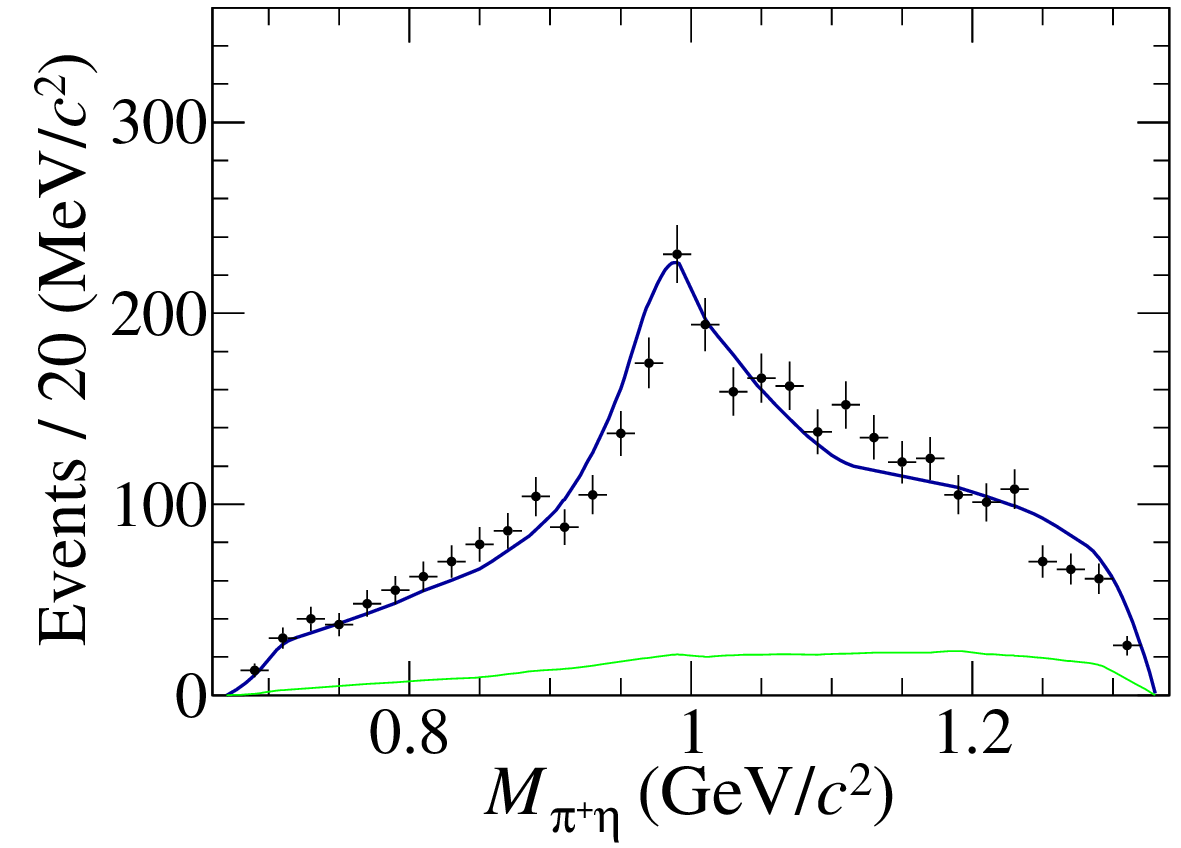}
%\put(-25,50){(a)}
\put(-85,70){I}
\end{minipage}
\begin{minipage}[b]{0.24\textwidth}
\epsfig{width=0.98\textwidth,clip=true,file=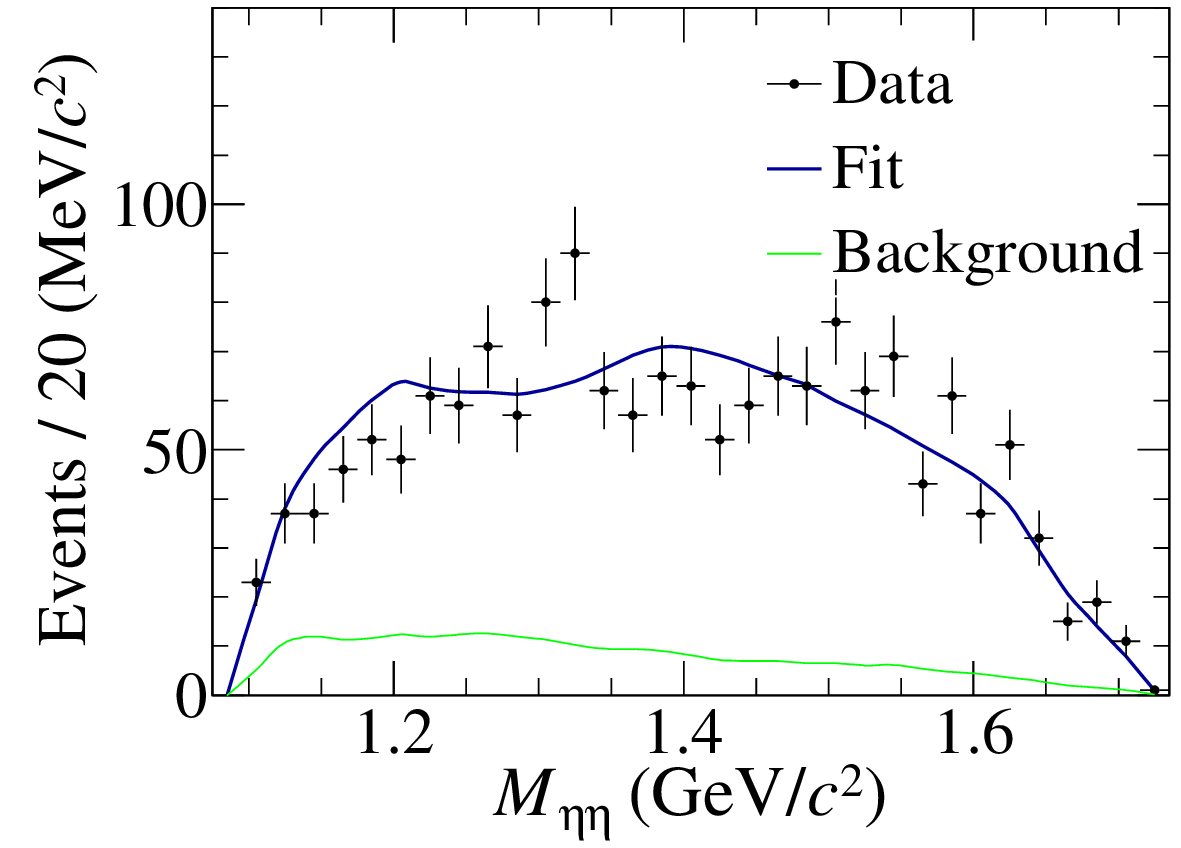}
%\put(-25,50){(b)}
\end{minipage}
\begin{minipage}[b]{0.24\textwidth}
\epsfig{width=0.98\textwidth,clip=true,file=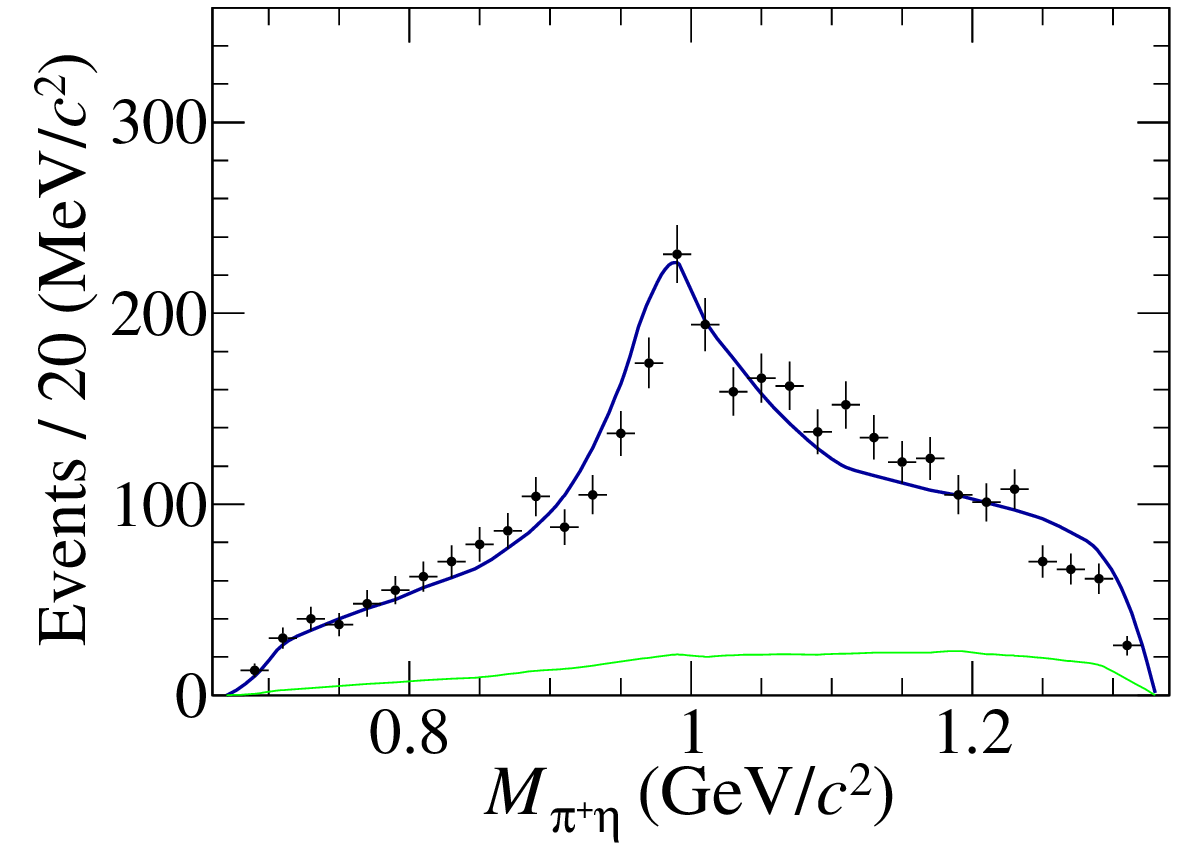}
%\put(-25,50){(c)}
\put(-85,70){II}
\end{minipage}
\begin{minipage}[b]{0.24\textwidth}
\epsfig{width=0.98\textwidth,clip=true,file=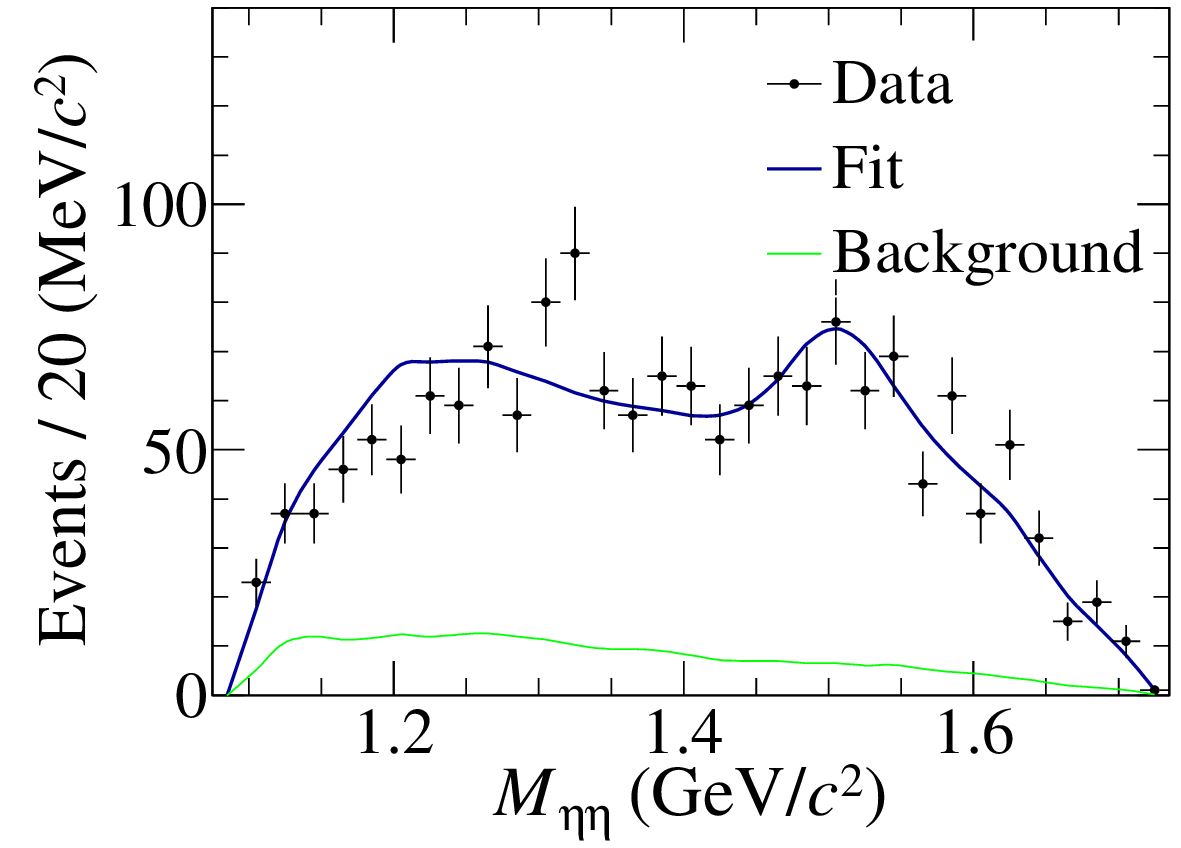}
%\put(-25,50){(d)}
\end{minipage}
\begin{minipage}[b]{0.24\textwidth}
\epsfig{width=0.98\textwidth,clip=true,file=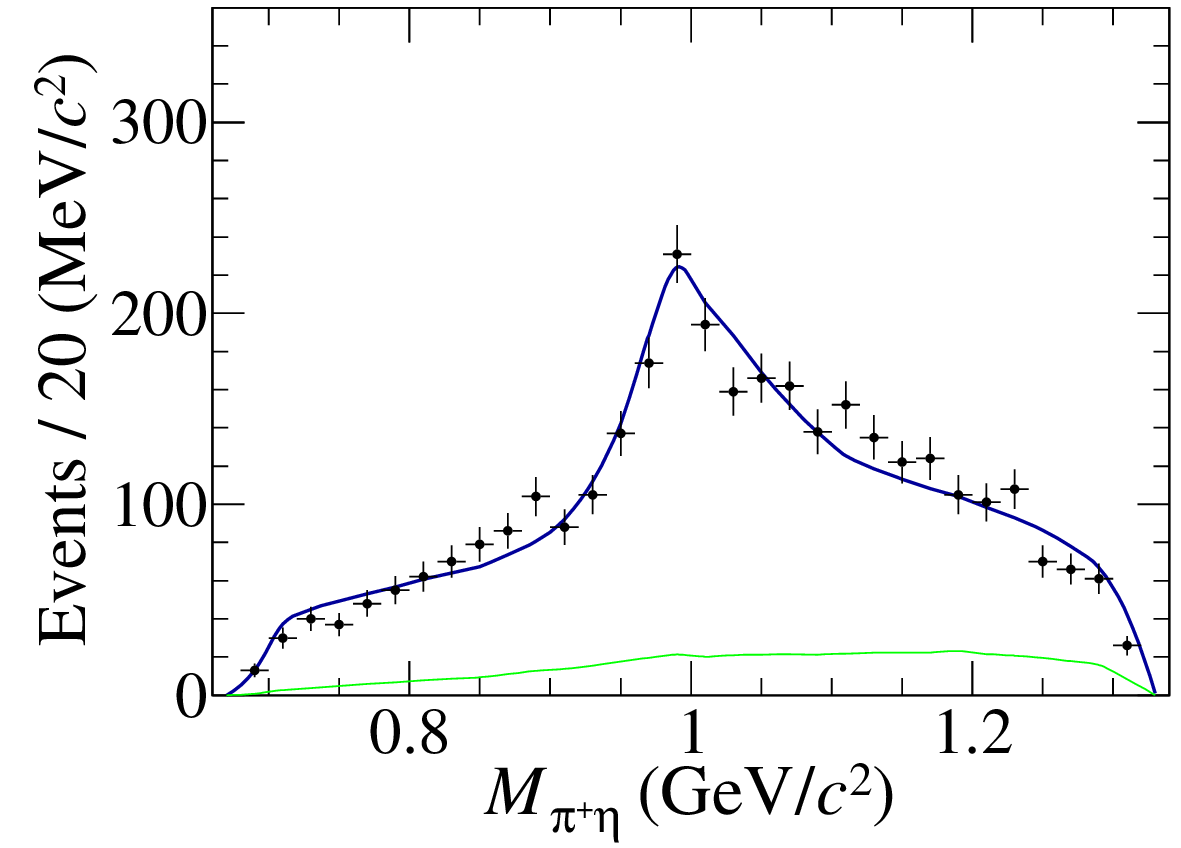}
%\put(-25,50){(e)}
\put(-85,70){III}
\end{minipage}
\begin{minipage}[b]{0.24\textwidth}
\epsfig{width=0.98\textwidth,clip=true,file=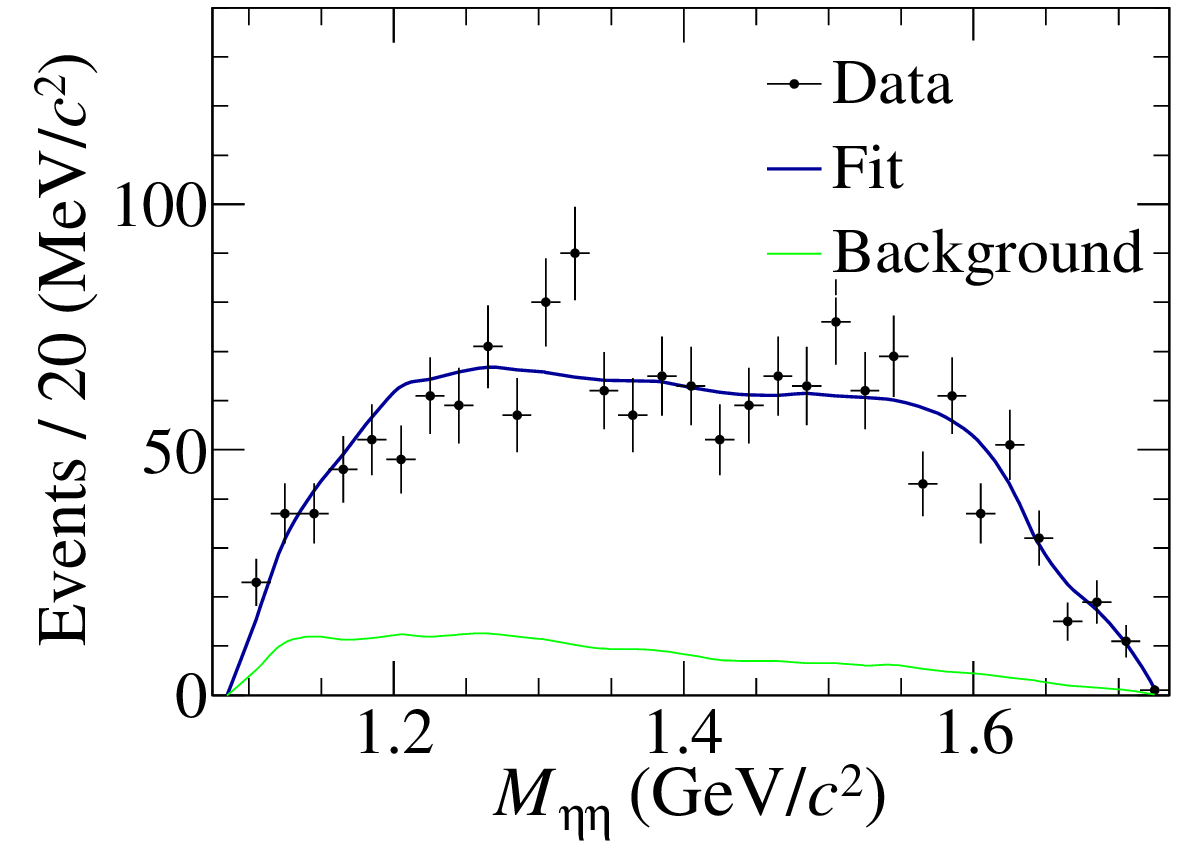}
%\put(-25,50){(f)}
\end{minipage}
\begin{minipage}[b]{0.24\textwidth}
\epsfig{width=0.98\textwidth,clip=true,file=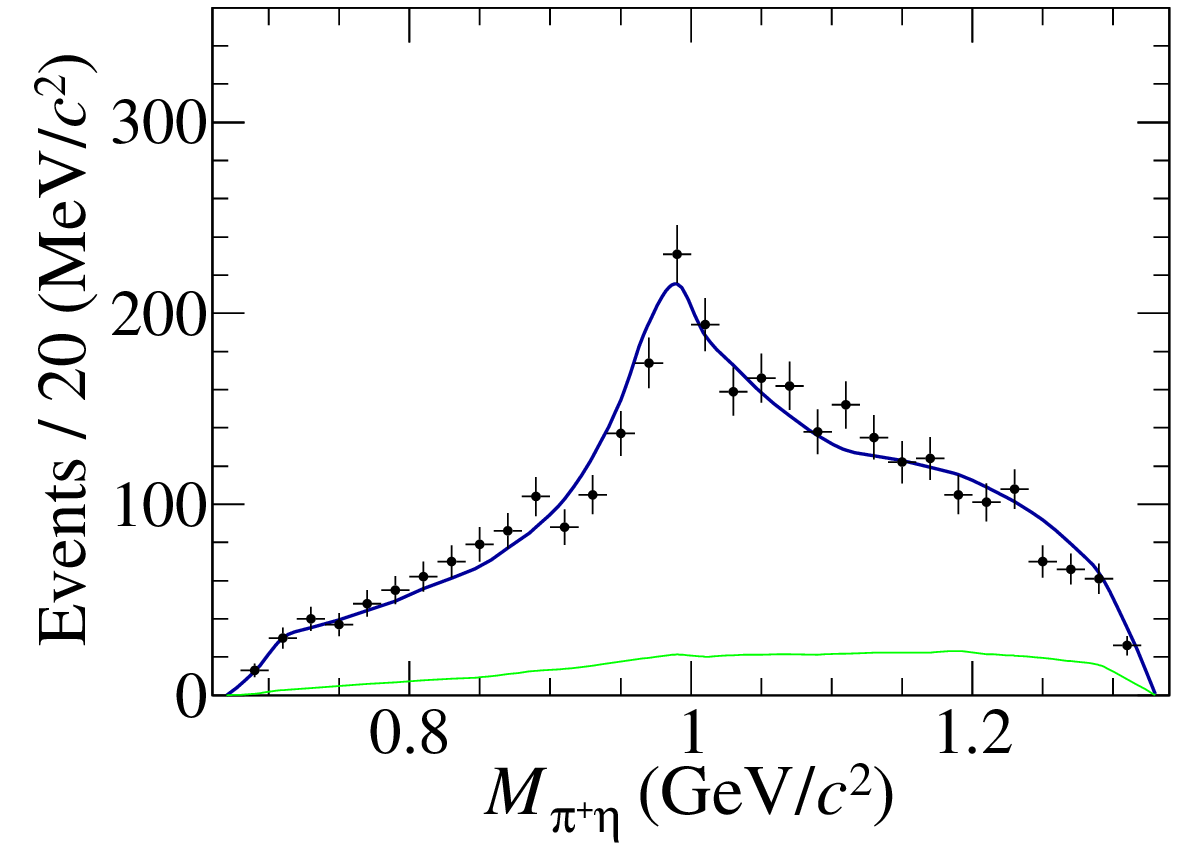}
%\put(-25,50){(g)}
\put(-85,70){IV}
\end{minipage}
\begin{minipage}[b]{0.24\textwidth}
\epsfig{width=0.98\textwidth,clip=true,file=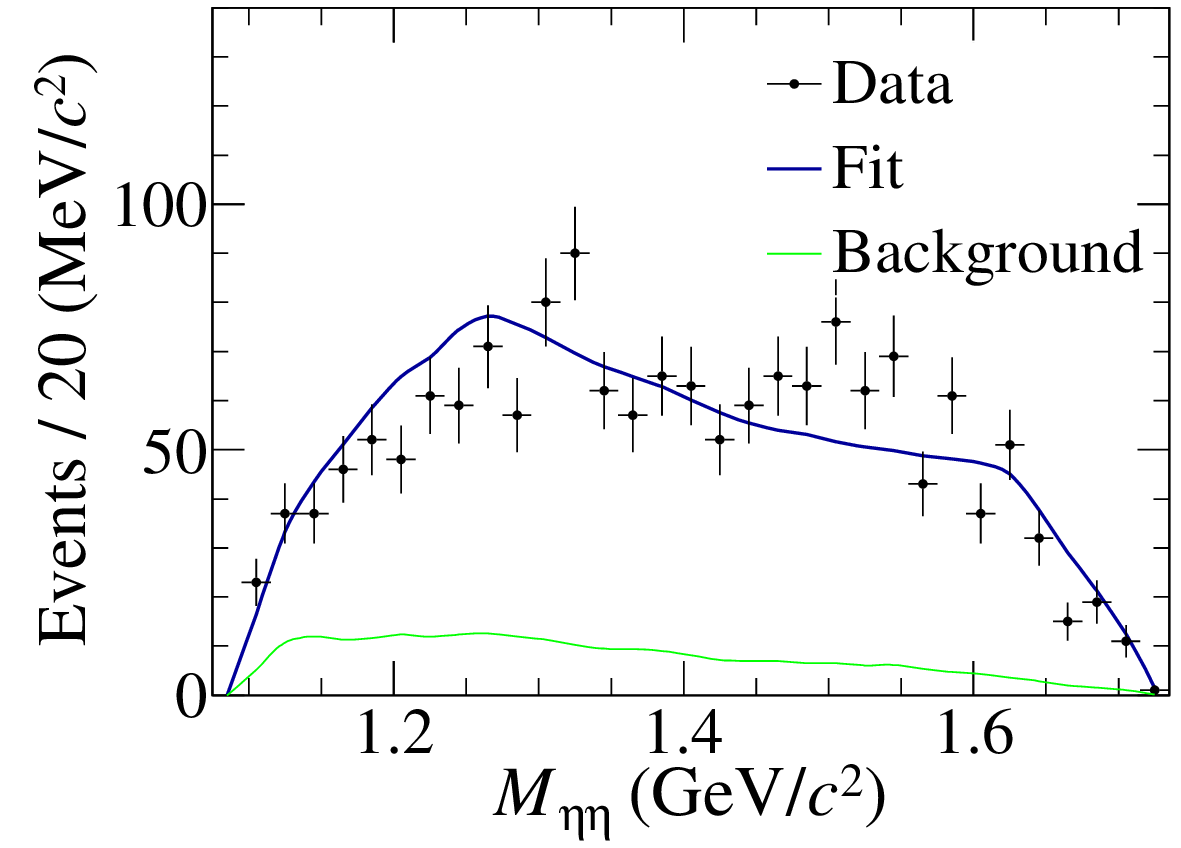}
%\put(-25,50){(h)}
\end{minipage}
\begin{minipage}[b]{0.24\textwidth}
\epsfig{width=0.98\textwidth,clip=true,file=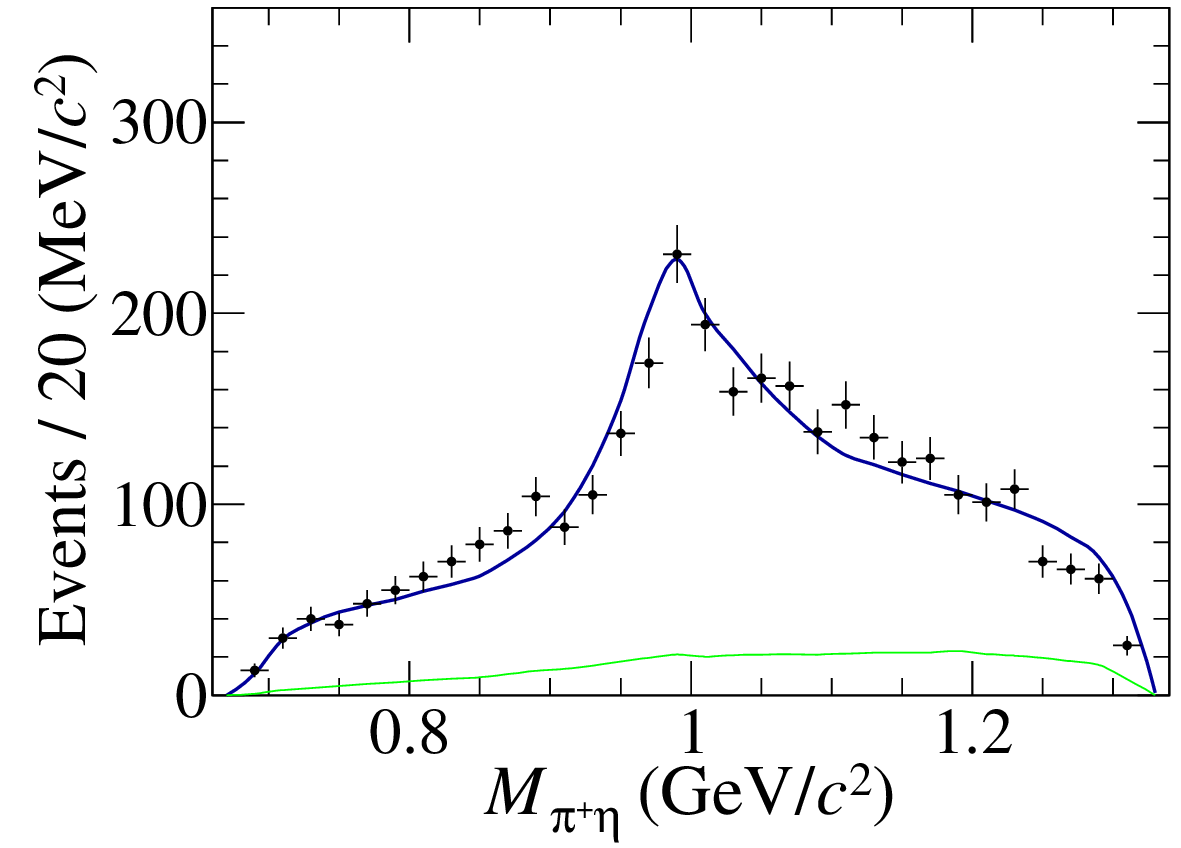}
%\put(-25,50){(i)}
\put(-85,70){V}
\end{minipage}
\begin{minipage}[b]{0.24\textwidth}
\epsfig{width=0.98\textwidth,clip=true,file=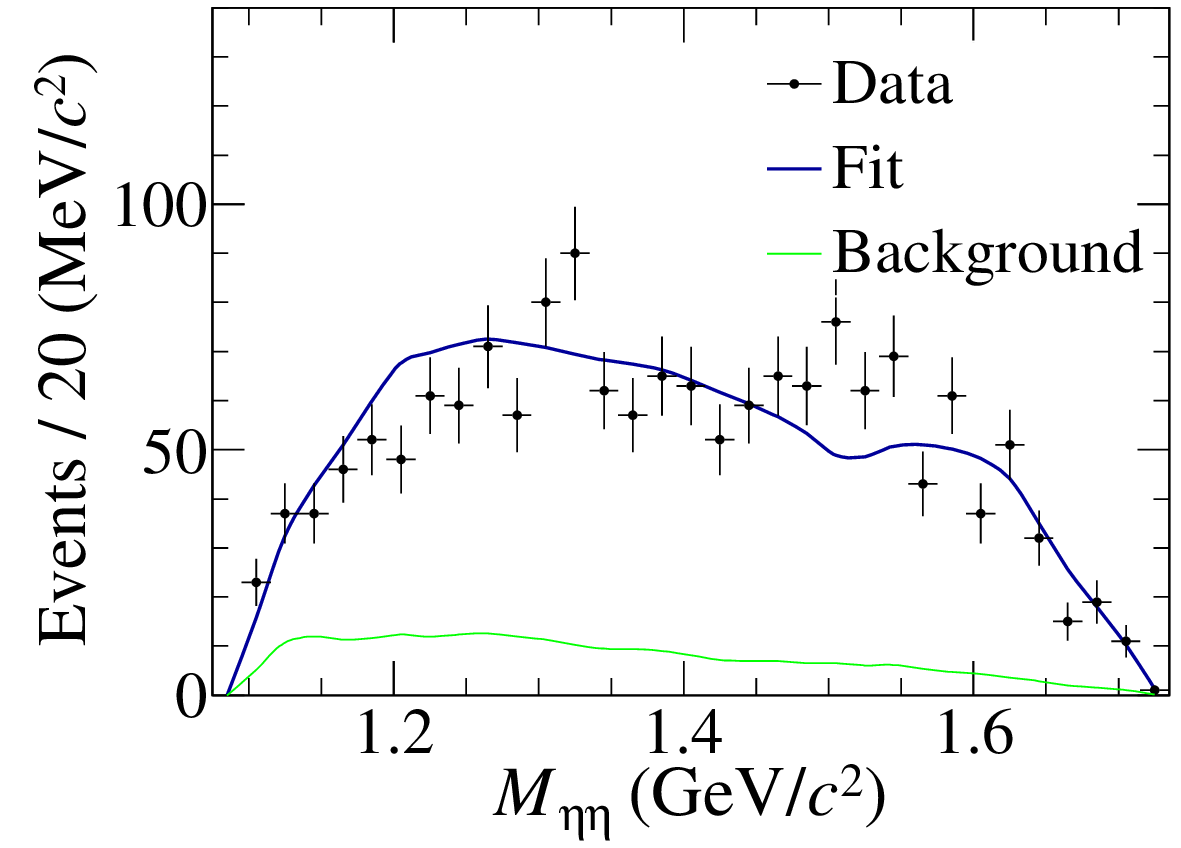}
%\put(-25,50){(j)}
\end{minipage}
\begin{minipage}[b]{0.24\textwidth}
\epsfig{width=0.98\textwidth,clip=true,file=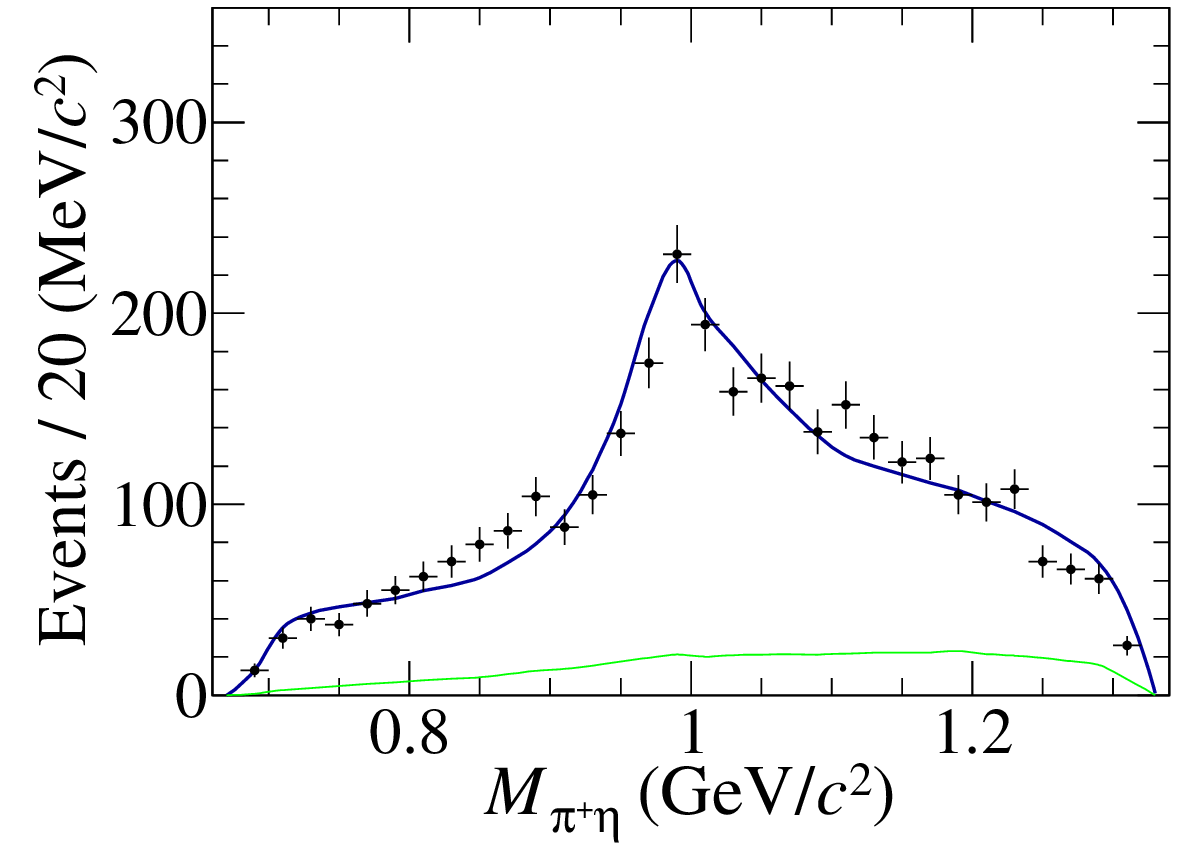}
%\put(-25,50){(k)}
\put(-85,70){VI}
\end{minipage}
\begin{minipage}[b]{0.24\textwidth}
\epsfig{width=0.98\textwidth,clip=true,file=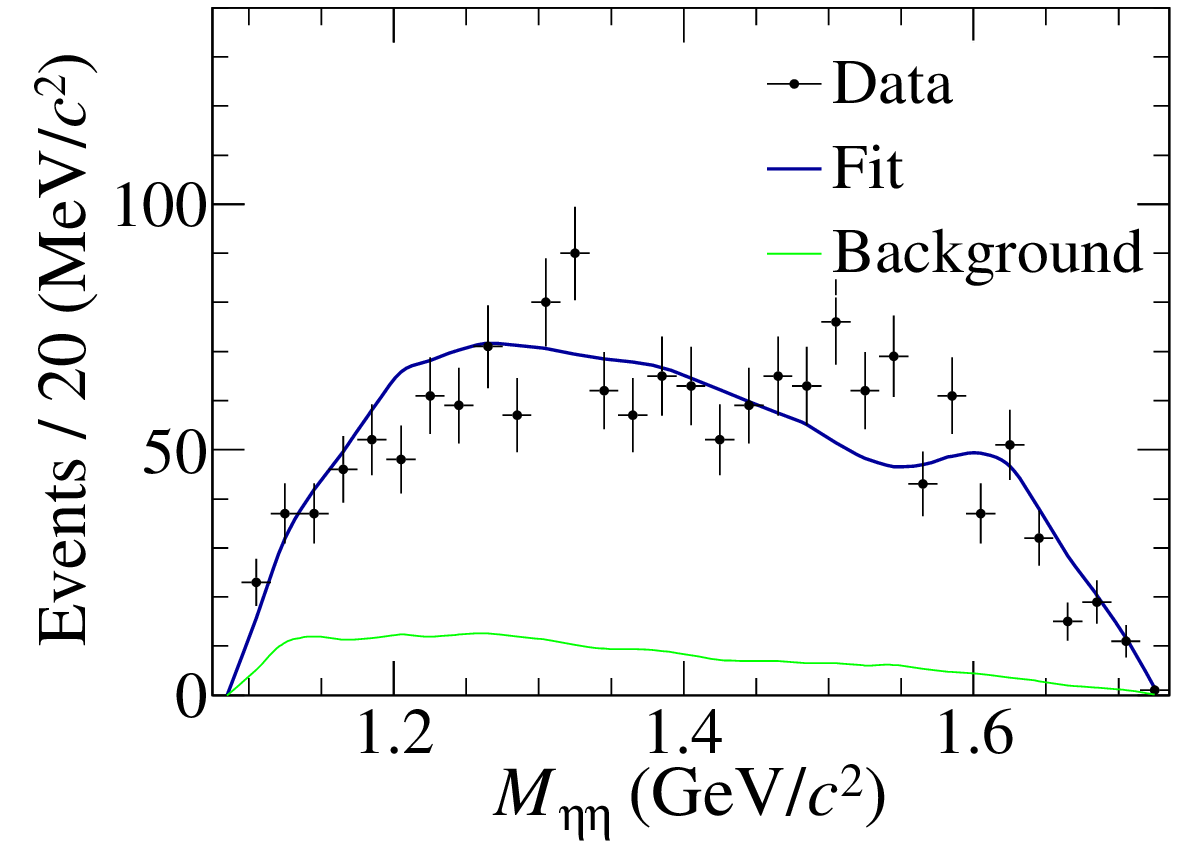}
%\put(-25,50){(l)}
\end{minipage}
\begin{minipage}[b]{0.24\textwidth}
\epsfig{width=0.98\textwidth,clip=true,file=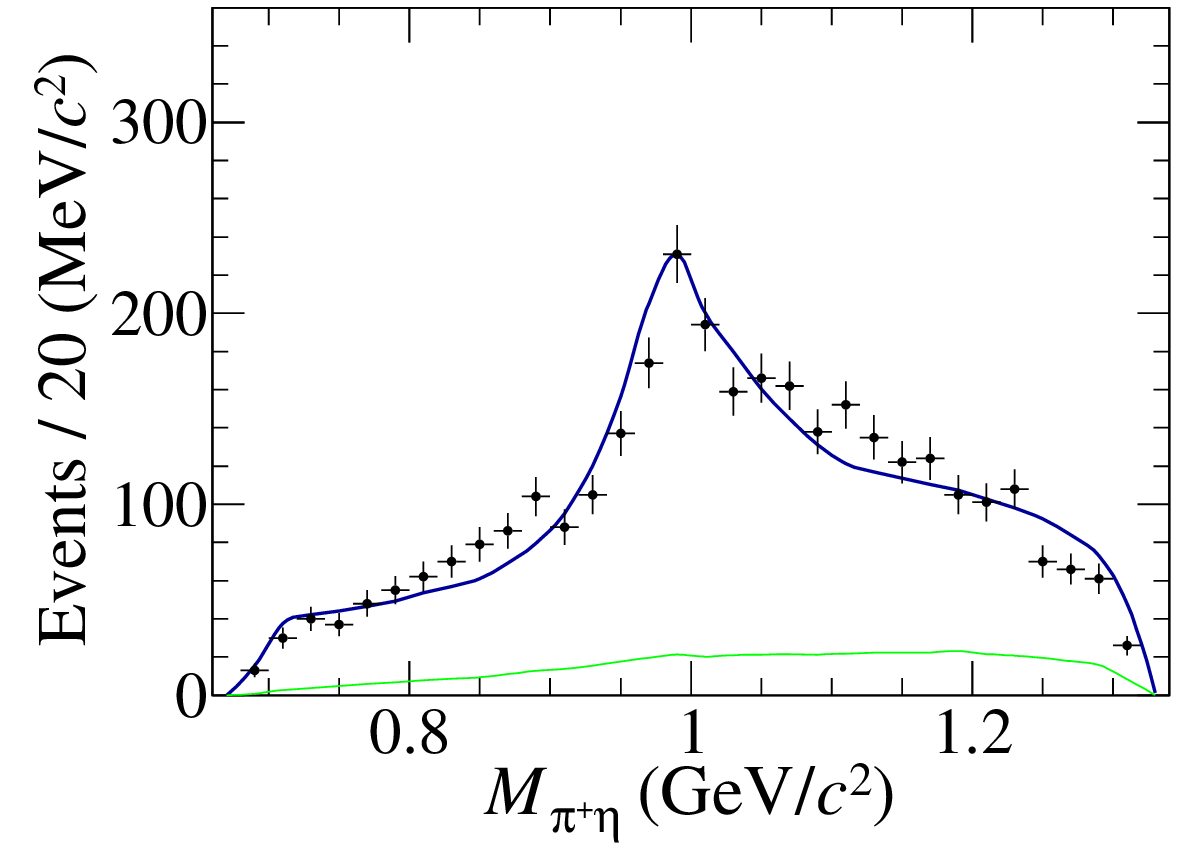}
%\put(-25,50){(m)}
\put(-85,70){VII}
\end{minipage}
\begin{minipage}[b]{0.24\textwidth}
\epsfig{width=0.98\textwidth,clip=true,file=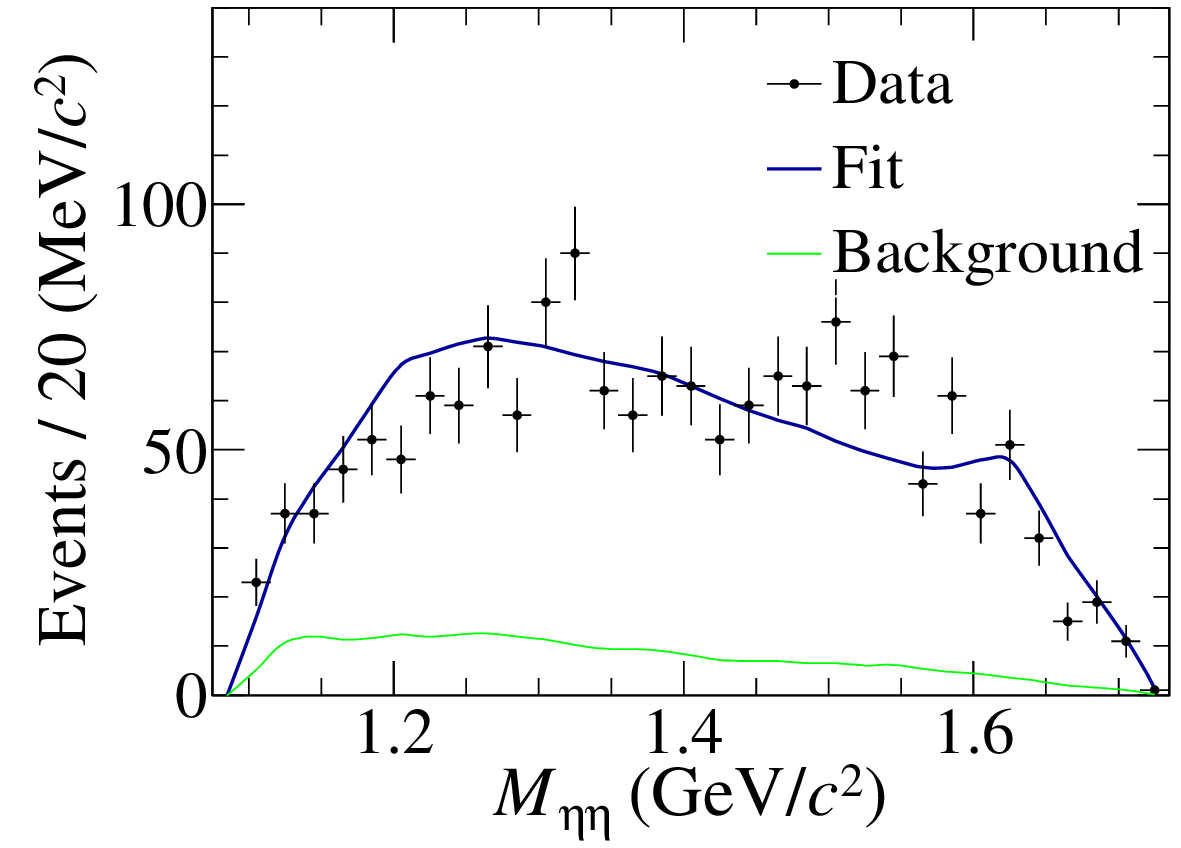}
%\put(-25,50){(n)}
\end{minipage}
\begin{minipage}[b]{0.24\textwidth}
\epsfig{width=0.98\textwidth,clip=true,file=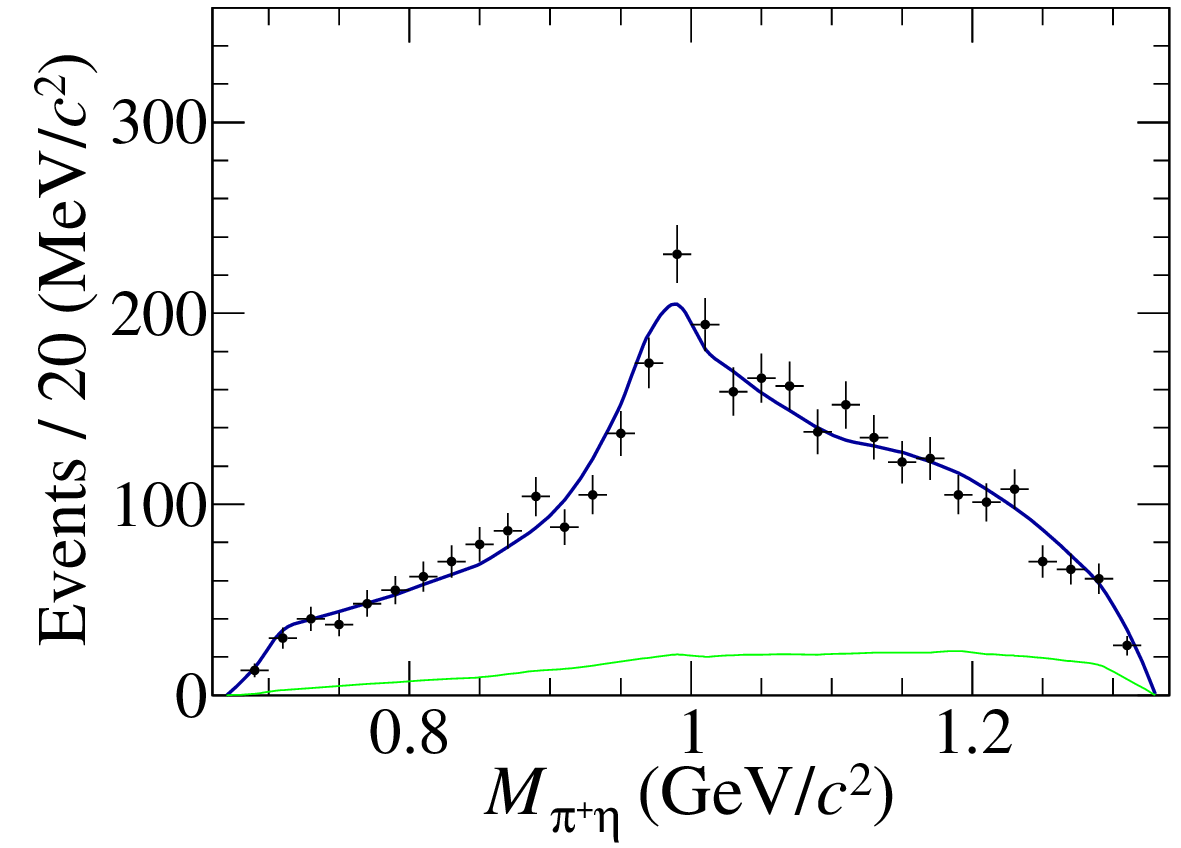}
%\put(-25,50){(o)}
\put(-85,70){VIII}
\end{minipage}
\begin{minipage}[b]{0.24\textwidth}
\epsfig{width=0.98\textwidth,clip=true,file=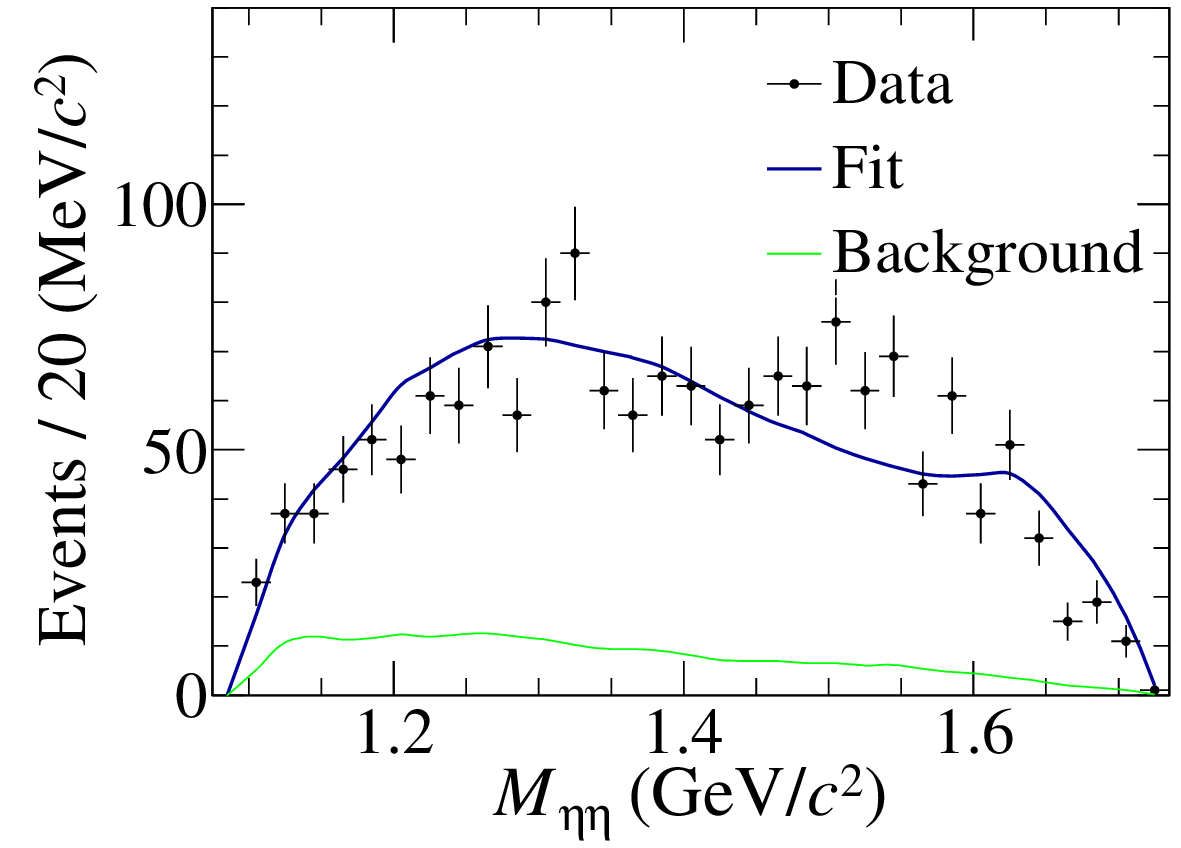}
%\put(-25,50){(p)}
\end{minipage}
\begin{minipage}[b]{0.24\textwidth}
\epsfig{width=0.98\textwidth,clip=true,file=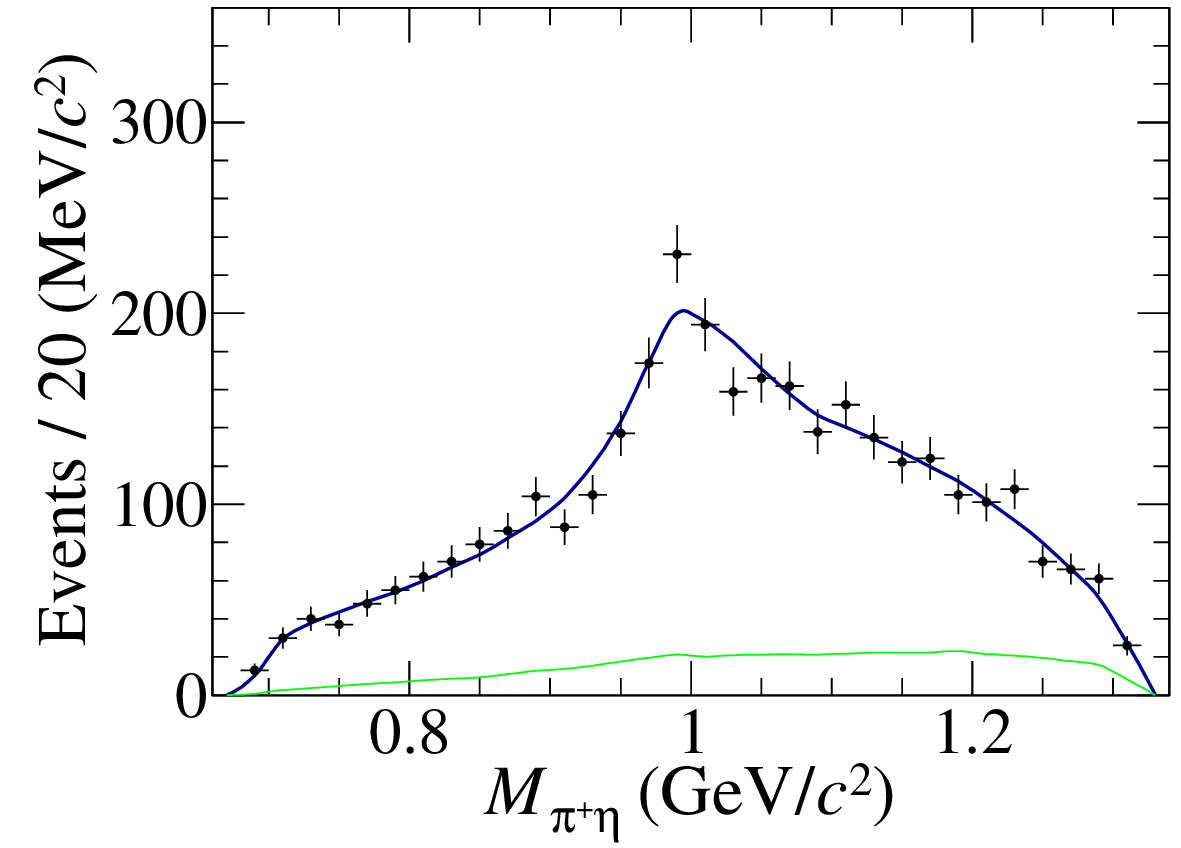}
%\put(-25,50){(q)}
\put(-85,70){IX}
\end{minipage}
\begin{minipage}[b]{0.24\textwidth}
\epsfig{width=0.98\textwidth,clip=true,file=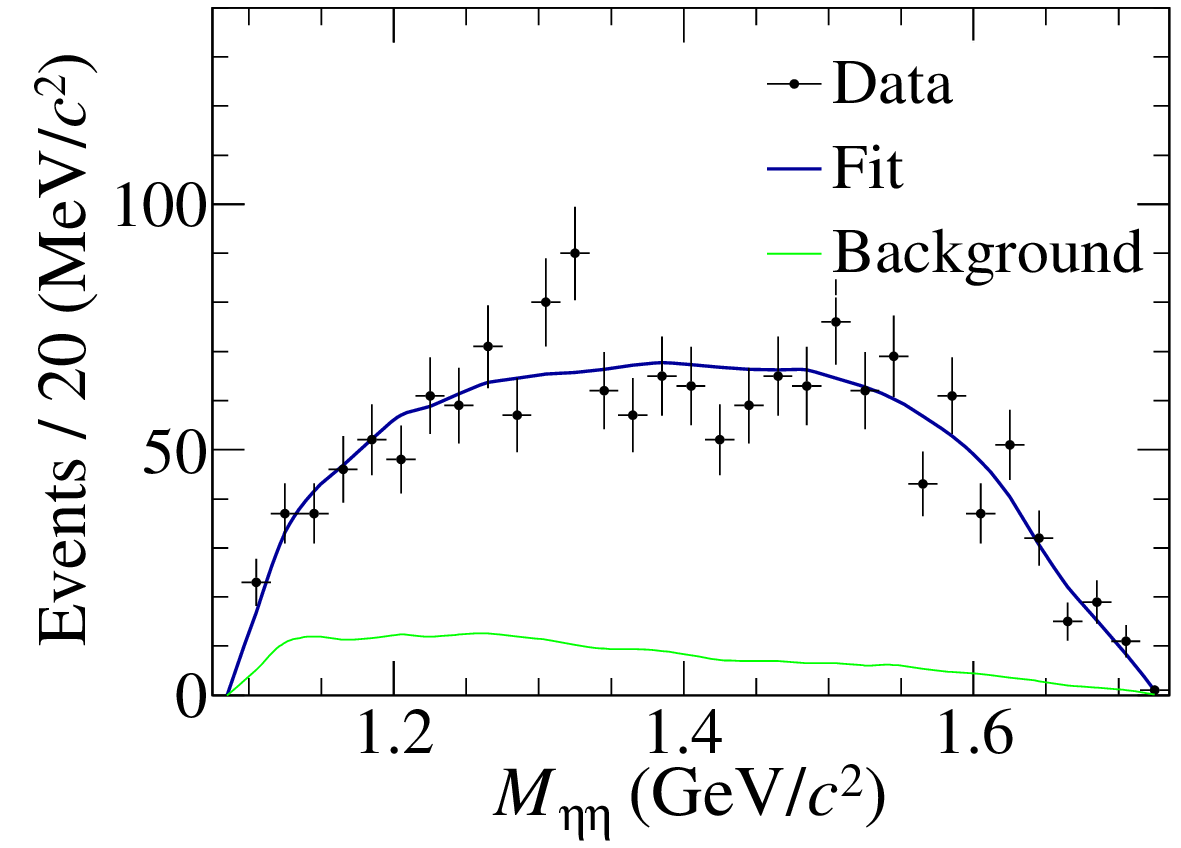}
%\put(-25,50){(r)}
\end{minipage}
\begin{minipage}[b]{0.24\textwidth}
\epsfig{width=0.98\textwidth,clip=true,file=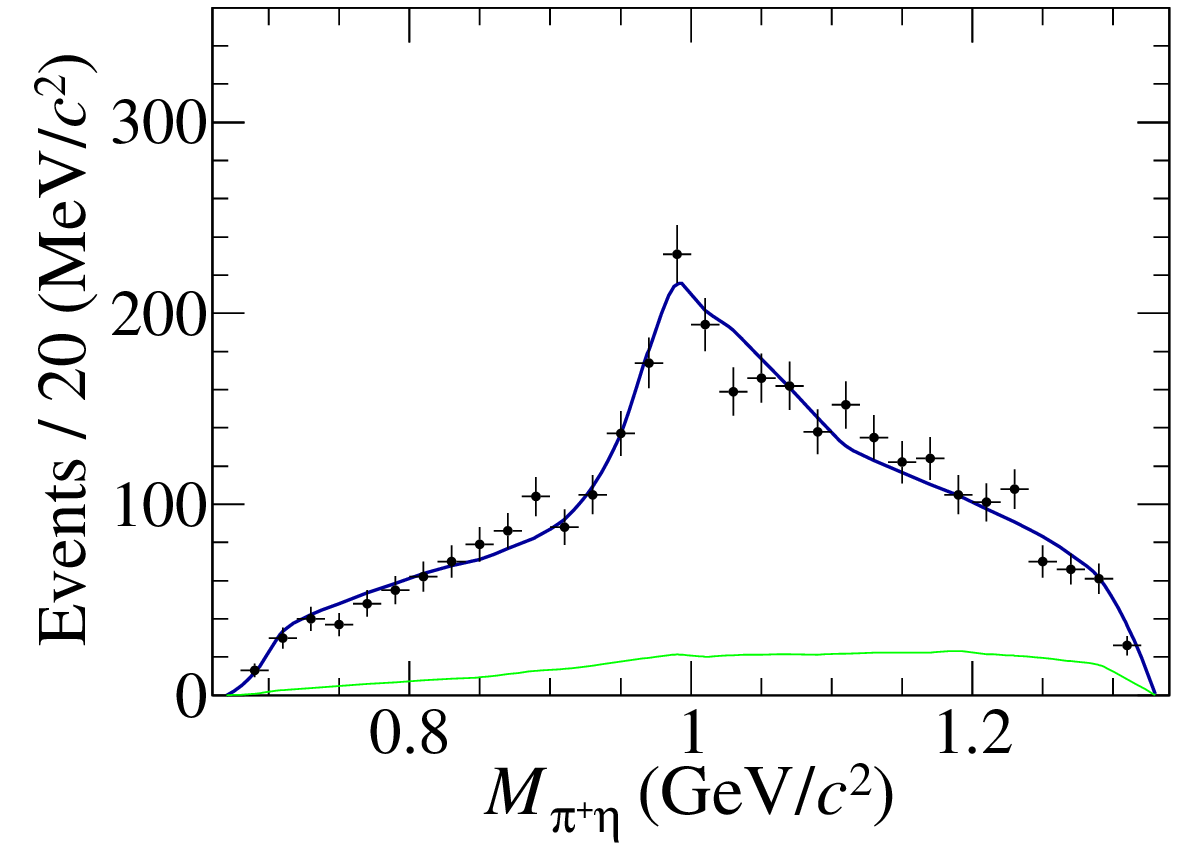}
%\put(-25,50){(s)}
\put(-85,70){X}
\end{minipage}
\begin{minipage}[b]{0.24\textwidth}
\epsfig{width=0.98\textwidth,clip=true,file=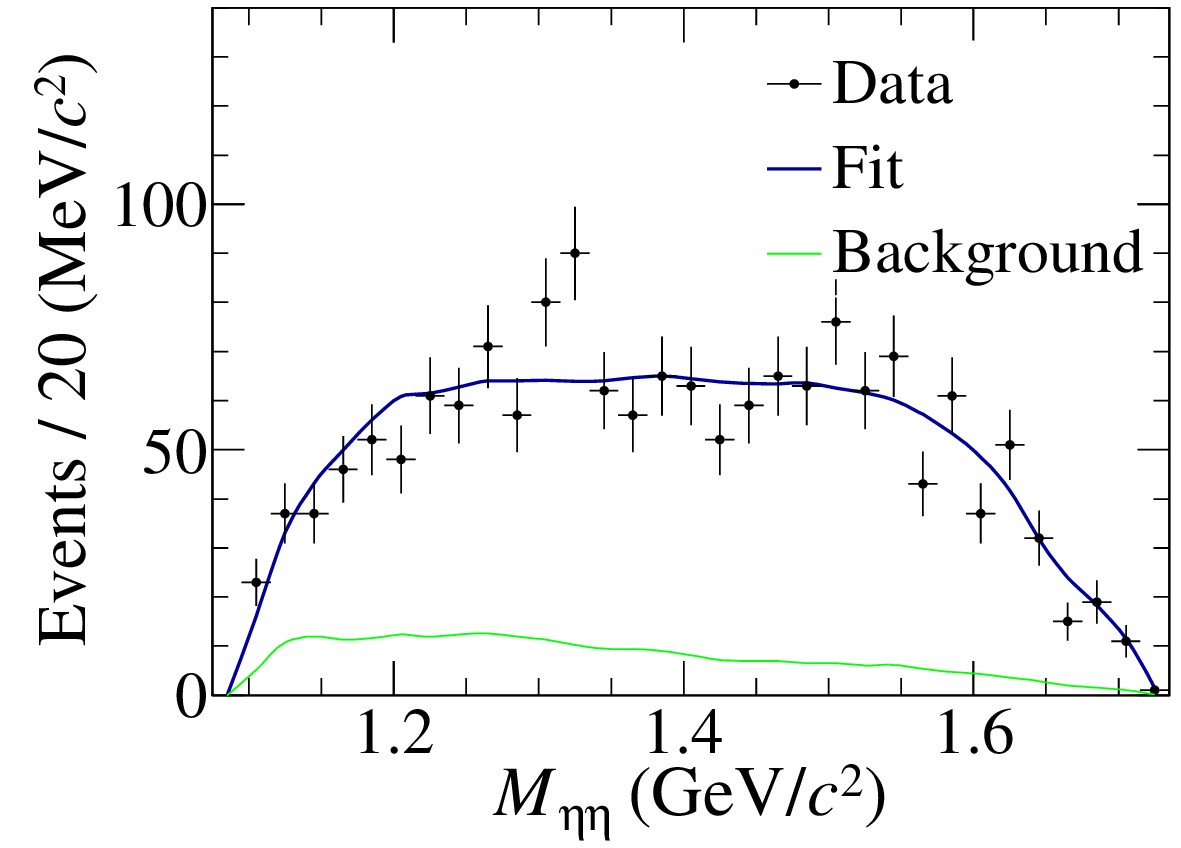}
%\put(-25,50){(t)}
\end{minipage}
\begin{minipage}[b]{0.24\textwidth}
\epsfig{width=0.98\textwidth,clip=true,file=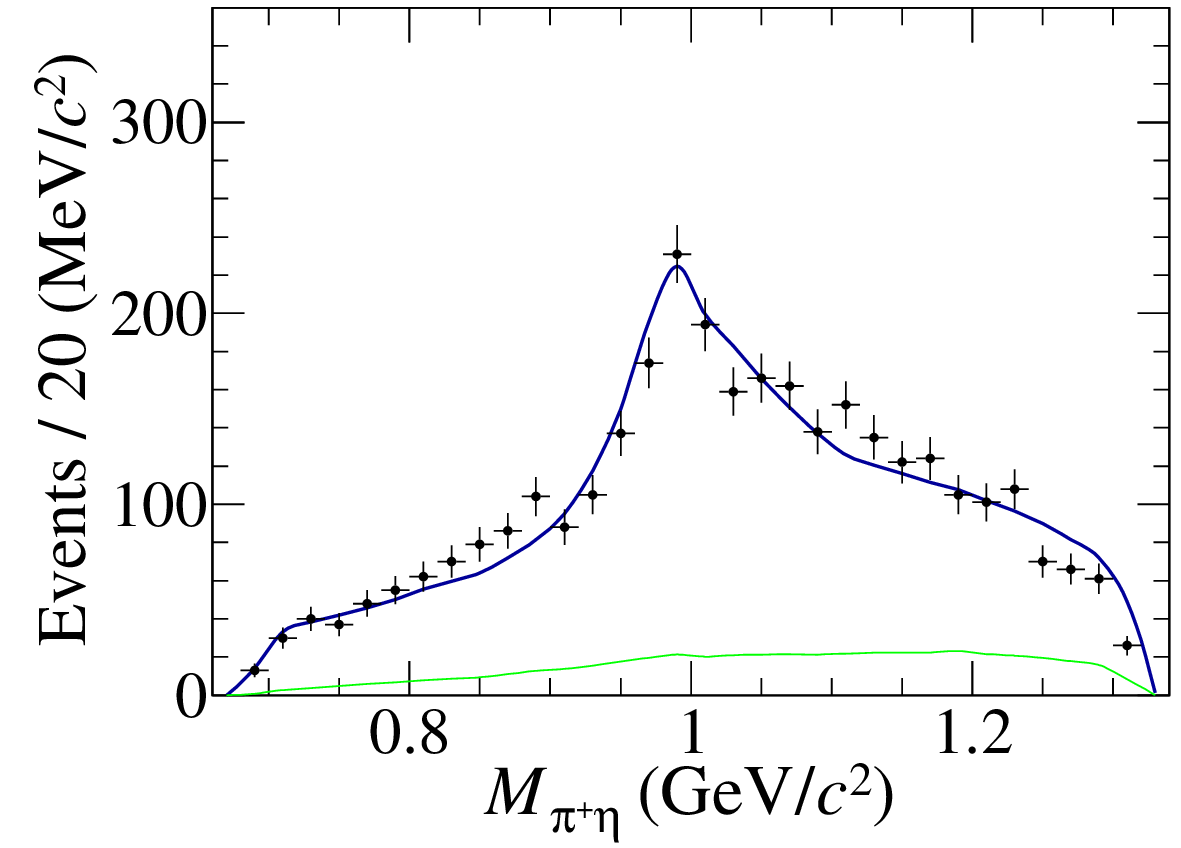}
%\put(-25,50){(u)}
\put(-85,70){XI}
\end{minipage}
\begin{minipage}[b]{0.24\textwidth}
\epsfig{width=0.98\textwidth,clip=true,file=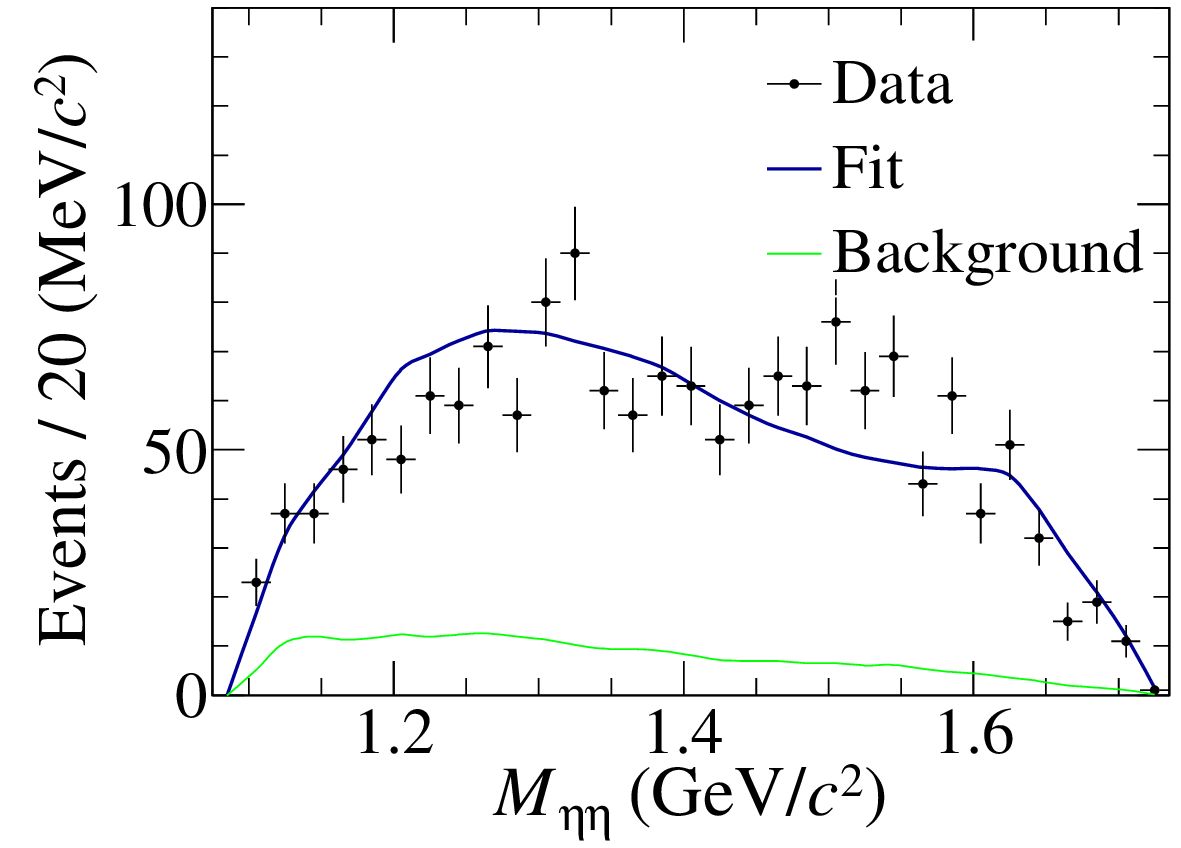}
%\put(-25,50){(v)}
\end{minipage}
\begin{minipage}[b]{0.24\textwidth}
\epsfig{width=0.98\textwidth,clip=true,file=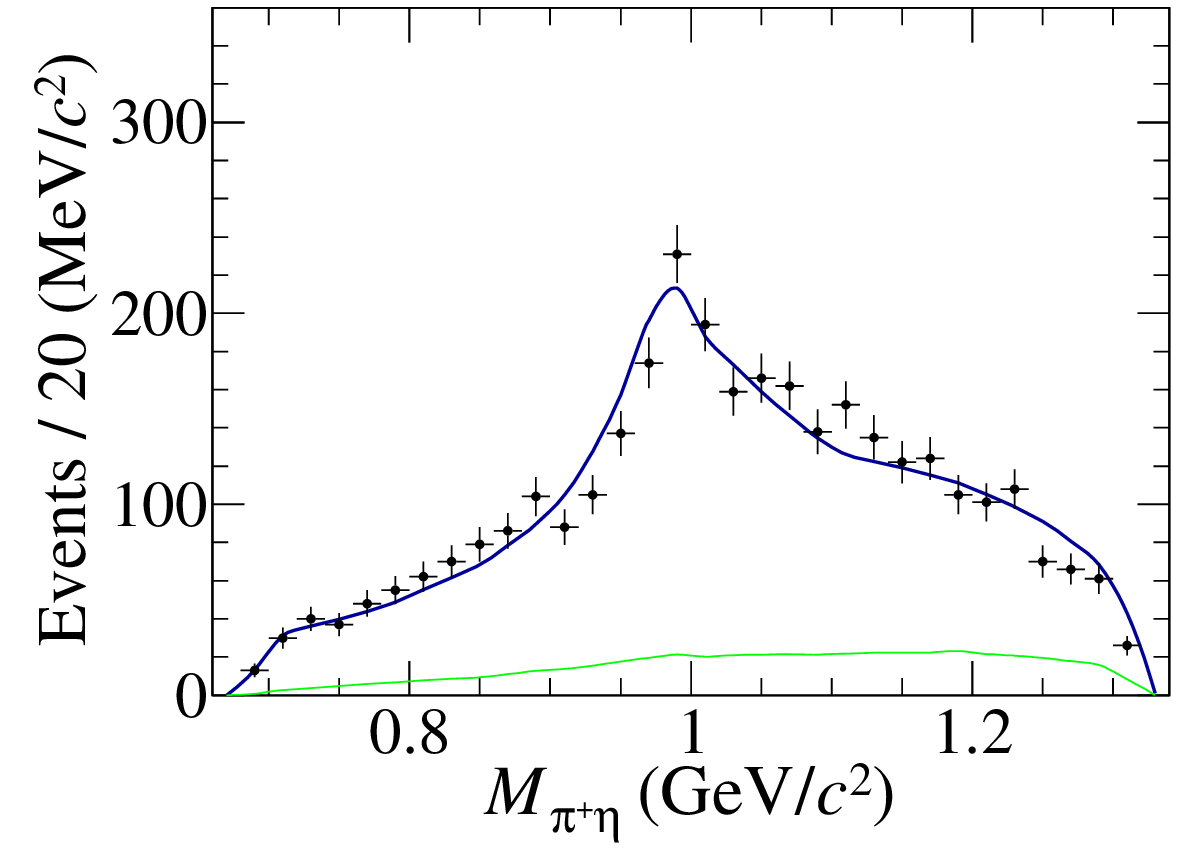}
%\put(-25,50){(w)}
\put(-85,70){XII}
\end{minipage}
\begin{minipage}[b]{0.24\textwidth}
\epsfig{width=0.98\textwidth,clip=true,file=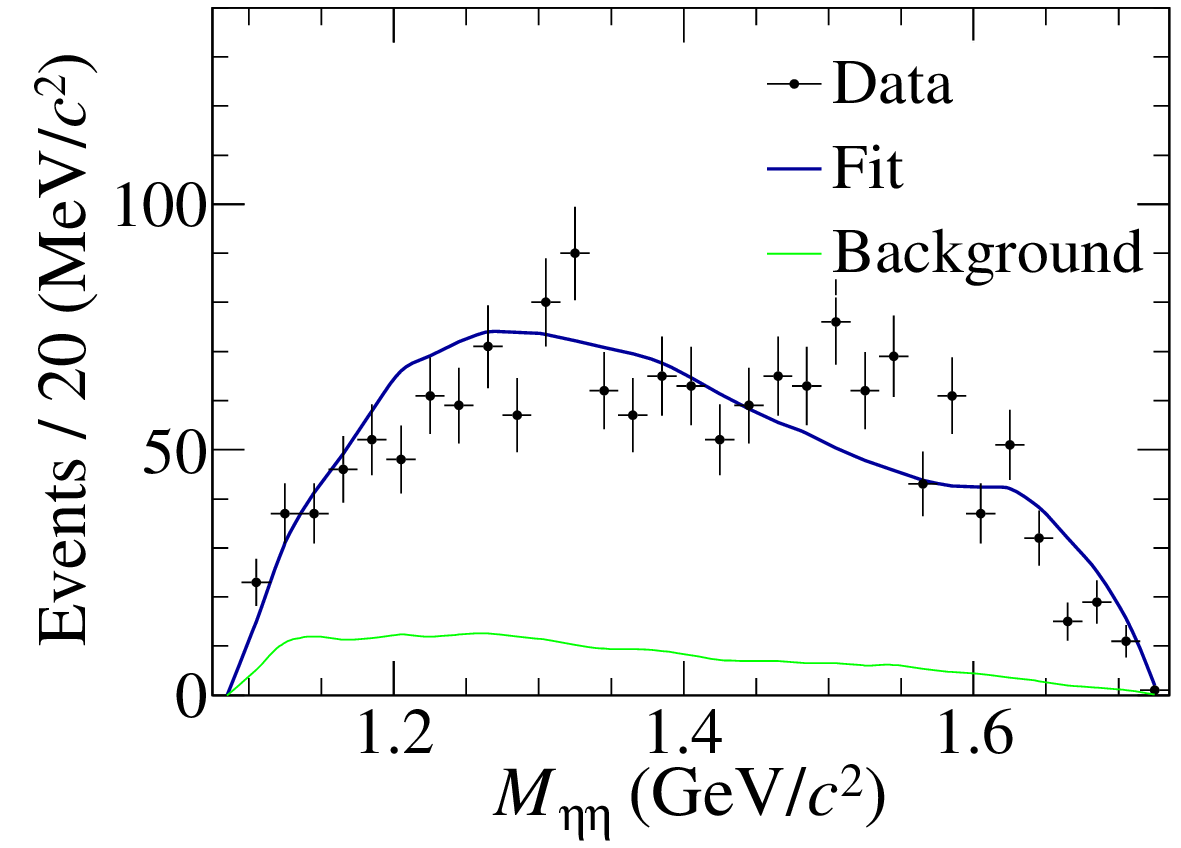}
%\put(-25,50){(x)}
\end{minipage}
\caption{The same as Fig.~\ref{fig:Flatteaddamp}, but for the models using the $K$-matrix formula for the $a_{0}(980)$ line shape. }
\label{fig:KMatrixaddamp}
\end{center}
\end{figure}

\clearpage
\newpage
\section{Additional tests with fixed additional-amplitude coefficients}
\label{app:fixed_addamp_tests}

This appendix gives the detailed results of additional tests in which the complex coefficients, magnitudes, or phases of possible additional amplitudes are fixed to the values obtained in the direct-production tests. 
These tests are performed to examine whether the shifted $a_0(980)$ pole
position can be restored by including conventional additional amplitudes with fixed coefficients. 
For all tables in this appendix, ${\rm NBin}=92$.
%The constant term is not included in these tests.

% \subsection{Flatt\'e parameterization}
% \label{app:fixed_addamp_flate}

\begin{table}[htbp]
\begin{center}
\caption{Fit results obtained by fixing the complex coefficient of each additional amplitude to the value obtained in Sec.~\ref{sec:testsetA}, while refitting the Flatt\'e parameters. %The negative $\Delta\ln\mathcal{L}$ and the worse fit quality indicate that these models are as good as the baseline. 
The $\Delta\ln\mathcal{L}$ is evaluated relative to the fit result for the baseline model with floating the $M_0$, $g_{\pi\eta}^2$, and $g_{K\bar{K}}^2$. 
%The constant term is not included. 
The units are ${\rm GeV}/c^2$ for $M_0$ and $\sqrt{s}_{\rm pole}$, and ${\rm GeV}^2/c^4$ for $g_{\pi\eta}^{2}$ and $g_{K\bar K}^{2}$.}
\begin{tabular}{c|cc|cccc} \hline 
Amplitude              
& $\Delta\ln\mathcal{L}$ 
& $\chi^2/{\rm NBin}$ 
& $M_{0}$ 
& $g_{\pi\eta}^{2}$ 
& $g_{K\bar{K}}^{2}$ 
& $\sqrt{s}_{\rm pole}$ \\ \hline
I      & $-7.4$    & $92.8/92$  & $1.050\pm0.009$ & $0.421\pm0.033$ & $0.305\pm0.081$ & $[(1.066\pm0.011)-i\,(0.081\pm0.017)]$\\
II     & $-7.4$    & $103.8/92$ & $1.047\pm0.009$ & $0.398\pm0.030$ & $0.338\pm0.081$ & $[(1.063\pm0.010)-i\,(0.069\pm0.016)]$\\
III    & $-8.7$    & $96.2/92$  & $1.039\pm0.008$ & $0.445\pm0.035$ & $0.243\pm0.072$ & $[(1.054\pm0.011)-i\,(0.101\pm0.016)]$\\
IV     & $-7.4$    & $96.3/92$  & $1.052\pm0.010$ & $0.421\pm0.034$ & $0.474\pm0.102$ & $[(1.070\pm0.010)-i\,(0.052\pm0.018)]$\\
V      & $-6.0$    & $98.0/92$  & $1.039\pm0.008$ & $0.367\pm0.024$ & $0.337\pm0.073$ & $[(1.055\pm0.008)-i\,(0.062\pm0.014)]$\\
VI     & $-8.3$    & $101.6/92$ & $1.030\pm0.007$ & $0.370\pm0.024$ & $0.303\pm0.067$ & $[(1.047\pm0.008)-i\,(0.069\pm0.013)]$\\  
VII    & $-5.6$    & $99.5/92$  & $1.032\pm0.007$ & $0.370\pm0.023$ & $0.265\pm0.062$ & $[(1.048\pm0.008)-i\,(0.076\pm0.013)]$\\
VIII   & $-5.3$    & $89.0/92$  & $1.043\pm0.008$ & $0.379\pm0.025$ & $0.298\pm0.065$ & $[(1.059\pm0.009)-i\,(0.071\pm0.015)]$\\
IX     & $-5.7$    & $93.6/92$  & $1.015\pm0.006$ & $0.331\pm0.025$ & $0.222\pm0.074$ & $[(1.030\pm0.007)-i\,(0.073\pm0.013)]$\\
${\rm IX^{\rm MI_1}}$ &  $+5.1$ & $88.4/92$ & $0.999\pm0.009$ & $0.368\pm0.032$ & $0.419\pm0.118$ & $[(1.027\pm0.008)-i\,(0.057\pm0.013)]$\\
${\rm IX^{\rm MI_2}}$ &  $+6.8$ & $85.5/92$ & $1.002\pm0.008$ & $0.364\pm0.024$ & $0.303\pm0.058$ & $[(1.024\pm0.006)-i\,(0.072\pm0.011)]$\\
X      & $-5.8$    & $88.4/92$  & $1.056\pm0.010$ & $0.427\pm0.034$ & $0.358\pm0.094$ & $[(1.073\pm0.012)-i\,(0.072\pm0.019)]$\\
XI     & $-5.7$    & $99.7/92$  & $1.041\pm0.007$ & $0.359\pm0.022$ & $0.318\pm0.062$ & $[(1.057\pm0.007)-i\,(0.062\pm0.013)]$\\
XII    & $-1.2$    & $87.5/92$  & $1.057\pm0.009$ & $0.385\pm0.026$ & $0.375\pm0.082$ & $[(1.073\pm0.010)-i\,(0.057\pm0.017)]$\\
\hline
\end{tabular}
\label{tab:flattefixcoeff}
\end{center}
\end{table}

\begin{table}[htbp]
\begin{center}
\caption{Fit results obtained by fixing the complex coefficient of each additional amplitude to the value obtained in Sec.~\ref{sec:testsetB}, while refitting the dispersive parameters. %The negative $\Delta\ln\mathcal{L}$ shows that these models are worse than the baseline. 
%The constant term is not included. 
The $\Delta\ln\mathcal{L}$ is evaluated relative to the fit result for the baseline model with floating the $M_0$, $g_{\pi\eta}^2$, and $g_{K\bar{K}}^2$. 
The units are ${\rm GeV}/c^2$ for $M_0$ and $\sqrt{s_{\rm pole}}$, and ${\rm GeV}^2/c^4$ for $g_{\pi\eta}^{2}$ and $g_{K\bar K}^{2}$. 
Here, ``NA'' indicates that no pole associated with the $a_0(980)$ is found on the $(-++)$ sheet.}
\begin{tabular}{c|cc|cccc} \hline 
Amplitude & $\Delta\ln\mathcal{L}$ & $\chi^2/{\rm NBin}$ 
& $M_0$ & $g_{\pi\eta}^{2}$ & $g_{K\bar K}^{2}$ & $\sqrt{s}_{\rm pole}$
\\ \hline
I      & $-6.7$   & $89.5/92$  & $1.032\pm0.006$ & $0.346\pm0.025$ & $0.322\pm0.059$ & $[(1.078\pm0.009)-i\,(0.042\pm0.018)]$\\
II     & $-3.1$   & $92.1/92$  & $1.035\pm0.006$ & $0.318\pm0.024$ & $0.367\pm0.058$ & $[(1.077\pm0.009)-i\,(0.024\pm0.015)]$\\
III    & $-7.9$   & $95.2/92$  & $1.021\pm0.006$ & $0.374\pm0.025$ & $0.267\pm0.056$ & $[(1.069\pm0.009)-i\,(0.067\pm0.016)]$\\
IV     & $-4.9$   & $94.9/92$  & $1.034\pm0.007$ & $0.324\pm0.025$ & $0.490\pm0.073$ &  NA \\
V      & $-5.2$   & $96.6/92$  & $1.032\pm0.006$ & $0.313\pm0.023$ & $0.406\pm0.059$ & $[(1.072\pm0.008)-i\,(0.016\pm0.014)]$\\
VI     & $-8.8$   & $104.0/92$ & $1.024\pm0.006$ & $0.333\pm0.023$ & $0.383\pm0.057$ & $[(1.068\pm0.008)-i\,(0.031\pm0.015)]$\\  
VII    & $-7.3$   & $103.6/92$ & $1.025\pm0.006$ & $0.343\pm0.023$ & $0.339\pm0.054$ & $[(1.071\pm0.008)-i\,(0.042\pm0.016)]$\\
VIII   & $-4.6$   & $88.4/92$  & $1.032\pm0.006$ & $0.334\pm0.025$ & $0.387\pm0.054$ & $[(1.077\pm0.009)-i\,(0.025\pm0.015)]$\\
IX     & $-7.0$   & $90.5/92$  & $1.009\pm0.005$ & $0.346\pm0.022$ & $0.256\pm0.058$ & $[(1.051\pm0.008)-i\,(0.066\pm0.014)]$\\
${\rm IX^{\rm MI_1}}$&$+3.0$&$88.9/92$&$1.000\pm0.007$&$0.318\pm0.026$&$0.416\pm0.064$&$[(1.049\pm0.010)-i\,(0.054\pm0.014)]$\\ 
${\rm IX^{\rm MI_2}}$&$+5.2$&$81.7/92$&$0.996\pm0.006$&$0.377\pm0.022$&$0.395\pm0.061$&$[(1.051\pm0.009)-i\,(0.056\pm0.012)]$\\
X      & $-5.9$   & $88.9/92$  & $1.031\pm0.006$ & $0.344\pm0.025$ & $0.296\pm0.055$ & $[(1.076\pm0.008)-i\,(0.047\pm0.017)]$\\
XI     & $-1.3$   & $89.7/92$  & $1.037\pm0.006$ & $0.300\pm0.022$ & $0.397\pm0.052$ & $[(1.076\pm0.008)-i\,(0.012^{+0.016}_{-0.007})]$\\
XII    & $+0.0$   & $84.0/92$  & $1.044\pm0.007$ & $0.318\pm0.025$ & $0.408\pm0.063$ & $[(1.086\pm0.009)-i\,(0.010^{+0.019}_{-0.006})]$\\
\hline
\end{tabular}
\label{tab:dispersivefixcoeff}
\end{center}
\end{table}

\begin{table}[htbp]
\footnotesize
\begin{center}
\caption{Fit results obtained by floating only $M_0$ while fixing the coupling constants to the reference values and fixing the full complex coefficient of each additional amplitude to the value obtained in Secs.~\ref{sec:testsetA} and~\ref{sec:testsetB}. 
The $\Delta\ln\mathcal{L}$ is evaluated relative to the baseline model in which only $M_0$ is floated.
%The negative $\Delta\ln\mathcal{L}$ shows that these models are worse than the baseline. 
%For the Flatt\'e parameterization, the coupling constants are fixed to the CLEO values; for the dispersively modified Flatt\'e parameterization, they are fixed to the BESIII values. 
%The constant term is not included. 
The units are ${\rm GeV}/c^2$ for $M_0$ and $M_{\rm pole}$.}
% \resizebox{\textwidth}{!}
{%
\begin{tabular}{c|cccc|cccc} \hline
\multirow{2}{*}{Amplitude}
& \multicolumn{4}{c|}{Flatt\'e}
& \multicolumn{4}{c}{Dispersively modified Flatt\'e} \\ \cline{2-9}
& $\Delta\ln\mathcal{L}$ 
& $\chi^2/{\rm NBin}$ 
& $M_0$ 
& $M_{\rm pole}$ 
& $\Delta\ln\mathcal{L}$ 
& $\chi^2/{\rm NBin}$ 
& $M_0$ 
& $M_{\rm pole}$ \\ \hline
I    & $-3.9$ & $95.6/92$   & $1.047\pm0.007$ & $1.062\pm0.006$ &$-5.6$ & $92.5/92$   & $1.033\pm0.006$ & $1.084\pm0.007$ \\
II   & $-2.6$ & $104.8/92$  & $1.043\pm0.007$ & $1.058\pm0.006$ &$-3.3$ & $98.2/92$   & $1.037\pm0.006$ & $1.088\pm0.006$ \\
III  & $-8.2$ & $102.7/92$  & $1.036\pm0.007$ & $1.053\pm0.006$ &$-7.0$ & $97.8/92$   & $1.022\pm0.006$ & $1.073\pm0.006$ \\
IV   & $-4.8$ & $101.1/92$  & $1.042\pm0.007$ & $1.057\pm0.006$ &$-5.9$ & $101.4/92$  & $1.030\pm0.006$ & $1.081\pm0.007$ \\
V    & $-0.5$ & $98.1/92$   & $1.037\pm0.007$ & $1.054\pm0.006$ &$-6.0$ & $104.3/92$  & $1.031\pm0.006$ & $1.082\pm0.006$ \\
VI   & $-2.7$ & $101.5/92$  & $1.030\pm0.007$ & $1.048\pm0.006$ &$-8.1$ & $108.7/92$  & $1.023\pm0.006$ & $1.074\pm0.006$ \\
VII  & $-0.3$ & $99.8/92$   & $1.033\pm0.007$ & $1.050\pm0.006$ &$-6.3$ & $107.7/92$  & $1.026\pm0.006$ & $1.076\pm0.006$ \\
VIII & $+0.3$ & $89.2/92$   & $1.043\pm0.007$ & $1.058\pm0.006$ &$-3.7$ & $91.4/92$   & $1.031\pm0.006$ & $1.082\pm0.007$ \\
IX   & $-1.1$ & $97.4/92$   & $1.014\pm0.007$ & $1.034\pm0.006$ &$-6.9$ & $96.4/92$   & $1.009\pm0.006$ & $1.060\pm0.006$ \\
${\rm IX^{\rm MI_1}}$&$+10.2$&$88.3/92$&$1.002\pm0.008$&$1.025\pm0.006$&$+4.6$&$89.1/92$&$1.001\pm0.006$& $1.051\pm0.006$ \\
${\rm IX^{\rm MI_2}}$&$+12.4$&$85.7/92$&$1.001\pm0.007$&$1.024\pm0.006$&$+6.7$&$81.5/92$&$0.996\pm0.006$& $1.047\pm0.006$ \\
X    & $-2.7$ & $91.7/92$   & $1.050\pm0.008$ & $1.064\pm0.006$ &$-5.6$ & $94.2/92$   & $1.032\pm0.006$ & $1.084\pm0.006$ \\
XI   & $-0.0$ & $99.7/92$   & $1.041\pm0.007$ & $1.056\pm0.006$ &$-3.8$ & $101.9/92$  & $1.037\pm0.006$ & $1.089\pm0.006$ \\
XII  & $+3.6$ & $87.7/92$   & $1.053\pm0.007$ & $1.067\pm0.006$ &$+0.0$ & $89.2/92$   & $1.044\pm0.007$ & $1.096\pm0.007$ \\
\hline
\end{tabular}
}
\label{tab:finaltests0_combined}
\end{center}
\end{table}

\begin{table}[htbp]
\footnotesize
\begin{center}
% \caption{Fit results obtained by floating only $M_0$ while fixing the coupling constants to the reference values and fixing only the magnitude of each additional amplitude to the value obtained in Secs.~\ref{sec:testsetA} and~\ref{sec:testsetB}. 
% The $\Delta\ln\mathcal{L}$ is relate to the fit result for the baseline model with only floating the $M_0$. 
\caption{Fit results obtained by floating $M_0$ and the phase of each additional amplitude while fixing the coupling constants to the reference values and fixing the magnitude of each additional amplitude to the value obtained in Secs.~\ref{sec:testsetA} and~\ref{sec:testsetB}.
The $\Delta\ln\mathcal{L}$ is evaluated relative to the baseline model in which only $M_0$ is floated.
%The negative $\Delta\ln\mathcal{L}$ shows that these models are worse than the baseline. 
%For the Flatt\'e parameterization, the coupling constants are fixed to the CLEO values; for the dispersively modified Flatt\'e parameterization, they are fixed to the BESIII values. 
%The constant term is not included. 
The units are ${\rm GeV}/c^2$ for $M_0$ and $M_{\rm pole}$.}
% \resizebox{\textwidth}{!}
{%
\begin{tabular}{c|cccc|cccc} \hline
\multirow{2}{*}{Amplitude}
& \multicolumn{4}{c|}{Flatt\'e}
& \multicolumn{4}{c}{Dispersively modified Flatt\'e} \\ \cline{2-9}
& $\Delta\ln\mathcal{L}$ 
& $\chi^2/{\rm NBin}$ 
& $M_0$ 
& $M_{\rm pole}$ 
& $\Delta\ln\mathcal{L}$ 
& $\chi^2/{\rm NBin}$ 
& $M_0$ 
& $M_{\rm pole}$ \\ \hline
I    & $+0.2$ & $99.0/92$   & $1.059\pm0.009$ & $1.072\pm0.008$ &$-4.9$ & $101.9/92$  & $1.040\pm0.011$ & $1.091\pm0.011$ \\
II   & $+0.1$ & $100.9/92$  & $1.055\pm0.010$ & $1.068\pm0.009$ &$+1.1$ & $91.5/92$   & $1.050\pm0.008$ & $1.102\pm0.008$ \\
III  & $+1.5$ & $92.7/92$   & $1.063\pm0.010$ & $1.076\pm0.009$ &$-6.4$ & $93.6/92$   & $1.026\pm0.008$ & $1.077\pm0.008$ \\
IV   & $-1.7$ & $88.5/92$   & $1.053\pm0.010$ & $1.067\pm0.009$ &$-5.7$ & $100.8/92$  & $1.030\pm0.006$ & $1.081\pm0.007$ \\
V    & $+1.2$ & $93.6/92$   & $1.044\pm0.010$ & $1.060\pm0.009$ &$-6.0$ & $104.4/92$  & $1.031\pm0.006$ & $1.082\pm0.006$ \\
VI   & $-2.4$ & $100.9/92$  & $1.031\pm0.007$ & $1.048\pm0.006$ &$-7.9$ & $107.7/92$  & $1.024\pm0.006$ & $1.075\pm0.007$ \\
VII  & $-0.2$ & $100.0/92$  & $1.033\pm0.007$ & $1.050\pm0.006$ &$-5.6$ & $104.8/92$  & $1.028\pm0.007$ & $1.079\pm0.007$ \\
VIII & $+0.1$ & $89.2/92$   & $1.043\pm0.007$ & $1.058\pm0.006$ &$-3.6$ & $91.0/92$   & $1.032\pm0.007$ & $1.083\pm0.007$ \\
IX   & $+1.7$ & $98.3/92$   & $1.034\pm0.013$ & $1.051\pm0.011$ &$-5.7$ & $100.1/92$  & $1.018\pm0.009$ & $1.068\pm0.009$ \\
${\rm IX^{\rm MI_1}}$&$+10.3$&$88.5/92$&$1.005\pm0.010$&$1.027\pm0.008$&$+4.7$&$89.1/92$&$1.003\pm0.008$& $1.053\pm0.008$ \\
${\rm IX^{\rm MI_2}}$&$+12.5$&$86.1/92$&$1.003\pm0.009$&$1.026\pm0.007$&$+6.7$&$81.6/92$&$0.996\pm0.006$& $1.047\pm0.006$ \\
X    & $-0.1$ & $92.1/92$   & $1.059\pm0.009$ & $1.072\pm0.008$ &$-0.4$ & $93.3/92$   & $1.047\pm0.008$ & $1.098\pm0.009$ \\
XI   & $+1.3$ & $96.4/92$   & $1.045\pm0.008$ & $1.061\pm0.007$ &$-2.3$ & $97.1/92$   & $1.042\pm0.007$ & $1.093\pm0.007$ \\
XII  & $+3.7$ & $86.6/92$   & $1.053\pm0.008$ & $1.067\pm0.006$ &$+0.0$ & $89.0/92$   & $1.044\pm0.007$ & $1.096\pm0.007$ \\
\hline
\end{tabular}
}
\label{tab:finaltests1_combined}
\end{center}    
\end{table}

\begin{table}[htbp]
\footnotesize
\begin{center}
% \caption{Fit results obtained by floating only $M_0$ while fixing the coupling constants to the reference values and fixing only the phase of each additional amplitude to the value obtained in Secs.~\ref{sec:testsetA} and~\ref{sec:testsetB}. 
% The negative $\Delta\ln\mathcal{L}$ shows that these models are worse than the baseline. 
\caption{Fit results obtained by floating $M_0$ and the magnitude of each additional amplitude while fixing the coupling constants to the reference values and fixing the phase of each additional amplitude to the value obtained in Secs.~\ref{sec:testsetA} and~\ref{sec:testsetB}.
The $\Delta\ln\mathcal{L}$ is evaluated relative to the baseline model in which only $M_0$ is floated.
%For the Flatt\'e parameterization, the coupling constants are fixed to the CLEO values; for the dispersively modified Flatt\'e parameterization, they are fixed to the BESIII values. 
%The constant term is not included. 
The units are ${\rm GeV}/c^2$ for $M_0$ and $M_{\rm pole}$.}
% \resizebox{\textwidth}{!}
{%
\begin{tabular}{c|cccc|cccc} \hline
\multirow{2}{*}{Amplitude}
& \multicolumn{4}{c|}{Flatt\'e}
& \multicolumn{4}{c}{Dispersively modified Flatt\'e} \\ \cline{2-9}
& $\Delta\ln\mathcal{L}$ 
& $\chi^2/{\rm NBin}$ 
& $M_0$ 
& $M_{\rm pole}$ 
& $\Delta\ln\mathcal{L}$ 
& $\chi^2/{\rm NBin}$ 
& $M_0$ 
& $M_{\rm pole}$ \\ \hline
I    & $+0.0$ & $92.3/92$  & $1.056\pm0.008$ & $1.070\pm0.007$ &$+0.0$ & $89.8/92$  & $1.044\pm0.007$ & $1.096\pm0.008$ \\
II   & $+0.5$ & $93.5/92$  & $1.052\pm0.009$ & $1.066\pm0.007$ &$+0.1$ & $89.2/92$  & $1.045\pm0.008$ & $1.097\pm0.008$ \\
III  & $+0.2$ & $92.0/92$  & $1.059\pm0.010$ & $1.072\pm0.009$ &$+0.2$ & $88.4/92$  & $1.041\pm0.008$ & $1.093\pm0.009$ \\
IV   & $+0.0$ & $92.1/92$  & $1.056\pm0.010$ & $1.069\pm0.008$ &$+0.1$ & $89.8/92$  & $1.043\pm0.007$ & $1.094\pm0.008$ \\
V    & $+2.6$ & $91.3/92$  & $1.047\pm0.008$ & $1.061\pm0.007$ &$+0.0$ & $89.7/92$  & $1.044\pm0.008$ & $1.096\pm0.008$ \\
VI   & $+1.5$ & $91.9/92$  & $1.045\pm0.009$ & $1.060\pm0.008$ &$+0.0$ & $89.6/92$  & $1.044\pm0.008$ & $1.096\pm0.008$ \\
VII  & $+0.8$ & $92.5/92$  & $1.049\pm0.009$ & $1.063\pm0.008$ &$+0.3$ & $90.5/92$  & $1.041\pm0.008$ & $1.093\pm0.008$ \\
VIII & $+2.4$ & $87.5/92$  & $1.049\pm0.008$ & $1.063\pm0.007$ &$+0.0$ & $89.4/92$  & $1.043\pm0.009$ & $1.094\pm0.009$ \\
IX   & $+1.9$ & $89.1/92$  & $1.035\pm0.012$ & $1.052\pm0.010$ &$+0.0$ & $89.1/92$  & $1.042\pm0.011$ & $1.094\pm0.011$\\
${\rm IX^{\rm MI_1}}$ &$+10.3$&$87.9/92$&$1.005\pm0.010$&$1.027\pm0.008$&$+5.8$&$85.9/92$&$1.026\pm0.010$&$1.077\pm0.011$\\
${\rm IX^{\rm MI_2}}$ &$+12.6$&$85.0/92$&$1.008\pm0.013$&$1.030\pm0.011$&$+6.7$&$81.6/92$&$0.998\pm0.014$&$1.048\pm0.014$\\
X    & $+0.1$ & $93.0/92$  & $1.057\pm0.009$ & $1.070\pm0.007$ &$+0.1$ & $90.2/92$  & $1.046\pm0.008$ & $1.097\pm0.008$ \\
XI   & $+4.2$ & $88.5/92$  & $1.050\pm0.008$ & $1.064\pm0.007$ &$+0.0$ & $89.9/92$  & $1.044\pm0.007$ & $1.096\pm0.007$ \\
XII  & $+3.8$ & $86.8/92$  & $1.053\pm0.008$ & $1.067\pm0.007$ &$+0.1$ & $89.3/92$  & $1.044\pm0.007$ & $1.096\pm0.007$ \\
\hline
\end{tabular}
}
\label{tab:finaltests2_combined}
\end{center}
\end{table}


\begin{thebibliography}{99}

\bibitem{BESIII:2019jjr} M.~Ablikim \textit{et al.} [BESIII], \href{https://doi.org/10.1103/PhysRevLett.123.112001}{Phys. Rev. Lett. \textbf{123}, 112001 (2019)}.

%\bibitem{Molina:2019udw} R.~Molina, J.~J.~Xie, W.~H.~Liang, L.~S.~Geng and E.~Oset, \href{https://doi.org/10.1016/j.physletb.2020.135279}{Phys. Lett. B \textbf{803}, 135279 (2020)}.

%\bibitem{Hsiao:2019ait} Y.~K.~Hsiao, Y.~Yu and B.~C.~Ke, \href{https://doi.org/10.1140/epjc/s10052-020-08468-9}{Eur. Phys. J. C \textbf{80}, no.9, 895 (2020)}.

\bibitem{BESIII:2021aza} M.~Ablikim \textit{et al.} [BESIII], \href{https://doi.org/10.1103/PhysRevD.104.L071101}{Phys. Rev. D \textbf{104}, L071101 (2021)}.

\bibitem{BESIII:2023htx} M.~Ablikim \textit{et al.} [BESIII], \href{https://doi.org/10.1103/PhysRevLett.132.131903}{Phys. Rev. Lett. \textbf{132}, 131903 (2024)}.

\bibitem{BESIII:2024tpv} M.~Ablikim \textit{et al.} [BESIII], \href{https://doi.org/10.1103/PhysRevD.110.L111102}{Phys. Rev. D \textbf{110}, L111102 (2024)}.

\bibitem{BESIII:2024mbf} M.~Ablikim \textit{et al.} [BESIII], \href{https://doi.org/10.1103/PhysRevLett.134.021901}{Phys. Rev. Lett. \textbf{134}, 021901 (2025)}.

\bibitem{ParticleDataGroup:2024cfk} S.~Navas \textit{et al.} [Particle Data Group], \href{https://pdg.lbl.gov/}{Phys. Rev. D \textbf{110}, 030001 (2024)}.

\bibitem{Ablikim:2013ntc} M.~Ablikim \textit{et al.} [BESIII], \href{https://doi.org/10.1088/1674-1137/37/12/123001}{Chin. Phys. C \textbf{37}, 123001 (2013)}.

\bibitem{BESIII:2015equ} M.~Ablikim \textit{et al.} [BESIII], \href{https://doi.org/10.1016/j.physletb.2015.11.043}{Phys. Lett. B \textbf{753}, 629-638 (2016)} [erratum: \href{https://doi.org/10.1016/j.physletb.2020.135982}{Phys. Lett. B \textbf{812}, 135982 (2021)}].

\bibitem{BESIII:2024lbn} M.~Ablikim \textit{et al.} [BESIII], \href{https://iopscience.iop.org/article/10.1088/1674-1137/ad70a0/pdf}{Chin. Phys. C 48, 123001 (2024)}. 

\bibitem{Flatte:1976xu} S.~M.~Flatte,
\href{https://doi.org/10.1016/0370-2693(76)90654-7}{Phys. Lett. B \textbf{63}, 224-227 (1976)}

\bibitem{Bugg:2008ig} D.~V.~Bugg,
\href{https://doi.org/10.1103/PhysRevD.78.074023}{Phys. Rev. D \textbf{78} (2008), 074023}.

\bibitem{Ikeno:2021kzf} N.~Ikeno, M.~Bayar and E.~Oset, \href{https://link.springer.com/article/10.1140/epjc/s10052-021-09174-w}{Eur. Phys. J. C \textbf{81}, 377 (2021)}.

\bibitem{Anisovich:1997pe}
V.~V.~Anisovich, A.~A.~Kondashov, Y.~D.~Prokoshkin, S.~A.~Sadovsky and A.~V.~Sarantsev, \href{https://doi.org/10.1134/1.1307464}{Phys. Atom. Nucl. \textbf{63}, 1410-1427 (2000)}.

\bibitem{BESIII:2021jnf}
M.~Ablikim \textit{et al.} [BESIII],
%``Amplitude analysis of Ds+{\textrightarrow}{\ensuremath{\pi}}+{\ensuremath{\pi}}-{\ensuremath{\pi}}+,''
\href{https://journals.aps.org/prd/abstract/10.1103/PhysRevD.106.112006}{Phys. Rev. D \textbf{106}, 112006 (2022)}.

\bibitem{CLEO:2011upl} G.~S.~Adams \textit{et al.} [CLEO], \href{https://doi.org/10.1103/PhysRevD.84.112009}{Phys. Rev. D \textbf{84}, 112009 (2011)}.

\bibitem{BESIII:2016tqo} M.~Ablikim \textit{et al.} [BESIII], \href{https://journals.aps.org/prd/abstract/10.1103/PhysRevD.95.032002}{Phys. Rev. D \textbf{95}, 032002 (2017)}.

\bibitem{Song:2025ofe} J.~Song, W.~H.~Liang and E.~Oset,
\href{https://arxiv.org/abs/2502.20160}{[arXiv:2502.20160 [hep-ph]]}.



\bibitem{CrystalBarrel:2019zqh} M.~Albrecht \textit{et al.} [Crystal Barrel],
%``Coupled channel analysis of ${\bar{p}p}\,\rightarrow \,\pi ^0\pi ^0\eta $, ${\pi ^0\eta \eta }$ and ${K^+K^-\pi ^0}$ at 900 MeV/c and of ${\pi \pi }$-scattering data,''
\href{https://doi.org/10.1140/epjc/s10052-020-7930-x}{Eur. Phys. J. C \textbf{80}, 453 (2020)}

\bibitem{detector} M. Ablikim {\it et al.} [BESIII], \href{https://linkinghub.elsevier.com/retrieve/pii/S0168900209023870}{Nucl. Instrum. Methods Phys. Res., Sect. A {\bf 614}, 345 (2010)}.

\bibitem{MRPC} X.~Wang {\it et al.}, \href{https://iopscience.iop.org/article/10.1088/1748-0221/11/08/C08009}{JINST {\bf 11}, C08009 (2016)}.

\bibitem{MARK-III:1985hbd} R.~M.~Baltrusaitis \textit{et al.} [MARK-III], \href{https://journals.aps.org/prl/abstract/10.1103/PhysRevLett.56.2140}{Phys. Rev. Lett. \textbf{56}, 2140 (1986)}.

\bibitem{BESIII:2023exq} M.~Ablikim \textit{et al.} [BESIII], \href{https://doi.org/10.1103/PhysRevD.109.072003}{Phys. Rev. D \textbf{109}, 072003 (2024)}.

\bibitem{sim} S. Agostinelli {\it et al.} [{\sc geant4}], \href{https://linkinghub.elsevier.com/retrieve/pii/S0168900203013688}{Nucl. Instrum. Meth. A \textbf{506}, 250-303 (2003)}.

\bibitem{EvtGen} D. J. Lange, \href{https://linkinghub.elsevier.com/retrieve/pii/S0168900201000894}{Nucl. Instrum. Meth. A \textbf{462}, 152-155 (2001)}; R. G. Ping, \href{https://iopscience.iop.org/article/10.1088/1674-1137/32/8/001}{Chin. Phys. C {\bf 32}, 599 (2008)}.

\bibitem{Chen:2000tv} J.~C.~Chen, G.~S.~Huang, X.~R.~Qi, D.~H.~Zhang and Y.~S.~Zhu,
%``Event generator for J / psi and psi (2S) decay,''
\href{https://journals.aps.org/prd/abstract/10.1103/PhysRevD.62.034003}{Phys. Rev. D \textbf{62}, 034003 (2000)}. 
R.~L.~Yang, R.~G.~Ping and H.~Chen,
%``Tuning and Validation of the Lundcharm Model with $J/\psi$ Decays,''
\href{https://doi.org/10.1088/0256-307X/31/6/061301}{Chin. Phys. Lett. \textbf{31}, 061301 (2014)}.

\bibitem{Hocker:2007ht} A.~Hocker {\it et al.}, {\it TMVA - Toolkit for Multivariate Data Analysis}, \href{https://arxiv.org/abs/physics/0703039}{arXiv: physics/0703039}; H.~Voss, A.~Hocker, J.~Stelzer and F.~Tegenfeldt,
%``TMVA, the Toolkit for Multivariate Data Analysis with ROOT,''
\href{https://doi.org/10.22323/1.050.0040}{PoS \textbf{ACAT}, 040 (2007)}.

\bibitem{Zou:2002ar} B.~S.~Zou and D.~V.~Bugg, \href{https://link.springer.com/article/10.1140/epja/i2002-10135-4}{Eur. Phys. J. A \textbf{16}, 537-547 (2003)}.

\bibitem{CLEO:2008msk} P.~Rubin \textit{et al.} [CLEO], \href{https://doi.org/10.1103/PhysRevD.78.072003}{Phys. Rev. D \textbf{78}, 072003 (2008)}.

\bibitem{EvangelhoVieira:2015ujl} D.~Evangelho Vieira, \href{https://inspirehep.net/files/a9edc1bd7959178b4a72bb01468756b5}{CERN-THESIS-2015-328}.

\bibitem{Zhang:2024qkg}
Z.~H.~Zhang and F.~K.~Guo,
%``Classification of coupled-channel near-threshold structures,''
\href{https://www.sciencedirect.com/science/article/pii/S0370269325001479?via%3Dihub}{Phys. Lett. B \textbf{863}, 139387 (2025)}.

\bibitem{Efron1979}
B.~Efron,
``Bootstrap methods: Another look at the jackknife,''
\href{https://doi.org/10.1214/aos/1176344552}{\textit{Ann. Statist.} \textbf{7}, 1--26 (1979)}.

\bibitem{EfronTibshirani1993}
B.~Efron and R.~J.~Tibshirani,
\textit{An Introduction to the Bootstrap},
\href{https://isbnsearch.org/isbn/9780412042317}{Chapman \& Hall/CRC, New York, 1993}.

\bibitem{BESIII:2024njj}
M.~Ablikim \textit{et al.} [BESIII],
%``Observation of D+{\textrightarrow}{\ensuremath{\eta}}'{\ensuremath{\mu}}+{\ensuremath{\nu}}{\ensuremath{\mu}} and First Experimental Study of D+{\textrightarrow}{\ensuremath{\eta}}'{\ensuremath{\ell}}+{\ensuremath{\nu}}{\ensuremath{\ell}} Decay Dynamics,''
\href{https://doi.org/10.1103/PhysRevLett.134.111801}{Phys. Rev. Lett. \textbf{134}, 111801 (2025)}.

\end{thebibliography}
\end{document}